%% file: main.tex
\newif\ifllncs
\llncsfalse

\newif\ifanon
\anonfalse

\newif\ifnotes
\notesfalse

\newif\iffull
\fulltrue

\ifllncs
\documentclass[envcountsame]{llncs}

\else
\documentclass[11pt,letter]{article}
\fi

\input{preamble}
\input{latexdefs}

\begin{document}

\ifllncs\else\pagenumbering{roman}\fi

\title{Doubly-Efficient Private Information Retrieval\\
from Modular Forms}

\ifanon
    \author{Submission \#}
    \date{}
\else
    \author{
    Alexandra Henzinger\thanks{\texttt{amh@cs.stanford.edu}. Supported by the NSF GRFP under Grant No. 2141064, a MIT EECS Great Educators fellowship, gifts from Apple, Google, and Meta, and NSF CNS-2054869.}\\Stanford \and
    Seyoon Ragavan\thanks{\texttt{sragavan@mit.edu}.  Supported by NSF CNS-2534400 and in part by Jane Street. This work was done in part while SR was a student researcher at Google Quantum AI. SR's access to ChatGPT Pro was supported by the UK AISI alignment project and the Simons Institute for the Theory of Computing, and his access to Claude Max was supported by Jane Street.}\\MIT
    }
    \ifllncs
    \institute{Massachusetts Institute of Technology}
    \fi
\fi 


\maketitle
\thispagestyle{plain}
\input{abstract}

\ifanon
\else
\newpage
\fi
   
\ifllncs
\else
\thispagestyle{empty}
\newpage
\thispagestyle{empty}
\tableofcontents
\newpage 
\pagenumbering{arabic} 
\fi 

\input{intro}
\input{rel}

\input{prelims}
\input{depir}

\input{math}

\section*{Statement on AI Use}
\phantomsection\addcontentsline{toc}{part}{Statement on AI Use}

\begin{flushleft}
{\centering
\renewcommand{\arraystretch}{1.3}
\begin{tabular}{@{}l@{\qquad}c@{}}
\textbf{Activity} & \textbf{AI used} \\
Asking the research question & No \\
Coming up with the approach & Yes \\
Proof development & Yes \\
Writing and exposition & Yes \\
Checking for bugs & Yes \\
Other supporting tasks & Yes \\
\end{tabular}}
\end{flushleft}

\begin{sloppypar}
\paragraph{Details (elaborating on the points above).}
The results of this paper were obtained via months-long conversations with ChatGPT (5.5--6 Astra), seeded with ideas of the authors and steered by the authors in its exploration of the search space. 
Initially, we gave prompts asking for any improvement on the tradeoffs of~\cite{HR26}, uploading our pre-existing research notes and specifying concrete ideas towards adapting matching-vector codes. After some back-and-forth, this led to the Track 1 results (\Cref{mainthm:track1}) and a different ad-hoc proof of the Track 3 result (\Cref{mainthm:track3}) in the 3-server case. We identified Fourier-sparse cone-delta functions as the key ingredient underlying these protocols.

Generalizing these constructions of Fourier-sparse cone-delta functions to infinitely many server counts (in fact, even one server count $> 3$) took several more weeks of iteration with ChatGPT. The first general constructions were along the lines of Track 2 (\Cref{mainthm:track2}), but their initial presentations were more mathematically involved than anything in this paper. We spent several more weeks steering ChatGPT towards results that were stronger and had less mathematically involved proofs, culminating in Track 3 (\Cref{mainthm:track3}) which only uses machinery from complex analysis and some number theory.

The aforementioned GPT models and Claude (Opus 5, Opus 5.5, and Fable 5) were all used for writing, in particular to draft and verify proofs. The authors are fully accountable for the contents and correctness of this paper. 
\end{sloppypar}

\section*{Acknowledgements}
\phantomsection\addcontentsline{toc}{part}{Acknowledgements}

We thank Vinod Vaikuntanathan, Henry Corrigan-Gibbs, and Dan Boneh for encouraging this work and for numerous useful discussions on PIR. We also thank Assaf Naor, Peter Sarnak, and Andrew Sutherland for teaching beautiful courses that helped with the presentation of this work.

\ifllncs
\bibliographystyle{splncs04}
\bibliography{refs}
\else
\bibliographystyle{alpha}
\bibliography{refs}
\fi

\appendix
\crefalias{section}{appendix}\crefalias{subsection}{subappendix}
\phantomsection\addcontentsline{toc}{part}{Appendices}
\input{depirx}
\input{mathapp}
\end{document}

%% file: preamble.tex
\usepackage[T1]{fontenc}
\usepackage{amssymb}
\usepackage{amsmath}
\usepackage{amsthm}
\makeatletter
\renewenvironment{proof}[1][\proofname]{%
  \par\pushQED{\qed}%
  \normalfont\topsep6\p@\@plus6\p@\relax
  \trivlist
  \item[\hskip\labelsep
    \itshape
    \if\relax\detokenize{#1}\relax\else
      #1\@addpunct{.}%
    \fi]\ignorespaces
}{%
  \popQED\endtrivlist\@endpefalse
}
\makeatother
\PassOptionsToPackage{hyphens}{url}

\usepackage[notquote]{hanging} 
\usepackage{mathtools}
\usepackage{amsmath}
\usepackage{pifont}
\ifllncs
\usepackage[pdfstartview=FitH,colorlinks,urlcolor=black,linkcolor=black,citecolor=black,pdfpagelabels,bookmarksdepth=2,bookmarksopen=true,bookmarksnumbered,breaklinks]{hyperref}
\usepackage[font={small}]{caption}
\else
\usepackage[utf8]{inputenc}
\usepackage{beton}
\usepackage{fullpage}
\usepackage[margin=1in]{geometry}
\usepackage{caption}
\usepackage{xcolor}
\definecolor{DarkBlue}{RGB}{0,0,150}
\definecolor{NotSoDarkBlue}{RGB}{15,15,210}
\definecolor{DarkRed}{RGB}{150,0,0}
\definecolor{DarkGreen}{RGB}{0,100,0}
\usepackage[pdfstartview=FitH,pdfpagemode=UseNone,colorlinks,allcolors=black,pagebackref,breaklinks]{hyperref}
\usepackage{microtype}      
\usepackage{libertine}
\usepackage{libertinust1math}

\usepackage{breakcites}
\usepackage[noadjust,sort,nocompress]{cite}

\usepackage{tikz,pgfplots}
\usepackage{tikz-cd}
\usetikzlibrary{calc}

\usepackage{mathtools, xparse}
\usepackage{graphicx}
\usepackage{rotating}
\usepackage{braket}
\usepackage{url}
\usepackage{bm}
\usepackage{float}
\usepackage{soul}
\usepackage{multirow}

\usepackage[ruled, algosection]{algorithm2e}
\usepackage{orcidlink}

\fi

\usepackage{booktabs}
\usepackage[table,dvipsnames]{xcolor}
\usepackage{adjustbox}

\usepackage{bbm}
\usepackage{mathtools}

\usepackage{framed}
\usepackage{multicol}
\ifllncs\else

\usepackage{amsthm}
\fi
\usepackage[capitalise]{cleveref}
\usepackage[inline]{enumitem}

\ifllncs
\spnewtheorem{informal-theorem}[theorem]{Informal Theorem}{\bfseries}{\itshape}
\spnewtheorem{experiment}{Experiment}[section]{\bfseries}{}
\spnewtheorem{fact}[theorem]{Fact}{\bfseries}{\itshape}
\spnewtheorem{defn}[theorem]{Definition}{\bfseries}{\itshape}
\spnewtheorem{const}[theorem]{Construction}{\bfseries}{}
\else 
\theoremstyle{plain}
\newtheorem{theorem}{Theorem}[section]{\bfseries}{\itshape}
\newtheorem{informal-theorem}[theorem]{Informal Theorem}{\bfseries}{\itshape}
{\bfseries}{}
\newtheorem{lemma}[theorem]{Lemma}{\bfseries}{\itshape}
\newtheorem{fact}[theorem]{Fact}{\bfseries}{\itshape}
\makeatletter
\def\th@remarkit{%
  \thm@headfont{\itshape}%
  \itshape
  \thm@preskip\topsep \divide\thm@preskip\tw@
  \thm@postskip\thm@preskip
}
\makeatother
\theoremstyle{remarkit}
\newtheorem{remark}[theorem]{Remark}{\itshape}{}
\theoremstyle{plain}
{\bfseries}{\itshape}
{\bfseries}{}
\newtheorem{corollary}[theorem]{Corollary}{\bfseries}{\itshape}
\newtheorem{claim}[theorem]{Claim}{\bfseries}{\itshape}
\fi 

\Crefname{informal-theorem}{Informal Theorem}{Informal Theorems}
\Crefname{prop}{Proposition}{Propositions}
\Crefname{fact}{Fact}{Facts}
\Crefname{game}{Game}{Games}
\Crefname{defn}{Definition}{Definitions}
\Crefname{remark}{Remark}{Remarks}
\Crefname{const}{Construction}{Constructions}


%% file: latexdefs.tex
\ifllncs
\else
\usepackage{amsthm}
\theoremstyle{plain}
\newtheorem{definition}[theorem]{Definition}
\newtheorem{proposition}[theorem]{Proposition}

\numberwithin{theorem}{section}
\numberwithin{conjecture}{section}
\numberwithin{problem}{section}

\newtheorem{maintheorem}[theorem]{Main Theorem}

\AddToHook{env/lemma/begin}{\crefalias{theorem}{lemma}}
\AddToHook{env/corollary/begin}{\crefalias{theorem}{corollary}}
\AddToHook{env/definition/begin}{\crefalias{theorem}{definition}}
\AddToHook{env/proposition/begin}{\crefalias{theorem}{proposition}}
\AddToHook{env/fact/begin}{\crefalias{theorem}{fact}}
\AddToHook{env/claim/begin}{\crefalias{theorem}{claim}}
\AddToHook{env/informal-theorem/begin}{\crefalias{theorem}{informal-theorem}}
\AddToHook{env/maintheorem/begin}{\crefalias{theorem}{maintheorem}}
\AddToHook{env/defn/begin}{\crefalias{theorem}{defn}}
\AddToHook{env/const/begin}{\crefalias{theorem}{const}}
\crefname{fact}{Fact}{Facts}
\crefname{claim}{Claim}{Claims}
\Crefname{claim}{Claim}{Claims}
\crefname{maintheorem}{Main Theorem}{Main Theorems}
\Crefname{maintheorem}{Main Theorem}{Main Theorems}
\crefname{informal-theorem}{Informal Theorem}{Informal Theorems}

\AddToHook{env/remark/begin}{\crefalias{theorem}{remark}}

\newcommand\numberthis{\addtocounter{equation}{1}\tag{\theequation}}

\fi

\newcommand{\ord}[1]{\ensuremath{#1^{\mathrm{th}}}}

\newcommand{\ans}{\mathsf{ans}}
\newcommand{\Query}{\mathsf{Query}}
\newcommand{\Recon}{\mathsf{Reconstruct}}
\newcommand{\Reconstruct}{\mathsf{Reconstruct}}

\newcommand{\supp}{\mathsf{supp}}

\newcommand{\query}{\mathsf{qu}}
\newcommand{\qu}{\query}
\newcommand{\state}{\mathsf{st}}

\DeclareMathOperator{\order}{ord}
\DeclareMathOperator{\GL}{GL}

\def\sfX{\mathsf{X}}
\def\sfZ{\mathsf{Z}}
\def\frakc{\mathfrak{c}}
\def\fraka{\mathfrak{a}}
\def\frakb{\mathfrak{b}}
\def\frakC{\mathfrak{C}}

\newcommand{\Preprocess}{\mathsf{Preprocess}}

\newcommand{\SAns}{\mathsf{Answer}}

\newcommand{\Answer}{\SAns}

\newcommand{\DB}{\mathsf{DB}}
\newcommand{\db}{\mathsf{DB}}

\newcommand{\T}{\mathsf{Table}}

\newcommand{\rgets}{\mathrel{\mathpalette\rgetscmd\relax}}
\newcommand{\rgetscmd}{\ooalign{$\leftarrow$\cr
    \hidewidth\raisebox{1.2\height}{\scalebox{0.5}{\ \rm R}}\hidewidth\cr}}%

\renewcommand{\vec}[1]{\mathbf{#1}}
\newcommand{\av}{\mathbf{a}} \newcommand{\cv}{\mathbf{c}}
 \newcommand{\ev}{\mathbf{e}}

\newcommand{\rv}{\mathbf{r}} 
\newcommand{\tv}{\mathbf{t}} \newcommand{\uv}{\mathbf{u}}
\newcommand{\vecv}{\mathbf{v}} \newcommand{\wv}{\mathbf{w}}
\newcommand{\xv}{\mathbf{x}} \newcommand{\yv}{\mathbf{y}}
\newcommand{\zv}{\mathbf{z}} \newcommand{\zerov}{\mathbf{0}}
 \newcommand{\BB}{\mathbf{B}}

\newcommand{\one}{\mathbbm{1}}

\newcommand{\itpara}[1]{\medskip\noindent\textit{#1}}

\renewcommand{\paragraph}[1]{\medskip\noindent\textbf{#1}}
\newcommand{\prmid}{\; \middle| \;}

\newcommand{\calA}{\ensuremath{\mathcal{A}}}

\newcommand{\calH}{\ensuremath{\mathcal{H}}}

\newcommand{\calL}{\ensuremath{\mathcal{L}}}

\newcommand{\calQ}{\ensuremath{\mathcal{Q}}}
\newcommand{\calR}{\ensuremath{\mathcal{R}}}

\newcommand{\calT}{\ensuremath{\mathcal{T}}}

\newcommand{\calW}{\ensuremath{\mathcal{W}}}

\newcommand{\Type}{\ensuremath{\mathsf{Type}}}

\def\kv{\mathbf{k}}

\def\bv{\mathbf{b}}

\newcommand{\boldG}{\ensuremath{\mathbf{G}}}

\DeclareMathOperator{\wt}{wt}
\newcommand{\zo}{\ensuremath{\{0,1\}}} 

\newcommand{\rcell}{\cellcolor{red!25}}
\newcommand{\gcell}{\cellcolor{green!25}}
\newcommand{\ycell}{\cellcolor{yellow!25}}
\newcommand{\graycell}{\cellcolor{gray!25}}

\newcommand{\gc}{\cellcolor{green!25}}

\newcommand{\CC}{\mathbb{C}}
\newcommand{\RR}{\mathbb{R}}
\newcommand{\QQ}{\mathbb{Q}}

\newcommand{\ZZ}{\mathbb{Z}}

\newcommand{\FF}{\mathbb{F}}
\newcommand{\F}{\mathbb{F}}
\newcommand{\N}{\mathbb{N}}
\newcommand{\bfone}{\mathbf{1}}
\newcommand{\poly}{\mathsf{poly}}
\newcommand{\polylog}{\mathsf{polylog}}

\DeclareMathOperator{\SL}{SL}
\DeclareMathOperator{\PSL}{PSL}

\newcommand{\pp}{\partial}

\allowdisplaybreaks

\DeclarePairedDelimiterX{\bigket}[1]{\bigg\lvert}{\bigg\rangle}{\,#1}
\DeclarePairedDelimiter{\norm}{\lVert}{\rVert}

\newcolumntype{R}[2]{%
  >{\adjustbox{angle=#1,lap=\width-(#2)}\bgroup}%
    c%
    <{\egroup}%
}

\newcommand{\squishheader}[1]{{{\rotatebox{0}{#1}}}}

\usepackage{xcolor}
\usepackage{amsmath}

\definecolor{bracegray}{gray}{0.7}

\let\originaloverbrace\overbrace
\let\originalunderbrace\underbrace

\RenewDocumentCommand{\overbrace}{m e{^}}{%
    {\color{bracegray}%
        \originaloverbrace{{\color{black}#1}}%
        \IfValueT{#2}{^{#2}}%
    }%
}

\RenewDocumentCommand{\underbrace}{m e{_}}{%
    {\color{bracegray}%
        \originalunderbrace{{\color{black}#1}}%
        \IfValueT{#2}{_{#2}}%
    }%
}

\newcommand{\ip}[2]{\langle #1, #2\rangle}

%% file: abstract.tex
\begin{abstract}
    We present new information-theoretic protocols for $s$-server, doubly-efficient private information retrieval (PIR) in sublinear server time and {\em one} extra bit of space on the servers, for every constant $s \geq 2$. On an $n$-bit database, each of the $s$ PIR servers first builds and stores a preprocessing data structure consisting of $n+1$ bits (which is unique to each server, and after which the server may discard the original database). Then, each server answers a PIR query in time $T_s(n)$, by reading at most $T_s$ bits of its data structure and sending them to the user. 
    
    In the two server setting ($s=2$), our PIR protocols achieve $T_2 = n^{0.82}$, matching the server time of the state-of-the-art schemes of Henzinger and Ragavan (EUROCRYPT 2026) while shrinking their storage from $O(n\sqrt{\log n})$ to $n+1$ bits. With $s \geq 3$ servers, our PIR protocols retain the $n+1$ bits of storage and additionally improve on the server time of all prior information-theoretic $s$-server schemes with $n^{1+o(1)}$ storage: for example, achieving $T_3 = n^{0.74}$ and $T_{10} = n^{0.61}$ (previously $n^{0.79}$ and $n^{0.75}$). Asymptotically, as the server count $s$ grows, we shrink the number of servers required to achieve a given server-work exponent from $s^{3.9}$ to $s$. 

    Our results draw inspiration from matching-vector techniques (Efremenko, STOC 2009), which we generalize to handle degree-two computations in the exponent. Using this generalization, we build doubly-efficient PIR from the new intermediate abstraction of Fourier-sparse cone-delta functions. Then, we construct such Fourier-sparse cone-delta functions via a connection to modular forms and complex analysis: if a Fourier-sparse cone-delta function does not exist, then there exists a modular form which has too many roots, contradicting the valence formula.
    
\end{abstract}

%% file: intro.tex
\section{Introduction}
\label{sec:intro}

A {\em private information retrieval} (PIR) protocol~\cite{CGKS95,KO97} runs between $s$ servers, who each hold a copy of an $n$-bit database $\DB \in \{0,1\}^n$, and a user, who holds an index $i \in [n]$. 
The protocol lets the user learn the $\ord{i}$ bit of the database, while each server learns  nothing about the index~$i$. 
More formally, each server cannot distinguish (either information-theoretically or computationally) between any two indices $i$ or $j$ that the user might be reading.
This privacy property does not come cheap: Beimel, Ishai, and Malkin~\cite{Beimel2004} prove that, in any PIR scheme where the $s$ servers store only the database $\db$, the servers must touch every bit of $\db$ to answer a PIR query---in effect, running in $\Omega(n)$ time---as otherwise they would learn which locations the user is not reading.

{\em Doubly-efficient} PIR escapes this limitation by letting the servers preprocess their database once, ahead of time~\cite{Beimel2004,LMW23}: the $s$~servers first transform the $n$-bit database $\db$ into a $\text{poly}(n)$-bit data structure of their choosing. Then, the servers use this data structure to answer each PIR query in $o(n)$ time. In this setting,  a much weaker version of Beimel, Ishai, and Malkin's lower bound applies:  if the servers store $b$ preprocessed bits, the PIR servers must run in $\Omega(n/b)$ time, as they could ``logically'' touch all bits of the database by probing only a far smaller number of bits in their preprocessed data structure~\cite{Beimel2004}.

At the same time, known doubly-efficient PIR schemes are far from  this ideal time-space tradeoff, either:
\vspace{-5pt}\begin{enumerate}[itemsep=2pt]
    \item  saving only polylogarithmic factors in  the servers' runtime~\cite{DBLP:conf/crypto/Lee26,lee26b,Lip09,WY05}, or
    \item incurring server storage that is  concretely enormous (in the computationally-private regime~\cite{LMW23,BIPW17,CHR17,okada2025algebraic}) or that grows asymptotically-faster than the database size (in the information-theoretic  one~\cite{Beimel2004,lazzaretti2024multi,DBLP:journals/iacr/GhoshalLMDS24,HR26}).
\end{enumerate}  
A puzzling question remains: to make a private data access, what is the right data structure for the job?

\subsection{Results}

In this paper, we introduce a new data structure for doubly-efficient PIR. 
Our approach draws inspiration from matching-vector PIR~\cite{Y08,Efr09,DG16,GKS25} and a long line of work on preprocessing PIR~\cite{DBLP:conf/crypto/BeimelIM00,lazzaretti2024multi,DBLP:journals/iacr/GhoshalLMDS24,HR26}.
While our protocols are simple to describe, their correctness is proven via a connection to modular forms and complex analysis. 
This yields the following results:
\begin{enumerate}
    \item We obtain the first doubly-efficient PIR schemes with a constant number of servers,  polynomially-sublinear server time $n^{1-\Omega(1)}$, and $n+1$ bits of storage per server. That is, our schemes incur {\em one bit} of overhead in server storage relative to schemes without preprocessing (although each server must store a different data structure, rather than an identical copy of the database).

    \item Among doubly-efficient PIR schemes with information-theoretic security and quasilinear (i.e., $n^{1+o(1)}$) server storage, we improve the communication and server time for all constant server counts $s \geq 3$ over the previous state of the art~\cite{HR26}.
    To be more precise, for a given server work budget, we shrink the number of required servers almost quartically (from $s^{3.9}$ to $s$) as the number of servers grows---all while achieving $n+1$ bits of storage.
\end{enumerate} 
As an added benefit, in our PIR schemes, the preprocessing data structure is simple and efficient to produce: it is a Fourier transform of the database with specific, server-dependent weights applied to each term,  computable in $n \cdot \polylog(n)$ time.

\input{table-server-counts}

We report the precise asymptotic performance of our doubly-efficient PIR, along with that of the prior state-of-the-art doubly-efficient schemes~\cite{HR26}, in \cref{table:server-counts}. Formal theorem statements follow in~\Cref{sec:mainthms}. 

\subsection{Guide to the Paper} 
\label{sec:guide}

Our new PIR constructions build on a shared   intermediate abstraction which we call {\em Fourier-sparse cone-delta functions}, a special case of the delta functions defined by Ghasemi and Kopparty in their study of matching-vector PIR~\cite{DBLP:conf/innovations/GhasemiK26}. This abstraction splits the paper into two parts: \Cref{sec:depir} builds doubly-efficient PIR from  Fourier-sparse cone-delta functions, and gives elementary (but limited) constructions of these functions.  \Cref{sec:math} constructs even sparser cone-delta functions from modular forms. 

Some of our designs and analysis of sparse cone-delta functions rely on tools from complex analysis and number theory. In the interest of readability and modularity, we organize these constructions into three ``reading tracks,'' according to the mathematics they use:

\begin{itemize}
    \item In~\Cref{sec:depir}, our elementary constructions of Fourier-sparse cone-delta functions, together with our compiler from cone-delta functions to doubly-efficient PIR,  yield the ``Track 1'' results displayed in~\Cref{table:server-counts}. These results include our  two-server doubly-efficient PIR protocol, as well as a three-server doubly-efficient PIR protocol with $n+1$ bits of server storage and  a server-time exponent better than the prior, quasilinear-storage state of the art~\cite{HR26}. (A reader interested in only our cryptographic protocols may focus exclusively on \cref{sec:depir}.)

    \item In~\Cref{sec:math}, we prove that even sparser cone-delta functions exist. Combining these results with the compiler from~\Cref{sec:depir} immediately gives improved PIR protocols.~\Cref{sec:math} further splits into two tracks, which draw on the theory of modular forms at various levels of complexity:
    \begin{itemize}
        \item Track 2 exposes the main conceptual ideas underlying the connection between modular forms and cone-delta functions (using, in particular, the theta functions of lattices).

        \item Track 3 gives our strongest results  with considerably more complexity (including the use of Jacobi forms).
    \end{itemize}
\end{itemize}
 \Cref{tab:guide} summarizes each track's PIR result, and two paths for reading it: a short one through the informal technical overview (``Path with overviews'') and a more detailed, full presentation (``Path with proofs''). We give a self-contained presentation of modular forms and their properties in~\Cref{sec:pt2bg}, as well as more informal intuitions in~\Cref{sec:overview:modforms}.

\input{tabguide}

\section{Technical Overview}
\label{sec:overview}

Next, we give an overview of our techniques. The high-level idea behind state-of-the art, information-theoretic doubly-efficient PIR schemes~\cite{Beimel2004,HR26} is to first build ``unbalanced'' PIR schemes with extremely short queries of length $\ell = (1+o(1)) \cdot \log n$ bits. Then, these schemes convert such an unbalanced PIR scheme into a doubly-efficient one by having the $s$ servers each build a data structure that stores a piece of information for every possible PIR query---i.e., consisting of $2^\ell = n^{1+o(1)}$ bits. Each possible PIR query determines a fixed pattern of $n^{1/2+\epsilon(s)}$ locations that the servers look up in this data structure and send back to the client, for constant $\epsilon(s)$ approaching $1 / \log s + o(1/\log s)$ as the server count $s$ grows.

We push this technique to its extreme: building PIR with queries of bitlength close to $\log n$ and 
having each server store a data structure consisting of $n+1$ bits. As before, each possible query determines a fixed pattern of $n^{1/2+\epsilon(s)}$ locations that the servers look up in this data structure and send back to the client; but now for improved constant $\epsilon(s)$ that approaches $0.255 / \log s + o(1/\log s)$ as $s$ grows. 

To make this design work, our starting point  is to adapt techniques from  {\em matching-vector codes}, which yield the most communication-efficient (but not doubly-efficient) PIR schemes to date~\cite{Y08,Efr09,DG16}. 
Towards modifying these techniques to produce short enough queries for preprocessing, we study the intermediate abstraction of {\em delta functions} that are sparse in the Fourier domain, a mathematical object introduced by Ghasemi and Kopparty in their study of matching-vector PIR~\cite{DBLP:conf/innovations/GhasemiK26}. We give new constructions of Fourier-sparse delta functions on the set $\left\{ \xv \in \ZZ_q^d: \langle \xv, \xv \rangle \equiv 0 \pmod{q}\right\}$, which we call \emph{the cone} (it is the zero set of the quadratic form $\langle \xv, \xv \rangle$, and closed under scaling). We refer to these functions as Fourier-sparse ``cone-delta'' functions, and we construct them via a connection to {\em modular forms}. Details on each of these steps follow.

\paragraph{Notation.} For a positive integer $n$, we write $[n] := \{1, \ldots, n\}$ and write $\one[E]$ to denote the indicator of an event~$E$. For an integer $q \geq 2$, we write $\ZZ_q$ to refer to the integers modulo~$q$; for vectors~$\uv, \vecv \in \ZZ_q^d$, we write their inner product as $\langle \uv, \vecv \rangle := \sum_{k \in [d]} u_k v_k \in \ZZ_q$. We use the notation $\rv \rgets \ZZ_q^d$ or $\rv \gets \ZZ_q^d$ to denote a vector sampled uniformly. Logarithms are base~$2$, unless otherwise specified.

\subsection{Inspiration from Matching-Vector PIR}\label{sec:overview:inspiration}
\label{sec:intro:mvc}

Given an $n$-bit database, the high-level idea behind $s$-server matching-vector PIR~\cite{Y08,Efr09,DG16}---which is communication-efficient, but not doubly-efficient---is to associate the indices in $[n]$ with distinct vectors $\uv_1, \ldots, \uv_n \in \ZZ_q^d$, for parameter $d$ that grows with $n$, and then rely on a simple, seemingly-tautological equation: 
\begin{equation}\label{eq:tautology}    
    \forall i \in [n], \qquad \DB_i = \sum_{j \in [n]} \DB_j \cdot \one[i = j] = \sum_{j \in [n]} \DB_j \cdot \one[\uv_i = \uv_j].
\end{equation}

To make this approach work, the vectors $\uv_1, \ldots, \uv_n \in \ZZ_q^d$ are selected to have suitable structure: namely, there exists a field $\F$ with a primitive \ord{q} root of unity $\gamma$ and an $s$-sparse polynomial $P \in \F[X]$ such that
\begin{equation}\label{eq:mvdecpoly}
    \forall i, j \in [n], \qquad P\left(\gamma^{\langle \uv_i, \uv_j \rangle}\right) = \one[\uv_i = \uv_j].
\end{equation}
Plugging this into equation~\eqref{eq:tautology} and writing $P(X) = \sum_{t \in [s]} \alpha_t \cdot X^{c_t}$ gives, for any desired index $i \in [n]$:
\begin{align*}
    \DB_i
    &= \sum_{j \in [n]} \DB_j \cdot P\left(\gamma^{\langle \uv_i, \uv_j \rangle}\right) \\
    &= \sum_{j \in [n]} \DB_j \cdot \left(\sum_{t \in [s]} \alpha_t \gamma^{\langle c_t \uv_i, \uv_j \rangle}\right) \\
    &= \sum_{t \in [s]} \left(\sum_{j \in [n]} \DB_j \cdot \alpha_t \gamma^{\langle c_t \uv_i, \uv_j \rangle}\right).
\end{align*}
This chain of substitutions suggests the following (not yet private) protocol: a user, who wants to read the database record at index $i \in [n]$, sends the vector $c_t \uv_i$ to each server~$t \in [s]$; server~$t$ in turn responds with the sum~$\sum_{j \in [n]} \DB_j \cdot \alpha_t \gamma^{\langle c_t \uv_i, \uv_j \rangle}$. Making this protocol private requires only a small tweak: the user instead samples a random vector $\rv \gets \ZZ_q^d$ and sends  the masked vector $\rv + c_t \uv_i$ to server $t \in [s]$. This masking does little damage to the above computation's correctness: namely,
\begin{equation*}
    \forall i \in [n], \qquad \underbrace{\DB_i \cdot \gamma^{\langle \rv, \uv_i \rangle}}_{\text{recovered by the user, who derives $\DB_i$}} = \sum_{t \in [s]} \underbrace{\left(\sum_{j \in [n]} \DB_j \cdot \alpha_t \gamma^{\langle \rv + c_t \uv_i, \uv_j \rangle}\right)}_{\text{server $t$'s PIR answer}}.
\end{equation*}
This recovers exactly the matching-vector PIR protocol of Efremenko~\cite{Efr09}. 
Why is this protocol not doubly-efficient? The reason is that equation~\eqref{eq:mvdecpoly}  forces the inner products $\langle \uv_i, \uv_j \rangle$ to be different when $i = j$ vs. $i \neq j$. This requirement, combined with the $s$-sparsity of polynomial~$P$,  limits how dense the collection of vectors $\{\uv_1, \ldots, \uv_n\}$ can be as a subset of $\ZZ_q^d$. In the best-known construction~\cite{Efr09}, these vectors have dimension~$d = \omega(\log n)$, making the queries too large for preprocessing.

Put differently, building a doubly-efficient PIR from the matching-vector template requires finding a replacement for equation~\eqref{eq:mvdecpoly} that both tests for equality between \emph{any} two vectors in $\ZZ_q^d$, and can be distributed across $s$ servers while retaining privacy, much as in the above protocol. 
The former requirement closely mirrors the motivation behind Ghasemi and Kopparty's study of Fourier-sparse delta functions on restricted sets~\cite{DBLP:conf/innovations/GhasemiK26}; our next insight is to construct such Fourier-sparse delta functions on a new set.

\subsection{Doubly-Efficient PIR from Sparse Cone-Delta Functions}\label{sec:depir:qpir}

To this end, we define the notion of {\em sparse cone-delta functions}, which is the new machinery at the core of our doubly-efficient PIR protocols. Informally, this is a function that is both sparse in its Fourier representation and that evaluates to the indicator function $\mathbbm{1}[\vec x = 0]$ on the cone (i.e., for all $\xv \in \ZZ_q^d$ with $\langle \xv, \xv \rangle \equiv 0 \pmod{q}$). (We defer a more detailed comparison with~\cite{DBLP:conf/innovations/GhasemiK26} to~\Cref{sec:bg}.) Looking ahead to our final construction, the reason for the cone is that we generalize matching-vector techniques from the previous subsection to handle a degree-two computation in the exponent of $\gamma$---allowing for direct cancellation of any inputs not on the cone. 

The rest of this section makes our construction more precise:

\begin{definition}[Fourier-sparse cone-delta function]\label{def:conedeltaoverview}\label{def:conedelta}
    Let $q$ be an odd prime and $\F$ a field containing a primitive \ord{q} root of unity $\gamma$.
    In this work, we say that a collection of coefficients $C = \left\{c_{\av} \in \F: \av \in \ZZ_q^d\right\}$ is a {\em $k$-sparse cone delta function} on $\ZZ_q^d$ if there are at most $k$ nonzero coefficients $c_{\av}$ in $C$ and  it holds that:
    \begin{equation}\label{eq:conedeltaoverview}
        \sum_{\av \in \ZZ_q^d} c_{\av} \gamma^{\langle \av, \xv \rangle} = \begin{cases}
            1,\text{ if }\xv = 0; \\
            0,\text{ if }\xv \neq 0\text{ lies on the cone (i.e., }\langle \xv, \xv \rangle \equiv 0 \pmod{q}\text{)}; \\
            \text{arbitrary, otherwise.}\footnotemark
        \end{cases}
    \end{equation}\footnotetext{Placing no restrictions on the sum whenever $\langle \xv, \xv \rangle \not\equiv 0 \pmod{q}$ is crucial; if this sum were forced to vanish for all non-zero $\xv$, the resulting cone-delta function could not be sparse.}
\end{definition}

Before presenting our construction of cone-delta functions in \cref{sec:overview:modforms}, we explain how these functions give rise to doubly-efficient PIR protocols. For any two vectors~$\uv, \vecv \in \ZZ_q^d$, we have:
\begin{align*}
    \one[\uv = \zerov] = \left(\frac{1}{q} \sum_{t \in \ZZ_q} \gamma^{t \langle \uv, \uv \rangle}\right) \cdot \left(\sum_{\av \in \ZZ_q^d} c_{\av} \gamma^{\langle \av, \uv \rangle}\right)
    \quad \Rightarrow \quad \one[\uv = \vecv] = \left(\frac{1}{q} \sum_{t \in \ZZ_q} \gamma^{t \langle \uv-\vecv, \uv-\vecv \rangle}\right) \cdot \left(\sum_{\av \in \ZZ_q^d} c_{\av} \gamma^{\langle \av, \uv-\vecv \rangle}\right).
\end{align*}
To make the cancellation more explicit: in this second equation, the term $\left(\frac{1}{q} \sum_{t \in \ZZ_q} \gamma^{t \langle \uv-\vecv, \uv-\vecv \rangle}\right)$ is the indicator that $\uv - \vecv$ lies on the cone (i.e., $\langle \uv - \vecv, \uv - \vecv \rangle \equiv 0 \pmod{q}$), since the element $\gamma$ has order $q$. Meanwhile, whenever $\uv - \vecv$ lies on the cone, the cone-delta property guarantees that the term $\left(\sum_{\av \in \ZZ_q^d} c_{\av} \gamma^{\langle \av, \uv-\vecv \rangle}\right)$ evaluates to $\mathbbm{1}[\vec u-\vec v = 0] = \mathbbm{1}[\vec u = \vec v]$. 

Our $q$-server, doubly-efficient PIR protocol then follows by plugging the above equations into the template from \cref{sec:intro:mvc} and in particular equation~\eqref{eq:tautology}. More precisely, letting the database consist of $n = q^d$ records $\DB_1, \ldots, \DB_n$, each in the field $\F$, and letting the collection of vectors $\uv_1, \ldots, \uv_n$ be  all of $\ZZ_q^d$:
\begin{align*}
    \DB_i &= \sum_{j \in [n]} \DB_j \cdot \one[\uv_i = \uv_j] \\
    &= \sum_{j \in [n]} \DB_j \cdot \left(\frac{1}{q} \sum_{t \in \ZZ_q} \gamma^{t \langle \uv_i-\uv_j, \uv_i-\uv_j \rangle}\right) \cdot \left(\sum_{\av \in \ZZ_q^d} c_{\av} \gamma^{\langle \av, \uv_i-\uv_j \rangle}\right) \\
    &= \frac{1}{q} \sum_{t \in \ZZ_q} \sum_{\av \in \ZZ_q^d} c_\av \sum_{j \in [n]} \DB_j \cdot \gamma^{t \langle \uv_i - \uv_j, \uv_i - \uv_j \rangle + \langle \av, \uv_i - \uv_j \rangle} \\
    &= \frac{1}{q} \sum_{t \in \ZZ_q} \sum_{\av \in \ZZ_q^d} \left(c_\av \gamma^{\langle \av, \uv_i \rangle+t \langle \uv_i, \uv_i \rangle} \left(\sum_{j \in [n]} \left(\DB_j \gamma^{t \langle \uv_j, \uv_j \rangle}\right) \cdot \gamma^{\langle -2t\uv_i-\av, \uv_j \rangle}\right)\right).
\end{align*}
Once again, masking the $i$-dependent terms on the RHS with a random vector is harmless:
\begin{equation}\label{eq:overviewrecons}
    \DB_i \cdot \gamma^{\langle \rv, \uv_i \rangle} = \frac{1}{q} \sum_{t \in \ZZ_q} \sum_{\av \in \ZZ_q^d} \left(c_\av \gamma^{\langle \av, \uv_i \rangle+t \langle \uv_i, \uv_i \rangle} \left(\sum_{j \in [n]} \left(\DB_j \gamma^{t \langle \uv_j, \uv_j \rangle}\right) \cdot \gamma^{\langle \rv-2t\uv_i-\av, \uv_j \rangle}\right)\right).
\end{equation}
The above identity then naturally points to the following $q$-server PIR:

\paragraph{Preprocessing.}
Each server $t \in \ZZ_q$ stores the truth table of the following function $f_t: \ZZ_q^d \to \FF$:
\begin{equation*}
    f_t(\xv) := \sum_{j \in [n]} \DB_j \gamma^{t \langle \uv_j, \uv_j \rangle + \langle \xv, \uv_j \rangle}.
\end{equation*}
This data structure consists of exactly $q^d$ elements in $\F$---matching the dimensions of the original database. A quirk is  that each server's data structure depends on its own index $t \in [q]$. To compute it, each server $t \in [q]$ applies a quadratic phase $\gamma^{t \langle \uv_j, \uv_j \rangle}$ to each entry of the database, then performs a Fourier transform.

\paragraph{Querying.} 
Suppose the user wants to recover the record $\DB_i$ for $i \in [n]$.
To do this, the user samples a random mask $\rv \gets \ZZ_q^d$ and sends server $t \in [q]$ the query vector $$\query_t = \vec r - 2t \cdot \uv_i \in \ZZ_q^d.$$
Security is immediate: each server's query is masked by the random vector $\vec r$.

\paragraph{Answering.} 
Given a query $\qu \in \ZZ_q^d$, the server replies with the set of evaluations~$f_t(\qu - \av)$ for all vectors $\av \in \ZZ_q^d$ such that the corresponding coefficient of the cone-delta function $c_{\av} \neq 0$. Rather than computing these values, the server can simply fetch them from its precomputed truth table.

\paragraph{Reconstructing.} The user combines the servers' answers according to equation~\eqref{eq:overviewrecons}, which tells us that:
\begin{equation*}
     \gamma^{-\langle \rv, \uv_i \rangle} \cdot \frac{1}{q} \sum_{t \in \ZZ_q} \sum_{\av \in \ZZ_q^d} c_{\av} \gamma^{\langle \av, \uv_i \rangle + t \langle \uv_i, \uv_i \rangle} \underbrace{f_t(\rv-2t\uv_i-\av)}_{\text{included in server }t\text{'s answer}} = \DB_i.
\end{equation*}

\paragraph{Efficiency.} Finally, we analyze the scheme's efficiency. The user sends $d \log q$ bits to each each server, who in turn responds with $k$ elements in $\F$ from its preprocessed table, where $k$ is the sparsity of the cone-delta function. Similarly, the server computation and the user's computation both are both $O(k)$ field operations.\footnote{Although the user's reconstruction is written as a sum over all $\av \in \ZZ_q^d$, they only need to handle the $k$ terms for which $c_{\av} \neq 0$.} 
We arrive at the following theorem (stated for databases whose records are field elements, though a standard transformation to bit-valued databases~\cite{DBLP:conf/stoc/DodisPT10}, specified in~\Cref{sec:bits}, applies):

\begin{informal-theorem}\label{thm:q-server-rate-one}
Let $q \geq 3$ be a constant prime number of servers, and $\F$ a constant field. 
If there exists a cone-delta function from $\ZZ_q^d \to \F$ with Fourier sparsity $k \leq q^{d(1-\Omega(1))}$, then there exists a $q$-server doubly-efficient PIR scheme operating over databases that consist of $q^d$ records in $\F$. This PIR scheme has information-theoretic security and the following efficiency:
\begin{itemize}[itemsep=0pt,topsep=3pt]
    \item each server stores a data structure that consists of $q^d$ records in $\F$ (and may then discard the original database),
    \item to make a PIR query, the user uploads $d\lceil \log q \rceil$ bits to each PIR server,
    \item to answer a PIR query, each server reads and sends back $k$ elements in $\F$ from its data structure, and
    \item to reconstruct, the user performs $O(k)$ field operations.
\end{itemize}
\end{informal-theorem}

We end this section on two remarks: first, the observation that equation~\eqref{eq:overviewrecons} carries out a computation that is degree-two in the vector $\uv_i$ in its exponent, which generalizes the approach of classical matching-vector PIR (\cref{sec:overview:inspiration}). The reason this works is that the degree-two term splits into a term $\gamma^{t \langle \uv_i, \uv_i \rangle}$ that the client applies during reconstruction, a term $\gamma^{t \langle \uv_j, \uv_j \rangle}$ that the server applies during preprocessing, and remaining linear terms that are handled as in matching-vector PIR. Put more abstractly, this degree-two function in the exponent is computable because its derivative is linear; Liu, Vaikuntanathan, and Wee use similar techniques in their protocols for conditional disclosure of secrets~\cite[Section~4.1]{LVW17}.

Second, looking ahead to \cref{sec:depir:spir}, the protocol for doubly-efficient PIR we just constructed  generalizes to arbitrary, non-prime server counts. To this end, we generalize of the notion of sparse, cone-delta functions, as will appear in \cref{def:genconedelta,sec:bg:sfd}.

\subsection{Sparse Cone-Delta Functions from Modular Forms}\label{sec:overview:modforms}

With that aside, the remaining piece of the doubly-efficient PIR construction is to show that Fourier-sparse cone-delta functions exist. 
For simple cases of $q \in \{2,3\}$, we present elementary constructions of sparse cone-delta functions (or related objects for $q = 2$) that only make use of Fourier analysis, yielding the two-server and three-server PIR results  given in~\Cref{table:server-counts,tab:guide} (``Track 1'').

Generalizing these constructions to Fourier-sparse cone-delta functions over larger $q$ (which yields doubly-efficient PIR with larger server counts), and even obtaining a stronger sparsity-bound for the case of $q = 3$, turns out to be far more involved and requires the use of modular forms. Before we dive into the proof, we give its high-level intuition: if a Fourier-sparse cone-delta function does {\em not} exist, then we will show that there exists a modular form which has too many roots, contradicting the valence formula.
(In fact, proofs of classic Reed--Muller PIR basically follows this same template, changing ``modular form'' and ``valence formula'' to ``polynomial'' and ``degree bound'' respectively. We reiterate this parallel in~\Cref{sec:overview:arch}.) 

We focus on the following warm-up result, which is the core of the ``Track 2'' results given in~\Cref{table:server-counts,tab:guide} and is proved formally in~\Cref{sec:gen1}:
\begin{informal-theorem}[Informal version of~\Cref{cor:gen1general}]\label{thm:informalwarmup}
    Let $q$ be an odd prime, and let $\calL \subseteq \RR^r$ be an even lattice\footnote{An even lattice is one where $\langle \yv, \yv \rangle \in 2\ZZ$ for all $\yv \in \calL$.} of determinant $\sqrt{q}$, with dual lattice $\calL^*$. For each coset $c$ of $\calL$ in $\calL^*$, let $w_\calL(c) := \frac{q}{2} \min_{\yv \in c} \langle \yv, \yv \rangle$ be the (scaled) squared length of a shortest vector in $c$. Then, as $d \to \infty$, there exists a $k$-sparse cone-delta functions on $\ZZ_q^d$ with $k = q^{(\eta_\calL + o(1))d}$, where
    \begin{equation*}
        \eta_\calL := \inf_{0 < t \leq 1} \log_q \Bigg( t^{-r(q+1)/24} \sum_{c \in \calL^*/\calL} t^{w_\calL(c)} \Bigg).
    \end{equation*}
\end{informal-theorem}
Roughly, the cone-delta functions we construct are supported on the frequency vectors $\av$ whose coordinates, identified appropriately\footnote{We may make this identification since $\det \calL = \sqrt{q} \Rightarrow |\calL^* / \calL| = q$, so the quotient group is cyclic of order $q$.} with cosets of $\calL$ in $\calL^*$, have total \emph{cost} $\sum_i w_\calL(a_i)$ at most $r(q+1)d/24$. The sum over $c$ is a generating function for these costs, and the infimum over $t$ is a Chernoff-type bound on the number of such frequency vectors. For primes $q \equiv 3 \pmod{4}$, instantiating~\Cref{thm:informalwarmup} with a suitable lattice of rank $2$ gives an exponent $\eta_q \leq \frac{1}{2} + \frac{0.467 + o(1)}{\log_2 q}$ as $q \to \infty$ (\Cref{cor:gen1specific}). The theorem takes the lattice as input, so a lattice with a smaller $\eta_\calL$ immediately gives sparser cone-delta functions and, through~\Cref{mainthm:track2}, faster PIR schemes; we discuss the search for such lattices in~\Cref{sec:open}. 

The warm-up is the simplest instance of our approach. It already gives $q$-server PIR schemes whose exponent is $\frac{1}{2} + \frac{0.467 + o(1)}{\log_2 q}$ for large $q$ (\Cref{mainthm:track2}), and it introduces every tool that the stronger construction of~\Cref{sec:gen4} uses to obtain smaller exponents and arbitrary server counts (\Cref{mainthm:track3}). (See~\Cref{table:server-counts} for a comparison of the ``Track 2'' and ``Track 3'' results.) Our proofs combine two well-known results---specifically, the transformation law of lattice theta functions under the Weil representation~\cite[Theorem~4.1]{Borcherds98} and the valence formula~\cite[Theorem~4.1.4]{Rankin77}---in a new way. Specifically, we use them to construct a dual witness to show that sparse cone-delta functions exist.\footnote{For $q = 3$ and our particular choice of rank-2 lattice (\Cref{def:gen1concrete}), the conclusion was already known: the support is contained in Hamming ball of an elementary construction of Alon and Lubetzky~\cite[proof of Theorem~2.2]{AlonLubetzky07}, though we offer an improved elementary construction in~\Cref{sec:q3elementary} (which shrinks the exponent to $0.7445$ from $0.7897$). These simpler results are all subsumed by our strongest 3-server result in~\Cref{cor:gen4exponent}, which uses heavier modular-form machinery.} We sketch the main ideas of the stronger construction at the end of~\Cref{sec:overview:mf}, deferring its overview to~\Cref{sec:gen4overview}. 

From here on, we take $\F = \CC$ and $\gamma = \zeta_q := e^{2\pi i/q}$.
The coefficients we construct in this subsection lie in $\QQ(\zeta_q)$; \Cref{sec:primefield} transfers them to the finite fields that our PIR schemes use. We describe each step informally, and leave details to the body of the paper.

\subsubsection{Connecting to the Weil Representation of $\SL_2(\ZZ)$}\label{sec:overview:weil}

Observe that the condition in equation~\eqref{eq:conedeltaoverview} can be compressed in terms of two matrices. The first is, unsurprisingly, the Fourier transform matrix $\rho_S \in \GL_q(\CC)$ defined by:
\begin{equation}
    \rho_S \ev_x := \sum_{y \in \ZZ_q} \zeta_q^{-2xy} \ev_y.
\end{equation}
Here $\{\ev_x\}_{x \in \ZZ_q}$ is the standard basis of $\CC^q$, and for $\xv \in \ZZ_q^d$ we write $\ev_\xv := \ev_{x_1} \otimes \cdots \otimes \ev_{x_d}$ for the standard basis of $(\CC^q)^{\otimes d} = \CC^{q^d}$. In these terms, we have:
\begin{align*}
    \sum_{\av \in \ZZ_q^d} c_{\av} \zeta_q^{\langle \av, \xv \rangle} = \sum_{\av \in \ZZ_q^d} c_{-2\av} \zeta_q^{\langle -2\av, \xv \rangle}
    = \cv^\top \rho_S^{\otimes d} \ev_\xv,\numberthis \label{eq:rhoSoverview}
\end{align*}
where we define $\cv \in \CC^{q^d}$ to collect the coefficients $c_{-2\av}$. 
Secondly, the condition that $\langle \xv, \xv \rangle \equiv 0\pmod{q}$ can be described in terms of the quadratic phase matrix $\rho_T \in \GL_q(\CC)$ defined by:
\begin{equation}
    \rho_T \ev_x := \zeta_q^{x^2} \ev_x.
\end{equation}
(The factor $-2$ in $\rho_S$ comes from the identity $(x+y)^2 - x^2 - y^2 = 2xy$, which ties $\rho_S$ to this quadratic phase.) In these terms, we have:
\begin{equation}\label{eq:rhoToverview}
    \langle \xv, \xv \rangle \equiv 0 \pmod{q} \Leftrightarrow \rho_T^{\otimes d} \ev_{\xv} = \ev_{\xv}.
\end{equation}
Putting equations~\eqref{eq:rhoSoverview} and~\eqref{eq:rhoToverview} together gives the following succinct description of a $k$-sparse cone-delta function: there should exist a $\cv \in \CC^{q^d}$ with at most $k$ nonzero entries such that:
\begin{equation}\label{eq:conedeltaintermsofweil}
    \cv^\top \rho_S^{\otimes d} \ev_\xv = \begin{cases}
        1,\text{ if }\xv = 0;\\
        0,\text{ if }\rho_T^{\otimes d} \ev_\xv = \ev_\xv \text{ and }\xv \neq 0;\\
        \text{arbitrary, otherwise}.
    \end{cases}
\end{equation}
One might hope that a conceptual understanding of the matrices $\rho_S$ and $\rho_T$ will let us reason about sparse cone-delta functions. 
Miraculously, these same two matrices arise in the representation theory of $\SL_2(\ZZ)$! We unpack this connection next.

The group $\SL_2(\ZZ)$ consists of all $2 \times 2$ matrices with integer entries and determinant 1; it is generated by $$S := \begin{bmatrix} 0 & -1 \\ 1 & 0 \end{bmatrix} \text{ and }T := \begin{bmatrix} 1 & 1 \\ 0 & 1 \end{bmatrix}.$$ 
One can intuitively think of this group as (up to  sign conventions) encoding the transformations used by the Euclidean algorithm for computing the greatest-common-divisor of two integers: $S$ corresponds to swapping and $T$ corresponds to adding one element to the other. It turns out that there exists a mapping $\rho: \SL_2(\ZZ) \to \GL_q(\CC)$ such that:
\begin{itemize}
    \item $\rho(S) = \rho_S$ and $\rho(T) = \rho_T$, 
    \item for all $g, h \in \SL_2(\ZZ)$, we have $\rho(gh) \propto \rho(g) \rho(h)$, by which we mean the two matrices are the same up to a nonzero scalar.\footnote{Readers familiar with quantum computing may recognize these matrices: up to a scalar and a relabeling of the basis, $\rho_S$ is the quantum Fourier transform on a single $q$-dimensional qudit, and $\rho_T$ is a diagonal quadratic phase gate. More generally, each $\rho(g)$ normalizes the group generated by the shift and clock operators $\sfX$ and $\sfZ$ (\Cref{def:heisenberg,lem:heisenbergweil}), so it is a qudit Clifford gate; for odd $q$, the Clifford group modulo phases is isomorphic to $\SL_2(\F_q) \ltimes \F_q^2$~\cite[Theorem~1]{Appleby05}. Clifford gates are only defined up to a global phase, which is why $\rho$ is multiplicative only up to scalars.}
\end{itemize}
This mapping $\rho$ is known as the \emph{Weil representation} of $\SL_2(\ZZ)$~\cite{Weil64,Gerardin77,Stromberg13}.
Another observation that will be useful for intuition is the following: for any element $g \in \SL_2(\ZZ)$ and vector $\vecv \in \CC^{q}$, we have:
\begin{equation}
    \cv^\top \rho(g)^{\otimes d} \vecv^{\otimes d} = \cv^\top \left(\rho(g) \vecv\right)^{\otimes d} = P_{\cv}\left(\rho(g) \vecv\right),
\end{equation}
for a homogeneous degree-$d$ polynomial $P_{\cv}: \CC^q \to \CC$ formed using the entries of $\cv$ as its coefficients. The intuitive takeaway of this section is thus the following: to study sparse cone-delta functions over $\CC$, a promising route is to study the matrices $\rho(g)^{\otimes d}$ for the Weil representation $\rho: \SL_2(\ZZ) \to \GL_q(\CC)$. This in turn boils down to reasoning about compositions $P \circ \rho(g)$ for homogeneous degree-$d$ $P: \CC^q \to \CC$.

Looking ahead, this motivates modular forms because they can be thought of as ``Fourier-analytic'' tools for studying representations of $\SL_2(\ZZ)$ (and in particular, the Weil representation).

\subsubsection{Proof Architecture Mirroring Reed--Muller PIR}\label{sec:overview:arch}

We prove that sparse cone-delta functions exist by a root-counting argument, which splits into four steps (with full definitions in~\Cref{sec:overview:mf}):
\begin{enumerate}
    \item \emph{Assume the contrary and dualize:} suppose that no cone-delta function has Fourier support in a set $S_d \subseteq \ZZ_q^d$ that we will specify. Then, by duality, there is a vector $\beta$, indexed by the points of the cone, with
    \begin{equation}\label{eq:betaoverview}
        \beta_{\zerov} = 1,\text{ and}\qquad \sum_{\xv} \beta_\xv \zeta_q^{\langle \xv, \zv \rangle} = 0\quad\forall \zv \in S_d.
    \end{equation}
    We build from its entries the coefficients of a homogeneous polynomial $P_\beta$ of degree $d$ in $q$ variables.
    \item \emph{Choose a curve through the high-dimensional space:} the theta function $\Theta_\calL: \calH \to \CC^q$ of a lattice $\calL$ of rank $r$, a vector-valued modular form on the upper half-plane $\calH$.
    \item \emph{Restrict the witness to the curve:} the function $P_\beta \circ \Theta_\calL: \calH \to \CC$ is a scalar modular form of weight $dr/2$.
    \item \emph{Count roots to get a contradiction:} when $S_d$ is chosen correctly, $P_\beta \circ \Theta_\calL$ is not identically zero, yet it vanishes at a cusp to higher order than the valence formula, a root-counting bound for modular forms, allows.
\end{enumerate}
In fact, this argument is qualitatively similar to how the reconstruction of classic Reed--Muller PIR~\cite{CGKS95,WY05} is shown, except vector-valued and scalar-valued modular forms are replaced with multivariate and univariate polynomials respectively.

\subsubsection{Connecting $\SL_2(\ZZ)$ to Modular Forms}\label{sec:overview:mf}

\paragraph{The M\"obius action of $\SL_2(\ZZ)$ on the upper half-plane.} To use all of the matrices $\rho(g)$ at once, we need a space on which $\SL_2(\ZZ)$ acts, for which we turn to lattices. Recall the intuition that $\SL_2(\ZZ)$ encodes the steps of the Euclidean algorithm, which repeatedly swaps two numbers and adds one to the other. Applied to a basis $(\omega_1, \omega_2)$ of a lattice in $\RR^2$, rather than to two integers, these steps produce new bases of the same lattice. More precisely, $g = \begin{bsmallmatrix} a & b \\ c & d_0 \end{bsmallmatrix} \in \SL_2(\ZZ)$ sends the basis $(\omega_1, \omega_2)$ to the basis $(c\omega_2 + d_0\omega_1, a\omega_2 + b\omega_1)$ of the same lattice, and the matrices in $\SL_2(\ZZ)$ are exactly the changes of basis that preserve the lattice and its orientation. (We write $d_0$ for the lower-right entry, since $d$ is the dimension.) This picture, and the rest of the background in this subsection, is classical; see, e.g.,~\cite[Chapter~VII]{Serre73} and~\cite{DiamondShurman05,Shimura71}.

Now identify $\RR^2$ with $\CC$ in the natural way, and regard two lattices as equivalent if one is a nonzero complex multiple of the other (i.e., a rotated and rescaled copy of it). Dividing by $\omega_1$ shows that every lattice is equivalent to one with a basis of the form $(1, \tau)$, and by ordering the basis we may take $\tau$ in the upper half-plane $\calH := \{\tau \in \CC : \Im \tau > 0\}$. Applying $g$ to this basis gives the basis $(c\tau + d_0, a\tau + b)$ of the same lattice, and dividing by $c\tau + d_0$ turns it into the basis $(1, g \ast \tau)$ of an equivalent lattice (\Cref{fig:lattice}), where
\begin{equation*}
    g \ast \tau := \frac{a\tau + b}{c\tau + d_0}.
\end{equation*}
That is, $\ZZ + \ZZ(g \ast \tau) = (c\tau + d_0)^{-1}(\ZZ + \ZZ\tau)$. Since $\Im(g \ast \tau) = \Im \tau / |c\tau + d_0|^2$, these M\"obius transformations map $\calH$ to itself, and this is the space we want: $\SL_2(\ZZ)$ acts on $\calH$ by $\tau \mapsto g \ast \tau$. In this picture, $S$ and $T$ are the two steps of the Euclidean algorithm: $S$ replaces the basis $(1, \tau)$ by $(\tau, -1)$, and $T$ replaces it by $(1, \tau + 1)$. As for the Euclidean algorithm, these two steps generate all of $\SL_2(\ZZ)$~\cite[Chapter~VII, Theorem~2]{Serre73}.

\input{figlattice}

A crucial ingredient for us will be extending this action to the point at $\infty$ (which can be thought of as the limit of $it$ as $t \to \infty$). This then also requires us to introduce the boundary points in $\QQ$, since $\frac{a\infty+b}{c\infty+d_0} = \frac{a}{c}$, informally speaking. We refer to these points in $\QQ \cup \{\infty\}$ as the \emph{cusps} of $\calH$ (\Cref{fig:cusps}); these will be the evaluation points at which we ultimately count roots to obtain a contradiction.

\input{figcusps}

\paragraph{Modular forms.} A \emph{modular form} is a holomorphic (i.e., complex differentiable) function from $\calH$ to $\CC^q$ that is ``compatible'' with the aforementioned action of $\SL_2(\ZZ)$ and its Weil representation in a precise sense (for such vector-valued modular forms, see, e.g.,~\cite[Section~4]{Borcherds98}).
This is analogous to Fourier analysis: a Fourier character is a function from $\ZZ_q$ to $\CC$ that turns translation of its argument into multiplication of its value by a scalar. We depict this analogy in~\Cref{fig:fourieranalogy}, and now turn our attention to defining modular forms more precisely.

\begin{definition}[Informal]
    A modular form of weight $k$ for a subgroup $\Gamma \subseteq \SL_2(\ZZ)$ and a representation $\rho: \Gamma \to \GL_\ell(\CC)$ is a holomorphic function $f: \calH \to \CC^\ell$ satisfying both of the following:
    \begin{itemize}
        \item it is compatible with the M{\"o}bius action and $\rho$, in the sense that:
        \begin{equation}\label{eq:vvmfoverview}
            f\left(\begin{bmatrix} a & b \\ c & d_0 \end{bmatrix} \ast \tau\right) \propto (c\tau+d_0)^k \rho(g) f(\tau) \quad\forall g \in \Gamma, \tau \in \calH,
        \end{equation}
        where $\propto$ hides a nonzero constant that may depend on $g$ but not on $\tau$; and

        \item very roughly, it does not blow up near any of the cusps in $\QQ \cup \{\infty\}$. In this case, we can define an order of vanishing of $f$ at each such cusp according to whether it approaches 0 near the cusp, and if so, how rapidly.
    \end{itemize}
    When $\ell = 1$, we refer to this as a \emph{scalar} modular form. 
    The differential terms $d\tau$ in~\Cref{fig:fourieranalogy} are obtained by noting that $\frac{d}{d\tau} (g \ast \tau) = \frac{1}{(c\tau+d_0)^2}$.
\end{definition}
There are two useful features of modular forms that we highlight here:
\begin{itemize}
    \item The constraint of equation~\eqref{eq:vvmfoverview} together with the requirement that $f$ is holomorphic imposes rigid structure on $f$ which limits how many roots $f$ can have (in analogy to the degree bound for low-degree polynomials). For modular forms, this is formally captured by the \emph{valence formula} (\Cref{thm:valence}), which our proofs make heavy use of.
    \item Secondly, composition with homogeneous polynomials gives a way to extract a scalar modular form from a vector-valued modular form. 
    Indeed, let $P$ be a homogeneous polynomial of degree $d$ on $\CC^\ell$ with $P(\rho(g)\vecv) \propto P(\vecv)$ for all $g$ in a subgroup $\Gamma$. Then,~\eqref{eq:vvmfoverview} and homogeneity give $P(\Theta(g \ast \tau)) \propto (c\tau + d_0)^{dk} P(\Theta(\tau))$ for all $g \in \Gamma$, so $P \circ \Theta$ is a scalar modular form of weight $dk$ for $\Gamma$. This is the scalarization step of \Cref{fig:fourieranalogy}(b), and an important ingredient of our proof. It is also how the classical correspondence between weight enumerators of codes and theta series of lattices arises~\cite{BroueEnguehard72,NebeRainsSloane06}.
\end{itemize}

\input{figfourieranalogy}

\paragraph{Our vector-valued modular form: a lattice theta function.} 
Modular forms are often defined by encoding suitable information about a lattice.
In our case, where $\rho$ is the Weil representation, we use \emph{theta functions} of lattices. Let $\calL \subseteq \RR^r$ be an even lattice (one with $\langle \yv, \yv \rangle \in 2\ZZ$ for all $\yv \in \calL$) of determinant $\sqrt{q}$; its rank $r$ is then even. The dual lattice $\calL^* := \{\xv \in \RR^r : \langle \xv, \yv \rangle \in \ZZ \text{ for all } \yv \in \calL\}$ contains $\calL$, and $\calL^*/\calL$ is cyclic of order $q$, so its elements are the cosets $x\av + \calL$, for $x \in \ZZ_q$ and a generator $\av + \calL$. The \emph{theta function} of $\calL$ is
\begin{equation*}
    \Theta_\calL(\tau) := \sum_{x \in \ZZ_q} \Bigg(\sum_{\yv \in x\av + \calL} e^{\pi i \langle \yv, \yv \rangle \tau}\Bigg) \ev_x \in \CC^q.
\end{equation*}
In the case where $\tau$ is purely imaginary, each coordinate of $\Theta_{\calL}(\tau)$ is the Gaussian mass of a different coset of $\calL$. 
One can check that $\Theta_{\calL}$ is modular for $\SL_2(\ZZ)$ and the Weil representation: for some intuition, equation~\eqref{eq:vvmfoverview} for $g = T$ follows from reindexing and for $g = S$ it follows from the Poisson summation formula (for the same reason that Poisson summation arises in lattice-based cryptography; see~\Cref{sec:rel}). 
This is well-known~\cite[Theorem~4.1]{Borcherds98}, but we provide a self-contained proof in~\Cref{sec:gen1step2}. Finally, as $\Im \tau \to \infty$, every term except the one for $\yv = \zerov$ decays, so $\Theta_\calL(\tau) \to \ev_0$.

\paragraph{Root counting for modular forms: the valence formula.}
The last ingredient for us to sketch is how we bound the root counts of modular forms, analogously to the degree bound for univariate polynomials. The \emph{valence formula} (\Cref{thm:valence}; see, e.g.,~\cite[Theorem~4.1.4]{Rankin77}) plays this role: a nonzero modular form of weight $k$ for $\Gamma$ has total order of vanishing (across equivalence classes of cusps under $\Gamma$) at most
\begin{equation*}
    \frac{k}{12}\left[\SL_2(\ZZ) : \Gamma\right].
\end{equation*}
Note that the dependencies on $k$ and $[\SL_2(\ZZ) : \Gamma]$ above are natural: $k$ can be thought of as a degree-type parameter so larger $k$ should allow for more roots, and larger $[\SL_2(\ZZ) : \Gamma]$ means $\Gamma$ is smaller and therefore equation~\eqref{eq:vvmfoverview} imposes fewer constraints on the modular form, thus allowing for more roots.

\paragraph{Putting the proof together.} We can now explain how the conditions on a cone-delta function from~\Cref{sec:overview:weil} reappear in the proof. The witness $\beta$ of~\Cref{sec:overview:arch} has the same structure, dually to equation~\eqref{eq:conedeltaintermsofweil}: it is supported on the cone, so it is fixed by $\rho_T^{\otimes d}$; it satisfies $\beta_\zerov = 1$; and by equation~\eqref{eq:betaoverview} it is orthogonal to $\rho_S^{\otimes d}\ev_\zv$ for every $\zv \in S_d$ (the sets $S_d$ we use are closed under scaling by nonzero residues, which absorbs the factor $-2$ in $\rho_S$). Each of these becomes a property of the scalar function $P_\beta \circ \Theta_\calL$:
\begin{itemize}
    \item \emph{The cone ($\rho_T$) gives modularity.} Every $g$ in the subgroup
    \begin{equation*}
        \Gamma_0(q) := \left\{\begin{bmatrix} a & b \\ c & d_0 \end{bmatrix} \in \SL_2(\ZZ) : c \equiv 0 \pmod{q}\right\}
    \end{equation*}
    acts on $\CC^q$, up to a scalar, by permuting the coordinates ($x \mapsto d_0x$) and multiplying them by quadratic phases. On the cone these phases cancel, just as for $\rho_T$, and after averaging $P_\beta$ over the permutations $x \mapsto ux$, the function $P_\beta \circ \Theta_\calL$ becomes a scalar modular form of weight $dr/2$ for $\Gamma_0(q)$. Since $[\SL_2(\ZZ) : \Gamma_0(q)] = q + 1$, the valence formula allows it a total order of vanishing of $dr(q+1)/24$.
    \item \emph{The normalization $\beta_\zerov = 1$ from~\eqref{eq:betaoverview} gives nonvanishing at the cusp at $\infty$.} As $\Im \tau \to \infty$, we have $\Theta_\calL(\tau) \to \ev_0$, so $P_\beta(\Theta_\calL(\tau))$ tends to a nonzero multiple of $\beta_\zerov$.
    \item \emph{The Fourier support ($\rho_S$) gives vanishing at the cusp at $0$.} Because $S \ast \infty = 0$, it turns out that the order of vanishing of $P_\beta \circ \Theta_{\calL}$ at 0 is governed by the lowest-order term of $\tau^{dr/2} P_\beta(\rho_S \Theta_\calL(\tau))$ that is nonzero. Expanded as a polynomial in the coordinates of $\Theta_\calL$, this has no monomial indexed by a frequency vector $\tv \in S_d$, by the orthogonality of $\beta$ given by equation~\eqref{eq:betaoverview}. The $t$th coordinate of $\Theta_\calL$ is a sum of Gaussians $e^{\pi i \langle \yv, \yv \rangle \tau}$ over a coset of $\calL$, so it vanishes at the cusp to order equal to the cost $w_\calL(t)$ of that coset; a monomial indexed by $\tv$ thus vanishes to order at least the total cost $\sum_i w_\calL(t_i)$.
\end{itemize}
This tells us how to choose $S_d$: we work backwards from the contradiction we want. The argument needs $P_\beta \circ \Theta_\calL$ to vanish at the cusp $0$ to order more than $dr(q+1)/24$, the total that the valence formula permits. The orthogonality of the witness removes the contributions of the frequency vectors in $S_d$, and every other frequency vector contributes at order at least its total cost. So the high vanishing order at $0$ is guaranteed as soon as $S_d$ contains every frequency vector of total cost at most $r(q+1)d/24$, and since we want $S_d$ to be small, we take it to consist of exactly these vectors (together with their multiples by nonzero residues). With this choice, $P_\beta \circ \Theta_\calL$ vanishes at $0$ to higher order than the valence formula allows, so no witness exists, and a cone-delta function with Fourier support in $S_d$ does. This is the support in~\Cref{thm:informalwarmup}: its budget $r(q+1)d/24$ is the total order of vanishing that the valence formula permits, and the lattice enters only through its rank $r$ (the weight of $\Theta_\calL$ is $r/2$) and the costs of its cosets. We give the details in~\Cref{sec:gen1overview}.

\paragraph{The stronger construction.} The stronger construction of~\Cref{sec:gen4} (``Track 3'' in~\Cref{table:server-counts,tab:guide}) follows the same high-level architecture, with a number of changes:
\begin{itemize}
    \item Firstly, instead of constructing Fourier-sparse cone-delta functions, we construct the more general quadratic-phase delta expansions alluded to at the end of~\Cref{sec:depir:qpir}. 
As a result, the dual witness is no longer supported on the cone, so $P_\beta \circ \Theta_\calL$ is no longer modular for $\Gamma_0(q)$. We use instead the smaller subgroup $\Gamma(q)$ of matrices congruent to the identity modulo $q$, on which the Weil representation is trivial. It has $\frac{q^2-1}{2}$ classes of cusps rather than two; we will use the analogous condition to equation~\eqref{eq:betaoverview} to establish high-order vanishing at $s(q-1)/2$ of these cusp classes (rather than just at the cusp at 0). 
    \item Next, we give the theta function an additional \emph{elliptic} variable, making it a vector of theta functions with characteristics~\cite[Chapter~I, \S3]{Mumford83} and, after composing with the witness polynomial, a Jacobi form~\cite{EichlerZagier85}. We extract two scalar modular forms from it rather than one: one from its Taylor expansion in the elliptic variable around $0$~\cite[Section~3]{EichlerZagier85}, and one from its values at the half-periods $\frac{1}{2}$, $\frac{\tau}{2}$ and $\frac{1+\tau}{2}$~\cite[Theorem~1.3]{EichlerZagier85}.
    \item The Fourier support set $S_d$ is chosen to consist of two components that respectively ensure high-order vanishing of these two scalar modular forms.

    To obtain the final contradiction, we again apply the valence formula to both of these, following prior analyses by~\cite{RichterWesterholtRaum15,Aoki22,LiWang25}.
\end{itemize}
We give a more detailed overview of this proof in~\Cref{sec:gen4overview}.

\subsection{Open Questions}
\label{sec:open}

We conclude the technical overview with open questions that call for further investigation.

\paragraph{Breaking the $\sqrt{n}$-server-work barrier?} Henzinger and Ragavan~\cite{HR26} explicitly ask whether there exists doubly-efficient PIR with $O(1)$ servers, $n^{1+o(1)}$ storage, and server work per query $n^{1/2-\Omega(1)}$. 
None of our new protocols meets this bar. Worse yet,~\Cref{thm:deltalb} suggests this may be inherent to our techniques: no cone-delta function can have low-enough Fourier sparsity to achieve such a result via our compiler in~\Cref{sec:depir:spir}. 
This state of affairs is particularly puzzling because of a complementary result: server work $n^{o(1)}$ per query is possible if the number of servers is permitted to grow with the database size~$n$, no matter how slowly~\cite{DBLP:journals/iacr/GhoshalLMDS24}.

More generally, for any constant server count $s \geq 2$, it remains open what the  smallest achievable exponent  $\alpha_s$ is that gives PIR with storage $n^{1+o(1)}$ and server work $n^{\alpha_s+o(1)}$ per query. 
One clean approach to this question would be to find better lattices to plug into Track 2.~\Cref{mainthm:track2} turns every even lattice $\calL$ of determinant $\sqrt{q}$ into a $q$-server PIR scheme with exponent $\eta_\calL$, so improving Track~2 reduces to a question about lattices alone: which even lattices of determinant $\sqrt{q}$ minimize $\eta_\calL$? (By~\Cref{thm:deltalb}, no lattice gives an exponent below $\frac{1}{2}$, but it is open whether some lattice matches or beats the stronger construction of~\Cref{sec:gen4}.)

\paragraph{Smooth storage-work tradeoffs?} 
Prior doubly-efficient PIR constructions based on Reed--Muller and multiplicity codes~\cite{Beimel2004,DBLP:journals/iacr/GhoshalLMDS24,HR26} have the appealing feature of supporting smooth tradeoffs between storage and server work per query. We do not know whether the new protocols in this work can be extended to such a tradeoff.

\paragraph{Other applications of modular forms?} 
Finally, throughout our results, modular forms play a role typically fulfilled by low-degree polynomials (e.g., in PIR).  Given the ubiquity of low-degree polynomials in theoretical computer science, as described next in \cref{sec:rel}, we are hopeful that the tools from modular forms developed in this work may have uses elsewhere in  coding theory and cryptography.

%% file: table-server-counts.tex
\begin{table}[ht]
\centering
\caption{Performance of information-theoretic doubly-efficient PIR with a constant number of servers $s \geq 2$, over a database of $n$ bits: prior state of the art~\cite{HR26} and our schemes. Each entry $n^\rho$ indicates that the PIR servers run in time $o(n^{\rho})$ per query, and that the PIR communication is $o(n^{\rho})$ bits; exponents are rounded up to four decimals. The last column reports the smallest number of servers needed to match the PIR exponent from this work using the prior state-of-the-art scheme~\cite{HR26} (restricted to the quasilinear-storage setting, given in their Corollary D.11); its last entry is approximately $s^{3.93}$. The last row gives the number of bits of per-server storage required by each PIR scheme, which stays constant as the server count $s$ grows. (One distinction is that, in our schemes, each server stores a different data structure, rather than an identical copy of the database.) For each $s$, the best server time/communication/storage is highlighted in green.
 }
\label{table:server-counts}
\smallskip
\begingroup
\renewcommand{\arraystretch}{1.15}
\begin{tabular}{cccccc}
 & \textbf{Prior work} & \multicolumn{3}{c}{\bf This work} & \multirow{3}{*}{\shortstack{Num. servers\\\cite{HR26} needs to\\match this work}} \\
\cmidrule(lr){2-2}\cmidrule(lr){3-5}
$s$ & \cite[D.11]{HR26} & Track~1 (Thm \ref{mainthm:track1})& Track~2 (Thm \ref{mainthm:track2}) & Track~3 (Thm \ref{mainthm:track3})& \\
\midrule
\multicolumn{4}{@{}l}{\em Server time and communication}\\[+3pt]
$2$ & \gcell $n^{0.8113}$ & \gcell $n^{0.8113}$ & --- & $n^{0.8204}$ & $2$ \\
$3$ & $n^{0.7897}$ & $n^{0.7445}$ & $n^{0.7897}$ & \gcell $n^{0.7327}$ & $11$ \\
$4$ & $n^{0.7897}$ & $n^{0.7445}$ & $n^{0.7897}$ & \gcell $n^{0.7025}$ & $19$ \\
$5$ & $n^{0.7628}$ & $n^{0.7445}$ & $n^{0.7897}$ & \gcell $n^{0.6535}$ & $73$ \\
$6$ & $n^{0.7628}$ & $n^{0.7445}$ & $n^{0.7897}$ & \gcell $n^{0.6521}$ & $79$ \\
$7$ & $n^{0.7456}$ & $n^{0.7445}$ & $n^{0.7095}$ & \gcell $n^{0.6198}$ & $307$ \\
$8$ & $n^{0.7456}$ & $n^{0.7445}$ & $n^{0.7095}$ & \gcell $n^{0.6198}$ & $307$ \\
$9$ & $n^{0.7456}$ & $n^{0.7445}$ & $n^{0.7095}$ & \gcell $n^{0.6198}$ & $307$ \\
$10$ & $n^{0.7456}$ & $n^{0.7445}$ & $n^{0.7095}$ & \gcell $n^{0.6078}$ & $587$ \\
$11$ & $n^{0.7239}$ & $n^{0.7445}$ & $n^{0.6609}$ & \gcell $n^{0.5901}$ & $2161$ \\
\midrule
$s \to \infty$ & $n^{\frac{1}{2} + \frac{1 + o(1)}{\log_2 s}}$ & $n^{0.7445}$ & $n^{\frac{1}{2} + \frac{0.467 + o(1)}{\log_2 s}}$ & \gcell $n^{\frac{1}{2} + \frac{0.255 + o(1)}{\log_2 s}}$ & $s^{2/\log_2(\pi e/6) + o(1)}$ \\
\midrule
\midrule
\multicolumn{4}{@{}l}{\em Server storage}\\[+3pt]
any & $n \cdot \polylog(n)$ & \gcell $n + 1$ & \gcell $n + 1$ & \gcell $n + 1$ & $n^{1+o(1)}$\\
\bottomrule
\end{tabular}
\endgroup
\end{table}

%% file: tabguide.tex
\begin{table}[t]
    \centering
    \small
    \renewcommand{\arraystretch}{1.25}
    \begin{tabular}{>{\raggedright\arraybackslash}p{0.12\textwidth} >{\raggedright\arraybackslash}p{0.255\textwidth} >{\raggedright\arraybackslash}p{0.255\textwidth} >{\raggedright\arraybackslash}p{0.255\textwidth}}
        \toprule
        & Track~1: Fourier analysis & Track~2: Theta functions & Track~3: Jacobi forms \\
        \midrule
        {Main result} & \Cref{mainthm:track1} & \Cref{mainthm:track2} & \Cref{mainthm:track3} \\ \addlinespace
        \multicolumn{4}{l}{\emph{Server time with $s$ servers}} \\
        \quad $s = 2$ & \gcell $n^{0.812}$ & --- & $n^{0.821}$ \\
        \quad $s = 3$ & $n^{0.745}$ & $n^{0.790}$ & \gcell $n^{0.733}$ \\
        \quad $s = 7$ & $n^{0.745}$ & $n^{0.710}$ & \gcell $n^{0.620}$ \\
        \quad $s \to \infty$ & $n^{0.745}$ & $n^{1/2 + (0.467 + o(1))/\log_2 s}$ & \gcell $n^{1/2 + (0.255 + o(1))/\log_2 s}$ \\ \addlinespace
        Mathematics & Fourier analysis over $\F_2$ and $\F_3$; polynomial interpolation & Modular forms for $\Gamma_0(q)$; the Weil representation; theta functions of lattices; the valence formula & All of Track~2, plus: Jacobi forms; the cusps of $\Gamma(q)$; Heisenberg operators; scalarization by Taylor coefficients and by values at 2-torsion points \\ \addlinespace
        Path with overviews & \S\S\ref{sec:overview:inspiration}--\ref{sec:depir:qpir} & \S\ref{sec:overview:modforms} and \S\ref{sec:gen1overview} & \S\ref{sec:gen4overview} \\ \addlinespace
        Path with proofs & \S\S\ref{sec:bg}--\ref{sec:elementary} and \S\ref{sec:fieldtobits} & \S\ref{sec:primefield}, \S\ref{sec:pt2bg} (except Def~\ref{def:heisenberg}, Lem~\ref{lem:heisenbergweil}, and Cor~\ref{lem:valencegammaq}), \S\ref{sec:gen1}; Appendices~\ref{sec:pt2appendix}--\ref{sec:gen1proofs} & the parts of \S\ref{sec:pt2bg} skipped in Track~2, \S\ref{sec:gen4}, and Appendix~\ref{sec:gen4proofs} \\
        \bottomrule
    \end{tabular}
    \caption{The three tracks of our results on constructing doubly-efficient PIR over $n$-bit databases. In every scheme, each server stores a server-dependent data structure consisting of $n+1$ bits. The server-time rows give the exponents of each track's own schemes: an entry $n^\rho$ means that the server time and the communication are $o(n^\rho)$. The best entry for each $s$ is highlighted in green. The ``reading path'' rows are cumulative: each column lists only what its track adds to the tracks preceding it.}
    \label{tab:guide}
\end{table}

%% file: figlattice.tex
\begin{figure}[ht]
    \centering
    \pgfmathsetmacro{\tx}{1.3}
    \pgfmathsetmacro{\ty}{0.6}
    \pgfmathsetmacro{\ga}{1}
    \pgfmathsetmacro{\gb}{-2}
    \pgfmathsetmacro{\gc}{1}
    \pgfmathsetmacro{\gd}{-1}
    \pgfmathsetmacro{\ax}{\gc*\tx + \gd}
    \pgfmathsetmacro{\ay}{\gc*\ty}
    \pgfmathsetmacro{\bx}{\ga*\tx + \gb}
    \pgfmathsetmacro{\by}{\ga*\ty}
    \pgfmathsetmacro{\nrm}{\ax*\ax + \ay*\ay}
    \pgfmathsetmacro{\sx}{(\bx*\ax + \by*\ay)/\nrm}
    \pgfmathsetmacro{\sy}{(\by*\ax - \bx*\ay)/\nrm}
    \begin{tikzpicture}[scale=2.1, >=stealth]
        \fill[blue!10] (0,0) -- (1,0) -- (1+\tx,\ty) -- (\tx,\ty) -- cycle;
        \fill[red!12] (0,0) -- (\ax,\ay) -- (\ax+\bx,\ay+\by) -- (\bx,\by) -- cycle;
        \draw[blue!70!black] (1,0) -- (1+\tx,\ty) -- (\tx,\ty);
        \draw[red!70!black, dashed] (\ax,\ay) -- (\ax+\bx,\ay+\by) -- (\bx,\by);
        \foreach \m in {-4,...,4} {
            \foreach \n in {-2,...,3} {
                \pgfmathsetmacro{\px}{\m+\n*\tx}
                \pgfmathsetmacro{\py}{\n*\ty}
                \pgfmathparse{(\px > -1.3) && (\px < 2.5) && (\py > -0.75) && (\py < 1.5) ? 1 : 0}
                \ifdim\pgfmathresult pt>0.5pt \fill (\px,\py) circle (0.7pt); \fi
            }
        }
        \draw[violet!80!black, densely dotted] (1,0) -- (1+\sx,\sy) -- (\sx,\sy);
        \draw[->, thick, violet!80!black] (0,0) -- (\sx,\sy) node[above] {$g \ast \tau$};
        \draw[->, thick, blue!70!black] (0,0) -- (1,0) node[below] {$1$};
        \draw[->, thick, blue!70!black] (0,0) -- (\tx,\ty) node[above right=-1pt] {$\tau$};
        \draw[->, thick, red!70!black] (0,0) -- (\ax,\ay) node[above right=-1pt] {$c\tau + d_0$};
        \draw[->, thick, red!70!black] (0,0) -- (\bx,\by) node[left] {$a\tau + b$};
        \draw[->, very thick, black!70, dash pattern=on 4pt off 2pt, shorten <=3pt, shorten >=3pt] (\ax,\ay) to[bend left=35] (1,0);
        \draw[->, very thick, black!70, dash pattern=on 4pt off 2pt, shorten <=3pt, shorten >=3pt] (\bx,\by) to[bend left=35] node[pos=0.3, left=3pt, black] {$\cdot\, (c\tau + d_0)^{-1}$} (\sx,\sy);
        \node[below] at (0,0) {$0$};
    \end{tikzpicture}
    \caption{The action of $\SL_2(\ZZ)$ through lattices. The dots form the lattice $\ZZ + \ZZ\tau$, with basis $(1, \tau)$ (blue). The matrix $g$ gives another basis $(c\tau+d_0, a\tau+b)$ of the same lattice (red). Rotating and scaling the plane by $(c\tau+d_0)^{-1}$ (dashed arrows) turns this basis into the basis $(1, g \ast \tau)$ of a similar lattice (purple), where $g \ast \tau = \frac{a\tau+b}{c\tau+d_0}$.}
    \label{fig:lattice}
\end{figure}
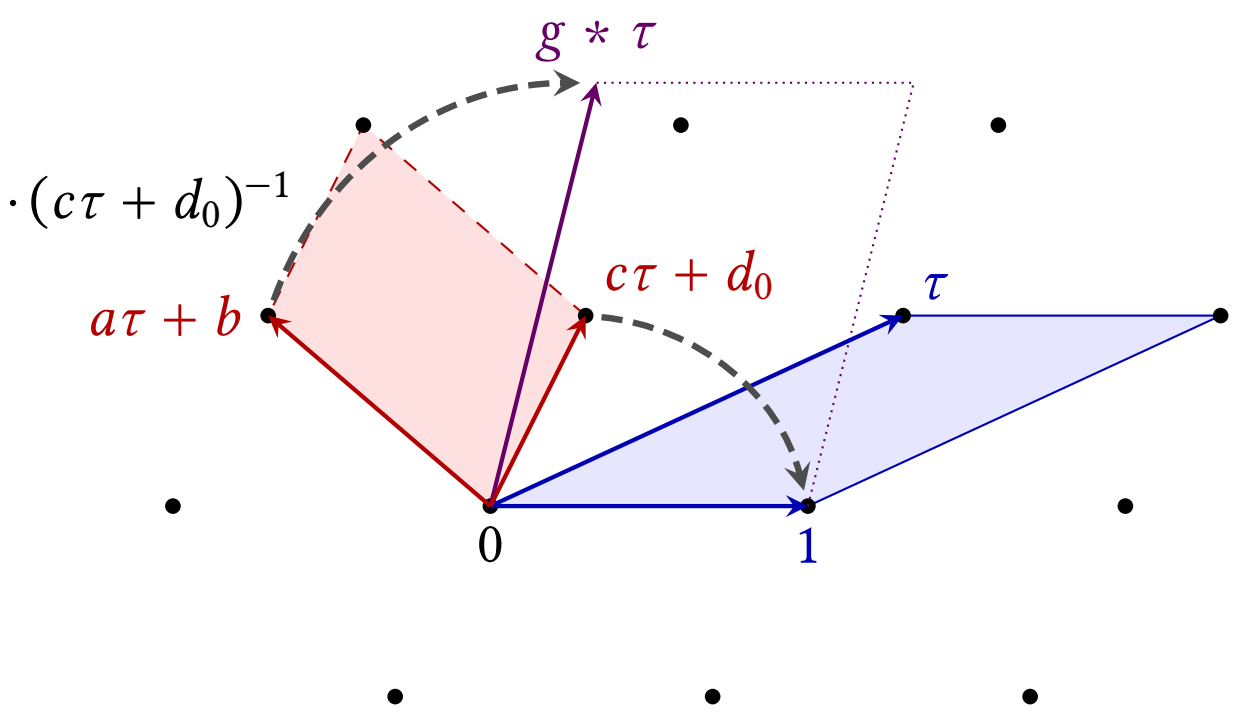

%% file: figcusps.tex
\begin{figure}[ht]
    \centering
    \begin{tikzpicture}[x=3.2cm, y=3.2cm, >=stealth]
        \fill[gray!8] (-1.55,0) rectangle (1.55,1.3);
        \draw[thick] (-1.55,0) -- (1.58,0) node[right] {$\RR$};
        \node at (-1.2,1.1) {$\calH$};
        \coordinate (inf) at (0,1.3);
        \filldraw[fill=white, draw=black, thick] (inf) circle (2pt);
        \node[above] at (inf) {$\infty$};
        \foreach \x/\lab/\g/\bend/\pos/\side in {
            -1/{-1}/{T^{-1}S}/20/0.5/left,
            -0.4/{-\frac{2}{5}}/{\begin{bsmallmatrix} -2 & -1 \\ 5 & 2 \end{bsmallmatrix}}/8/0.72/left,
            0/0/{S}/0/0.55/right,
            1/1/{TS}/-20/0.5/right} {
            \filldraw[fill=white, draw=black, thick] (\x,0) circle (2pt);
            \node[below=2pt] at (\x,0) {$\lab$};
            \draw[->, thick, black!70, shorten <=3pt, shorten >=3pt] (inf) to[bend right=\bend] node[pos=\pos, \side, black] {$\g$} (\x,0);
        }
    \end{tikzpicture}
    \caption{The upper half-plane $\calH$, with the cusp $\infty$ (approached as $\Im \tau \to \infty$) and some of the cusps in $\QQ$. Every cusp is the image $g \ast \infty = \frac{a}{c}$ of $\infty$ under some $g = \begin{bsmallmatrix} a & b \\ c & d_0 \end{bsmallmatrix} \in \SL_2(\ZZ)$; the arrows show one such $g$ for each cusp drawn, where $TS = \begin{bsmallmatrix} 1 & -1 \\ 1 & 0 \end{bsmallmatrix}$ and $T^{-1}S = \begin{bsmallmatrix} -1 & -1 \\ 1 & 0 \end{bsmallmatrix}$.}
    \label{fig:cusps}
\end{figure}
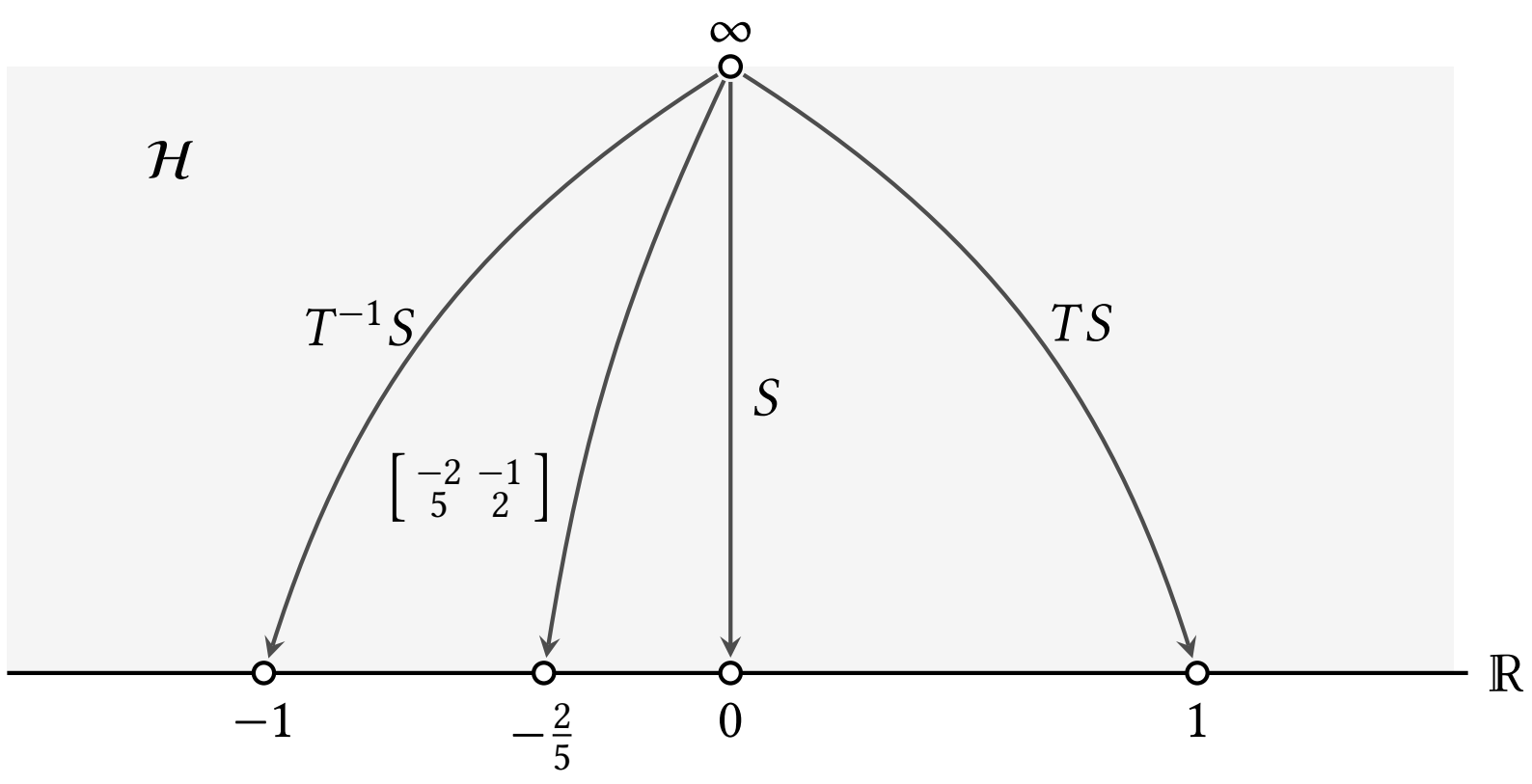

%% file: figfourieranalogy.tex
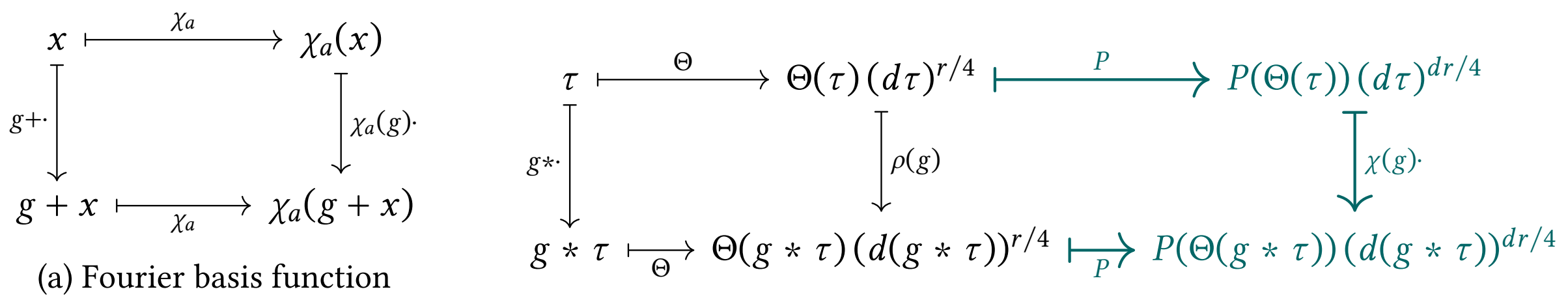
\begin{figure}[t]
    \centering
    \begin{minipage}[b]{0.32\textwidth}
        \centering
        \begin{tikzcd}[column sep=large, row sep=large]
            x \arrow[r, mapsto, "\chi_a"] \arrow[d, mapsto, "g + \cdot"'] & \chi_a(x) \arrow[d, mapsto, "\chi_a(g) \cdot"] \\
            g + x \arrow[r, mapsto, "\chi_a"'] & \chi_a(g + x)
        \end{tikzcd}

        \vspace{0.5em}
        {\small (a) Fourier basis function $\chi_a(x) = \zeta_q^{ax}$;\\ $g \in \ZZ_q$ acts on $\ZZ_q$ by translation}
    \end{minipage}%
    \hfill
    \begin{minipage}[b]{0.66\textwidth}
        \centering
        \begin{tikzcd}[column sep=1.8em, row sep=large]
            \tau \arrow[r, mapsto, "\Theta"] \arrow[d, mapsto, "g \ast \cdot"'] & \Theta(\tau)\,(d\tau)^{r/4} \arrow[r, mapsto, "P", teal!80!black, thick] \arrow[d, mapsto, "\rho(g)"] & |[text=teal!80!black]| P(\Theta(\tau))\,(d\tau)^{dr/4} \arrow[d, mapsto, "\chi(g) \cdot", teal!80!black, thick] \\
            g \ast \tau \arrow[r, mapsto, "\Theta"'] & \Theta(g \ast \tau)\,(d(g \ast \tau))^{r/4} \arrow[r, mapsto, "P"', teal!80!black, thick] & |[text=teal!80!black]| P(\Theta(g \ast \tau))\,(d(g \ast \tau))^{dr/4}
        \end{tikzcd}

        \vspace{0.5em}
        {\small (b) Vector-valued modular form $\Theta$ of weight $\frac{r}{2}$, scalarized by $P$;\\ $g \in \Gamma \subseteq \SL_2(\ZZ)$ acts on $\calH$ by M\"obius maps}
    \end{minipage}
    \caption{Fourier basis functions and modular forms both intertwine the action of a group on their domain with a linear action on their values. The scalarization step (teal) has no counterpart in (a), since every representation of $\ZZ_q$ splits into characters, while the Weil representation of $\SL_2(\ZZ)$ does not.}
    \label{fig:fourieranalogy}
\end{figure}

%% file: rel.tex
\section{Related Work}
\label{sec:rel}

\input{table}

\paragraph{Doubly-efficient PIR.} Doubly-efficient private information retrieval has been studied in two settings: information-theoretic schemes, which require several non-colluding servers (as in this work), and single-server schemes, which require cryptographic assumptions. In either case, despite decades of study, only a handful of constructions are known, each with severe efficiency limitations shown in \cref{table:rel}.
All prior information-theoretic doubly-efficient PIR schemes~\cite{Beimel2004,lazzaretti2024multi,DBLP:journals/iacr/GhoshalLMDS24,HR26} store an $\omega(n)$-bit data structure on each server. 
All  single-server  doubly-efficient PIR schemes~\cite{BIPW17,LMW23,okada2024towards,okada2025algebraic} with polynomially-sublinear server time per query store a data structure of size  $\Omega(\poly(\lambda) \cdot \polylog(n)\cdot n )$ bits~\cite{chen2025secret}, which is concretely enormous. In contrast, our construction stores $n+1$ bits on each server---with no hidden constants.

In response to the elusiveness of doubly-efficient PIR, a weaker notion emerged in the literature: {\em secret-key} doubly-efficient PIR, which lets a client  (using its own secret key) construct the database encoding that is later on stored and served by the server. The first candidates~\cite{BIPW17,CHR17} relied on new permuted-code assumptions~\cite{BHW19,BHMW21,BW21}. Building on this framework, recent work by Chen, Ishai, Jain, Mour, Rosen, and Xing~\cite{DBLP:journals/iacr/ChenIJMRX26} gives candidate doubly-efficient secret-key PIR schemes with $O(n)$ storage and $n^{o(1)}$ communication and online server work.
The server's storage can approach the database size, in suitable parameter regimes. Unlike in these schemes, our work does {\em not} require the PIR data structure to depend on the client's secret key; the servers can preprocess the database  once and for all and then use the resulting encoding to answer queries from {\em all} clients.

\paragraph{Decoding polynomials and Fourier-sparse delta functions.} In the matching-vector PIR of Efremenko~\cite{Efr09}, the number of servers is equal to the number of monomials of an $S$-decoding polynomial.
Two  lines of work reduce this number of servers by (A)~finding sparser $S$-decoding polynomials~\cite{Efr09,Chee2011,GR26}, and (B)~designing more server-efficient ways of using them in PIR~\cite{DG16,GKS25,ABL25}. Ghasemi and Kopparty~\cite{DBLP:conf/innovations/GhasemiK26} recast the former task as a question about delta functions: how sparse in the Fourier basis can a function
be, if it equals $1$ at the origin and vanishes on the rest of the boolean cube $\{0,1\}^r$?  Their lower bounds show that any such function must have Fourier sparsity $\geq r+1$, as later shown to be tight (under a number-theoretic conjecture) for carefully chosen composite moduli~\cite{GR26}.
Our lower bound on the sparsity of cone-delta functions (\Cref{thm:deltalb}) adapts the multiplication argument of~\cite[Lemma~3.7]{DBLP:conf/innovations/GhasemiK26}.

\paragraph{Applications of modular forms.} Modular forms have found a number of uses throughout theoretical computer science and combinatorics.

\itpara{Explicit constructions of Ramanujan expander graphs~\cite{LPS88,Margulis88}.}
These explicit constructions are Cayley graphs on $\SL_2(\FF_p)$ for prime $p$. Their expansion is proven through bounds on the Fourier coefficients of certain modular forms, whereas our proofs use bounds on their orders of vanishing.

\itpara{Distance bounds for special binary codes.} The weight enumerator of a doubly-even self-dual binary code is invariant under a finite group, which confines it to a small ring of polynomials, and substituting theta functions into such invariant polynomials gives modular forms~\cite{BroueEnguehard72,NebeRainsSloane06}. Mallows and Sloane~\cite{MallowsSloane73} used this rigidity to upper-bound the minimum distance of such codes: beyond a point determined by the length, no such enumerator can have all of its low-weight coefficients vanish, which is the analog of a modular form vanishing to too high an order. Our proofs also substitute theta functions into an invariant polynomial and show that it cannot vanish as much as required, but the polynomial comes from a dual witness rather than from a code.

\itpara{Sphere packing.} 
The problem of nailing down the densest possible sphere packing in 8-dimensional space was resolved by
Viazovska~\cite{Viazovska17} using modular forms; soon after, Cohn, Kumar, Miller, Radchenko, and Viazovska~\cite{CKMRV17} did the same for dimension $24$. 
Very roughly, upper bounding the density can be boiled down to finding a solution to a dual linear program~\cite{CohnElkies03,Delsarte73}; the aforementioned works used modular forms to construct optimal dual witnesses. We note that this is in contrast to our results, where modular forms are used to rule out --- rather than construct --- a dual witness to a particular linear system.

\paragraph{Discrete Gaussians in cryptography.} 
The lattice theta function $\Theta_{\calL}(\tau)$ we use in~\Cref{sec:gen1} is closely related to discrete Gaussian masses of lattice cosets; indeed, they coincide when $\tau \in \calH$ is purely imaginary. 
Discrete Gaussian masses are central to lattice-based cryptography; the use of the Poisson summation formula to prove a theta function's transformation under $S$ is analogous to its use for proving the lattice smoothing lemma in cryptography~\cite{Ban93,MR07,R09}. 
The smoothing lemma can be interpreted in our language as a statement about the rate of convergence of $\Theta_{\calL}(\tau)$ (after appropriate normalization) to the all-ones vector as $\tau$ approaches the cusp at 0.

\paragraph{Vanishing bounds for Jacobi forms.}
Our analysis in~\Cref{sec:gen4} connects closely to work bounding how much a nonzero Jacobi form can vanish at a cusp~\cite{Aoki22,LiWang25}.
Useful tools include extracting modular forms from expansions in the elliptic variable and from evaluations at torsion points~\cite{EichlerZagier85,RichterWesterholtRaum15,Aoki22,LiWang25}, then applying modular-form vanishing bounds.
We use this strategy with a Taylor expansion at the origin and evaluations at the three nonzero half-periods.
The resulting inequalities rule out Jacobi forms arising from hypothetical dual witnesses, establishing the existence of our quadratic-phase delta expansions.

\paragraph{Elliptic curves.} 
Modular forms are intimately related to elliptic curves, which are solution sets over fields to equations of the form $y^2 = x^3+ax^2+bx+c$. 
One intuition for this connection is that sublattices of $\CC$ (which we have already connected to modular forms) correspond in an extremely tight way to elliptic curves over $\CC$:
\begin{itemize}[nosep,leftmargin=*]
    \item Each $\tau \in \calH$ is associated to an elliptic curve $E_\tau$ over $\CC$.
    \item $\tau, \tau'$ are related by some element of $\SL_2(\ZZ)$ (under the M{\"o}bius action) if and only if $E_{\tau}$ and $E_{\tau'}$ are isomorphic.
    \item The points on $E_\tau$ form an abelian group whose structure is identical to that of the torus $\CC/(\ZZ+\ZZ\tau)$ (see~\Cref{sec:gen4overview,fig:gen4twotorsion}).
\end{itemize}
The arithmetic relationship between elliptic curves and modular forms is central to the proof of Fermat's Last Theorem~\cite{Frey86,Serre87,Ribet90,Wiles95,TaylorWiles95}.
Elliptic-curve groups are also ubiquitous in cryptography~\cite{miller1985use,koblitz1987elliptic,menezes1993elliptic,koblitz2000state}.

%% file: table.tex
\begin{table}[t]
\centering
\caption{PIR schemes that use preprocessing to reduce server computation, on a database size $n \in \N$ and security parameter $\lambda \in \N$. We hide doubly-logarithmic factors in $n$, polylogarithmic factors in $\lambda$, and lower-order terms; we use $c \in \N$ to denote an arbitrary constant.  Communication costs (in column~3) exclude setup and one-time communication. In all schemes, the communication costs can be made $\poly(\lambda) \cdot \log (n)$, without changing the asymptotic computation or storage costs, by additionally assuming LWE (or RLWE). Concrete storage overheads (in column 6) are measured or estimated on hundreds-of-megabyte to gigabyte-sized databases. $^*$VBB denotes virtual black-box obfuscation, an idealized form of program obfuscation for which no concretely-efficient candidate constructions exist to date.}
\label{table:rel}
\smallskip
\resizebox{\linewidth}{!}{%
\begin{tabular}{@{}lccccc}
& & & \multicolumn{3}{c}{Server costs} \\
\cmidrule(lr){4-6}
\squishheader{\parbox{9em}{Scheme\\[-2pt](crypto assumptions)}} &
  \squishheader{\parbox{2.05em}{Num.\\[-2pt]servers}} &
  \squishheader{\parbox{6em}{Communication\\[-2pt](asymptotic)}} &
  \squishheader{\parbox{5.5em}{Computation\\[-2pt](asymptotic)}} &
  \squishheader{\parbox{4.15em}{Storage\\[-2pt](asymptotic)}} &
  \squishheader{\parbox{8em}{Storage\\[-2pt] (concrete overhead)}} \\
\midrule
\multicolumn{6}{@{}l}{\textbf{{Linear-time} PIR}}\\
\enspace \citen{SimplePIR,frodopir} (LWE) &   $1$ & $\log(n) \cdot \sqrt{n}$ & \rcell $n$ & \gcell $\lambda \cdot \sqrt{n}$ &  \gcell $\approx1\times$ \\
\addlinespace[4pt]
\multicolumn{6}{@{}l}{\textbf{PIR with {polylogarithmic} time savings}}\\
\enspace \citen{DBLP:conf/crypto/Lee26} (LWE) &  $1$ & $\poly(\lambda) \cdot \sqrt{n}$ & \ycell $n/\log^2 n$ & \ycell $n^{1+o(1)}$ & \graycell ? \\
\enspace \citen{lee26b} (QR) &  $1$ & $\lambda \cdot \sqrt{n}$ & \ycell $n/\log^2 n$ & \ycell $\poly(\lambda) \cdot n^{1+1/c}$ & \graycell ? \\
\enspace \citen{WY05} &  2 & $n^{1/3}$ & \ycell $n/\log^c n$ & \rcell $\poly(n)$ & \graycell ? \\
\addlinespace[4pt]
\multicolumn{6}{@{}l}{\textbf{Doubly-efficient PIR with {\color{red!70!black}{super-constant}} servers}}\\
\enspace \citen{lazzaretti2024multi} &  $\omega(\log n)$ & $n^{o(1)}$ & \gcell $n^{o(1)}$ & \ycell $n^{1+o(1)}$ & \graycell ?  \\
\enspace \citen{DBLP:journals/iacr/GhoshalLMDS24} &  $\omega(1) $ & $n^{o(1)}$ & \gcell $n^{o(1)}$ & \ycell $n^{1+o(1)}$ & \graycell ? \\
\addlinespace[4pt]
\multicolumn{6}{@{}l}{\textbf{Doubly-efficient PIR with constant servers}}\\
\enspace \citen{BIPW17,CHR17} (\textbf{\color{red!70!black}{VBB}}$^*$)&  $1$ & $\poly(\lambda) \cdot n^{o(1)}$ & \gcell $\poly(\lambda) \cdot n^{o(1)}$ & \ycell $\poly(\lambda) \cdot n^{1+o(1)}$ & \graycell ?\\
\enspace \citen{LMW23,okada2024towards}/\hyperlink{cite.okada2025algebraic}{25} (RL\rlap{WE)} &  $1$ & $\poly(\lambda) \cdot n^{o(1)}$ & \gcell $\poly(\lambda) \cdot n^{o(1)}$ & \ycell $\poly(\lambda) \cdot n^{1+o(1)}$& \rcell $\approx 10^8-10^9\times$\\
\enspace \citen{DBLP:conf/crypto/BeimelIM00,DBLP:journals/iacr/GhoshalLMDS24} &  $2$ & $n^{0.82}$ & \gcell $n^{0.82}$ & \rcell $n^{1.48}$ & \rcell $\approx 10^5-10^8\times$\\
\enspace \citen{HR26} &  $2$ & $n^{0.82}$ & \gcell $n^{0.82}$ & \ycell $\sqrt{\log n} \cdot n$ & \ycell $\approx10-500\times$ \\

 \enspace {\em This work} &  $2$ & $n^{0.82}$ & \gcell $n^{0.82}$ & \gcell $n+1$ & \gcell $\approx1\times$ \\

\midrule
\addlinespace[2pt]
\enspace \citen{HR26} &  $c $ & $n^{1/2 + 1/\log_2 c + o(1)}$ & \gcell $n^{1/2 + 1/\log_2 c + o(1)}$ & \ycell $\polylog(n) \cdot n$ & \graycell ? \\
%
\enspace {\em This work} &  $c $ & $n^{1/2 + 0.255/\log_2 c + o(1)}$ & \gcell $n^{1/2 + 0.255/\log_2 c + o(1)}$ & \gcell $n+1$ & \gcell $\approx 1\times$ \\
\bottomrule
\bigskip
\end{tabular}%
\medskip
}
\end{table}

%% file: prelims.tex
\section{Preliminaries}
\label{sec:bg}

\subsection{Private Information Retrieval}
\label{sec:bg:pir}

\paragraph{Definition.} Following prior work~\cite{Beimel2004,HR26}, for any $s \geq 2$ and ring $\calR$, we define an {\em $s$-server PIR} protocol over a database in $\calR^n$  to be a tuple of four routines (which each may take a security parameter as an implicit input). Crucially, our definition allows each server to store a distinct preprocessed data structure.

\medskip

\begin{hangparas}{3ex}{1}
$\Preprocess(j, \db) \to \T_j$.
Given a server index $j \in [s]$ and a database $\db \in \calR^n$, output a preprocessed data structure $\T_j$ to be stored on the \ord{j} server.

\smallskip

$\Query(i) \to (\state, \qu_1, \dots, \qu_s)$.
Given an index $i \in [n]$, 
output a secret client state  and $s$ PIR  queries  to be sent to each of the servers.

\smallskip

$\Answer^{\T}(\qu) \to \ans$.
Given a client query $\qu$, along with oracle access to the   preprocessed database structure $\T$, output a PIR answer $\ans$.

\smallskip

$\Recon(\state, \ans_1, \dots, \ans_s) \to r$. 
Given a secret client state  and 
$s$ PIR answers, output the recovered database record $r \in \calR$.
\end{hangparas}

\bigskip

We require a PIR scheme to meet two properties, correctness and security, as we define next.

\itpara{Correctness.} At the end of the protocol, the user successfully learns the record of the database that it wants to read: that is, on database size $n \in \N$, for every database $\db \in \calR^n$ and every index $i \in [n]$,  
    \begin{align*}
        \Pr \left[ r = \db_i \prmid \begin{array}{r@{\ }l}
        \text{for }j \in [s], \, \T_j &\gets \Preprocess(j, \db)\\
        \state, \qu_1, \dots, \qu_s &\gets \Query(i)\\
        \text{for }j \in [s], \, \ans_j &\gets \Answer^{\T_j}(\qu_j)\\
        r &\gets  \Reconstruct(\state, \ans_1, \dots, \ans_s)
        \end{array} 
    \right] = 1.
    \end{align*}

\itpara{Security.} At the end of the protocol, each server learns {\em nothing}%
about what record the user read. Specifically, we define an $s$-server PIR protocol to have \emph{information-theoretic privacy} if, on database size $n \in \N$ and for all pairs of indices $i, j \in [n]$, each server's view is identical whether the user is querying for index $i$ or  $j$. Formally, for each $k \in [s]$, it holds that:
$$\left\{ \qu_k : \state, \qu_1, \dots,  \qu_s \gets \Query(i) \right\}  \equiv \left\{ \qu_k : \state, \qu_1,  \dots, \qu_s \gets \Query(j) \right\}.$$

We define an $s$-server PIR protocol to have \emph{computational privacy} if,  on security parameter $\lambda \in \N$ and on a polynomially-bounded database size $n = N(\lambda) \in \N$, for all pairs of indices $i, j \in [n]$,  the above two distributions are computationally indistinguishable.

\paragraph{Efficiency metrics.} We measure the efficiency of an $s$-server PIR scheme over a database in $\calR^n$ as in prior work~\cite{Beimel2004,LMW23,HR26}:
\begin{itemize}[leftmargin=*]
\item {\em Communication.} The PIR scheme's communication is the worst-case bit length of all $s$ queries and answers exchanged between the user and the $s$ servers, on any database in $\calR^n$.

\medskip

\item {\em Server storage.} The PIR scheme's server storage is the worst-case bit length of any of the $s$ preprocessed data structures $\T_1, \dots, \T_s$, produced on any database in $\calR^n$.

\medskip

\item {\em Server time.} The PIR scheme's server time is the worst-case runtime of $\Answer^{\T_j}(\cdot)$, summed across all $j$ from $1$ to $s$ and measured in terms of the number of word operations (including but not limited to RAM accesses performed) on any database in $\calR^n$.
\end{itemize}

Finally, following Lin, Mook, and Wichs, we define the following family of PIR schemes with double-efficiency properties~\cite{LMW23}: a PIR scheme is {\em doubly-efficient} if each of the $s$ servers stores a polynomially-large data structure and  answers  queries in sublinear time. That is, on database size $n \in \N$, the server storage is $\poly(n)$ bits and the server time is $o(n)$.

\subsection{Fourier-Sparse Cone-Delta Functions}
\label{sec:bg:sfd}

We next define the mathematical abstractions at the core of our new template for doubly-efficient PIR. The following two definitions specialize those of Ghasemi and Kopparty~\cite{DBLP:conf/innovations/GhasemiK26} to the abelian group $\ZZ_q^d$:

\begin{definition}[{Fourier sparsity~\cite[Definition~3.1]{DBLP:conf/innovations/GhasemiK26}}]
    Let $q$ be a prime and $\F$ be a field containing a primitive \ord{q} root of unity $\gamma$.
    For any function $f: \ZZ_q^d\to \F$, there exists a unique function $\hat{f}: \ZZ_q^d\to \F$ such that:
        $f(\xv) = \sum_{\vec w \in \ZZ_q^d} \hat{f}(\vec w) \cdot  \gamma^{\langle \vec w, \xv \rangle}.$
    The function~$\hat{f}$ is the Fourier representation of $f$.
    Then, the {\em Fourier sparsity} of the function~$f$ is the size of its Fourier representation's support, $$\supp \left(\hat{f}\right) = \left\{\vec w \in \ZZ_q^d: \hat{f}(\vec w) \neq 0\right\}.$$
\end{definition}

\begin{definition}[{Delta function~\cite[Definition~2.1]{DBLP:conf/innovations/GhasemiK26}}]
    For any subset $S \subseteq \ZZ_q^d$, a function $\delta: \ZZ_q^d\to \F$ is a \emph{delta function on $S$} if it holds that $$\delta(\xv) = \begin{cases} 1,\text{ if }\xv = 0 \\ 0,\text{ if }\xv \in S \backslash \{0\}.\end{cases}$$
\end{definition}

In these terms, a $k$-sparse cone-delta function, as per~\Cref{def:conedeltaoverview}, is exactly a delta function for the cone (i.e., the set $\left\{\xv \in \ZZ_q^d: \langle \xv, \xv \rangle \equiv 0 \pmod{q} \right\}$) with Fourier sparsity $\leq k$. 
However, our protocols work with the following more general definition that allows us to accommodate arbitrary server counts $s \leq q$ (rather than only prime server counts $q$).

\begin{definition}[Quadratic-phase delta expansion]\label{def:genconedelta}
    A collection of coefficients $\left\{c_{\vec w, t} \in \F: \vec w \in \ZZ_q^d, t \in \ZZ_q\right\}$ is a \emph{quadratic-phase delta expansion} over $\F$, of dimension $d$, and of order $q$, if it holds that
    \begin{equation}
        \sum_{t \in \ZZ_q} \sum_{\vec w \in \ZZ_q^d} c_{\vec w, t} \cdot \gamma^{\langle \vec w, \xv \rangle + t \cdot \langle \xv, \xv \rangle} = \begin{cases}
            1,\text{ if }\xv = 0 \\
            0,\text{ if }\xv \in \ZZ_q^d\backslash \{0\}.
        \end{cases}
    \end{equation}
    We say this expansion has support $\calW \times \calT$ if there exist subsets $\calW \subseteq \ZZ_q^d$ and $\calT \subseteq \ZZ_q$ such that $\supp \left\{c_{\vec w, t}\right\} \subseteq \calW \times \calT$. Finally, we say this expansion is {\em $(k, s)$-sparse} if  $|\calW| \leq k$ and $|\calT| \leq s$.
\end{definition}

We call a set $\calW \subseteq \ZZ_q^d$ \emph{permutation-invariant} if it is invariant under permuting the $d$ coordinates. All the Fourier supports that we construct are permutation-invariant.

\medskip
Looking ahead to \cref{sec:depir},
in all our PIR schemes that build on a $(k, s)$-sparse quadratic-phase delta expansion, the Fourier sparsity $k$ governs the PIR scheme's communication, while the parameter $s$ determines the required number of non-colluding servers. Finally, we note that the notions in Definitions~\ref{def:conedelta} and~\ref{def:genconedelta} are existentially equivalent when $s = q$, for an odd prime $q$:
\begin{lemma}\label{lem:conedelta-qpde}
    For any odd prime $q$, if $f$ is a cone-delta function on $\ZZ_q^d$ with Fourier sparsity $k$, then there exists a $(k, q)$-sparse quadratic-phase delta expansion of order $q$, of dimension $d$, and over  the same field, with support $\supp(\hat{f}) \times \ZZ_q$.
     Conversely, for any $s$, if there exists a $(k, s)$-sparse quadratic-phase delta expansion of order $q$, dimension $d$, and over a field $\F$, then there exists a  cone-delta function with Fourier sparsity $k$ over the same field.
\end{lemma}
\begin{proof}
    First, consider a cone-delta function $f$ with Fourier sparsity $k$.
    Then, consider the function
    $$g(\xv) := \left(\frac{1}{q} \sum_{t \in \ZZ_q} \gamma^{t\langle \xv, \xv \rangle}\right) \cdot f(\xv).$$
    This is a delta function on $\ZZ_q^d$, since:
    \begin{enumerate}
        \item at 0, both terms evaluate to 1, and
        \item at a nonzero point we either have $\langle \xv, \xv \rangle \not\equiv 0 \pmod{q}$ in which case the first term vanishes, or $\langle \xv, \xv \rangle \equiv 0 \pmod{q}$ in which case the second term vanishes by the cone-delta guarantee.
    \end{enumerate} 
    Expanding $f(\xv)$ out in its Fourier basis, we get:
    $$g(\xv) = \sum_{t \in \ZZ_q} \sum_{\vec w \in \ZZ_q^d} {\frac{\hat{f}(\vec w)}{q}}
    \cdot \gamma^{\langle \vec w, \xv \rangle + t\langle \xv, \xv \rangle},$$
    so the coefficients $c_{\vec w, t} := \hat{f}(\vec w)/q$ (for all $t \in \ZZ_q$) form a $(k, q)$-sparse quadratic-phase delta expansion, completing the argument.

    Conversely, consider a $(k, s)$-sparse quadratic-phase delta expansion $\{c_{\vec w, t}\}$, and let $g(\xv)$ denote the left-hand side of its defining identity, so that $g$ is the delta function on $\ZZ_q^d$.
    Then, whenever $\langle \xv, \xv \rangle \equiv 0 \pmod{q}$, we have:
    \begin{align*}
        g(\xv) &= \sum_{\vec w \in \ZZ_q^d} {\left(\sum_{t \in \ZZ_q} c_{\vec w, t}\right)}
        \cdot \gamma^{\langle \vec w, \xv \rangle} =: f(\xv),
    \end{align*}
    and the function $f$ so defined on all of $\ZZ_q^d$ agrees with the delta function $g$ on the cone, so it is a cone-delta function; its Fourier coefficients $\sum_t c_{\vec w, t}$ are supported on $\calW$, so its Fourier sparsity is at most $k$.
\end{proof}

%% file: depir.tex
\part{Doubly-Efficient PIR from Fourier-Sparse Cone-Delta Functions}
\label{sec:depir}

Next, we show that the mathematical objects just defined (Fourier-sparse, cone-delta functions and their generalization to quadratic-phase delta expansions, as in \cref{def:conedelta,def:genconedelta}) give new constructions of doubly-efficient PIR, which are the first to achieve $n+1$ bits of storage on each server.
We state the resulting theorems first (\Cref{sec:mainthms}). Then, we present the doubly-efficient PIR constructions for varying server counts (\cref{sec:depir:2pir,sec:depir:spir}). Finally, we give an elementary construction of the  needed, new mathematical objects (\cref{sec:elementary}), while deferring more complex constructions to \cref{sec:math} of the paper.

\label{sec:depir:rateone}

\section{Full Doubly-Efficient PIR Results}
\label{sec:mainthms}

We state our results on doubly-efficient PIR as three theorems, one for each track in~\Cref{tab:guide}. All three construct the following kind of scheme:

\begin{definition}\label{def:exponent}
    For an integer $s \geq 2$ and a real number $\rho \in [0, 1]$, an \emph{$s$-server doubly-efficient PIR scheme with exponent $\rho$} consists of an $s$-server PIR scheme with information-theoretic security and the following efficiency, on database of $n \in \N$ bits:
    \begin{itemize}[itemsep=0pt,topsep=3pt]
        \item Each server stores a data structure consisting of $n + 1$ bits (and may then discard the original database). The data structure is unique to each server and computable in $n \cdot \polylog(n)$ time, given public parameters that are shared by the servers and the user and can be sampled in expected $\polylog(n)$ time.
        \item To make a PIR query, the user uploads $O(\log^2 n)$ bits to each server.
        \item To answer a PIR query, each server reads and sends back at most $n^{\rho + o(1)}$ bits of its data structure, resulting in  server time and  communication $n^{\rho + o(1)}$.
    \end{itemize}
\end{definition}

We give the proofs associated with each theorem statement, which combine results proved in the next sections, in \cref{sec:depirx}.

\begin{maintheorem}[Track~1: Fourier analysis]\label{mainthm:track1}
    There is a two-server doubly-efficient PIR scheme with 
    exponent $H_2(1/4) < 0.8113$, where $H_2$ is the binary entropy function, and a three-server doubly-efficient PIR scheme with exponent $\eta_{\mathrm{ends}} < 0.7445$, where $\eta_{\mathrm{ends}} = 0.7444\ldots$ is the constant of~\Cref{cor:q3ends}.
\end{maintheorem}

\begin{maintheorem}[Track~2: theta functions of lattices]\label{mainthm:track2}
    Let $q$ be an odd prime, and let $\calL \subseteq \RR^r$ be an even lattice of determinant $\sqrt{q}$ (\Cref{sec:highdimlattices}). For each coset $c \in \calL^*/\calL$, let $w_\calL(c) := \frac{q}{2} \min_{\yv \in c} \langle \yv, \yv \rangle$, and let
    \begin{equation*}
        \eta_\calL := \frac{1}{\ln q} \inf_{0 < t \leq 1} \Bigg(\ln \sum_{c \in \calL^*/\calL} t^{w_\calL(c)} - \frac{r(q+1)}{24} \ln t\Bigg).
    \end{equation*}
    Then there is a $q$-server scheme with exponent $\eta_\calL$.

    In particular, for every prime $q \equiv 3 \pmod 4$, there is a $q$-server scheme with exponent $\eta_q$, the constant of~\eqref{eq:gen1exp}, where $\eta_3 < 0.7897$, $\eta_7 < 0.7095$, $\eta_{11} < 0.6609$, and $\eta_q \leq \frac{1}{2} + \frac{0.467 + o(1)}{\log_2 q}$ as $q \to \infty$. Consequently, as $s \to \infty$, there are $s$-server schemes with exponent $\frac{1}{2} + \frac{0.467 + o(1)}{\log_2 s}$.
\end{maintheorem}

The exponent of Track~3 is the value of an explicit two-parameter optimization problem, which we define first.

\begin{definition}[The Track~3 exponent]\label{def:track3exponent}
    Let $q$ be an odd prime and $s \in [2, q]$ an integer. For $\vartheta \in [0, 1]$, define the cost $w_{q, \vartheta}: \ZZ_q \to \RR$ and its generating function by
    \begin{equation*}
        w_{q, \vartheta}(j) := \min_{\substack{k \in \ZZ \\ k \equiv j \pmod q}} \big(k(k-1) + 2\vartheta k\big), \qquad M_\vartheta(t) := \sum_{j \in \ZZ_q} t^{w_{q, \vartheta}(j)} \quad (0 < t \leq 1),
    \end{equation*}
    and write $w_q := w_{q, 1/2}$, so that $w_q(j) = \min_{k \equiv j \pmod q} k^2$. Call a pair $(\vartheta, \nu)$ of real numbers \emph{feasible} (for $q$ and $s$) if
    \begin{equation*}
        0 \leq \vartheta \leq \min\left\{1, \frac{q+1}{6s}\right\} \qquad\text{and}\qquad 0 \leq \nu \leq \frac{q^2-1}{8s},
    \end{equation*}
    and for such a pair let
    \begin{equation*}
        \gamma(\vartheta, \nu) := \frac{q-1}{2}\,\vartheta + \left(1 - \frac{6s}{q+1}\vartheta\right)\left(\frac{q^2-1}{4s} - 2\nu\right).
    \end{equation*}
    Finally, define
    \begin{equation}\label{eq:gen4exp}
    \begin{aligned}
        E^{\mathrm{sq}}_{q}(\nu) &:= \frac{1}{\ln q} \inf_{0 < t \leq 1} \left(\ln M_{1/2}(t) - \nu \ln t\right), \\
        E^{\mathrm{edge}}_{q, s}(\vartheta, \nu) &:= \frac{1}{\ln q} \inf_{0 < t \leq 1} \left(\ln M_{\vartheta}(t) - \gamma(\vartheta, \nu) \ln t\right), \\
        \eta_{q, s}(\vartheta, \nu) &:= \max\left(E^{\mathrm{sq}}_{q}(\nu),\ E^{\mathrm{edge}}_{q, s}(\vartheta, \nu)\right), \\
        \eta_{q, s} &:= \inf \left\{\eta_{q, s}(\vartheta, \nu) : (\vartheta, \nu)\ \text{feasible}\right\}.
    \end{aligned}
    \end{equation}
\end{definition}

As in Track~2, these are Chernoff bounds on the number of frequency vectors of small total cost. The expansions behind~\Cref{mainthm:track3} are supported on the vectors whose total cost is small for one of two costs on $\ZZ_q$, together with a few affine images of these vectors (\Cref{def:gen4aprimesupport}): the square cost $w_q$, with budget $\nu$ per coordinate, and the tilted edge cost $w_{q, \vartheta}$, with budget $\gamma(\vartheta, \nu)$. The two costs come from the two scalar modular forms in the proof (\Cref{sec:gen4overview}). The parameters $\vartheta$ and $\nu$ trade the two budgets against each other, and the feasibility constraints are the conditions that the proof needs.

\begin{maintheorem}[Track~3: Jacobi forms]\label{mainthm:track3}
    Let $q$ be an odd prime and $s \in [2, q]$ an integer. For every feasible pair $(\vartheta, \nu)$ (\Cref{def:track3exponent}), there is an $s$-server scheme with exponent $\eta_{q, s}(\vartheta, \nu)$; in particular, there is an $s$-server scheme with exponent $\eta_{q, s}$. More concretely, $\eta_{3, 2} < 0.8204$, $\eta_{3, 3} < 0.7327$, $\eta_{5, 4} < 0.7025$, $\eta_{5, 5} < 0.6535$, $\eta_{7, 6} < 0.6521$, $\eta_{7, 7} < 0.6198$, $\eta_{11, 10} < 0.6078$, $\eta_{11, 11} < 0.5901$, and $\eta_{q, q} \leq \frac{1}{2} + \frac{0.255 + o(1)}{\log_2 q}$ as $q \to \infty$. Consequently, as $s \to \infty$, there are $s$-server schemes with exponent $\frac{1}{2} + \frac{0.255 + o(1)}{\log_2 s}$.
\end{maintheorem}

\begin{remark}[Reducing the communication with fully-homomorphic encryption]
    In each of these PIR schemes, a standard transformation (e.g., in~\cite{HR26}) uses fully-homomorphic encryption to preserve their efficiency guarantees while reducing the total communication to $\poly(\lambda, \log n)$ and downgrading their security guarantee to a computational (rather than information-theoretic) one. The storage and preprocessing time are  unaffected, while the server time and upload communication obtain additional $\poly(\lambda)$ factors.
\end{remark}

\section{Doubly-Efficient PIR with Two Servers}
\label{sec:depir:2pir}
\input{fig2server}

In this section, we give a stand-alone presentation of two-server doubly-efficient PIR scheme in which each server stores exactly as many field elements as the database has records. Our specific construction works over databases consisting of $2^d$ record in $\F_p$, for parameters $d$ and $p$ that we specify. The full construction appears in \cref{fig:2pir}; it works much in the same way as the scheme presented in the technical overview (\cref{sec:depir:qpir}), but uses other quadratic functions than used in our cone-delta definition (\cref{def:conedelta}) that work over the special field $\F_2$ used in the exponent. 

Applying a standard transformation of Dodis, P{\u{a}}tra{\c{s}}cu, and Thorup~\cite{DBLP:conf/stoc/DodisPT10}, which we review in \cref{sec:fieldtobits}, then gives a doubly-efficient PIR scheme that works over arbitrary $n$-bit databases. Each PIR server stores an $(n+1)$-bit data structure and runs in $o(n^{0.82})$ time per query, proving the two-server result in \Cref{mainthm:track1} (``Track 1'').  Compared to the prior, doubly-efficient PIR of Henzinger and Ragavan~\cite{HR26}, this new two-server PIR scheme improves the server-side storage from $1.5 \sqrt{\log n} \cdot n$ bits all the way down to $n+1$ bits, while asymptotically matching its per-query server work and communication.

In the remainder of this section, we show that \cref{fig:2pir}'s PIR scheme proves:

\begin{theorem}\label{thm:2-server-rate-one}
For every integer $d \geq 2$ with $d \equiv 2 \pmod4$ and every odd prime $p > d$, there is a two-server PIR scheme with information-theoretic security that operates over any database consisting of $2^d$ records in $\F_{p}$, with the following efficiency:
\begin{itemize}[itemsep=0pt,topsep=3pt]
    \item each server stores a server-dependent data structure that consists of $2^d$ elements of $\F_p$, computed from the database in $O(d 2^d)$ operations over $\F_p$ (and may then discard the original database),
    \item to make a PIR query, the user uploads $d$ bits to each PIR server, and
    \item to answer a PIR query, each server reads and sends back at most $2^{H_2(1/4)(d+1)}$ elements of $\F_p$ in its data structure, where $H_2$ is the binary entropy function.
\end{itemize}
\end{theorem}

\subsection{Setup}

On a database which consists of $2^d$ records in $\F_{p}$, for an integer $d \geq 2$ with $d \equiv 2 \pmod 4$ and an odd prime $p > d$, the PIR scheme is parametrized by:
\begin{enumerate}
    \item A {\em sparse decoding set $\calW$} defined, for the integer $\ell = (d - 2)/4$, as \begin{align*}
 \mathcal W=
 \{\vec w\in\F_2^d:|\vec w|\le \ell\}
 \,\cup\,
 \{\vec w\in\F_2^d:|\vec w|\ge d-\ell+1\}.
\end{align*}
In other words, the set~$\calW$ contains all vectors of weight at most $\ell$, together with the complements of all vectors of weight at most $\ell-1$. These vectors are distinct, giving  $|\mathcal W|=\sum_{j=0}^\ell\binom{d+1}{j}$.

\item A {\em set of public recovery coefficients $c_{\vec w} \in \F_{p}$} for each vector~$\vec w \in \calW$, defined to be the Fourier coefficients of the following polynomial, using the parity function $\pi(\vec x) = |\vec x | \pmod 2$:
\begin{align*}
 G(\vec x) &=
 \prod_{j=1}^\ell\left(1-\frac{|\vec x|+\pi(\vec x)}{4j}\right) \in \F_p, \qquad \text{  for all }
  \vec x\in\F_2^d\\
 c_{\vec w} &= \frac{1}{2^{d}} \cdot \sum_{\vec v\in\F_2^d}
 G(\vec v) \cdot (-1)^{\ip{\vec w}{\vec v}},  \qquad \text{  for all }
  \vec w\in \calW
\end{align*}
These are well defined, since $p$ is odd and $p > d > 4\ell$, so that $2^d$ and $4j$ for $j \in [\ell]$ are invertible in $\F_p$.
\end{enumerate}
We additionally define the quadratic function $Q: \F_2^d \to \F_2$  as $ Q(\vec x) = \sum_{1\leq i\leq j\leq d}x_ix_j.$

\paragraph{Properties.} 
These parameter choices generalize the earlier Fourier-sparse delta-function framework, as they use the quadratic cone given by $Q(\vec x)=0$,  rather than direct vector inner-products $\langle \vec x, \vec x \rangle \equiv 0 \pmod{q}$ as in \cref{def:conedelta}. (This is necessary since $\ip{\vec x}{\vec x}$ is linear over $\F_2$.)
In particular, mirroring the cone-delta definition, these choices offer the following property, which will lie at the core of our PIR scheme: for all vectors $\vec x \in \F_2^d$ with $Q(\vec x) = 0$, 
\begin{equation}\label{eq:decoder}
\sum_{\vec w \in \calW} c_{\vec w}\cdot (-1)^{Q(\vec x + \vec w) + Q(\vec w)} = \begin{cases}1,\text{ if }\vec x = \vec 0\\0,\text{ otherwise.}\end{cases}
\end{equation}

\noindent We give a proof of this property next.

\begin{proof} 
Consider any nonzero vector $\vec x \in \F_2^d$ with $Q(\vec x)=0$. Then, as $Q(\vec x)=0$ implies that $|\vec x| \equiv 0,3 \pmod 4$, the integer $|\vec x|+\pi(\vec x)$ must be a multiple of $4$. Its possible values are $4,8,\ldots,4\ell$, as $d = 4\ell+2$.
Therefore, the function $G(\vec x)$, which is defined as $$G(\vec x) = \prod_{j=1}^\ell\left(1-\frac{|\vec x|+\pi(\vec x)}{4j}\right) \in \F_p,$$ must evaluate to $1$ at input $\vec x = \vec 0$ and evaluate to 0 at every other point
$\vec x \in \F_2^d$ with $Q(\vec x)=0$.

Next, we examine the representation of $G(\vec x)$ in the Fourier domain. Applying the identity that, for any bit $b \in \{0,1\}$ it holds that $b = (1 - (-1)^b) / 2$, we see that 
\begin{align*}
|\vec x| &= \sum_{i=1}^d x_i = \frac{d - \sum_{i=1}^d (-1)^{x_i}}{2} &&\text{ and}\\
\pi(\vec x) &= \frac{1-(-1)^{\pi(\vec x)}}{2} &&\text{ giving that}\\
 |\vec x|+\pi(\vec x) &= \frac{d+1-\sum_{i=1}^d (-1)^{x_i}-(-1)^{\pi(\vec x)}}{2}.
\end{align*}
As a result, the function $G(\vec x)$ can be expressed as a multilinear polynomial of degree at most $\ell$ in the $d+1$ variables $X_1 = (-1)^{x_1}, \dots, X_d = (-1)^{x_d}$, and $Y = (-1)^{\pi(\vec x)}$, using $X_i^2=Y^2=1$. In this polynomial, any monomial without the term $Y = (-1)^{\pi(\vec x)}$ can be expressed as $(-1)^{\langle \vec w, \vec x \rangle}$, for some vector $\vec w$ of norm at most $\ell$.  Any monomial with the term $Y = (-1)^{\pi(\vec x)}$
can be expressed as $(-1)^{\langle \mathbf{\vec 1 + \vec w}, \vec x \rangle}$ for some vector $\vec w$ of norm at most $\ell-1$ and where $\vec 1$ is the all-ones vector. Here, $|\vec 1 + \vec w| = d - |\vec w| \geq d - \ell + 1$, so the vector $\vec 1 + \vec w$ has norm at least $d-\ell+1$. Each of these vectors is in the sparse-decoding set $\calW$,  implying that the Fourier support of function $G$ lies in~$\mathcal W$. That is, the public recovery coefficients $c_{\vec w}$ include all the non-zero coefficients of the Fourier representation of $G$.

Finally, using this observation and using that $Q(\vec x + \vec w) + Q(\vec w)=
\ip{\vec w}{\vec x+\pi(\vec x) \cdot \mathbf1}$ whenever $Q(\vec x)=0$, it must also be that
\begin{align*}
 \sum_{\vec w\in\calW}c_{\vec w} \cdot (-1)^{Q(\vec x + \vec w) + Q(\vec w)} &= \sum_{\vec w \in \calW} \left( \frac{1}{2^{d}} \cdot \sum_{\vec v\in\F_2^d} G(\vec v) \cdot (-1)^{\ip{\vec w}{\vec v}} \right) \cdot (-1)^{Q(\vec x + \vec w) + Q(\vec w)}\\
 &= \frac1{2^d}\sum_{\vec v\in\F_2^d}G(\vec v)
 \sum_{\vec w\in\F_2^d}
 (-1)^{\ip{\vec w}{\vec v+\vec x+\pi(\vec x) \cdot \mathbf1}}\\
 &=G(\vec x+\pi(\vec x) \cdot \mathbf1)\\
 &=
 \begin{cases}
  1,&\vec x=\vec0,\\
  0,&\vec x\ne\vec0.
 \end{cases}
\end{align*}
The last equality follows because, if vector $\vec x$ has even weight, then $G(\vec x+ \pi(\vec x) \cdot \mathbf 1) = G(\vec x)$. On the other hand, if vector $\vec x$ has odd weight and $Q(\vec x)=0$, then $|\vec x|\equiv3\pmod4$, so its complement has positive
weight $d-|\vec x|\equiv3\pmod4$ and also satisfies that $Q(\mathbf1+\vec x) = 0$. This completes the proof.
\qedhere
\end{proof}

\subsection{Protocol}

We next sketch the two-server, doubly-efficient PIR protocol, with servers indexed by $t\in\{0,1\}$. 

\paragraph{Preprocessing.} The servers split the database into $2^d$ records in $\F_p$, each indexed by the vectors in $\F_2^d$. Server $t \in \zo$ then stores the truth table of the following function $f_t: \F_2^d \to \F_p$:
\[
 f_t(\vec x)=\sum_{\vec v \in \F_2^d} \DB_{\vec v} \cdot
 (-1)^{Q(\vec x + \vec v) + t \cdot Q(\vecv)}.
\]
Here, arithmetic in the exponent happens mod $2$ (since $-1$ has order 2), while arithmetic in the base happens mod $p$---a setup mirroring matching-vector PIR~\cite{DG16}.

\paragraph{Querying.} Letting $\vec i \in \F_2^d$ be the address of the requested record, the user samples vector $\vec r \rgets \F_2^d$, then sends query $\query_0= \vec r$ to server $0$ and query $\query_1= \vec r+ \vec i$ to server $1$.

\paragraph{Answering.} For $t \in \zo$ and given query $\query_t$, server $t$ returns all of the evaluations $f_t(\query_t +  \vec w)$ for each vector $\vec w \in \calW$ to the user, by looking up each value in its precomputed truth table. 

\paragraph{Reconstructing.} The user recovers the chosen \ord{\vec i} database record, which is a $\F_p$-element, as
\begin{align*}
 \mathsf{record} = \frac12 \sum_{\vec w\in\mathcal W} c_{\vec w} \cdot
 \bigl((-1)^{Q(\query_1 + \vec w)} \cdot f_0(\query_0 +  \vec w)
 +(-1)^{Q(\query_0 + \vec w)+Q(\vec i)} \cdot f_1(\query_1 +  \vec w)\bigr)
 \pmod p.
\end{align*}

\subsection{Analysis}

We give a formal proof of \cref{thm:2-server-rate-one}.

\begin{proof}[]
\paragraph{Correctness.} Consider a user who aims to read the \ord{\vec i} database element, for $\vec i \in \F_2^d$. Expanding out the recovery equation and using the quadratic identity $$Q(\vec x+\vec y+\vec z)=Q(\vec x+\vec y)+Q(\vec x+\vec z)+Q(\vec y+\vec z)+Q(\vec x)+Q(\vec y)+Q(\vec z)\pmod2,$$ we see that:
\begin{align*}
  \mathsf{record} &= \frac12 \sum_{\vec w\in\mathcal W} c_{\vec w} \cdot
 \bigl((-1)^{Q(\vec r + \vec w + \vec i)} \cdot f_0(\query_0 +  \vec w)
 +(-1)^{Q(\vec r + \vec w)+Q(\vec i)} \cdot f_1(\query_1 +  \vec w)\bigr)\\
 &= \frac12 \sum_{\vec w\in\mathcal W} c_{\vec w} \cdot
 \sum_{\vec v \in \F_2^d} \DB_{\vec v} \cdot
 (-1)^{Q(\vec r + \vec w + \vec i)+Q(\vec r + \vec w + \vec v)}
 \cdot \bigl(1+(-1)^{Q(\vecv + \vec i)}\bigr)\\
 &= \frac12 \sum_{\vec v \in \F_2^d} \DB_{\vec v} \cdot \bigl( \sum_{\vec w\in\mathcal W} c_{\vec w} \cdot 
  (-1)^{Q(\vec w + \vec v + \vec i)+Q(\vec w)} \bigr) \cdot (-1)^{Q(\vec r + \vec v)+Q(\vec r + \vec i)} \cdot \bigl(1
 + 
 (-1)^{ Q(\vecv + \vec i)}\bigr).
\end{align*}
In this last equation, if $Q(\vecv + \vec i) \neq 0$, then the term $\bigl(1
 + 
(-1)^{ Q(\vecv + \vec i)}\bigr)$ is zero and the corresponding part of the sum cancels. On the other hand, if $Q(\vecv + \vec i) =0$, then the relation on the public recovery coefficients in Equation~\eqref{eq:decoder} implies that the term $\bigl(\sum_{\vec w \in \calW} c_{\vec w} \cdot (-1)^{Q(\vec w + \vec v + \vec i)+Q(\vec w)} \bigr)$ is 1 if $\vec v = \vec i$ and 0 otherwise. As a result, the whole equation simplifies out to
 \[ 
 \mathsf{record} = \frac12 \cdot \DB_{\vec i}  \cdot \bigl(1
 + 
 (-1)^{ Q(\vec0)}\bigr) = \DB_{\vec i}.
 \]
That is, the user correctly recovers its desired database element.

\paragraph{Security.} PIR security is immediate: every query sent to each server is distributed uniformly in $\F_2^d$.

\paragraph{Efficiency.}  Each preprocessed truth table consists of $2^d$ entries in $\F_p$, exactly matching the size of the original database. On each PIR query, the user sends each server $d$ bits. Each of the servers reads and returns
$|\calW|$ elements in $\F_p$ to the user, using $O(d)$ operations per element to compute the point $\query_t + \vec w$. Since $\ell \leq (d+1)/4$, the binomial bound below, together with the fact that $H_2$ is increasing on $[0, \frac{1}{2}]$, gives
\[
 | \calW | =\sum_{j=0}^\ell\binom{d+1}{j}
 \leq 2^{H_2(1/4)(d+1)}.
\]

To construct its truth table, each server performs at most $2^d$ sign multiplications followed by one unnormalized Walsh--Hadamard transform, taking $O(d2^d)$ operations in $\F_p$ in total. The signs are determined by
$Q(\vec v)=|\vec v| \cdot (|\vec v|+1)/2\pmod2$.  The public coefficients can be generated in the same asymptotic time.
\end{proof}

\section{Doubly-Efficient PIR with $s \geq 3$ Servers}
\label{sec:depir:spir}

Next, we construct doubly-efficient PIR with $n+1$ bits of storage for any number of servers $s \geq 3$. As in the technical overview (\cref{sec:depir:qpir}) and in \cref{sec:depir:2pir}, each server stores a Fourier transform of the database, whose entries are weighted linear combinations of the database records. Moreover, much as in \cref{sec:depir:2pir}, our specific construction works over databases consisting of $q^d$ records in $\F_p$, for parameters $q$, $d$, and $p$ that we specify. The full PIR construction appears in \cref{fig:spir}; it makes use of a set of public coefficients that correspond to a quadratic-phase delta expansion, as in \cref{def:genconedelta}.

Applying the standard transformation~\cite{DBLP:conf/stoc/DodisPT10} (given in \cref{sec:fieldtobits}) then yields a $s$-server doubly-efficient PIR scheme that works over arbitrary $n$-bits databases, with $n+1$ bits of storage per server. This PIR construction is  key to the results in \cref{mainthm:track1,mainthm:track2,mainthm:track3}, which differ only in the quadratic-phase delta expansion that the protocol uses.

\input{figsserver}

\paragraph{Overview.} The main difference from the prime-server PIR scheme (given in the technical overview) is how recovery works. In the prime-server scheme with $q$ servers, cancellation happens in two stages: first, summing across all servers' answers removes any database values labeled with a vector $\vec v \in \F_q^d$ such that $Q(\vecv-\vec i) \neq 0$, where $\vec i$ is the address of the requested record and $Q$ is a quadratic form (in \cref{sec:depir:qpir}, we take $Q(\vec x) = \langle \vec x, \vec x \rangle$). Then, applying the coefficients of a Fourier-sparse, cone-delta function (as per \cref{def:conedelta}) removes any database values labeled with a vector $\vec v \in \F_q^d$ such that $Q(\vecv-\vec i) = 0$ {\em except} the \ord{\vec i} database record. Via this two-step process, the user recovers exactly the desired database record.

This approach, which sums across all $q$ servers, appears to break when fewer than $q$ servers are available. However, in this section, we show that such a PIR scheme can work using $s < q$ servers by choosing server-dependent recovery coefficients such that both types of cancellation happen together, in the same step. This technique is inspired by sparse decoding polynomials, which reduce the number of servers in classic (not doubly-efficient) matching-vector PIR~\cite{Efr09,GKS25,ABL25,BO26,GR26}.
The resulting PIR  proves that:

\begin{theorem}\label{thm:s-server-rate-one}
For any number of servers $s\geq2$ and prime $q\geq\max\{3,s\}$, if there exists a $(k, s)$-sparse quadratic-phase delta expansion of order $q$, dimension $d$, and over a field~$\F_p$ (\Cref{def:genconedelta}),
then there is an $s$-server PIR scheme that operates over any database consisting of $q^d$ entries in $\F_p$, with information-theoretic security and the following efficiency:
\begin{itemize}[itemsep=0pt,topsep=3pt]
    \item each server stores a server-dependent data structure that consists of $q^d$ entries in $\F_p$, computed from the database in $O(d q^d)$ operations over $\F_p$ (and may then discard the original database),
    \item to make a PIR query, the user uploads $\lceil d \log_2 q\rceil$ bits to each PIR server, and
    \item to answer a PIR query, each server reads and sends back at most $k$ $\F_p$-elements in its data structure, using $O(d)$ operations per element.
\end{itemize}
\end{theorem}

\begin{remark}\label{rem:smallerdims}
    A $(k, s)$-sparse quadratic-phase delta expansion over $\F_p$ on $\ZZ_q^d$ also gives a $(k, s)$-sparse quadratic-phase delta expansion over $\F_p$ on $\ZZ_q^{d'}$ for any $d' < d$, by a similar padding argument to~\Cref{cor:roundd}. So~\Cref{thm:s-server-rate-one} then gives PIR schemes as above over databases of $q^{d'}$ records in $\F_p$, for every $d' \leq d$.
\end{remark}

\subsection{Protocol}

The scheme is parameterized by a prime $q\geq\max\{3,s\}$, an integer $d \geq 1$, and a prime $p$ such that 
\begin{enumerate}[nosep]
    \item $\F_p$ contains an element $\gamma$ of order $q$ and
    \item there exists a $(k, s)$-sparse quadratic-phase delta expansion of order $q$, of dimension $d$, and over field $\F_p$, as per \cref{def:genconedelta}. 
\end{enumerate} 
The support of this expansion is contained in $\calW \times \calT$ for some sets $\calW \subseteq \F_q^d$ and $\calT \subseteq \ZZ_q$ with $|\calW| \leq k$ and $|\calT| \leq s$. Since $s \leq q$, we may enlarge $\calT$ so that $|\calT| = s$, and we index the $s$ servers by the elements $t \in \calT$. That is, we have public coefficients $\{c_{\vec w,t}\in\F_p:\vec w\in\calW,\ t\in\calT\}$ such that, for every $\vec x\in\F_q^d$,
\begin{equation}\label{eq:s-decoder}
 \sum_{t\in\calT}\sum_{\vec w\in\mathcal W}
 c_{\vec w,t}\cdot
 \gamma^{\langle\vec w,\vec x\rangle+t \cdot \ip{\vec x}{\vec x}}
 = \begin{cases} 1,\text{ if }\vec x = \vec 0\\ 
 0, \text{ else}.\end{cases}
\end{equation}

\noindent Then, given any database that consists of $q^d$ elements in $\F_p$, the PIR scheme works as follows:

\paragraph{Preprocessing.}
The servers split the database into $q^d$ records in $\F_p$, each indexed by the vectors in $\F_q^d$. Each server $t\in\calT$ stores the truth-table of the following function $f_t: \F_q^d \to \F_p$:
$$f_t(\vec x) = \sum_{\vec v \in\F_q^d}\db_{\vec v}\cdot
 \gamma^{\langle\vec x,\vec v\rangle+t\cdot \langle \vec v, \vec v \rangle}.$$
Mirroring the prime-server construction from \cref{sec:depir:qpir}, arithmetic in the exponent happens modulo $q$; arithmetic in the base happens modulo $p$.

\paragraph{Querying.}
The user maps its index $i$ into a query point $\vec i\in\F_q^d$ such that the record $\DB_{\vec i}$ contains the database value he aims to read. Then, the user samples a random mask $\vec r\rgets\F_q^d$ and sends server $t\in\calT$ the query vector
\[
 \query_t=\vec r-2t\cdot\vec i \in \F_q^d.
\]

\paragraph{Answering.}
Given query point $\query_t$, server $t$ responds with all evaluations $f_t(\query_t+\vec w)$ for $\vec w\in\calW$, by looking up each value in its precomputed truth table. 

\paragraph{Reconstructing.}
The user combines the servers' answers, weighting them by the public coefficients of the quadratic-phase delta expansion, to recover and output the following $\F_p$-element:
\begin{equation}\label{eq:s-recon}
 \mathsf{record} =
 \sum_{t\in\calT}\sum_{\vec w\in\mathcal W}
 c_{\vec w,t} \cdot f_t (\query_t+\vec w)\cdot
 \gamma^{\langle t \cdot \vec i - \vec r-\vec w,\vec i\rangle}.
\end{equation}

\subsection{Analysis}

We prove each of the scheme's requirements in sequence.

\begin{proof}[] 
\paragraph{Correctness.}
The above reconstruction routine exactly outputs the \ord{\vec i} record in the database since, by expanding expressions and applying the identity $\langle \vec v-\vec i, \vec v - \vec i \rangle =\langle \vec v, \vec v \rangle + \langle \vec i, \vec i \rangle -2\langle\vec i,\vec v\rangle$, we see that
\begin{align*}
 \mathsf{record}
 &=
 \sum_{t\in\calT}\sum_{\vec w\in\mathcal W}
 c_{\vec w,t} \cdot \left( \sum_{\vec v \in\F_q^d}\db_{\vec v}\cdot
 \gamma^{\langle\vec r -2t \cdot \vec i + \vec w,\vec v\rangle+t\cdot \langle \vec v, \vec v \rangle} \right) \cdot
 \gamma^{\langle t \cdot \vec i - \vec r - \vec w,\vec i\rangle}\\
 &= \sum_{\vec v\in\F_q^d}
 \db_{\vec v}
 \cdot \gamma^{\langle\vec r,\vec v-\vec i\rangle}
 \left(
 \sum_{t\in\calT}\sum_{\vec w\in\mathcal W}
 c_{\vec w,t}\cdot
 \gamma^{\langle\vec w,\vec v-\vec i\rangle
       +t\langle \vec v-\vec i, \vec v-\vec i\rangle}
 \right)
 = \db_{\vec i}.
\end{align*}
The final equality here follows directly from property~\eqref{eq:s-decoder}; this proves the scheme's correctness.

\paragraph{Security.} Security is immediate: each query is masked by the uniformly random vector $\vec r \in \F_q^d$.

\paragraph{Efficiency.} The scheme's efficiency properties follow by inspection. Each server stores $q^d$ entries in $\F_p$ in its preprocessed table. The user sends each server $\lceil d\log_2 q\rceil$ bits. Each server reads and returns $|\mathcal W| \leq k$ elements in $\F_p$, using $O(d)$ operations per element to compute the point $\query_t + \vec w$.

Each preprocessing table is an entry-wise multiplication by
$\gamma^{t \cdot \langle \vec v, \vec v \rangle}$, followed by a $d$-dimensional Fourier transform.  Applying the transform one coordinate at a time takes $O(dq^d)$ operations, as $q$ is fixed.
\end{proof}

\section{Elementary Results on Sparse Cone-Delta Functions}\label{sec:elementary}

\subsection{Simple Constructions for $q = s = 3$ from Hamming Slices}\label{sec:q3elementary}

For $q=3$, we can construct sparse cone-delta functions without using modular forms (whereas \cref{sec:math} will use modular forms). We improve on an elementary Hamming-ball construction of Alon and Lubetzky~\cite{AlonLubetzky07}, by using Hamming shells from both ends of the weight range, obtaining Fourier sparsity $3^{(0.7444\ldots+o(1))d}$  (\Cref{lem:q3ends,cor:q3ends}); it yields a $3$-server doubly-efficient PIR scheme with server work $n^{0.7445}$ (\Cref{mainthm:track1}). 
(For comparison, the more sophisticated construction in \cref{sec:gen4} gives the exponent $0.7327$ for $q = 3$ shown in \cref{table:server-counts}; by a lower bound given next, no cone-delta function achieves an exponent below $1/2$.)

\paragraph{Setup.} Throughout this subsection, $\F$ is a field containing a primitive cube root of unity $\gamma$, and $\wt(\xv)$ denotes the number of nonzero coordinates of $\xv \in \ZZ_3^d$. Every nonzero element of $\ZZ_3$ squares to $1$, so $\langle \xv, \xv \rangle \equiv \wt(\xv) \pmod{3}$, and the cone $\{\xv : \langle \xv, \xv \rangle \equiv 0 \pmod{3}\}$ consists exactly of the points of weight $3r$ for $r = 0, 1, \ldots, \lfloor d/3 \rfloor$. We look for cone-delta functions of the form $f(\xv) = \sum_{\av \in \ZZ_3^d} c_{\wt(\av)} \gamma^{\langle \av, \xv \rangle}$, whose Fourier coefficients depend only on the weight of the frequency vector; such an $f$ is a combination of the \emph{shell sums} $\sum_{\wt(\av) = j} \gamma^{\langle \av, \xv \rangle}$, and its Fourier sparsity is at most $\sum_{j : c_j \neq 0} \binom{d}{j} 2^j$.

\begin{lemma}[{Shell sums and ternary Krawtchouk polynomials~\cite[Chapter~5]{MacWilliamsSloane77}}]\label{lem:q3shells}
    For $0 \leq j \leq d$ and $\xv \in \ZZ_3^d$ of weight $w$, the sum $\sum_{\wt(\av) = j} \gamma^{\langle \av, \xv \rangle}$ depends only on $w = \wt(\xv)$. Hence we abuse notation and denote this sum by $K_j(w)$. Then, we have:
    \begin{align*}
        \sum_{j=0}^{d} K_j(w)\, z^j &= (1 + 2z)^{d-w} (1 - z)^w, \\
        K_j(w) &= \sum_{n=0}^{j} (-3)^n\, 2^{j-n} \binom{d-n}{j-n} \binom{w}{n} &&\text{for all $j \in [0, d]$},\\
        K_{d-i}(w) &= 2^{d-w} (-1)^w \sum_{n=0}^{i} \left(-\frac{3}{2}\right)^n 2^{-(i-n)} \binom{d-n}{i-n} \binom{w}{n}&&\text{for all $i \in [0, d]$}.
    \end{align*}
    The second identity above holds over $\ZZ$ and the third identity holds over $\ZZ[1/2]$.
\end{lemma}
\noindent
So $f$ is a cone-delta function exactly when:
\begin{equation}\label{eq:krawtchouk}
    \sum_j c_j K_j(3r) = \one[r = 0],\quad\text{for } r = 0, 1, \ldots, \lfloor d/3 \rfloor. 
\end{equation}
To understand this condition, associate to each $j \in [0, d]$ the vector \begin{equation}\label{eq:vjq3}
    \vecv_j := (K_j(0), K_j(3), \ldots, K_j(3\lfloor d/3 \rfloor)).
\end{equation}
Then~\eqref{eq:krawtchouk} says that there is a cone-delta function supported on the Hamming shells from a set $J \subseteq \{0, 1, \ldots, d\}$ if and only if the vectors $\vecv_j$ for $j \in J$ span $(1, 0, \ldots, 0)$. 
For each Hamming shell $j$, we pay a cost of $\binom{d}{j} 2^j$ in Fourier sparsity, and we would like to select a collection of shells $J$ that has total cost as low as possible, while satisfying the spanning condition. 
We plot the cost of shell $j$ as a function of $j$ in~\Cref{fig:q3shells}: the cost increases to a peak at $j = 2d/3$ and then decreases again.

Bearing this intuition in mind, we now restate the existing construction of cone-delta functions implicit in prior work by Alon and Lubetzky~\cite{AlonLubetzky07}. It proceeds by simply taking $J = \{0, 1, \ldots, \lfloor d/3 \rfloor\}$. 
We omit the proof of~\cite{AlonLubetzky07} and also note that the below is a special case of~\Cref{lem:q3ends} which we will prove.

\begin{lemma}[{Hamming ball, adapted from~\cite[proof of Theorem~2.2]{AlonLubetzky07}}]\label{lem:q3ball}
    Suppose that $\F$ has characteristic $0$ or greater than $\max(\lfloor d/3 \rfloor, 2)$. Then there is a cone-delta function on $\ZZ_3^d$ over $\F$ whose Fourier support is contained in the Hamming ball $\{\av : \wt(\av) \leq \lfloor d/3 \rfloor\}$. Its Fourier sparsity is at most $\sum_{j \leq \lfloor d/3 \rfloor} \binom{d}{j} 2^j = 3^{(\eta_{\mathrm{ball}} + o(1))d}$, where $\eta_{\mathrm{ball}} := 1 - \frac{1}{3}\log_3 2 = 0.7896\ldots$.
\end{lemma}

However, the picture in~\Cref{fig:q3shells} points to a way to improve on this construction. 
Intuitively, we want to span a vector in dimension $\lfloor d/3 \rfloor + 1$, so hopefully it suffices to greedily take the $\lfloor d/3 \rfloor + 1$ lowest-cost shells i.e., some shells close to the $j = 0$ end and some shells close to the $j = d$ end. 
Fortunately for us, these shells actually do have the desired spanning, and this gives us the improved $0.744\ldots$ exponent. We formally check this next:

\begin{lemma}[Both ends]\label{lem:q3ends}
    Let $\ell, \ell' \geq 0$ be integers with $\ell + \ell' = \lfloor d/3 \rfloor$, and suppose that $\F$ has characteristic $0$ or a prime characteristic $p > \max\{\lfloor d/3 \rfloor, 2\}$. Then the vectors $\vecv_j$ (defined in~\eqref{eq:vjq3}) for $j \in J := \{0, 1, \ldots, \ell\} \cup \{d-\ell'+1, \ldots, d\}$ are linearly independent. In particular, they span $(1, 0, \ldots, 0)$.

    Consequently, there is a cone-delta function on $\ZZ_3^d$ over $\F$ whose Fourier coefficients depend only on the weight of the frequency vector, and whose Fourier support is contained in $\{\av : \wt(\av) \leq \ell\} \cup \{\av : \wt(\av) \geq d - \ell' + 1\}$.
\end{lemma}
\begin{proof}
    For $0 \leq n \leq \lfloor d/3 \rfloor$, let $B_n(r) := \binom{3r}{n} = \frac{1}{n!}\prod_{i=0}^{n-1}(3r - i)$, a polynomial in $r$ over $\F$ of degree exactly $n$ (its leading coefficient $3^n/n!$ is invertible, as the characteristic of $\F$ is $0$ or exceeds $\lfloor d/3 \rfloor \geq n$ and is not $3$). 
    Define $\lambda := -\frac{1}{8} \in \F$ (which is defined since the characteristic is not $2$). By~\Cref{lem:q3shells} with $w = 3r$, so that $2^{d-w}(-1)^w = 2^d \lambda^r$:
    \begin{itemize}
        \item for $j \leq \ell$, the function $K_j(3r)$ is a combination of $B_0, \ldots, B_j$ with coefficient $(-3)^j \neq 0$ on $B_j$; and
        \item for $i < \ell'$, the function $K_{d-i}(3r)$ is $2^d \lambda^r$ times a combination of $B_0, \ldots, B_i$ with coefficient $(-\frac{3}{2})^i \neq 0$ on $B_i$.
    \end{itemize}
    To parse the above, define:
    \begin{align*}
        \bv_j &:= (B_j(0), \ldots, B_j(\lfloor d/3 \rfloor)),\quad j \in \{0, 1, \ldots, \ell\}; \\
        \bv'_j &:= (0^j, \ldots, (\lfloor d/3 \rfloor)^j),\quad j \in \{0, 1, \ldots, \ell\};\\
        \bv_{d-i} &:= (\lambda^0 B_i(0), \lambda^1 B_i(1), \ldots, \lambda^{\lfloor d/3 \rfloor} B_i(\lfloor d/3 \rfloor)),\quad i \in \{0, 1, \ldots, \ell'-1\} \\
        \bv'_{d-i} &:= (\lambda^0 0^i, \lambda^1 1^i, \ldots, \lambda^{\lfloor d/3 \rfloor} (\lfloor d/3 \rfloor)^i),\quad i \in \{0, 1, \ldots, \ell'-1\}.
    \end{align*}
    Then it follows that there exists an upper triangular matrix $\mathbf{M}$ with nonzero diagonal such that:
    \begin{equation*}
        \begin{bmatrix}
            \vecv_0 & \ldots & \vecv_{\ell} & \vecv_{d} & \ldots & \vecv_{d-\ell'+1}
        \end{bmatrix} = \begin{bmatrix}
            \bv_0 & \ldots & \bv_\ell & \bv_{d} & \ldots & \bv_{d-\ell'+1}
        \end{bmatrix}
        \mathbf{M}.
    \end{equation*}
    Since $\mathbf{M}$ is invertible, to show that the $\vecv_j$'s are linearly independent it suffices to show that the $\bv_j$'s are linearly independent. 
    A similar argument, now applied to the $\bv_j$'s, reduces this to showing that the $\bv'_j$'s are linearly independent (in words, this replaces each evaluation $B_j(r)$ with the leading-order monomial $r^j$). 
    In other words, we want to show that the matrix:
    \begin{align*}
        V &:= \begin{bmatrix} \bv'_0 & \ldots & \bv'_\ell & \bv'_d & \ldots & \bv'_{d-\ell'+1} \end{bmatrix} \\
        &= \begin{bmatrix}
            1 & \ldots & 0 & \lambda^0 & \ldots & 0 \\
            1 & \ldots & 1 & \lambda^1 & \ldots & \lambda^1 1^{\ell'-1} \\
            \vdots & \ddots & \vdots & \vdots & \ddots & \vdots \\
            1 & \ldots & (\lfloor d/3 \rfloor)^{\ell} & \lambda^{\lfloor d/3 \rfloor} &\ldots & \lambda^{\lfloor d/3 \rfloor} (\lfloor d/3 \rfloor)^{\ell'-1}
        \end{bmatrix},
    \end{align*}
    is invertible. 
    This is a \emph{confluent Vandermonde matrix,} and~\cite[Theorem~21, equation~(3.2)]{Krattenthaler99} tells us that its determinant is 
    \[
        \det V = \lambda^{\binom{\ell'}{2}} (\lambda - 1)^{(\ell+1)\ell'}
        \left(\prod_{j=0}^{\ell} j!\right)
        \left(\prod_{j=0}^{\ell'-1} j!\right),
    \]
    which is nonzero under our characteristic assumptions: $\lambda = -1/8$ and $\lambda - 1 = -9/8$ are nonzero, and all factorials have indices at most $\lfloor d/3 \rfloor$. The desired invertibility follows, completing our proof.
\end{proof}

\begin{corollary}\label{cor:q3ends}
    Let $H_2(t) := -t\log_2 t - (1-t)\log_2(1-t)$ be the binary entropy function. For every $\alpha \in [0, \frac{1}{3}]$, taking $\ell = \lfloor \alpha d \rfloor$ in~\Cref{lem:q3ends} gives Fourier sparsity $3^{(\eta(\alpha) + o(1))d}$, where
    \begin{equation*}
        \eta(\alpha) := \frac{1}{\log_2 3} \max\left\{H_2(\alpha) + \alpha,\ H_2\left(\tfrac{1}{3} - \alpha\right) + \tfrac{2}{3} + \alpha\right\}.
    \end{equation*}
    The two terms are equal at $\alpha^* = 0.2994\ldots$, and $\eta_{\mathrm{ends}} := \eta(\alpha^*) = 0.7444\ldots$; in particular, $\eta(0.2994) < 0.7445$.
\end{corollary}
\begin{proof}
    The sparsity is at most $\sum_{j \leq \ell} \binom{d}{j} 2^j + \sum_{i < \ell'} \binom{d}{i} 2^{d-i}$, where $\ell' = N - \ell$. In the first sum, the ratio of consecutive terms is $\frac{2(d-j)}{j+1} \geq 1$ for $j \leq \frac{2d-1}{3}$, so its terms increase and it is at most $(d+1) \binom{d}{\ell} 2^{\ell}$, as $\ell \leq \frac{d}{3}$. In the second sum, the ratio of consecutive terms is $\frac{d-i}{2(i+1)} \geq 1$ for $i \leq \frac{d-2}{3}$, so it is at most $(d+1)\binom{d}{\ell'} 2^{d - \ell'}$, as $\ell' - 1 \leq \frac{d}{3} - 1$. Finally, $\binom{d}{k} \leq 2^{d H_2(k/d)}$ for $0 \leq k \leq d$ (as $\binom{d}{k} t^k (1-t)^{d-k} \leq 1$ for $t = k/d$). With $\ell/d \to \alpha$ and $\ell'/d \to \frac{1}{3} - \alpha$, and $H_2$ continuous, the two sums are $2^{(H_2(\alpha) + \alpha + o(1))d}$ and $2^{(H_2(1/3 - \alpha) + 2/3 + \alpha + o(1))d}$. The numerical values are a direct computation; see~\Cref{fig:q3shells}.
\end{proof}

\input{figq3shells}

\begin{remark}[Relation to~\Cref{sec:math}]\label{remark:q3parttwo}
    The construction of~\Cref{lem:q3ball} has two features. First, the Fourier support consists of the frequency vectors $\av$ of small total cost $\sum_i w(a_i)$, where the cost $w$ of a residue in $\ZZ_3$ is $0$ for the zero residue and $1$ otherwise. Second, the vanishing on the cone comes from interpolation in the single variable $\wt(\xv)$, and the degree it needs is dictated by root counting: a univariate polynomial that equals $1$ at $0$ and vanishes at $N$ prescribed points must have degree at least $N$, and this degree suffices. The second feature does not extend to $q \geq 5$, because $\langle \xv, \xv \rangle \bmod q$ is then no longer a function of $\wt(\xv)$: for $q = 5$, the points $(1, 0, \ldots, 0)$ and $(2, 0, \ldots, 0)$ both have weight $1$, but $\langle \xv, \xv \rangle$ equals $1$ and $4$ respectively. The constructions of~\Cref{sec:math} keep the first feature, with the support again consisting of the frequency vectors of small total cost, for a cost function on $\ZZ_q$ that we construct from a lattice (\Cref{def:prices}), and they replace interpolation in $\wt(\xv)$ by a root-counting argument for modular forms. At $q = 3$, the resulting cost is $1$ on both nonzero residues, and the support of the warm-up construction (\Cref{sec:gen1}) is exactly the Hamming ball of~\Cref{lem:q3ball}.

    The construction of~\Cref{lem:q3ends} gives up the first feature, as the shells of weight close to $d$ have high cost. The support of the stronger construction of~\Cref{sec:gen4} also reaches both ends of the weight range: it contains the shifts $\zv + b\bfone$ of low-cost vectors $\zv$ (\Cref{def:gen4aprimesupport}), and for $q = 3$ and $b \neq 0$ these have weight at least $d - \wt(\zv)$.
\end{remark}

\subsection{Lower Bound}

Every exponent in~\Cref{table:server-counts} is above $1/2$. We now show that this is unavoidable for our approach: a cone-delta function over any field must have Fourier sparsity at least $q^{d/2-O(1)}$, and by~\Cref{lem:conedelta-qpde}, the same bound applies to every quadratic-phase delta expansion. So our doubly-efficient PIR constructions from these objects cannot achieve a server work exponent below $1/2$. The main idea is that a cone-delta function must be a delta function on a large subspace, which forces it to have large Fourier support.

\begin{theorem}\label{thm:deltalb}
    Let $q$ be an odd prime, and let $\F$ be any field with a primitive \ord{q} root of unity $\gamma$. Assume there exists a cone delta function with Fourier sparsity $k$. Then, we have $k \geq q^{2 \lfloor d/4 \rfloor}$.
\end{theorem}

To set up the proof of this theorem, we use the following intermediate lemmas:

\begin{proposition}\label{prop:sos-1}
    There exist residues $a, b \in \ZZ_q$ such that $a^2+b^2 \equiv -1\pmod{q}$.
\end{proposition}
\begin{proof}
    The sets $\{a^2: a \in \ZZ_q\}$ and $\{-1-b^2: b \in \ZZ_q\}$ both have size $(q+1)/2$, so they must intersect. The conclusion follows.
\end{proof}

\begin{lemma}\label{lem:v0}
    There exists a subspace $V$ of $\ZZ_q^4$ of dimension 2 such that $\langle \xv, \xv \rangle \equiv 0\pmod{q}$ for all $\xv \in V$.
\end{lemma}
\begin{proof}
    Take the $a, b$ given by~\Cref{prop:sos-1} and consider the subspace of $\ZZ_q^4$ spanned by $(a, b, 1, 0)$ and $(b, -a, 0, 1)$. 
    For any $x, y \in \ZZ_q$, we have $x(a, b, 1, 0) + y(b, -a, 0, 1) = (ax+by, bx-ay, x, y)$. 
    Finally, we have $(ax+by)^2 + (bx-ay)^2 + x^2 + y^2 = (a^2+b^2+1)(x^2+y^2) \equiv 0 \pmod{q}$.
\end{proof}

\begin{corollary}\label{cor:V}
    There exists a subspace $V$ of $\ZZ_q^d$ of dimension $2\lfloor d/4 \rfloor$ such that $\langle \xv, \xv \rangle \equiv 0 \pmod{q}$ for all $\xv \in V$.
\end{corollary}
\begin{proof}
    For the $V_0$ given in~\Cref{lem:v0}, we will take the subspace $V := \bigoplus_{i \in [\lfloor d/4 \rfloor]} V_0$, which clearly has dimension $2\lfloor d/4 \rfloor$. In other words, we split the $d$ coordinates into $\lfloor d/4 \rfloor$ blocks of 4 coordinates each and let $V$ be the set of vectors in $\ZZ_q^d$ such that the restriction to any of these blocks yields an element of $V_0$, and any coordinate outside these blocks is zero. The conclusion follows.
\end{proof}

\begin{lemma}\label{lem:fouriersupport}
    Suppose $f: \ZZ_q^d \to \F$ is a delta function on $V$ for a subspace $V \subseteq \ZZ_q^d$. Then $\hat{f}$ is nonzero in at least $q^{\dim V}$ positions.
\end{lemma}
\begin{proof}
    This is the multiplication argument of~\cite[Lemma~3.7 and Theorem~3.9]{DBLP:conf/innovations/GhasemiK26}, applied with the indicator of $V$.
    Define $g(\xv) := \left(\frac{1}{q^{d-\dim V}} \sum_{\yv \in V^\perp} \gamma^{\langle \xv, \yv \rangle}\right) \cdot f(\xv)$. The first term is 1 if $\xv \in V$ and 0 otherwise, so $g$ is the delta function on all of $\ZZ_q^d$. Consequently, it must have full Fourier support $q^d$. 
    On the other hand, by definition its Fourier support has size at most $|V^\perp| \cdot |\supp \hat{f}| = q^{d-\dim V} \cdot |\supp \hat{f}|$. Putting these together forces $\hat{f}$ to have support size $\geq q^{\dim V} = |V|$ as claimed.
\end{proof}

\begin{proof}[Proof of~\Cref{thm:deltalb}]
    If $f$ is a delta function on the cone, it must in particular be a delta function on the subspace $V$ given by~\Cref{cor:V} of size $q^{2\lfloor d/4 \rfloor}$. Then~\Cref{lem:fouriersupport} implies that $\hat{f}$ must have support size at least $q^{2\lfloor d/4 \rfloor}$, as desired.
\end{proof}

\section{From Expansions to $n$-Bit PIR}\label{sec:bits}

The constructions of~\Cref{sec:math} give quadratic-phase delta expansions over the field $\QQ(\zeta_q)$, while the PIR schemes of~\Cref{sec:depir:2pir,sec:depir:spir} operate over databases of records in a finite field $\F_p$. In this section, we first transfer expansions from $\QQ(\zeta_q)$ to $\F_p$, for every sufficiently large prime $p \equiv 1 \pmod q$ (\Cref{sec:primefield}). We then convert PIR schemes over databases of $\F_p$-records into PIR schemes over databases of $n$ bits, with $n + 1$ bits of storage per server (\Cref{sec:fieldtobits}). The proofs of the main theorems (\Cref{sec:mainthms}) combine these two steps.

\subsection{From Characteristic 0 to Finite Fields}\label{sec:primefield}

In the following, we provide some building blocks to transition from our cone-delta results over $\QQ(\zeta_q)$ to usable cone deltas over finite fields that can be explicitly computed; these are necessary for the PIR protocols of~\Cref{sec:depir:spir}. 
To do this, the main observation is that the Fourier support sets used in~\Cref{thm:gen1,thm:gen4aprime} are symmetric by definition; they are invariant under permutations of the $d$ coordinates. This allows us to impose the same symmetry on the coefficients:

\begin{lemma}\label{lem:symmetrize}
    Assume that there exists a quadratic-phase delta expansion on $\ZZ_q^d$ for $\F = \QQ(\zeta_q)$ with support $\calW \times \calT$, such that the set $\calW$ is invariant under permutations of the $d$ coordinates. 
    Then there also exists a quadratic-phase delta expansion supported on the same $\calW \times \calT$ with the same symmetry invariance i.e., $c'_{\av, t} = c'_{\av', t}$ if $\av, \av'$ are permutations of each other. Going forward, we refer to such a quadratic-phase delta expansion as \emph{symmetric.}
\end{lemma}
\begin{proof}
    By averaging: define $c'_{\av, t} := \frac{1}{d!} \sum_{\pi\text{ permutation on }[d]} c_{\pi(\av), t}$. This automatically has the stated symmetry. 
    To show that this works, we have:
    \begin{align*}
        \sum_{t \in \ZZ_q} \sum_{\av \in \ZZ_q^d} c'_{\av, t} \zeta_q^{\langle \av, \xv \rangle + t \langle \xv, \xv \rangle} &= \frac{1}{d!} \sum_{t \in \ZZ_q} \sum_{\pi} \sum_{\av \in \ZZ_q^d} c_{\pi(\av), t} \zeta_q^{\langle \av, \xv \rangle + t\langle \xv, \xv \rangle} \\
        &= \frac{1}{d!} \sum_{t \in \ZZ_q} \sum_{\pi} \sum_{\av \in \ZZ_q^d} c_{\pi(\av), t} \zeta_q^{\langle \pi(\av), \pi(\xv) \rangle + t\langle \pi(\xv), \pi(\xv) \rangle} \\
        &= \frac{1}{d!} \sum_{\pi} \sum_{t \in \ZZ_q} \sum_{\av \in \ZZ_q^d} c_{\av, t} \zeta_q^{\langle \av, \pi(\xv) \rangle + t\langle \pi(\xv), \pi(\xv) \rangle} \\
        &= \frac{1}{d!} \sum_\pi \one[\pi(\xv) = 0] \\
        &= \one[\xv = 0],
    \end{align*}
    as desired.
\end{proof}

\begin{lemma}\label{lem:cramer}
    Assume there exists a symmetric quadratic-phase delta expansion over $\QQ(\zeta_q)$ supported on $\calW \times \calT$.  
    Then there exists a symmetric quadratic-phase delta expansion supported on the same $\calW \times \calT$ and a positive integer $M \leq \exp(\poly(d))$ such that all coefficients in the delta expansion take the form $\frac{1}{M}(a_0 + a_1\zeta_q + a_2 \zeta_q^2 + \ldots + a_{q-2} \zeta_q^{q-2})$, for integers $a_0, \ldots, a_{q-2} \leq \exp(\poly(d))$.
\end{lemma}
\begin{proof}[Proof sketch]
    For a vector $\av \in \ZZ_q^d$, define:
    \begin{equation*}
        \Type(\av) := \left(|\{i: a_i = j\}|\right)_{j \in \ZZ_q}.
    \end{equation*}
    It is easy to see that the quadratic-phase delta expansion condition can be written as a system of linear equations over $\QQ(\zeta_q)$, where:
    \begin{itemize}
        \item There is one variable $c_{\av, t}$ for each $t \in \calT$ and each $\Type(\av)$ that appears in $\calW$; and
        \item There is one equation for each $\Type(\xv)$.
    \end{itemize}
    The number of variables and equations is thus bounded above by $q(d+1)^q = \poly(d)$. Moreover, each coefficient is a sum of at most $q^d$ powers of $\zeta_q$.
    
    We can further unroll this into a linear system over $\QQ$ by writing every element of $\QQ(\zeta_q)$ as a $\QQ$-linear combination of $1, \zeta_q, \ldots, \zeta_q^{q-2}$, at the expense of a factor of $q-1$ in the number of variables. With this, it is easy to see that all coefficients have absolute value at most $q^d$.
    
    The resulting linear system is solvable, so it has a solution arising from a full-rank square submatrix. Moreover, Cramer's rule tells us that this solution will be an integer divided by the determinant of this submatrix. Since this is a submatrix of some dimension $d' \leq \poly(d)$ with entries at most $q^d$ in magnitude, it follows that its determinant is $\leq d'! \cdot q^{dd'} \leq \exp(\poly(d))$. The upper bound on the numerators also follows from Cramer's rule.
\end{proof}

\begin{lemma}
\label{lem:finite-field-specialization}
Assume there exists a symmetric quadratic-phase delta expansion over $\QQ(\zeta_q)$ with support $\calW \times \calT$, where $\calW \subseteq \ZZ_q^d$ and $\calT \subseteq \ZZ_q$. 
Then there is a positive integer $M \leq M_d$, where $M_d := D! \cdot q^{dD}$ for $D := q^2 (d+1)^q$ (so $M_d \leq \exp(\poly(d))$), such that, for every prime $p \equiv 1 \pmod{q}$ with $p > M$ and every primitive $q$th
root of unity $\gamma\in\mathbb F_p$, there is a symmetric
quadratic-phase delta expansion over $\mathbb F_p$
supported on $\calW\times \calT$. Moreover, given $p$, $\gamma$, $\calT$ and the list of types of the vectors in $\calW$, the coefficients of such an expansion can be computed in $\poly(d, \log p)$ time.
\end{lemma}

\begin{proof}
Let $M$ be as given by~\Cref{lem:cramer}; its proof bounds $M$ by $d'! \cdot q^{dd'}$ for some $d' \leq (q-1) \cdot q(d+1)^q \leq D$, so $M \leq M_d$. Let $p$ and $\gamma$ be as above, so that $p \nmid M$.
The symmetric quadratic-phase delta expansion over $\QQ(\zeta_q)$ can now be mapped into $\FF_p$ by replacing $1/M$ with $M^{-1}\bmod{p}$, reducing all coefficients, modulo $p$, and mapping $\zeta_q \mapsto \gamma$. More formally, we are applying a homomorphism from $\ZZ[\zeta_q, 1/M] \to \FF_p$ given by $\zeta_q \mapsto \gamma$ and $1/M \mapsto M^{-1}\bmod{p}$. 

For the last part, replace $\zeta_q$ by $\gamma$ in the linear system of the proof of~\Cref{lem:cramer}: it has one unknown $c_{\tau, t}$ for each type $\tau$ of the vectors in $\calW$ and each $t \in \calT$, and one equation for each type $\sigma$, so $\poly(d)$ unknowns and equations. A solution over $\F_p$ exists (the expansion above is one), and every solution gives a symmetric quadratic-phase delta expansion over $\F_p$ supported on $\calW \times \calT$, as both sides of the defining identity depend only on the type of $\xv$. For $\xv$ of type $\sigma$, the coefficient of $c_{\tau, t}$ in the equation for $\sigma$ is $\gamma^{t\langle \xv, \xv \rangle}$ times the coefficient of $\prod_{j} Y_j^{\tau_j}$ in the polynomial $\prod_{r \in \ZZ_q} \big(\sum_{j \in \ZZ_q} \gamma^{jr} Y_j\big)^{\sigma_r}$, which has $\poly(d)$ monomials and can be expanded with $\poly(d)$ operations over $\F_p$. Gaussian elimination then finds a solution with $\poly(d)$ further operations over $\F_p$, each taking $\poly(\log p)$ time.
\end{proof}

\subsection{From Field Elements to Bits}\label{sec:fieldtobits}

In this section, we give a compiler that transforms a PIR scheme that operates over databases whose entries are field elements, to databases whose entries are bits.
The main workhorse is the following theorem by~\cite{DBLP:conf/stoc/DodisPT10}:

\begin{theorem}[\cite{DBLP:conf/stoc/DodisPT10}]\label{thm:dpt}
    Let $K$ be an alphabet size and $N$ a number of records. Then, there is an absolute constant $c > 0$ such that, for all integers $K, N \geq 2$, there exists an integer
    $ L \le
        \lceil N\log_2 K + cN/\sqrt{K} \rceil
    $
    and deterministic algorithms
    $\mathsf{Encode}_{K,N}: [K]^N \to \{0,1\}^{L}$ and 
    $\mathsf{Decode}_{K,N}: [N]\times\{0,1\}^{L}\to[K]$,
    such that:
    \begin{itemize}
        \item For every $\vecv\in[K]^N$ and $i\in[N]$,
        $
            \mathsf{Decode}_{K,N}
            (i,\mathsf{Encode}_{K,N}(\vecv))=v_i.
        $
        \item Given random access to the encoded string,
        $\mathsf{Decode}_{K,N}$ uses
        $\poly(\log N, \log K)$ bit operations.
        \item $\mathsf{Encode}_{K,N}$ uses $N \cdot \poly(\log N, \log K)$
        bit operations.
    \end{itemize}
\end{theorem}

To encode the servers' tables of $\F_p$-elements compactly with~\Cref{thm:dpt}, we need a prime $p$ that is very close to a power of $2$. The following lemma finds one in a given residue class; we use it both for odd primes $q$ and for $q = 2$, in which case the condition $p \equiv 1 \pmod 2$ just says that $p$ is odd.

\begin{lemma}[Primes just above a power of two]\label{lem:goodprime}
Fix a prime $q$. There is a randomized algorithm that takes as input a positive integer $M$ and a sufficiently large integer $n$, and constructs an integer $B$ and a prime $p \equiv 1 \pmod{q}$ such that:
\[
    p > M,\qquad
    p\ge n^4,\qquad
    2^B<p\le 2^B\left(1+\frac1{4n}\right),
    \qquad
    B = O(\log M + \log^3 n).
\]
This algorithm runs in expected time $\poly(B)$.
\end{lemma}

\begin{proof}
Set
\[
    B:=
    \max\left(
        \lceil\log_2(M+1)\rceil,\,
        (1+\lceil\log_2 n\rceil)^3
    \right),
    \qquad
    x:=2^B,
    \qquad
    H:=\left\lfloor\frac{x}{4n}\right\rfloor.
\]
Then $x>M$ and $x\ge n^4$. 
Recalling that $q$ is fixed, it is straightforward to check that the prime number theorem in arithmetic
progressions, with its error term $O(x e^{-c\sqrt{\ln x}})$ for a fixed modulus~\cite[Chapter~22, p.~133]{Davenport80}, implies that:
\[
    \#\{p\in(x,x+H]:
          p\text{ prime},\ p\equiv1\pmod q\}
    =
    (1+o(1))\frac{H}{(q-1)\ln x}.
\]
Thus, among the integers in this interval congruent to
$1$ modulo $q$, a fraction $\Theta_q(1/B)$ are prime. 
As a result, we can obtain a prime $p$ with $x < p \leq x+H$ and $p \equiv 1 \pmod{q}$ by rejection sampling. For this $p$, we have $p > M$ and:
\[
    n^{\omega(1)} \le 2^B < p\le2^B\left(1+\frac1{4n}\right).
\]
\end{proof}

In the following lemma, we show how to handle databases of $n$-bits by generically reducing to handling databases of $\FF_p$-records where $p$ is very close to a power of 2. The lemma takes as input PIR schemes over databases of $b^j$ records, for a constant $b$: we use $b = q$ for the schemes of~\Cref{thm:s-server-rate-one}, and $b = 16$ for the two-server scheme of~\Cref{thm:2-server-rate-one} (see the proof of~\Cref{mainthm:track1}).

\begin{lemma}[PIR for $n$-bit databases]\label{lem:bit-compiler}
Fix integers $s, b \geq 2$. 
Let $n$ be a sufficiently large database size (in bits), and let $d$ be the largest integer with $b^d \leq n$.
Suppose there are integers $B,p$ with $p$ a prime, such that:
\[
    p \ge n^4,
    \qquad
    2^B\le p\le 2^B\left(1+\frac{1}{4n}\right),
    \qquad
    B, \log p \leq \polylog(n).
\]
Assume additionally that, for every $j \in [d]$, there is an information-theoretically private $s$-server PIR scheme over databases of $b^j$ records in $\FF_p$, where each server stores $b^j$ elements of $\FF_p$, computed in $b^j \cdot \polylog(n)$ time, the user uploads $O(\log n)$ bits, and each server reads and sends back at most $k$ elements of $\FF_p$, in $k \cdot \polylog(n)$ time.
Then, there is an information-theoretically private
$s$-server PIR scheme for arbitrary $n$-bit databases
with:
\begin{itemize}[nosep]
    \item at most $n+1$ stored bits per server;
    \item $n \cdot \polylog (n)$ preprocessing time;
    \item $O(\log^2 n)$ upload communication; and
    \item $k \cdot \polylog(n)$ server time and
    download communication.
\end{itemize}
In particular, if $k\le n^{\eta+o(1)}$ for a constant
$\eta$, this is an $s$-server scheme with exponent $\eta$ (\Cref{def:exponent}).
\end{lemma}

\begin{proof}
First, it suffices to deal with the case where $n$ is divisible by $B$: we can use such a protocol for the first $B \lfloor n/B \rfloor$ bits of the database, while the remaining $\leq B = \polylog(n)$ bits can just be sent regardless of the client's query.

Bearing this in mind, let $n' = n/B$. Interpret the database as $n'$ records in $\{0, 1, \ldots, 2^B-1\}$, which can trivially be lifted to $n'$ records in $\FF_p$. Now we use the base-$b$ decomposition of $n'$ to chunk the database into mini-databases of size $b^{d_1} \geq b^{d_2} \geq \ldots \geq b^{d_r}$, with $d_1, \ldots, d_r \geq 1$, together with fewer than $b$ leftover records (from the last digit), which can also be sent regardless of the client's query. This is not a new idea e.g., see~\cite[Corollary 3.3]{HR26}. It is easy to see that we have $r \leq O(\log n)$ and $d_1, \ldots, d_r \leq d$. We will apply the given PIR schemes to each of these mini-databases. (The client only needs a record from one of the mini-databases, but they will send dummy queries for the others to ensure privacy.) A few efficiency claims become clear already:
\begin{itemize}
    \item The upload is $O(r\log n) = O(\log^2 n)$.
    \item The download communication is at most $O(kr \log p) + O(b \log p + B) \leq k \cdot \polylog(n)$.
\end{itemize}
It remains to address storage and server time: if we concatenate the data structures of the given schemes for each mini-database, together with the leftover records, each server needs to store $n'$ records over $\FF_p$. To encode this into bits, each server uses the encoding given by~\Cref{thm:dpt} to store their data structure. The resulting encoding has size at most:
\begin{align*}
    \lceil n'\log_2 p + O(n'/\sqrt{p}) \rceil &\leq \lceil n'B + n'\log(1+1/(4n)) + O(n'/n^2)\rceil \\
    &= n + \left\lceil \frac{n}{B}\log(1+1/(4n)) + O(1/(nB))\right\rceil \\
    &\leq n+\left\lceil \frac{1}{4B\ln 2} + O\left(\frac{1}{nB}\right)\right\rceil \\
    &= n+1,
\end{align*}
for sufficiently large $n$. Here, we are using the fact that the conditions force $B \geq \omega(1)$. 
This implies the claimed storage. To access a stored $\FF_p$-element from this encoding, each server needs to make $\poly(\log n, \log p) = \polylog(n)$ accesses to the encoding, giving the claimed $k \cdot \polylog(n)$ server time. Finally, preprocessing computes the data structures of the mini-databases in $n' \cdot \polylog(n)$ time, and then their encoding in $n \cdot \polylog(n)$ time (\Cref{thm:dpt}).
\end{proof}

%% file: fig2server.tex
\begin{figure}[]
{\FrameSep6pt
\begin{framed}
{\small
\textbf{Two-server doubly-efficient PIR.}
\label{const:2pir}%
Parameterized by  positive integer $d \equiv 2 \pmod4$ and  odd prime~$p > d$. The PIR scheme operates over databases consisting of $2^d$ records in $\F_{p}$. Define the set~$\calW\subseteq\F_2^d$ and the coefficients~$\{c_{\vec w}\}_{\vec w\in\calW}$ as  in \cref{sec:depir:2pir}. For vectors~$\vec x\in\F_2^d$, define the quadratic function in $\F_2^d \to \F_2$ as:
\vspace{-0.5em}\[
 Q(\vec x) = \sum_{1\leq i\leq j\leq d}x_ix_j.%
\]

\vspace{-0.5em}\hrulefill\vspace{-1em}
\setlength{\columnsep}{0.5cm}
\begin{multicols*}{2}

$\Preprocess(\DB\in \F_{p}^{2^d})\to\T_0,\T_1$
\begin{itemize}[leftmargin=4mm,itemsep=0.35em]
 \item Split $\DB$ into $2^d$ records
 $\{\DB_{\vec v}\}_{\vec v\in\F_2^d}$ each in $\F_{p}$.
 \item For $t\in\{0,1\}$, let the function $f_t: \F_2^d \to \F_{p}$ be
 $$f_t(\vec x)=\sum_{\vec v\in\F_2^d}
 \DB_{\vec v}\cdot
 (-1)^{Q(\vec x + \vec v)  + t \cdot Q(\vecv)}.$$
 \item Set $\T_t$ to be the truth table of the function $f_t$.
\end{itemize}

\bigskip

$\Query(i \in [2^d])\to\state,\query_0,\query_1$
\begin{itemize}[leftmargin=4mm,itemsep=0.35em]
\item Let $\vec i\in\F_2^d$ be the bit-vector corresponding to index~$i$.
 \item Sample $\vec r\rgets\F_2^d$ and set $\query_0=\vec r$ and $\query_1=\vec r+\vec i$.
 \item Output $\state=(\vec i,\vec r)$, $\query_0$, and $\query_1$.
\end{itemize}

\vfill\null
\columnbreak

$\Answer^{\T_t}(\query_t\in\F_2^d)\to\ans_t \in \F_{p}^{|\calW|}$
\begin{itemize}[leftmargin=4mm,itemsep=0.35em]
 \item Output $\ans_t = \left\{\T_t[\query_t+\vec w] \right\}_{\vec w \in \calW}$.
\end{itemize}

\bigskip

$\Recon(\state,\ans_0,\ans_1)\to \F_{p}$
\begin{itemize}[leftmargin=4mm,itemsep=0.35em]
 \item Parse $(\vec i,\vec r)$ from $\state$ and $\T_t[\query_t+\vec w]$ from each $\ans_t$.
 \item Output the $\F_{p}$-element  computed as
 \begin{align*}
 \frac12 \sum_{t \in \zo}\sum_{\vec w\in\calW} c_{\vec w} \,
 (-1)^{Q(\vec r + (t+1) \cdot \vec i +\vec w) + t \cdot Q(\vec i)} \, \T_t[\query_t+\vec w].
 \end{align*}
\end{itemize}

\vfill\null
\end{multicols*}
}
\vspace{-2em}
\end{framed}}
\caption{Two-server doubly-efficient PIR, over database consisting of $2^d$ records in $\F_p$.}
\label{fig:2pir}
\end{figure}

%% file: figsserver.tex
\begin{figure}[!t]
\begingroup
\setlength{\FrameSep}{6pt}
\begin{framed}
\small
\setlength{\abovedisplayskip}{5pt}
\setlength{\belowdisplayskip}{5pt}
\setlength{\abovedisplayshortskip}{3pt}
\setlength{\belowdisplayshortskip}{3pt}
\noindent\textbf{Doubly-efficient PIR with $s$ servers.}
Parameterized by a prime $q\geq\max\{3,s\}$, an integer $d$, and a prime $p$ such that the field $\F_p$ contains an element $\gamma$ of order $q$. The PIR scheme operates over databases consisting of $q^d$ records in $\F_p$. Define the sets $\calW$ and $\calT$ (with $|\calT| = s$) and the coefficients $\left\{ c_{\vec w, t} \in \F_p: \vec w \in \calW, t \in \calT \right\}$ as in \cref{sec:depir:spir}. They satisfy, for any $\vec x\in\F_q^d$,
\[
 \sum_{t\in\calT}\sum_{\vec w\in\mathcal W}
 c_{\vec w,t}\cdot \gamma^{\langle\vec w,\vec x\rangle+t \cdot \langle \vec x, \vec x \rangle}
 = \begin{cases}1,\text{ if }\vec x = \vec 0\\0, \text{ else.}\end{cases}
\]
\hrule\medskip
\noindent
\begin{minipage}[t]{0.485\linewidth}
\textbf{$\mathsf{Preprocess}(\db \in \F_p^{q^d})\to(\T_t)_{t\in\calT}$}
\begin{itemize}[leftmargin=4mm,itemsep=0.45em,topsep=0.4em]
 \item Split $\db$ into $q^d$ records $\left\{ \db_{\vec v} \right\}_{\vec v \in \F_q^d}$, each in $\F_p$.
 
 \item For $t\in\calT$, let the function $f_t: \F_q^d \to \F_p$ be
 \[
 f_t(\vec x)=\sum_{\vec v\in\F_q^d}\db_{\vec v}\cdot
 \gamma^{\langle\vec x,\vec v\rangle+
              t \cdot \langle \vec v, \vec v \rangle}.
 \]

 \item Set $\T_t$ to be the truth table of the function $f_t$.
\end{itemize}

\bigskip
\textbf{$\mathsf{Query}(i \in [q^d])\to\mathsf{st},(\query_t)_{t \in \calT}$}
\begin{itemize}[leftmargin=4mm,itemsep=0.45em,topsep=0.4em]
 \item Let $\vec i\in\F_q^d$ be the vector corresponding to index $i$.

 \item Sample $\vec r\rgets\F_q^d$ and set $\query_t=\vec r-2t\cdot\vec i$ for $t\in\calT$.

 \item Output $\state=(\vec i,\vec r)$ and $(\query_t)_{t\in\calT}$.
\end{itemize}
\end{minipage}\hfill
\begin{minipage}[t]{0.485\linewidth}
\textbf{$\mathsf{Answer}^{\T_t}(\query_t \in \F_q^d)\to\mathsf{ans}_t \in \F_p^{|\calW|}$}
\begin{itemize}[leftmargin=4mm,itemsep=0.45em,topsep=0.4em]
 \item Output $\ans_t 
 = \left\{\T_t[\query_t+\vec w]\right\}_{\vec w \in \calW}.$
\end{itemize}

\bigskip
\textbf{$\mathsf{Reconstruct}(\state,(\mathsf{ans}_t)_{t \in \calT}) \to \F_p$}
\begin{itemize}[leftmargin=4mm,itemsep=0.45em,topsep=0.4em]
 \item Parse $(\vec i,\vec r)$ from $\state$
 and $\T_t[\query_t+\vec w]$ from each $\ans_t$.

 \item Output the $\F_p$-element computed as
 \[
 \begin{aligned}
 \sum_{t\in\calT}\sum_{\vec w\in\mathcal W}
 c_{\vec w,t} \cdot \T_t[\query_t+\vec w] \cdot
 \gamma^{\langle t \cdot \vec i-\vec r-\vec w,\vec i\rangle}.
 \end{aligned}
 \]
\end{itemize}
\end{minipage}
\end{framed}
\endgroup
\caption{Doubly-efficient PIR with $s$ servers,  over databases consisting of $q^d$ records in $\F_p$.}
\label{fig:spir}
\end{figure}

%% file: figq3shells.tex
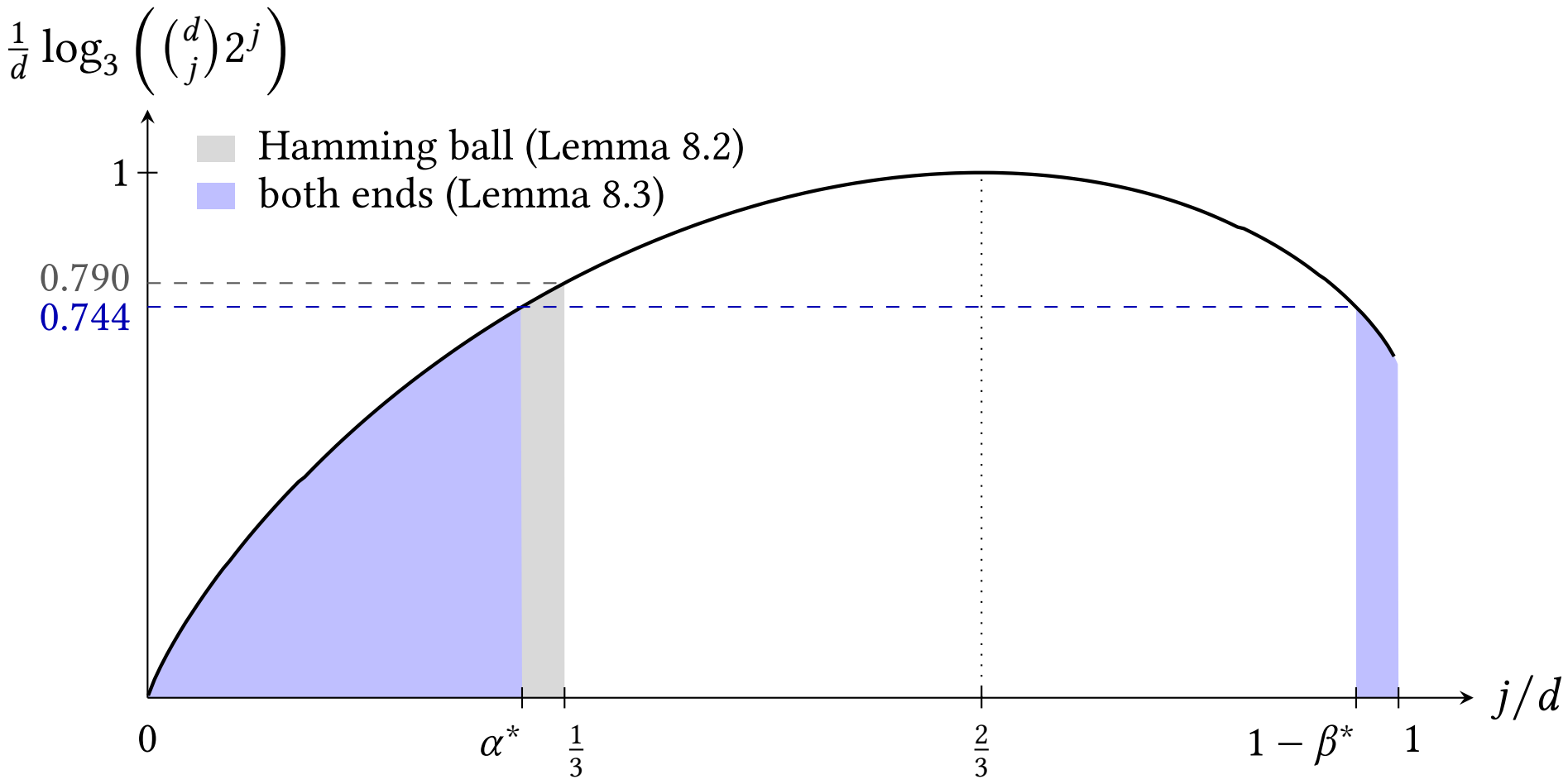
\begin{figure}[ht]
    \centering
    \begin{tikzpicture}[x=10cm, y=4.2cm, >=stealth,
        declare function={g(\t) = ((-\t*ln(\t) - (1-\t)*ln(1-\t))/ln(2) + \t)/1.5849625;}]
        \def\as{0.2994}
        \def\bs{0.9661}
        \fill[gray!30] (0,0) -- plot[domain=0.0005:0.33333, samples=80] (\x, {g(\x)}) -- (0.33333,0) -- cycle;
        \fill[blue!25] (0,0) -- plot[domain=0.0005:\as, samples=80] (\x, {g(\x)}) -- (\as,0) -- cycle;
        \fill[blue!25] (\bs,0) -- plot[domain=\bs:0.9995, samples=40] (\x, {g(\x)}) -- (1,0) -- cycle;
        \draw[->] (0,0) -- (1.06,0) node[right] {$j/d$};
        \draw[->] (0,0) -- (0,1.12) node[above] {$\frac{1}{d}\log_3 \left(\binom{d}{j} 2^j\right)$};
        \draw[thick] plot[domain=0.0005:0.9995, samples=200] (\x, {g(\x)});
        \draw[gray!70!black, dashed] (0,0.78969) -- (0.33333,0.78969);
        \draw[blue!70!black, dashed] (0,0.74444) -- (\bs,0.74444);
        \node[left, gray!70!black] at (0,0.80) {\small $0.790$};
        \node[left, blue!70!black] at (0,0.725) {\small $0.744$};
        \node[left] at (0,1) {\small $1$};
        \draw (-0.008,1) -- (0.008,1);
        \foreach \t in {\as, 0.33333, 0.66667, \bs, 1} {
            \draw (\t,0.02) -- (\t,-0.02);
        }
        \node[anchor=north east, xshift=4pt] at (\as,-0.02) {\small $\alpha^*$};
        \node[anchor=north west, xshift=-4pt] at (0.33333,-0.02) {\small $\frac{1}{3}$};
        \node[below] at (0.66667,-0.02) {\small $\frac{2}{3}$};
        \node[anchor=north east, xshift=4pt] at (\bs,-0.02) {\small $1 - \beta^*$};
        \node[anchor=north west, xshift=-3pt] at (1,-0.02) {\small $1$};
        \node[below] at (0,-0.02) {\small $0$};
        \draw[dotted] (0.66667,0) -- (0.66667,1);
        \fill[gray!30] (0.04,1.02) rectangle (0.07,1.07);
        \node[right] at (0.075,1.045) {\small Hamming ball (\Cref{lem:q3ball})};
        \fill[blue!25] (0.04,0.93) rectangle (0.07,0.98);
        \node[right] at (0.075,0.955) {\small both ends (\Cref{lem:q3ends})};
    \end{tikzpicture}
    \caption{Sizes of the Hamming shells of $\ZZ_3^d$. The shell of weight $j = td$ has $\binom{d}{j} 2^j = 3^{(g(t) + o(1))d}$ points, where $g(t) := (H_2(t) + t)/\log_2 3$; the profile $g$ peaks at $g(\frac{2}{3}) = 1$. Both constructions use $\lfloor d/3 \rfloor + 1$ shells, and their Fourier sparsity is governed by the largest shell they use. The Hamming ball of~\Cref{lem:q3ball} uses the weights up to $\frac{d}{3}$ (gray), where the shells reach $3^{0.790d}$. The construction of~\Cref{lem:q3ends} trades the gray shells between $\alpha^* d$ and $\frac{d}{3}$ for the top $\beta^* d$ shells (blue), where the shells are small again; balancing the two ends ($\alpha^* \approx 0.2994$, $\beta^* = \frac{1}{3} - \alpha^* \approx 0.0339$) lowers the largest shell to $3^{0.744d}$ (\Cref{cor:q3ends}).}
    \label{fig:q3shells}
\end{figure}

%% file: math.tex
\part{Fourier-Sparse Cone-Delta Functions from Modular Forms}
\label{sec:math}

This part proves the existence results that power the PIR schemes of Part~I: cone-delta functions and quadratic-phase delta expansions (\Cref{def:conedelta,def:genconedelta}) over $\ZZ_q^d$ whose Fourier supports are exponentially smaller than the trivial $q^d$. Throughout this part, $\calQ := \{\xv \in \ZZ_q^d : \langle \xv, \xv \rangle \equiv 0 \pmod{q}\}$ denotes the cone, the set on which cone-delta functions are constrained (\Cref{def:conedelta}). We prove two such results: a warm-up (\Cref{cor:gen1specific}), producing the cone-delta functions behind $q$-server PIR (\Cref{mainthm:track2}), and a stronger theorem (\Cref{thm:gen4aprime}, with its support bounds in~\Cref{cor:gen4exponent}), which trades the number of servers $s$ against the support exponent.

Both proofs follow the architecture of the technical overview (\Cref{sec:overview:arch}), which mirrors the reconstruction argument for Reed--Muller PIR: if the desired function does not exist, duality provides a witness; composing a polynomial built from the witness with a vector-valued theta function produces scalar modular forms; and the witness forces these forms to vanish at the cusps to higher order than the valence formula allows. Recall the dual form of the Reed--Muller argument. There, the database is encoded as a polynomial $F$ of degree at most $D$ in several variables. To read $F(\xv)$, the user picks a uniformly random direction $\mathbf{h}$, sends the points $\xv + j\mathbf{h}$ for $j \in [s]$ to $s \geq D + 1$ servers, and recovers $F(\xv)$ by interpolating $F$ along the line $j \mapsto \xv + j\mathbf{h}$. Interpolation could fail only if some polynomial $\Delta$ of degree at most $D$ had $\Delta(\xv) = 1$ while vanishing at all $s$ query points; restricted to the line, such a $\Delta$ would be a univariate polynomial of degree at most $D$ that is nonzero at $j = 0$ yet has $s > D$ roots, which is impossible. \Cref{tab:rmanalogy} lines up the two arguments in the case of the warm-up, and~\Cref{fig:gen1flow,fig:gen4flow} show how the steps fit together in each of our two proofs. The stronger construction adds an elliptic variable to the theta function, which yields a second scalarization and finer bounds on the orders at the cusps, so the number of servers can drop below $q$ (\Cref{tab:gen4comparison}).
\begin{table}[t]
    \centering
    \small
    \renewcommand{\arraystretch}{1.3}
    \begin{tabular}{p{0.17\textwidth} p{0.33\textwidth} p{0.40\textwidth}}
        \toprule
        Step & Reed--Muller PIR & Warm-up (\Cref{sec:gen1}) \\
        \midrule
        1. Witness & $\Delta$ of degree $\leq D$, with $\Delta(\xv) = 1$ and $\Delta(\xv + j\mathbf{h}) = 0$ for $j \in [s]$ & $\beta$ on the cone, with $\beta_\zerov = 1$ and Fourier sums vanishing on $S_d$; the polynomial $P_\beta$ \\
        2. Curve & the line $j \mapsto \xv + j\mathbf{h}$ & the theta function $\tau \mapsto \Theta_\calL(\tau) \in \CC^q$, for $\tau \in \calH$ \\
        3. Restriction & $g(j) = \Delta(\xv + j\mathbf{h})$, of degree $\leq D$ & $P_\beta \circ \Theta_\calL$, a modular form of weight $dr/2$ \\
        4. Nonvanishing & $g(0) = 1$ & a nonzero multiple of $\beta_\zerov = 1$ as $\tau \to i\infty$ \\
        \phantom{4.} Vanishing & at $j = 1, \ldots, s$ & to order more than $\frac{dr(q+1)}{24}$ at the cusp $0$ \\
        \phantom{4.} Root bound & at most $D$ roots & valence formula: total order $\frac{dr(q+1)}{24}$ \\
        \bottomrule
    \end{tabular}
    \caption{The reconstruction argument for Reed--Muller PIR, in the dual form recalled at the start of~\Cref{sec:math}, side by side with our proof that sparse cone-delta functions exist, in the case of the warm-up. On the left, $\Delta$ is a polynomial of degree at most $D$, $\xv$ is the queried point, $\mathbf{h}$ is the user's random direction, and the $s \geq D + 1$ servers receive the points $\xv + j\mathbf{h}$. On the right, $\calL$ is a lattice of rank $r$, and $S_d$ is the set of frequency vectors of small total cost.}
    \label{tab:rmanalogy}
\end{table}

\paragraph{Relation to the literature.} Almost all of the results on modular forms that we use in this part are classical. We restate them, often with proofs, only to keep the presentation self-contained for readers outside the area, and we point to sources along the way. This includes the background of~\Cref{sec:pt2bg} (see, e.g.,~\cite{Serre73,DiamondShurman05,Shimura71}); the Weil representation and its action on the finite Heisenberg group~\cite{Weil64,Gerardin77,Stromberg13}; the transformation laws of lattice theta functions~\cite[Theorem~4.1]{Borcherds98} and of theta functions with characteristics~\cite[Chapter~I, \S\S3 and~7]{Mumford83}; and the facts that restricting a Jacobi form to a torsion point, or taking the first nonvanishing coefficient of its Taylor expansion in the elliptic variable, produces modular forms~\cite[Theorem~1.3 and Section~3]{EichlerZagier85}. Combining such restrictions and Taylor coefficients with norms and the valence formula in order to bound orders of vanishing is also an established technique~\cite{RichterWesterholtRaum15,Aoki22,LiWang25}. Likewise, substituting a vector of theta functions into a suitably invariant polynomial (\Cref{lem:scalarisation}) is also how the classical correspondence between weight enumerators of codes and theta series of lattices arises~\cite{BroueEnguehard72,NebeRainsSloane06}.

What we believe is new is the connection itself: the reduction of the existence of Fourier-sparse delta functions to the impossibility of a dual witness, and the encoding of the witness's orthogonality constraints as orders of vanishing of modular forms at cusps --- in the stronger construction, via the symmetrized witness, the per-class quantities attached to the selected cusp classes, and the analysis of the edge polynomials in Step~4. To our knowledge, modular forms have not previously been used to construct Fourier-sparse delta functions. The closest arguments we know of are the classical bounds on the minimum distance of self-dual codes~\cite{MallowsSloane73}, which also substitute theta functions into an invariant polynomial and bound its order of vanishing; there the polynomial is a weight enumerator, whereas ours comes from a dual witness (\Cref{sec:rel}). For $q = 3$, the conclusion of the warm-up also follows from an elementary construction of Alon and Lubetzky (\Cref{sec:gen1}).

\Cref{sec:pt2bg} collects the background: $\SL_2(\ZZ)$ and its congruence subgroups, the Weil representation, modular forms and their cusp expansions, and the valence formula in the form we use. \Cref{sec:gen1} carries out the warm-up via even lattices (for primes $q \equiv 3 \pmod 4$), and~\Cref{sec:gen4} proves the stronger theorem. Standard complex-analysis facts and routine or lengthy verifications are deferred to the appendix, beginning with~\Cref{sec:complexanalysis}.

\section{Background: $\SL_2(\ZZ)$, the Weil Representation, and Modular Forms}\label{sec:pt2bg}

Everything in this section is standard; we restate it to fix notation and normalizations, and to keep the presentation self-contained. For modular forms for congruence subgroups, see, e.g.,~\cite{Serre73,DiamondShurman05,Shimura71}; for vector-valued modular forms for the Weil representation, see, e.g.,~\cite[Section~4]{Borcherds98}. Standard facts from complex analysis that we use freely are collected in~\Cref{sec:complexanalysis}.

For readers coming from the technical overview, this section makes precise the objects sketched in~\Cref{sec:overview:weil,sec:overview:mf}, and we point back to the figures there along the way. Each piece has a fixed role in the proofs of~\Cref{sec:gen1,sec:gen4} (see the flow charts in~\Cref{fig:gen1flow,fig:gen4flow}). The Weil representation (\Cref{sec:weil}) is the symmetry under which both of our theta functions transform. The upper half-plane and its cusps (\Cref{sec:upperhalfplane}) are where these functions live, and the expansions at the cusps (\Cref{sec:cuspinfty,sec:modularforms}) define the orders of vanishing that our proofs bound. The composition laws (\Cref{sec:compositionlaws}) contain the scalarization step, and the valence formula (\Cref{sec:valenceformula}) is the root count that ends both proofs.

\subsection{$\SL_2(\ZZ)$ and Congruence Subgroups}

\begin{definition}
    The group $\SL_2(\ZZ)$ consists of matrices $\begin{bmatrix} a & b \\ c & d_0 \end{bmatrix}$ with integer entries and determinant $ad_0-bc = 1$. Throughout this part, we write a generic element of $\SL_2(\ZZ)$ as $\begin{bmatrix} a & b \\ c & d_0 \end{bmatrix}$, reserving the letter $d$ for dimension and degree parameters. We single out the two elements
    $$S := \begin{bmatrix} 0 & -1 \\ 1 & 0\end{bmatrix} \qquad\text{and}\qquad T := \begin{bmatrix} 1 & 1 \\ 0 & 1 \end{bmatrix}.$$
\end{definition}

\begin{fact}\label{fact:generation}
    $\SL_2(\ZZ)$ is generated by $S$ and $T$. (Serre~\cite[Chapter~VII, Theorem~2]{Serre73} proves this for $\PSL_2(\ZZ) = \SL_2(\ZZ)/\{\pm I\}$; the statement for $\SL_2(\ZZ)$ follows since $-I = S^2 \in \langle S, T \rangle$.) Moreover, direct computation shows that $S^2 = (ST)^3 = -I$; in particular, $S$ and $T$ obey the relations $S^4 = I$ and $(ST)^3 = S^2$.
\end{fact}

Geometrically, $S$ and $T$ are the two steps of the Euclidean algorithm, applied to a basis of a lattice in $\CC$ (\Cref{sec:overview:mf} and~\Cref{fig:lattice}); this is how $\SL_2(\ZZ)$ will act on the upper half-plane in~\Cref{sec:upperhalfplane}.

Of particular interest to us will be considering particular subgroups of $\SL_2(\ZZ)$:
\begin{definition}\label{def:subgroups}
    For an odd prime $q$, we define the following subgroups of $\SL_2(\ZZ)$:
    \begin{align*}
        \Gamma(q) &= \left\{\begin{bmatrix} a & b \\ c & d_0 \end{bmatrix} \equiv \begin{bmatrix} 1 & 0 \\ 0 & 1 \end{bmatrix} \pmod{q}\right\} \\
        \Gamma_0(q) &= \left\{\begin{bmatrix} a & b \\ c & d_0 \end{bmatrix} \equiv \begin{bmatrix} * & * \\ 0 & * \end{bmatrix} \pmod{q}\right\}.
    \end{align*}
    A \emph{congruence subgroup of $\SL_2(\ZZ)$ of level $q$} is any subgroup that contains $\Gamma(q)$; in particular, $\SL_2(\ZZ)$ itself and $\Gamma_0(q)$ are congruence subgroups of level $q$. Equivalently, a subgroup $\Gamma$ is a congruence subgroup of level $q$ if and only if a matrix's membership in $\Gamma$ only depends on the values of its entries modulo $q$: if $\Gamma \supseteq \Gamma(q)$ and $g \equiv g' \pmod{q}$, then $g'g^{-1} \in \Gamma(q)$ and so $g \in \Gamma \Leftrightarrow g' \in \Gamma$; conversely, a subgroup with the latter property contains every matrix congruent to $I$.
\end{definition}

The two subgroups play complementary roles for us. The Weil representation of~\Cref{sec:weil} is trivial on $\Gamma(q)$, and on $\Gamma_0(q)$ it merely permutes the coordinates of $\CC^q$ and multiplies them by phases (\Cref{fact:weilrep}). So $\Gamma_0(q)$ is the group of the warm-up (\Cref{sec:gen1}), where the phases cancel on the cone, and $\Gamma(q)$ is the group of the stronger construction (\Cref{sec:gen4}), which works with this smaller group at the cost of many more cusps (\Cref{fig:gen4cusps}).

\begin{remark}\label{remark:minusI}
    We do \emph{not} require a congruence subgroup to contain $-I$; indeed, $-I \notin \Gamma(q)$. Because $-I$ will act trivially on all the geometric objects introduced below, some statements (most importantly the valence formula,~\Cref{thm:valence}) are stated for subgroups containing $-I$; to apply them to $\Gamma(q)$, we will pass to the congruence subgroup
    $$\pm\Gamma(q) := \langle \Gamma(q), -I \rangle = \Gamma(q) \cup (-I)\Gamma(q).$$
\end{remark}

\begin{lemma}[Indices]\label{lem:indices}
    Let $q$ be an odd prime. The entrywise-reduction map $\pi: \SL_2(\ZZ) \to \SL_2(\ZZ_q)$ is a surjective group homomorphism with kernel $\Gamma(q)$, and $|\SL_2(\ZZ_q)| = q^3-q$. Consequently:
    \begin{itemize}
        \item $[\SL_2(\ZZ) : \Gamma(q)] = q^3-q$;
        \item $[\SL_2(\ZZ) : \pm\Gamma(q)] = \frac{q^3-q}{2}$; and
        \item $[\SL_2(\ZZ) : \Gamma_0(q)] = q+1$.
    \end{itemize}
\end{lemma}
\begin{proof}
    The surjectivity of $\pi$ and the order $|\SL_2(\ZZ_q)| = q^3 - q$ are standard; see, e.g.,~\cite[Section~1.2]{DiamondShurman05} (stated there for general level). The indices follow at once: $\Gamma(q) = \pi^{-1}(I)$ is the kernel, so $[\SL_2(\ZZ) : \Gamma(q)] = |\SL_2(\ZZ_q)|$; the group $\pm\Gamma(q)$ is the disjoint union of the two cosets $\Gamma(q)$ and $(-I)\Gamma(q)$; and $\Gamma_0(q) = \pi^{-1}(B)$ for the subgroup $B$ of upper-triangular matrices $\begin{bsmallmatrix} a & b \\ 0 & a^{-1} \end{bsmallmatrix}$ in $\SL_2(\ZZ_q)$, which number $q(q-1)$, so that (by surjectivity of $\pi$, which puts the cosets of $\Gamma_0(q)$ and of $B$ in bijection) the index is $(q^3-q)/(q(q-1)) = q+1$.
\end{proof}

\subsection{The Weil Representation}\label{sec:weil}

We first fix some notation that will be in force throughout this part: $\zeta_q := e^{2\pi i/q}$, so that $\zeta_q^t$ is well-defined for $t \in \ZZ_q$, and $\{\ev_x\}_{x \in \ZZ_q}$ denotes the standard basis of $\CC^q$, indexed by $\ZZ_q$.

\begin{definition}[Representation]\label{def:representation}
    A \emph{representation} of a group $G$ on $\CC^N$ is a map $\rho: G \to \GL_N(\CC)$ satisfying $\rho(gh) = \rho(g)\rho(h)$ for all $g, h \in G$. (It follows that $\rho(1) = I$ and $\rho(g^{-1}) = \rho(g)^{-1}$.) When $N = 1$, we identify $\GL_1(\CC)$ with $\CC \setminus \{0\}$ and call $\rho$ \emph{one-dimensional}.
\end{definition}

\begin{definition}[Legendre symbol]\label{def:legendre}
    For $x \in \ZZ_q$, define $\left(\frac{x}{q}\right) := 1$ if $x$ is a nonzero square in $\ZZ_q$, $-1$ if $x$ is not a square in $\ZZ_q$, and $0$ if $x = 0$. For an integer $x$, define $\left(\frac{x}{q}\right)$ by first reducing $x$ modulo $q$.
\end{definition}

\begin{fact}\label{fact:legendre}
    The Legendre symbol is multiplicative, i.e., $\left(\frac{xy}{q}\right) = \left(\frac{x}{q}\right)\left(\frac{y}{q}\right)$. (This is a standard consequence of the cyclicity of $\ZZ_q^*$; see, e.g.,~\cite[Chapter~5]{IrelandRosen90}.)
\end{fact}

\begin{fact}[Quadratic Gauss sums]\label{fact:gausssum}
    Let $q$ be an odd prime and let $\mu$ be an integer with $\mu \not\equiv 0 \pmod{q}$. Then there exists $\epsilon_{\mu, q} \in \{\pm 1, \pm i\}$ (so in particular $\epsilon_{\mu, q}^4 = 1$ and $|\epsilon_{\mu, q}| = 1$) such that
    $$\sum_{y \in \ZZ_q} \zeta_q^{\mu y^2} = \epsilon_{\mu, q}^{-1} \sqrt{q}.$$
    (This is classical: \cite[Chapter~6]{IrelandRosen90} shows that the sum equals $\left(\frac{\mu}{q}\right) g$ for a constant $g = \sum_{x \in \ZZ_q} \left(\frac{x}{q}\right) \zeta_q^{x}$ independent of $\mu$, with $g^2 = \left(\frac{-1}{q}\right) q$; so the square of the sum is $\pm q$, which gives the claim.)
\end{fact}

\begin{definition}\label{def:weil}
    Fix an integer $\mu$ with $\mu \not\equiv 0 \pmod{q}$.
    Define the matrices $\rho_S, \rho_T \in \CC^{q \times q}$ (depending on $q$ and $\mu$) as follows:
    \begin{itemize}
        \item $\rho_T$ is diagonal, mapping
        $$\ev_x \mapsto \zeta_q^{\mu x^2} \ev_x.$$

        \item $\rho_S$ is a multiple of the $q$th order discrete Fourier transform, mapping:
        $$\ev_x \mapsto \frac{\epsilon_{\mu, q}}{\sqrt{q}} \sum_{y \in \ZZ_q} \zeta_q^{-2\mu xy} \ev_y.$$
    \end{itemize}
\end{definition}

The technical overview (\Cref{sec:overview:weil}) explains why these two matrices are the natural ones for us: $\rho_T$ multiplies by the quadratic phase that defines the cone, and $\rho_S$ is a Fourier transform, so the two conditions defining a cone-delta function are statements about $\rho_T$ and $\rho_S$. There we used the matrices without the constant $\frac{\epsilon_{\mu, q}}{\sqrt{q}}$ and with $\mu = 1$; with this constant and parameter $\mu$, the following is a representation, not only a representation up to scalars. (\Cref{fig:fourieranalogy} compares the Weil representation with the characters of Fourier analysis.)

The matrices $\rho_T$ and $\rho_S$ extend to a representation of $\SL_2(\ZZ)$, the Weil representation, which is trivial on $\Gamma(q)$ and monomial on $\Gamma_0(q)$. This is classical, and we restate it in the form we need. In the language of finite quadratic modules, $\rho$ is the Weil representation attached to the quadratic module $(\ZZ_q, x \mapsto \mu x^2/q)$, and~\Cref{def:weil} is exactly the normalization of Str\"omberg~\cite[Definition~5.1]{Stromberg13} (Scheithauer~\cite{Scheithauer09} works with the dual representation); see also~\cite[Section~4]{Borcherds98} and, for the finite-field setting, \cite{Weil64,Gerardin77}.

\begin{fact}[{The Weil representation~\cite[Definition~5.1 and Lemmas~5.12 and~5.13]{Stromberg13}}]\label{fact:weilrep}
    Let $\mu \not\equiv 0 \pmod{q}$ be as in~\Cref{def:weil}. There is a unique representation $\rho$ of $\SL_2(\ZZ)$ on $\CC^q$ with $\rho(T) = \rho_T$ and $\rho(S) = \rho_S$; we call it the \emph{Weil representation} with parameter $\mu$. Moreover:
    \begin{itemize}
        \item $\rho(g) = I$ for every $g \in \Gamma(q)$. Consequently, $\rho(g)$ depends only on the reduction of $g$ modulo $q$, and we write $\rho(\bar{g}) := \rho(g)$ for $\bar{g} \in \SL_2(\ZZ_q)$ and any lift $g \in \SL_2(\ZZ)$ of $\bar{g}$ (which exists by~\Cref{lem:indices}).
        \item For every $g = \begin{bmatrix} a & b \\ c & d_0 \end{bmatrix} \in \Gamma_0(q)$ and every $x \in \ZZ_q$:
        \begin{equation}\label{eq:weilformula}
            \rho(g)\, \ev_x = \left(\frac{a}{q}\right) \zeta_q^{\mu b d_0 x^2}\, \ev_{d_0 x}.
        \end{equation}
    \end{itemize}
\end{fact}

Finally, we record how the Weil representation interacts with the \emph{Heisenberg operators}, the cyclic shift and the diagonal phase matrix on $\CC^q$. This is used only in~\Cref{sec:gen4}, where translations of an elliptic variable act through these operators (\Cref{lem:thetatranslation}).

\begin{definition}[Heisenberg operators]\label{def:heisenberg}
    Define the Heisenberg operators $\sfX, \sfZ \in \CC^{q \times q}$ by $\sfX \ev_r = \ev_{(r-1)\bmod{q}}$ and $\sfZ \ev_r = \zeta_q^r \ev_r$. (In coordinates: $(\sfX \yv)_j = y_{(j+1) \bmod q}$ and $(\sfZ \yv)_j = \zeta_q^j\, y_j$.)
\end{definition}

The Weil representation interacts with these operators in a simple way: conjugating by $\rho(g)$ permutes the operators $\sfX^m\sfZ^n$ up to roots of unity.

\begin{lemma}[{Weil conjugation of the Heisenberg operators~\cite[n\textsuperscript{o}~10, Th\'eor\`eme~1]{Weil64}, \cite[Theorem~2.4(a)]{Gerardin77}}]\label{lem:heisenbergweil}
    Let $\rho$ be the Weil representation with parameter $\mu$ (\Cref{fact:weilrep}). For every $g \in \SL_2(\ZZ)$ and all integers $m, n$, there exist integers $m', n', c$ (depending on $g$, $m$, $n$, and $\mu$) such that:
    \begin{equation*}
        \rho(g)\, \sfX^m \sfZ^n = \zeta_q^{c}\, \sfX^{m'} \sfZ^{n'}\, \rho(g).
    \end{equation*}
\end{lemma}

\subsection{Lattices in $\CC$ and the Upper Half-Plane}\label{sec:upperhalfplane}

All of our theta functions are sums of Gaussian-type terms $e^{\pi i \tau \langle \yv, \yv \rangle}$ over the points of a lattice. The parameter $\tau$ controlling the Gaussian lies in the upper half-plane, and the symmetries used in our proofs come from an action of $\SL_2(\ZZ)$ on it. This subsection unpacks that action, the automorphy factor $G(g, \tau)$ through which every transformation law below will be expressed, and the \emph{cusps} (the rational numbers together with $\infty$) --- the points, besides those of $\calH$, at which the valence formula will count our forms' zeros.

The technical overview explains this action through lattices (\Cref{sec:overview:mf}): $\tau$ stands for the lattice $\ZZ + \ZZ\tau$, the matrix $g$ changes its basis to $(c\tau + d_0, a\tau + b)$, and dividing by $c\tau + d_0$ turns this into a basis of the form $(1, g \ast \tau)$ (\Cref{fig:lattice}). In this picture, the automorphy factor $G(g, \tau) = c\tau + d_0$ defined below is exactly the number we divide by, and the chain rule (\Cref{prop:chainrule}) says that performing two changes of basis in turn multiplies these factors. \Cref{fig:cusps} shows the cusps, and how elements of $\SL_2(\ZZ)$ move $\infty$ to other cusps.

\begin{definition}
    For a complex number $z \in \CC$, let $\Re z$ and $\Im z$ denote its real and imaginary parts.
    We let $\calH \subset \CC$ denote the strict upper half-plane, i.e., $\{z \in \CC: \Im z > 0\}$.
    Throughout, we will use $\tau$ to denote a complex number in $\calH$.
\end{definition}

\begin{fact}
    $\SL_2(\ZZ)$ acts on $\calH$ under the following group action:
    $$\begin{bmatrix} a & b \\ c & d_0 \end{bmatrix} \ast \tau := \frac{a\tau+b}{c\tau+d_0}.$$
    (Note that under this convention, $-I$ fixes every $\tau$ so this is actually an action of $\SL_2(\ZZ)/(\pm I) = \PSL_2(\ZZ)$.)
\end{fact}

\begin{definition}
    Define $G(g, \tau) := c\tau+d_0$, where $g = \begin{bmatrix} a & b \\ c & d_0 \end{bmatrix} \in \SL_2(\ZZ)$. (Note in particular that $G(I, \tau) = 1$ for all $\tau$.)
\end{definition}

\begin{proposition}[{Chain rule~\cite[Lemma~1.2.2(a)]{DiamondShurman05}}]\label{prop:chainrule}
    For any $g, h \in \SL_2(\ZZ)$ and $\tau \in \calH$, we have:
    $$G(gh, \tau) = G(g, h \ast \tau) G(h, \tau).$$
\end{proposition}

\begin{definition}[Cusps]\label{def:cusps}
    The \emph{cusps} are the elements of $\QQ \cup \{\infty\}$. The action of $\SL_2(\ZZ)$ extends to them by the same formula:
    \begin{equation*}
        \begin{bmatrix} a & b \\ c & d_0 \end{bmatrix} \ast x := \frac{ax+b}{cx+d_0} \quad (x \in \QQ), \qquad \begin{bmatrix} a & b \\ c & d_0 \end{bmatrix} \ast \infty := \frac{a}{c},
    \end{equation*}
    where a fraction with denominator $0$ is understood as $\infty$. (Intuitively, $\infty$ stands for the point $i\infty$ at the top of the upper half-plane, and the rational cusps are points on its boundary, the real line.)
    The cusps will be essential to our proofs: the valence formula (\Cref{thm:valence}) counts a modular form's zeros both at points of $\calH$ and at cusps, and our arguments work by forcing high-order vanishing precisely at the cusps.
\end{definition}

\subsection{Modular Forms}

\subsubsection{Weakly Modular Functions}

In the following, let $\Gamma$ be a congruence subgroup of $\SL_2(\ZZ)$ of level $q$ (\Cref{def:subgroups}), and let $\rho: \Gamma \to \GL_N(\CC)$ be a representation (\Cref{def:representation}) for a positive integer $N > 0$. We additionally assume throughout that $\rho$ is of \emph{finite order}, meaning that every matrix $\rho(g)$ in its image has finite order. This assumption is what gives weakly modular functions well-behaved expansions at the cusps (\Cref{sec:cuspinfty}), and it holds in all our applications: most importantly, the Weil representation factors through the finite group $\SL_2(\ZZ_q)$ by~\Cref{fact:weilrep}, so it --- and every representation obtained from it by restriction or conjugation --- has finite order.

\begin{definition}[Weakly modular functions]\label{def:weakmodular}
    Let $k$ be a non-negative integer.
    A holomorphic $F: \calH \to \CC^N$ is weakly modular of weight $k$ for $\Gamma$ with representation $\rho$ if:
    \begin{align}
        F(g \ast \tau) &= (c\tau+d_0)^k \rho(g) F(\tau),\text{ or more explicitly,} \label{eq:transformationlaw}\\
        F\left(\frac{a\tau+b}{c\tau+d_0}\right) &= (c\tau+d_0)^k \rho\left(\begin{bmatrix} a & b \\ c & d_0 \end{bmatrix}\right) F(\tau),\nonumber
    \end{align}
    for all $g = \begin{bmatrix} a & b \\ c & d_0 \end{bmatrix} \in \Gamma$ and $\tau \in \calH$.
    Note that the types here are consistent: $\rho(g) \in \CC^{N \times N}$ and it acts on $F(\tau) \in \CC^N$ to produce some other vector in $\CC^N$.
\end{definition}

This is the precise form of the informal definition~\eqref{eq:vvmfoverview} of the technical overview, and~\Cref{fig:fourieranalogy}(b) draws the transformation law~\eqref{eq:transformationlaw} as a commutative diagram.

\noindent\textbf{Proportionality.} For two functions $F_1, F_2$ of the same complex variables (in our applications, some of $\tau$, $w$, and $z$, or the coordinates of a vector $\vecv \in \CC^N$), with values in $\CC$ or in $\CC^N$, we write $F_1 \propto F_2$ if $F_1 = \lambda F_2$ for a nonzero constant scalar $\lambda \in \CC$, and $F_1 \propto_\zeta F_2$ if moreover $\lambda$ can be taken to be a root of unity. Here ``constant'' means independent of the complex variables; $\lambda$ may depend on everything else in sight, such as a group element $g$. We stress that $\lambda$ is always a \emph{scalar}: these relations never absorb a matrix, so in vector-valued identities such as the transformation law~\eqref{eq:transformationlaw}, the matrix $\rho(g)$ (like any Heisenberg operator below) is always written out explicitly. Both relations are symmetric and transitive, are preserved when we substitute functions of other variables for the variables (the constant does not change), and are compatible with products, powers, and inverses, as well as with applying a homogeneous polynomial $P$ of degree $D$ to both sides (if $F_1 = \lambda F_2$ then $P(F_1) = \lambda^D P(F_2)$); for $\propto_\zeta$ this uses that the roots of unity form a group. This lets us suppress constants that our arguments never need to know exactly. What they do need is that, for a scalar function, such constants assemble into a representation of finite order. For one-dimensional representations, finite order in the sense above (every value is a root of unity) already implies the uniform statement $\chi^j = 1$:

\begin{lemma}[Characters of finite order]\label{lem:finiteorderchar}
    Let $\Gamma$ be a congruence subgroup and $\chi: \Gamma \to \CC^\times$ a one-dimensional representation each of whose values $\chi(g)$ is a root of unity. Then $\chi^j = 1$ for some positive integer $j$. (So for one-dimensional representations of congruence subgroups, the two senses of ``finite order'' --- every value a root of unity, and $\chi^j = 1$ for some $j \geq 1$ --- agree, and we use them interchangeably.)
\end{lemma}
\begin{proof}
    $\Gamma$ has finite index in $\SL_2(\ZZ)$ (it contains $\Gamma(q)$, which has index $q^3 - q$ by~\Cref{lem:indices}), and $\SL_2(\ZZ)$ is generated by $S$ and $T$ (\Cref{fact:generation}), so $\Gamma$ is finitely generated~\cite[Chapter~II, Proposition~4.2]{LyndonSchupp77}. Hence $\chi(\Gamma)$ is generated by finitely many roots of unity, so it is a finite cyclic group, and $\chi^j = 1$ for $j := |\chi(\Gamma)|$.
\end{proof}

\begin{lemma}[Scalar transformation laws]\label{lem:scalarcharacter}
    Let $\Gamma$ be a congruence subgroup, $k$ a non-negative integer, and $f: \calH \to \CC$ a holomorphic function that is not identically zero, such that $f(g \ast \tau) \propto G(g, \tau)^k f(\tau)$ (as functions of $\tau$) for every $g \in \Gamma$; that is, $f(g \ast \tau) = \chi(g)\, G(g, \tau)^k f(\tau)$ for a nonzero constant $\chi(g)$ (uniquely determined, as $f \not\equiv 0$). Then $\chi: \Gamma \to \CC^\times$ is a one-dimensional representation, so $f$ satisfies the transformation law~\eqref{eq:transformationlaw} with representation $\chi$. If moreover $f(g \ast \tau) \propto_\zeta G(g, \tau)^k f(\tau)$ for every $g \in \Gamma$, then $\chi^j = 1$ for some positive integer $j$; in particular $\chi$ has finite order, so $f$ is weakly modular of weight $k$ for $\Gamma$ with representation $\chi$.
\end{lemma}
\begin{proof}
    For $g_1, g_2 \in \Gamma$, applying the transformation law twice and then~\Cref{prop:chainrule} gives
    \begin{equation*}
        \chi(g_1g_2)\, G(g_1g_2, \tau)^k f(\tau) = f(g_1 g_2 \ast \tau) = \chi(g_1)\, G(g_1, g_2 \ast \tau)^k\, \chi(g_2)\, G(g_2, \tau)^k f(\tau) = \chi(g_1)\chi(g_2)\, G(g_1g_2, \tau)^k f(\tau);
    \end{equation*}
    evaluating at any $\tau$ with $f(\tau) \neq 0$ gives $\chi(g_1g_2) = \chi(g_1)\chi(g_2)$. For the second claim, every value $\chi(g)$ is then a root of unity, so~\Cref{lem:finiteorderchar} applies.
\end{proof}

\noindent
To informally understand what~\eqref{eq:transformationlaw} is saying, note that:
\begin{align*}
    \frac{d(g*\tau)}{d\tau} &= \frac{d}{d\tau} \frac{a\tau+b}{c\tau+d_0} \\
    &= \frac{(c\tau+d_0)a - c(a\tau+b)}{(c\tau+d_0)^2} \\
    &= \frac{1}{(c\tau+d_0)^2}.
\end{align*}
This allows us to rewrite~\eqref{eq:transformationlaw} as:
\begin{align*}
    F(g \ast \tau) &= \left(\frac{d(g \ast \tau)}{d\tau}\right)^{-k/2} \rho(g) F(\tau) \\
    \Leftrightarrow F(g \ast \tau) (d(g \ast \tau))^{k/2} &= \rho(g) F(\tau) (d\tau)^{k/2}.
\end{align*}
In words, $F$ intertwines two actions of $g$: on the upper half-plane, $g$ moves the point $\tau$ to $g \ast \tau$ (rescaling $(d\tau)^{k/2}$ along the way), while on $\CC^N$, $\rho(g)$ acts linearly. This is the left square of the commutative diagram in~\Cref{fig:fourieranalogy}(b), which is drawn there for the theta function $\Theta$, of weight $k = \frac{r}{2}$. From this, we intuitively deduce two things: firstly, that~\Cref{prop:chainrule} from earlier is indeed the standard chain rule from calculus in disguise; and secondly, to check~\eqref{eq:transformationlaw} it suffices to check it for $g$ ranging over a set that generates $\Gamma$. Nevertheless, we check this more formally:
\begin{proposition}\label{prop:generatingisenough}
    If~\eqref{eq:transformationlaw} holds for $(g_1, \tau)$ and $(g_2, \tau)$ for all $\tau \in \calH$, then it also holds for $(g_1^{-1}, \tau)$ and $(g_1g_2, \tau)$.
\end{proposition}
\begin{proof}
    For the product: $F(g_1 g_2 \ast \tau) = G(g_1, g_2 \ast \tau)^k \rho(g_1) F(g_2 \ast \tau) = G(g_1, g_2 \ast \tau)^k G(g_2, \tau)^k \rho(g_1) \rho(g_2) F(\tau)$, which equals $G(g_1g_2, \tau)^k \rho(g_1g_2) F(\tau)$ by~\Cref{prop:chainrule}. For the inverse: substitute $\tau \mapsto g_1^{-1} \ast \tau$ into the law for $g_1$ and rearrange, using $G(g_1, g_1^{-1} \ast \tau)\, G(g_1^{-1}, \tau) = G(I, \tau) = 1$ (\Cref{prop:chainrule} again).
\end{proof}

\subsubsection{The Cusp at Infinity}\label{sec:cuspinfty}

An important task for us is to understand when a weakly modular function can be extended to include the cusps in $\QQ \cup \{\infty\}$ in a reasonable way. This is also where we define the orders of vanishing at cusps, which are the quantities that all of our root counts bound.
By ``reasonable'', we mean that the resulting function will be holomorphic, and this will give us a modular form.
This is most naturally done by first addressing the cusp at $\infty$, so we begin there.

\begin{definition}[Width of the cusp at $\infty$]
    The \emph{width} of the cusp at infinity is the minimal positive integer $h$ such that $T^h = \begin{bmatrix} 1 & h \\ 0 & 1 \end{bmatrix} \in \Gamma$.
    If $\Gamma$ is a congruence subgroup of level $q$, then the width is well-defined and divides $q$ (since $T^q \in \Gamma(q) \subseteq \Gamma$); as $q$ is prime, this means $h \in \{1, q\}$, and $h$ need not equal $q$.
\end{definition}
Letting $h$ be the width of the cusp at $\infty$, we have:
\begin{equation}\label{eq:periodicity1}
    F(\tau+h) = \rho\left(\begin{bmatrix} 1 & h \\ 0 & 1 \end{bmatrix}\right) F(\tau).
\end{equation}
To analyze this periodicity condition, we first put the matrix $A := \rho(T^h)$ appearing in it --- the matrix describing how $F$ transforms over one period --- in a canonical form, using the standing finite-order assumption on $\rho$:
\begin{proposition}[{\cite[Section~6.4, Theorem~6]{HoffmanKunze71}}]\label{prop:finiteorder}
    If $A \in \GL_N(\CC)$ has finite order, then $A$ is diagonalizable and all of its eigenvalues are roots of unity.
\end{proposition}
\noindent
By~\Cref{prop:finiteorder}, we may fix a basis $\vecv_1, \ldots, \vecv_N$ of $\CC^N$ consisting of eigenvectors of $A$, with dual basis $\uv_1, \ldots, \uv_N$ (characterized by $\uv_i^* \vecv_j = \one[i = j]$), and unique real numbers $\alpha_1, \ldots, \alpha_N \in [0, 1)$ such that $A \vecv_j = e^{2\pi i \alpha_j} \vecv_j$ for each $j$. Define:
\begin{equation}\label{eq:eigendecomp}
    M := \sum_{j \in [N]} \alpha_j \vecv_j \uv_j^*,
\end{equation}
so that $\exp(2\pi i M) = \sum_{j \in [N]} e^{2\pi i \alpha_j} \vecv_j \uv_j^* = A$. (Here and below we use that the $\vecv_j \uv_j^*$ are idempotents with $\vecv_i \uv_i^* \cdot \vecv_j \uv_j^* = \one[i=j]\, \vecv_j \uv_j^*$ and $\sum_j \vecv_j \uv_j^* = I$, so that $M^m = \sum_j \alpha_j^m \vecv_j \uv_j^*$ for all $m \geq 1$ and the exponential series simplifies as displayed.)
This allows us to write~\Cref{eq:periodicity1} as a periodicity condition: since $\exp(-2\pi i M(\tau+h)/h) = \exp(-2\pi i M \tau/h)\exp(-2\pi i M)$ (the two exponents, being multiples of $M$, commute) and $\exp(-2\pi i M) A = I$, we have
\begin{align*}
    \exp(-2\pi iM(\tau+h)/h) F(\tau+h) &= \exp(-2\pi i M \tau/h) F(\tau).
\end{align*}
In other words, the holomorphic function $\exp(-2\pi i M \tau/h) F(\tau)$ is $h$-periodic. (Geometrically: the map $\tau \mapsto e^{2\pi i \tau/h}$ sends $\calH$ onto the punctured unit disk, identifying $\tau$ with $\tau + h$; an $h$-periodic holomorphic function of $\tau$ is exactly a holomorphic function of this disk coordinate, and its behavior ``at $\tau = i\infty$'' becomes its behavior at the puncture; see~\Cref{fig:cuspinfty}.)

\input{figcuspinfty}

By~\Cref{fact:fourierexp}, it therefore admits a unique infinite series expansion:
\begin{equation}\label{eq:qexpansion}
    \exp(-2\pi i M\tau/h) F(\tau) = \sum_{n = -\infty}^{\infty} \cv_n e^{2\pi i \tau n/h},
\end{equation}
where each $\cv_n \in \CC^N$, and the series converges absolutely and locally uniformly on $\calH$.
Expanding this further to recover $F(\tau)$, we have:
\begin{align*}
    F(\tau) &= \exp(2\pi i M\tau/h) \sum_{n = -\infty}^{\infty} \cv_n e^{2\pi i \tau n/h} \\
    &= \left(\sum_{j \in [N]} e^{2\pi i \alpha_j \tau/h} \vecv_j \uv_j^*\right) \sum_{n = -\infty}^{\infty} \cv_n e^{2\pi i \tau n/h} \text{ (substituting in~\Cref{eq:eigendecomp})} \\
    &= \sum_{n = -\infty}^\infty \sum_{j \in [N]} e^{2\pi i (n+\alpha_j)\tau/h} (\uv_j^* \cv_n) \vecv_j
\end{align*}
\begin{definition}[Holomorphic at $\infty$]\label{def:holoinfty}
    We say that $F$ is holomorphic at the cusp at $\infty$ if and only if we have $\cv_n = 0$ for all $n < 0$.
    In this event, we define:
    \begin{equation}
        \order_\infty(F) := \min(n+\alpha_j: \uv_j^* \cv_n \neq 0).
    \end{equation}
    We take the convention $\order_\infty(0) = +\infty$.
\end{definition}

\begin{remark}\label{remark:orderwelldef}
    Two conventions make~\Cref{def:holoinfty} well-posed. First, because each $\alpha_j$ lies in $[0, 1)$, the exponent $n + \alpha_j$ determines both $n$ (its integer part) and $\alpha_j$ (its fractional part), so the minimum ranges over an unambiguous set of exponents. Second, nothing depends on how the eigenbasis was chosen: for each eigenvalue $e^{2\pi i \alpha}$ of $A$, the spectral projector $P_\alpha := \sum_{j: \alpha_j = \alpha} \vecv_j \uv_j^*$ depends only on $A$ (it is the projection onto $\ker(A - e^{2\pi i\alpha} I)$ along the sum of the other eigenspaces); hence so does $M = \sum_\alpha \alpha P_\alpha$ and thus the coefficients $\cv_n$; and $\uv_j^* \cv_n \neq 0$ holds for some $j$ with $\alpha_j = \alpha$ if and only if $P_\alpha \cv_n \neq 0$.
\end{remark}

\subsubsection{Arbitrary Cusps}\label{sec:modularforms}

Next, we characterize what it means for $F$ to be holomorphic at cusps other than $\infty$.
The idea is to act with an element of $\SL_2(\ZZ)$ that sends $\infty$ to this cusp and then adapt our methods from~\Cref{sec:cuspinfty} (\Cref{fig:cusps} shows such elements for a few cusps).
To this end, consider a cusp $\fraka \in \QQ$.
Using~\Cref{def:cusps}, we know there must exist a group element $g_\fraka \in \SL_2(\ZZ)$ such that $g_\fraka \ast \infty = \fraka$.
Accordingly, define:
\begin{equation}\label{eq:pullback}
    F_\fraka(\tau) := G(g_\fraka, \tau)^{-k} F(g_\fraka \ast \tau).
\end{equation}
In addition to sending the cusp at $\infty$ to $\fraka$, this transformation gives us a function that is weakly modular of weight $k$ with respect to the \emph{conjugate} subgroup $\Gamma_\fraka := g_\fraka^{-1} \Gamma g_\fraka$ and the conjugate representation $\rho_\fraka: g_\fraka^{-1} g g_\fraka \mapsto \rho(g)$.
($\Gamma_\fraka$ is again a congruence subgroup of the same level: $\Gamma(q)$ is normal in $\SL_2(\ZZ)$, being the kernel of the reduction map of~\Cref{lem:indices}, so $\Gamma_\fraka \supseteq g_\fraka^{-1} \Gamma(q) g_\fraka = \Gamma(q)$. Moreover, $\rho_\fraka$ is again a representation of finite order, since its image coincides with that of $\rho$; so the results of~\Cref{sec:cuspinfty} apply to $F_\fraka$.) 
To see this, note that:
\begin{align*}
    F_\fraka(g_\fraka^{-1} g g_\fraka \ast \tau) &= G(g_\fraka, g_\fraka^{-1} g g_\fraka \ast \tau)^{-k} F(gg_\fraka \ast \tau) \text{ (by~\Cref{eq:pullback})} \\
    &= G(g, g_\fraka \ast \tau)^k G(g_\fraka, g_\fraka^{-1} g g_\fraka \ast \tau)^{-k} \rho(g) F(g_\fraka \ast \tau) \text{ (by~\Cref{def:weakmodular})} \\
    &= G(g, g_\fraka \ast \tau)^k G(g_\fraka, g_\fraka^{-1} g g_\fraka \ast \tau)^{-k} G(g_\fraka, \tau)^k \rho(g) F_\fraka(\tau) \\
    &= G(g_\fraka^{-1}gg_\fraka, \tau)^k \rho(g) F_\fraka(\tau),
\end{align*}
as claimed.
The last step above can be seen from two applications of~\Cref{prop:chainrule}:
\[
    G(g_\fraka, g_\fraka^{-1}gg_\fraka \ast \tau) G(g_\fraka^{-1}gg_\fraka, \tau) = G(gg_\fraka, \tau) = G(g, g_\fraka \ast \tau) G(g_\fraka, \tau).
\]
With this transformation, the definition of holomorphy at the cusp at $\fraka$ is automatic:
\begin{definition}\label{def:holocusp}
    The width of the cusp of $F$ at $\fraka$ for $\Gamma$ is defined to be the width of the cusp of $F_\fraka$ at $\infty$ for $\Gamma_\fraka$ (more directly, it is the smallest integer $h > 0$ such that $g_\fraka \begin{bmatrix} 1 & h \\ 0 & 1 \end{bmatrix} g_\fraka^{-1} \in \Gamma$).
    $F$ is holomorphic at the cusp $\fraka$ if and only if $F_\fraka$ is holomorphic at $\infty$.
    Finally, we define $\order_\fraka(F) = \order_\infty(F_\fraka)$.
\end{definition}

\begin{lemma}[Well-definedness of cusp data]\label{lem:cuspwelldef}
    The width of the cusp at $\fraka$, whether $F$ is holomorphic at $\fraka$, and $\order_\fraka(F)$ do not depend on the choice of $g_\fraka \in \SL_2(\ZZ)$ with $g_\fraka \ast \infty = \fraka$. Moreover, if $\frakb = \gamma \ast \fraka$ for some $\gamma \in \Gamma$, then the width of the cusp at $\frakb$ equals that at $\fraka$, and $F$ is holomorphic at $\frakb$ if and only if it is holomorphic at $\fraka$, in which case $\order_\frakb(F) = \order_\fraka(F)$.
\end{lemma}
\begin{proof}
    See~\Cref{sec:cuspwelldefproof}. For scalar forms, the statement about orders is~\cite[Theorem~4.1.2(i)]{Rankin77}.
\end{proof}

\noindent
Accordingly, we will regard cusps in the same $\Gamma$-orbit as equivalent, and speak of the width, holomorphy, and order of $F$ at an equivalence class of cusps.

Finally, we can define a modular form:
\begin{definition}
    Let $k$ be a non-negative integer.
    A weakly modular $F: \calH \to \CC^N$ of weight $k$ for $\Gamma$ with representation $\rho$ is referred to as a \emph{modular form} if it is holomorphic at all cusps.
\end{definition}

\subsubsection{Composition Laws}\label{sec:compositionlaws}

The following lemma turns a vector-valued modular form into a scalar one by substituting it into a polynomial that transforms by a character under the representation. This is also how the classical correspondence between invariant polynomials, such as weight enumerators of self-dual codes, and modular forms arises: by substituting theta functions into them~\cite{BroueEnguehard72,NebeRainsSloane06}. In our proofs, it is the step labeled ``scalarization via $P_\beta$'' in~\Cref{fig:gen1flow,fig:gen4flow}, drawn in teal in~\Cref{fig:fourieranalogy}(b). We include a proof for completeness.

\begin{lemma}[Scalarization]\label{lem:scalarisation}
    Consider a modular form $F: \calH \to \CC^N$ of weight $k$ for $\Gamma$ with representation $\rho: \Gamma \to \GL_N(\CC)$.
    Suppose we also have a homogeneous polynomial $P: \CC^N \to \CC$ of degree $D$, not the zero polynomial, such that $P(\rho(g)\vecv) \propto P(\vecv)$ as functions of $\vecv \in \CC^N$, for every $g \in \Gamma$; that is, for each $g \in \Gamma$ there is a nonzero constant $\chi(g)$ (uniquely determined, as $P \neq 0$) with
    \begin{equation}\label{eq:scalarisation}
        P(\rho(g) \vecv) = \chi(g) P(\vecv) \qquad \text{for all } \vecv \in \CC^N.
    \end{equation}
    Then $\chi: \Gamma \to \CC^\times$ is a one-dimensional representation with $\chi^j = 1$ for some positive integer $j$, and the composed function $f := P \circ F$ is a scalar modular form of weight $kD$ for $\Gamma$ with representation $\chi$.
\end{lemma}
\begin{proof}
    Fix $\vecv_0 \in \CC^N$ with $P(\vecv_0) \neq 0$. For $g_1, g_2 \in \Gamma$, applying~\eqref{eq:scalarisation} twice gives $\chi(g_1g_2) P(\vecv_0) = P(\rho(g_1)\rho(g_2)\vecv_0) = \chi(g_1)\chi(g_2) P(\vecv_0)$, so $\chi$ is a one-dimensional representation. Moreover, every value of $\chi$ is a root of unity: for $g \in \Gamma$, if $\rho(g)^n = I$ (recall that $\rho$ has finite order), then $\chi(g)^n P(\vecv_0) = P(\rho(g)^n \vecv_0) = P(\vecv_0)$, so $\chi(g)^n = 1$. Hence $\chi^j = 1$ for some positive integer $j$, by~\Cref{lem:finiteorderchar}.

    Next, $f = P \circ F$ is holomorphic on $\calH$ (a polynomial composed with a holomorphic function), and it is weakly modular: for any $\tau \in \calH$ and $g = \begin{bmatrix} a & b \\ c & d_0 \end{bmatrix} \in \Gamma$, homogeneity of $P$ gives
    \begin{align*}
        f(g \ast \tau) = P(F(g \ast \tau)) = P\left((c\tau+d_0)^k \rho(g) F(\tau)\right) = (c\tau+d_0)^{kD}\, P(\rho(g) F(\tau)) = (c\tau+d_0)^{kD} \chi(g) f(\tau).
    \end{align*}

    Finally, $f$ is holomorphic at every cusp. The idea is that $P$ turns the cusp expansion of $F$ into one for $f$, and the covariance~\eqref{eq:scalarisation} keeps the exponents in the right residue class; the verification is routine and deferred to~\Cref{sec:scalarisationcuspsproof}.
\end{proof}

\begin{lemma}[{Products~\cite[Section~4.3]{Rankin77}}]\label{lem:product}
    Let $f_1, f_2: \calH \to \CC$ be scalar modular forms for $\Gamma$ of weights $k_1, k_2$ with one-dimensional representations $\chi_1, \chi_2$. Then $f_1 f_2$ is a scalar modular form of weight $k_1 + k_2$ for $\Gamma$ with representation $\chi_1 \chi_2$, and at every cusp $\fraka$ and every $\tau_0 \in \calH$:
    \begin{equation*}
        \order_\fraka(f_1 f_2) = \order_\fraka(f_1) + \order_\fraka(f_2), \qquad \order_{\tau_0}(f_1 f_2) = \order_{\tau_0}(f_1) + \order_{\tau_0}(f_2),
    \end{equation*}
    where $\order_{\tau_0}$ denotes the order of vanishing at an interior point (\Cref{def:interiororder} below); if either factor is identically zero, both equations hold in $[0, +\infty]$ with the convention $\order(0) = +\infty$.
\end{lemma}

\begin{lemma}[Restriction]\label{lem:restriction}
    Let $F: \calH \to \CC^N$ be a modular form of weight $k$ for $\Gamma$ with representation $\rho$, and let $\Gamma' \subseteq \Gamma$ be a congruence subgroup of level $q$. Then $F$ is a modular form of weight $k$ for $\Gamma'$ with representation $\rho|_{\Gamma'}$.
\end{lemma}
\begin{proof}
    See~\Cref{sec:restrictionproof}.
\end{proof}

\subsubsection{The Valence Formula}\label{sec:valenceformula}

The engine of our lower-bound arguments is the valence formula: a nonzero modular form has a fixed total number of zeros --- counted over the $\Gamma$-equivalence classes of points of $\calH$ and of cusps, with appropriate weights --- determined only by its weight and the index of $\Gamma$. It plays the role that the degree bound plays for univariate polynomials, as in the technical overview (\Cref{sec:overview:mf}): the weight $k$ is a degree-like parameter, and a smaller group $\Gamma$ imposes fewer constraints and so allows more zeros. We first define the order of vanishing at interior points.

\begin{definition}\label{def:interiororder}
    For a holomorphic $f: \calH \to \CC$ that is not identically zero and a point $\tau_0 \in \calH$, let $\order_{\tau_0}(f)$ denote the order of vanishing of $f$ at $\tau_0$, i.e., the index of the first nonzero coefficient in the Taylor expansion of $f$ at $\tau_0$. We take the convention $\order_{\tau_0}(0) = +\infty$.
\end{definition}

\begin{remark}\label{remark:interiororder}
    If $f \not\equiv 0$ is weakly modular of weight $k$ for $\Gamma$ with a one-dimensional representation $\chi$, then $\order_{\tau_0}(f)$ depends only on the $\Gamma$-orbit of $\tau_0$: for $g \in \Gamma$, the function $\tau \mapsto f(g \ast \tau) = (c\tau + d_0)^k \chi(g) f(\tau)$ differs from $f$ by a nonvanishing holomorphic factor, and $\tau \mapsto g \ast \tau$ is a biholomorphism of $\calH$, so $f$ vanishes at $g \ast \tau_0$ to the same order as at $\tau_0$. We also note that the stabilizer $\{g \in \Gamma : g \ast \tau_0 = \tau_0\}$ of any point of $\calH$ is finite (see, e.g.,~\cite[Chapter~2]{DiamondShurman05}).
\end{remark}

We now state the valence formula, in the form in which we will use it. It is classical, also for nontrivial one-dimensional representations (multiplier systems), and we cite it.

\begin{theorem}[{Valence formula~\cite[Theorem~4.1.4]{Rankin77}}]\label{thm:valence}
    Let $\Gamma$ be a congruence subgroup of level $q$ containing $-I$, and let $f: \calH \to \CC$ be a modular form of weight $k$ for $\Gamma$ with a one-dimensional representation $\chi$ satisfying $\chi^j = 1$ for some positive integer $j$, with $f$ not identically zero. Then $f$ has finitely many $\Gamma$-equivalence classes of zeros in $\calH$, and:
    \begin{equation}\label{eq:valence}
        \sum_{\text{cusps }\fraka} \frac{\order_\fraka(f)}{2} + \sum_{\tau \in \calH/\Gamma} \frac{\order_\tau(f)}{|\{g \in \Gamma: g \ast \tau = \tau\}|} = \frac{k}{24} \left[\SL_2(\ZZ) : \Gamma\right].
    \end{equation}
    Here the first sum ranges over the $\Gamma$-equivalence classes of cusps (well-defined by~\Cref{lem:cuspwelldef}), with orders as in~\Cref{def:holoinfty,def:holocusp}, and the second sum ranges over the $\Gamma$-equivalence classes of zeros of $f$ in $\calH$ (well-defined by~\Cref{remark:interiororder}).
    (At each application of this theorem we will verify, or explicitly defer, the hypothesis that some power of $\chi$ is trivial. Note also that taking $g = -I$ in the transformation law gives $f(\tau) = (-1)^k \chi(-I) f(\tau)$, so necessarily $\chi(-I) = (-1)^k$.)
\end{theorem}

\begin{corollary}[Valence formula for $\Gamma(q)$]\label{lem:valencegammaq}
    Let $f: \calH \to \CC$ be a modular form of weight $k$ for $\Gamma(q)$ with a one-dimensional representation $\chi$ satisfying $\chi^j = 1$ for some positive integer $j$, with $f$ not identically zero. Then:
    \begin{equation}\label{eq:valencegammaq}
        \sum_{\text{cusps }\frakc} \frac{\order_\frakc(f)}{2} + \sum_{\tau \in \calH/\Gamma(q)} \frac{\order_\tau(f)}{2\,|\{g \in \Gamma(q): g \ast \tau = \tau\}|} = \frac{k}{24} \cdot \frac{q^3-q}{2},
    \end{equation}
    where the first sum ranges over the $\Gamma(q)$-equivalence classes of cusps, with widths and orders computed with respect to $\Gamma(q)$, and the second over the $\Gamma(q)$-equivalence classes of zeros of $f$ in $\calH$.
\end{corollary}
\begin{proof}[Proof sketch]
    By extending $f$ from $\Gamma(q)$ to $\pm\Gamma(q) = \Gamma(q) \cup (-I)\Gamma(q)$ and applying~\Cref{thm:valence} there.
\end{proof}

\section{Warm-Up}\label{sec:gen1}

This section proves our warm-up result, \Cref{cor:gen1specific}: for primes $q \equiv 3 \pmod 4$, a cone-delta function on $\ZZ_q^d$ with Fourier sparsity $q^{(\eta_q + o(1))d}$. It already produces the objects behind $q$-server PIR (\Cref{mainthm:track2}), and its proof introduces, in the simplest setting, every tool the stronger construction of~\Cref{sec:gen4} will reuse. The plan: after an overview of the proof (\Cref{sec:gen1overview}) and the setup of even lattices (\Cref{sec:highdimlattices}), we state the general result (\Cref{thm:gen1}), its consequence for Fourier sparsity, which holds for every lattice the theorem allows (\Cref{cor:gen1general}), and its concrete rank-$2$ instantiation (\Cref{cor:gen1specific}); the support-size bounds behind both corollaries are proved in~\Cref{sec:instantiationgen1}.

The ingredients of this argument are classical: the transformation law of lattice theta functions under the Weil representation and the valence formula. What is new is their use to show that sparse cone-delta functions exist (see the start of~\Cref{sec:math}). Its conclusion is not new for $q = 3$: the instantiation of~\Cref{sec:instantiationgen1} uses the $A_2$ lattice, the set $S_d$ of~\Cref{thm:gen1} is the Hamming ball of radius $d/3$ (as $w_\calL$ takes the value $1$ at both nonzero residues, by~\Cref{lem:gen1weights}), and a cone-delta function with Fourier support in this ball is also given by an elementary polynomial in the proof of~\cite[Theorem~2.2]{AlonLubetzky07}, specialized to the zero vector.

\subsection{Overview of the Proof}\label{sec:gen1overview}

This subsection assembles the proof of~\Cref{thm:gen1} and its corollaries from the ideas of the technical overview (\Cref{sec:overview:weil,sec:overview:arch,sec:overview:mf}); we give the details in the rest of this section.

\paragraph{The dual object.} By duality (Step~1, \Cref{sec:gen1step1}), it suffices to rule out a \emph{witness}: a vector $(\beta_\xv)_{\xv \in \calQ}$, indexed by the points of the cone $\calQ$, with
\begin{equation}\label{eq:gen1witness}
    \beta_\zerov = 1 \qquad\text{and}\qquad \sum_{\xv \in \calQ} \beta_\xv \zeta_q^{\langle \av, \xv \rangle} = 0 \quad\text{for all } \av \in S_d
\end{equation}
(these are~\eqref{eq:beta0} and~\eqref{eq:betafourier} in Step~1). As with the polynomial $P_\cv$ of the technical overview (\Cref{sec:overview:weil}), we package the witness as the homogeneous polynomial $P_\beta(\vecv) := \sum_{\xv \in \calQ} \beta_\xv\, v_{x_1} v_{x_2} \cdots v_{x_d}$ of degree $d$ on $\CC^q$. The conditions on $\beta$ become three statements about $P_\beta$:
\begin{enumerate}
    \item $P_\beta(\ev_0) = \beta_\zerov = 1$, since only the term $\xv = \zerov$ survives.
    \item $P_\beta(\rho_T \vecv) = P_\beta(\vecv)$ for every $\vecv$: substituting $\rho_T\vecv$ multiplies the term for $\xv$ by $\zeta_q^{\mu\langle \xv, \xv \rangle}$, which is $1$ on the cone.
    \item $P_\beta(\rho_S \vecv)$, as a polynomial in $\vecv$, has no monomial $v_{z_1} v_{z_2} \cdots v_{z_d}$ with $\zv \in S_d$: up to the normalizing constant of $\rho_S$, the coefficient of this monomial is a sum of the sums $\sum_{\xv \in \calQ} \beta_\xv \zeta_q^{\langle -2\mu\zv', \xv \rangle}$ over the rearrangements $\zv'$ of $\zv$, which vanish by~\eqref{eq:gen1witness} as long as $S_d$ is closed under permuting coordinates and under multiplying by nonzero residues (recall that $\mu \not\equiv 0 \pmod{q}$, by~\Cref{prop:munonzero}). 
    
    Looking ahead, our choice of $S_d$ is precisely to ensure a large vanishing order of $P_\beta \circ \Theta_{\calL}$ at the cusp at 0 via this observation.
\end{enumerate}
The first and third statements are the analogues of $\Delta(\xv) = 1$ and of $\Delta$ vanishing at the queried points in Reed--Muller PIR (\Cref{tab:rmanalogy}). We now evaluate $P_\beta$ along the theta function $\Theta_\calL$ of an even lattice $\calL$ of rank $r$ and determinant $\sqrt{q}$, which is a vector-valued modular form for the Weil representation~\cite[Theorem~4.1]{Borcherds98}.

\paragraph{Scalarization.} By the second statement, $P_\beta \circ \Theta_\calL$ is invariant under $\tau \mapsto \tau + 1$. Every other $g \in \Gamma_0(q)$ acts on $\CC^q$, up to a scalar, by rescaling the indices $x \mapsto d_0 x$ and multiplying by quadratic phases (\Cref{fact:weilrep}). The phases equal $1$ on the cone, as in the second statement. To absorb the rescaling, we replace $P_\beta$ by its average over the rescalings $x \mapsto ux$ for $u \in \ZZ_q^*$ (\Cref{sec:gen1step2}); the average still satisfies the three statements, except that its value at $\ev_0$ is $q - 1$. With this change, $P_\beta(\rho(g)\vecv) \propto P_\beta(\vecv)$ for all $g \in \Gamma_0(q)$ (\Cref{lem:gen1scalarisation}), so $f := P_\beta \circ \Theta_\calL$ is a scalar modular form of weight $dr/2$ for $\Gamma_0(q)$. Substituting theta functions into an invariant polynomial in this way is also how the classical correspondence between weight enumerators of codes and theta series of lattices arises~\cite{BroueEnguehard72,NebeRainsSloane06}.

\paragraph{The cusp at $\infty$.} As $\Im \tau \to \infty$, we have $\Theta_\calL(\tau) \to \ev_0$, so $f(\tau) \to P_\beta(\ev_0) \neq 0$ (\Cref{lem:gen1nonzero}). In particular, $f$ is not identically zero.

\paragraph{The cusp at $0$.} To expand $f$ at the cusp $0$, we move it to $\infty$: by the transformation of $\Theta_\calL$ under $S$ and the homogeneity of $P_\beta$,
\begin{equation*}
    f_0(\tau) := \tau^{-dr/2} f(-1/\tau) \propto P_\beta\big(\rho_S\, \Theta_\calL(\tau)\big).
\end{equation*}
Write $\Theta_\calL(\tau) = \sum_{t \in \ZZ_q} \theta_t(\tau)\, \ev_t$, where $\theta_t(\tau) := \sum_{\yv \in t\av + \calL} e^{\pi i \langle \yv, \yv \rangle \tau}$. By the third statement, when we expand $P_\beta(\rho_S \Theta_\calL(\tau))$ as a polynomial in the $\theta_t(\tau)$, only the monomials $\theta_{t_1}(\tau) \cdots \theta_{t_d}(\tau)$ with $\tv \notin S_d$ survive. Each of them is a power series in $e^{2\pi i \tau/q}$:
\begin{equation*}
    \theta_{t_1}(\tau) \cdots \theta_{t_d}(\tau) = \sum_{\substack{\yv_1, \ldots, \yv_d: \\ \yv_i \in t_i\av + \calL}} e^{2\pi i n \tau/q}, \qquad\text{where } n := \frac{q}{2} \sum_{i=1}^d \langle \yv_i, \yv_i \rangle \geq \sum_{i=1}^d w_\calL(t_i\av),
\end{equation*}
and $w_\calL(x\av) := \frac{q}{2} \min_{\yv \in x\av + \calL} \langle \yv, \yv \rangle$ (\Cref{def:prices}). Each exponent $n$ is a nonnegative integer (\Cref{prop:evendistances}). We call $w_\calL(x\av)$ the \emph{cost} of $x \in \ZZ_q$. Now take $S_d$ to be the set of $\tv \in \ZZ_q^d$ with $\sum_i w_\calL(ut_i\av) \leq Bd$ for some $u \in \ZZ_q^*$; the union over $u$ makes $S_d$ closed under multiplying by nonzero residues, as the third statement requires. Every $\tv \notin S_d$ then has $\sum_i w_\calL(t_i\av) > Bd$, so every surviving exponent exceeds $Bd$, and $f$ vanishes to order more than $Bd$ at the cusp $0$ (\Cref{sec:gen1step3}).

\paragraph{Counting roots.} By the valence formula (\Cref{thm:valence}; see, e.g.,~\cite[Theorem~4.1.4]{Rankin77}), the nonzero modular form $f$, of weight $dr/2$ for $\Gamma_0(q)$, has total order of vanishing $\frac{dr(q+1)}{24}$. All orders are nonnegative, so its order at the cusp $0$ is at most $\frac{dr(q+1)}{24}$. Taking $B = \frac{r(q+1)}{24}$ contradicts the previous paragraph. Hence no witness exists, and there is a cone-delta function with Fourier support in $S_d$ (\Cref{thm:gen1}). \Cref{tab:rmanalogy}, in the introduction to this part, summarizes the analogy with the Reed--Muller PIR argument of~\Cref{sec:overview:arch}, and~\Cref{fig:gen1flow} shows how the pieces fit together, in the format we reuse for the stronger construction (\Cref{fig:gen4flow}).

\input{figgen1flow}

\paragraph{The size of the support.} It remains to bound $|S_d|$, which is a union of $q - 1$ sets of the form $\{\zv \in \ZZ_q^d : \sum_i w(z_i) \leq Bd\}$ for a cost function $w$ on $\ZZ_q$. A moment-generating-function (Markov) bound controls the size of each such set (\Cref{lem:sublevel}), and optimizing over its parameter gives $|S_d| \leq q^{(\eta_\calL + o(1))d}$ for an explicit exponent $\eta_\calL$ (\Cref{cor:gen1general}). For the warm-up, we use a lattice of rank $r = 2$, which exists exactly when $q \equiv 3 \pmod{4}$ (\Cref{remark:qmod4}). Its cost function is $w_\calL(x\av) = x^2$ for $|x| \leq (q+1)/4$ (\Cref{lem:gen1weights}), and the budget is $B = (q+1)/12$ per coordinate. Heuristically, $S_d$ is then close to the set of integer vectors in a Euclidean ball of radius about $\sqrt{qd/12}$, which has $q^{d/2 + O(d/\ln q)}$ points. The precise bound is $\eta_q \leq \frac{1}{2} + \frac{0.467 + o(1)}{\log_2 q}$ (\Cref{cor:gen1specific}).

\subsection{Lattices in $\RR^r$}\label{sec:highdimlattices}

\begin{definition}[Lattices, determinants]
    Let $r$ be an even positive integer.
    For a basis matrix $\BB \in \RR^{r \times r}$ of full rank, we associate to it the lattice $\calL(\BB) = \BB \ZZ^r \subseteq \RR^r$.
    We define the dual lattice $\calL^*$ to be $\BB^{-\top}\ZZ^r$; equivalently, it can be defined by:
    $$\calL^* = \left\{\xv \in \RR^r: \langle \xv, \yv \rangle \in \ZZ \quad \forall \yv \in \calL\right\}.$$
    The determinant of $\calL$ is defined to be $|\det \BB|$, which we will use interchangeably with $\det \calL$.
    We can also associate to a lattice its Gram matrix $\boldG = \BB^\top \BB$.
\end{definition}

\begin{definition}
    A lattice $\calL$ is said to be \emph{even} if for all $\xv \in \calL$ we have $\langle \xv, \xv \rangle \in 2\ZZ$.
\end{definition}

\begin{proposition}\label{prop:evencontainment}
    If $\calL$ is even, then we have $\calL \subseteq \calL^*$. (Note that the converse need not be true e.g., take $\calL = \ZZ^r$.)
    Moreover, we have $|\calL^* / \calL| = (\det \calL)^2$.
\end{proposition}
\begin{proof}
    For any $\xv, \yv \in \calL$, we have:
    $$\langle \xv, \yv \rangle = \frac{1}{2} \langle \xv+\yv, \xv+\yv \rangle - \frac{1}{2} \langle \xv, \xv \rangle - \frac{1}{2} \langle \yv, \yv \rangle \in \ZZ.$$
    This implies that $\calL \subseteq \calL^*$.
    Secondly, we have $|\calL^*/\calL| = \det \calL/\det \calL^* = (\det \calL)^2$.
\end{proof}

\begin{proposition}\label{prop:evendistances}
    Let $q$ be an odd prime. Suppose $\calL$ is even with $\det \calL = \sqrt{q}$.
    Then:
    \begin{itemize}
        \item For any $\xv \in \calL^*$ and $\yv \in \calL$, we have $\frac{1}{2} \langle \xv+\yv, \xv+\yv \rangle \equiv \frac{1}{2} \langle \xv, \xv \rangle \pmod{1}$.
        \item For any $\xv \in \calL^*$, we have $\langle \xv, \xv \rangle \in \frac{2}{q} \ZZ$.
    \end{itemize}
\end{proposition}
\begin{proof}
    See~\Cref{sec:evendistancesproof}.
\end{proof}

\begin{definition}\label{def:prices}
    Let $q$ be an odd prime and let $\calL \subseteq \RR^r$ be an even lattice of determinant $\sqrt{q}$. For a coset $\av + \calL \in \calL^*/\calL$, we define:
    \begin{equation}
        w_\calL(\av) = w_\calL(\av+\calL) = \frac{q}{2} \min_{\xv \in \av+\calL} \langle \xv, \xv \rangle.
    \end{equation}
    (The minimum is attained because every ball contains only finitely many points of the coset, as in the proof of~\Cref{lem:latticegaussian}.) By~\Cref{prop:evendistances}, we have $w_\calL(\av) \in \ZZ_{\geq 0}$.
\end{definition}

\begin{definition}\label{def:phases}
    Let $q$ be an odd prime and let $\calL \subseteq \RR^r$ be an even lattice of determinant $\sqrt{q}$. The group $\calL^*/\calL$ then has order $(\det \calL)^2 = q$ (\Cref{prop:evencontainment}), so it is cyclic.
    Let $\av + \calL$ be an arbitrary generator of the group (so $\av \in \calL^*$).

    For an input $x \in \ZZ_q$, define $\phi(x) \in \RR/\ZZ$ as follows: take an arbitrary representative $\yv \in x\av + \calL$, and output $\frac{1}{2} \langle \yv, \yv \rangle \bmod{1}$.
    By Proposition~\ref{prop:evendistances}, $\phi(x)$ is well-defined (i.e., it does not depend on the chosen representative $\yv$), and by taking $\yv = x\av$, we can write
    $$\phi(x) = \frac{1}{2} \langle \av, \av \rangle x^2 \bmod{1} = \frac{\mu x^2}{q} \bmod{1},$$
    where $\mu := \frac{q}{2} \langle \av, \av \rangle$ is an integer by~\Cref{prop:evendistances}. We call $\mu$ the \emph{phase coefficient} associated to $\calL$ and $\av$.

    Similarly, for inputs $x, y \in \ZZ_q$, we define $\psi(x, y) \in \RR/\ZZ$ as follows: take arbitrary representatives $\xv \in x\av+\calL, \yv \in y\av + \calL$, and output $\langle \xv, \yv \rangle \bmod{1}$.
    This is well-defined by definition of the dual lattice and again we can take:
    $$\psi(x, y) = \langle \av, \av \rangle xy \bmod{1} = \frac{2\mu xy}{q} \bmod{1}.$$
\end{definition}

\begin{proposition}\label{prop:munonzero}
    In the setting of~\Cref{def:phases}, the phase coefficient satisfies $\mu \not\equiv 0 \pmod{q}$.
\end{proposition}
\begin{proof}
    See~\Cref{sec:munonzeroproof}.
\end{proof}

The phase coefficient is further constrained by the rank of the lattice, via a classical formula of Milgram; we will use this to pin down the Gauss-sum constant in the theta transformation law.

\begin{fact}[{Milgram's formula, specialized~\cite[Appendix~4, p.~127]{MilnorHusemoller73}}]\label{fact:milgram}
    In the setting of~\Cref{def:phases}, the Gauss-sum constant of the phase coefficient $\mu$ --- defined by~\Cref{fact:gausssum}, which applies since $\mu \not\equiv 0 \pmod{q}$ by~\Cref{prop:munonzero} --- satisfies
    \begin{equation*}
        \epsilon_{\mu, q} = (-i)^{r/2}.
    \end{equation*} (Indeed, since $t \mapsto t\av + \calL$ enumerates $\calL^*/\calL$ and every $\yv \in t\av + \calL$ has $\frac{1}{2}\langle \yv, \yv \rangle \equiv \phi(t) \equiv \mu t^2/q \pmod{1}$, we have $\sum_{t \in \ZZ_q} \zeta_q^{\mu t^2} = \sum_{\yv + \calL \in \calL^*/\calL} e^{\pi i \langle \yv, \yv \rangle}$. Milgram's formula evaluates the latter sum, for any even lattice, as $\sqrt{|\calL^*/\calL|}\, e^{2\pi i \sigma/8}$, where $\sigma$ is the signature of the inner product; our form is positive definite, so $\sigma = r$ and the sum is $\sqrt{q}\, i^{r/2}$. Comparing with $\sum_{t} \zeta_q^{\mu t^2} = \epsilon_{\mu, q}^{-1} \sqrt{q}$ from~\Cref{fact:gausssum} gives $\epsilon_{\mu, q} = i^{-r/2} = (-i)^{r/2}$, an ordinary integer power since $r$ is even.)
\end{fact}

\subsection{Main Result}

The main theorem of this section is:
\begin{theorem}\label{thm:gen1}
    Let $q$ be an odd prime, and let $\calL \subseteq \RR^r$ be an even lattice of determinant $\sqrt{q}$.
    The group $\calL^* / \calL$ has order $q$, so it is cyclic with some generator $\av+\calL$ for a generator $\av \in \calL^*$.
    Let $d > 0$ be an arbitrary positive integer, and define the set:
    \begin{equation}
        S_d = \bigcup_{u \in \ZZ_q \backslash \{0\}} \left\{\zv \in \ZZ_q^d: \sum_{i = 1}^d w_\calL(uz_i\av) \leq \frac{r(q+1)}{24} d\right\}.
    \end{equation}
    Then there exists a cone-delta function $f: \ZZ_q^d \to \QQ(\zeta_q)$ such that $\hat{f}$ is supported on $S_d$.
\end{theorem}

(Throughout this part, we use the notions of~\Cref{sec:bg:sfd} over the field $\QQ(\zeta_q)$, with $\gamma = \zeta_q$.)

To turn~\Cref{thm:gen1} into a bound on Fourier sparsity, it remains to bound $|S_d|$. The only information about the lattice that this bound uses is the distribution of the costs $w_\calL$ over the $q$ cosets of $\calL$ in $\calL^*$, which we record in the generating function
\begin{equation}\label{eq:gen1mgf}
    M_\calL(t) := \sum_{\yv + \calL \in \calL^*/\calL} t^{w_\calL(\yv)} \qquad (0 < t \leq 1).
\end{equation}
\begin{corollary}\label{cor:gen1general}
    Let $q$ be an odd prime, and let $\calL \subseteq \RR^r$ be an even lattice of determinant $\sqrt{q}$. For $d$ a growing parameter, the set $S_d$ of~\Cref{thm:gen1} has size $|S_d| \leq q^{(\eta_\calL+o(1))d}$; hence there exists a cone-delta function $f: \ZZ_q^d \to \QQ(\zeta_q)$ whose Fourier support is contained in $S_d$, and so has Fourier sparsity at most $q^{(\eta_\calL+o(1))d}$, where
    \begin{equation}\label{eq:gen1generalexp}
        \eta_\calL := \frac{1}{\ln q} \inf_{0 < t \leq 1} \left(\ln M_\calL(t) - \frac{r(q+1)}{24} \ln t\right).
    \end{equation}
\end{corollary}
Note that $\eta_\calL$ depends only on the lattice $\calL$, and not on the choice of the generator $\av + \calL$ in~\Cref{thm:gen1}.

We instantiate this with the following lattice, for which $M_\calL$, and hence $\eta_\calL$, can be computed explicitly.

\begin{definition}[Concrete instantiation of~\Cref{thm:gen1}]\label{def:gen1concrete}
    Assume $q$ is a prime $\equiv 3\pmod{4}$. Let $r = 2$ and $\mathbf{B} := \begin{bmatrix} \sqrt{2} & 1/\sqrt{2} \\ 0 & \sqrt{q/2} \end{bmatrix}$. 
    One can then verify $\mathbf{G} = \mathbf{B}^\top \mathbf{B} = \begin{bmatrix} 2 & 1 \\ 1 & (q+1)/2\end{bmatrix}$. This has determinant $q$ and moreover it is even on the diagonal and integral in all entries so the resulting lattice $\calL(\mathbf{B})$ is even and has determinant $\sqrt{q}$. We choose our generator $\av + \calL$ of $\calL^*/\calL$ to have representative $\av = (0, \sqrt{2/q})$.
\end{definition}

\begin{remark}\label{remark:qmod4}
    The condition $q \equiv 3 \pmod 4$ in~\Cref{def:gen1concrete} is necessary: an even lattice $\calL \subseteq \RR^2$ of determinant $\sqrt{q}$ exists if and only if $q \equiv 3 \pmod 4$. Indeed, as $\calL$ is even, it is contained in $\calL^*$ (\Cref{prop:evencontainment}), so its Gram matrix has integer entries and an even diagonal: it has the form $\begin{bmatrix} 2a & b \\ b & 2c \end{bmatrix}$ for integers $a, b, c$. Its determinant is $(\det \calL)^2 = q$, so $q = 4ac - b^2 \equiv -b^2 \pmod 4$. As $q$ is odd, $b$ is odd, so $q \equiv -1 \pmod 4$. Conversely, for $q \equiv 3 \pmod 4$, \Cref{def:gen1concrete} gives such a lattice.
\end{remark}

This gives the following; here and below, $K := (q+1)/4$ (an integer for the primes $q \equiv 3 \pmod{4}$ considered).
\begin{corollary}\label{cor:gen1specific}
    For any fixed prime $q \equiv 3 \pmod{4}$ and $d$ a growing parameter, there exists a cone-delta function $f: \ZZ_q^d \to \QQ(\zeta_q)$ with Fourier sparsity at most $q^{(\eta_q+o(1))d}$, where
    \begin{equation}\label{eq:gen1exp}
    \eta_q := \frac{1}{\ln q} \inf_{0 < t \leq 1} \ln \left(t^{-(q+1)/12}\left(1 + 2\sum_{u = 1}^K t^{u^2} + 2\sum_{u = K+1}^{(q-1)/2} t^{u^2-qu+qK}\right)\right).
    \end{equation}
    In particular, $\eta_3 \leq 0.789691$, $\eta_7 \leq 0.709410$, and $\eta_{11} \leq 0.660807$. As $q \to \infty$ along primes $q \equiv 3 \pmod{4}$, we have $\eta_q \leq \frac{1}{2} + \frac{C_{\mathrm{warmup}} + o(1)}{\ln q} \leq \frac{1}{2} + \frac{0.467 + o(1)}{\log_2 q}$, where $C_{\mathrm{warmup}} \leq 0.3234$ is the constant of~\Cref{lem:gen1asymptotic}.
\end{corollary}

We prove~\Cref{cor:gen1general,cor:gen1specific} in~\Cref{sec:instantiationgen1}: the first by a Chernoff bound on the number of frequency vectors of small total cost, and the second by computing the costs of the cosets of the lattice of~\Cref{def:gen1concrete}.

\paragraph{Other lattices.} \Cref{cor:gen1general} is a modular piece of our approach: it accepts any even lattice $\calL$ of determinant $\sqrt{q}$, and through~\Cref{mainthm:track2} every such lattice yields a $q$-server PIR scheme with exponent $\eta_\calL$. A lattice with a smaller $\eta_\calL$ therefore immediately gives a better scheme, with no change to the rest of the argument. By~\eqref{eq:gen1generalexp}, $\eta_\calL$ is small when the shortest vectors in the nonzero cosets of $\calL$ in $\calL^*$ are long relative to the rank $r$, which sets the budget $r(q+1)/24$. The lattice of~\Cref{def:gen1concrete} is one choice, whose costs we can compute in closed form for every prime $q \equiv 3 \pmod 4$. Other lattices also cover primes that it does not. For instance, for $q = 5$, where no such lattice of rank $2$ exists (\Cref{remark:qmod4}), consider the root lattice $A_4 = \{\xv \in \ZZ^5 : \sum_i x_i = 0\}$, of rank $4$. It is even with determinant $\sqrt{5}$, and its nonzero cosets in $A_4^*$ consist of the vectors with coordinate sum $0$ whose coordinates are all congruent to $\frac{c}{5}$ modulo $1$, for $c \in \{1, 2, 3, 4\}$. The shortest such vector takes the value $\frac{c-5}{5}$ in $c$ coordinates and $\frac{c}{5}$ in the other $5 - c$, so it has squared length $\frac{c(5-c)}{5}$. The costs are therefore $w_{A_4} = 2$ on two cosets and $3$ on the other two, so $M_{A_4}(t) = 1 + 2t^2 + 2t^3$, and the infimum in~\eqref{eq:gen1generalexp}, attained at $t = \frac{1}{2}$, gives $\eta_{A_4} = \log_5 \frac{7}{2} < 0.7784$. This yields $5$-server PIR schemes with a smaller exponent than the $q = 3$ case of~\Cref{cor:gen1specific} ($0.7897$, which is what the Track~2 column of~\Cref{table:server-counts} lists for five servers), though not smaller than those of Tracks~1 and~3. Whether some lattice improves on~\Cref{def:gen1concrete}, or even on the stronger construction of~\Cref{sec:gen4}, is open; we discuss this in~\Cref{sec:open}.

\subsection{Proof Step 1: Duality}\label{sec:gen1step1}

Now we begin attacking~\Cref{thm:gen1}. Unwinding the definitions, the theorem asserts the solvability of a finite linear system. Indeed, a function with $\hat{f}$ supported on $S_d$ is exactly a combination $f(\xv) = \sum_{\zv \in S_d} \hat{f}(\zv)\, \zeta_q^{\langle \xv, \zv \rangle}$, and being a cone-delta function constrains its values only on the cone $\calQ$: the value must be $1$ at $\xv = \zerov$ and $0$ at the other points of $\calQ$. So, with the following vectors indexed by elements $\xv \in \calQ$:
\begin{itemize}
    \item the delta vector, which is 1 when $\xv = 0$ and 0 everywhere else; and
    \item for every $\zv \in S_d$, the vector $\vecv_\zv$ with entry $\zeta_q^{\langle \xv, \zv \rangle}$ in index $\xv$,
\end{itemize}
\Cref{thm:gen1} asserts precisely that the vectors $\vecv_\zv$ span the delta vector, with coefficients (namely, the Fourier coefficients $\hat{f}(\zv)$) in the field $\QQ(\zeta_q)$.

Suppose for the sake of contradiction that they do not. Two pieces of finite-dimensional linear algebra convert this assumption into the dual object we will work with for the rest of the proof. First, \emph{the base field does not matter}: the vectors in question have entries in $\QQ(\zeta_q)$, and the solvability of a finite linear system over a field is determined by a comparison of matrix ranks, which is unchanged under field extension; so the delta vector is not in the span of the $\vecv_\zv$ even with coefficients in $\CC$. Second, \emph{duality}: if a vector of $\CC^{\calQ}$ lies outside a subspace, then some linear functional vanishes on the subspace but not on that vector (complete a basis of the subspace by the vector, and further to a basis of $\CC^\calQ$, and take the corresponding dual-basis functional). Writing such a functional as $\wv \mapsto \sum_{\xv \in \calQ} \beta_\xv w_\xv$ for a coefficient vector $\beta \in \CC^{\calQ}$, normalized so that its value on the delta vector is $1$, we have:
\begin{align}
    \beta_0 &= 1 \label{eq:beta0}\\
    \sum_{\xv \in \calQ} \beta_{\xv} \zeta_q^{\langle \xv, \zv \rangle} &= 0 \quad\forall \zv \in S_d.\label{eq:betafourier}
\end{align}
Going forward, our goal is to show that the above two equations imply a contradiction. 

\subsection{Proof Step 2: Vector and Scalar Theta Modular Forms}\label{sec:gen1step2}

In the following, we build on the notation in~\Cref{sec:highdimlattices,sec:weil}. In particular, let $\av \in \calL^*$ be an arbitrary generator of $\calL^*/\calL$. Throughout the proof, $\mu \in \ZZ$ denotes the phase coefficient associated to $\calL$ and $\av$ in~\Cref{def:phases}, and $\rho, \rho_S, \rho_T$ denote the Weil representation of~\Cref{def:weil,fact:weilrep} with this parameter $\mu$. (Recall that~\Cref{def:weil} requires $\mu \not\equiv 0 \pmod{q}$, which~\Cref{prop:munonzero} guarantees.)

\begin{definition}
    Define $\Theta_{\calL}: \calH \to \CC^q$ by:
    \begin{equation}\label{eq:thetadef}
        \Theta_{\calL}(\tau) = \sum_{x \in \ZZ_q} \left(\sum_{\yv \in x\av + \calL} \exp\left(\pi i \langle \yv, \yv \rangle \tau\right)\right) \ev_x.
    \end{equation}
\end{definition}

We want to show that $\Theta_{\calL}$ is a vector-valued modular form of weight $r/2$ for $\SL_2(\ZZ)$ with the Weil representation $\rho$ (note $r/2 \in \ZZ$, as $r$ is even). The modularity of lattice theta functions with respect to the Weil representation is classical: for a positive definite even lattice of even rank $r$, it is the special case $p = 1$, $\alpha = \beta = 0$ of~\cite[Theorem~4.1]{Borcherds98}. (There, the scalar prefactor of $\rho(S)$ is written as $\sqrt{i}^{\,-r} = (-i)^{r/2}$, which equals our $\epsilon_{\mu, q}$ by Milgram's formula, \Cref{fact:milgram}.) Our setting nevertheless admits a short self-contained verification, which we give. We proceed in steps: convergence and holomorphy on $\calH$ (\Cref{lem:thetaconverges}), the transformation laws under the generators $T$ and $S$ (\Cref{lem:gen1T,lem:gen1S}), and holomorphy at the cusps together with the behavior at $\infty$ (\Cref{lem:thetacusp}).

\begin{lemma}\label{lem:thetaconverges}
    The sum in~\eqref{eq:thetadef} is well-defined at every $\tau$ and $\Theta_{\calL}(\tau)$ is holomorphic on $\calH$.
\end{lemma}
\begin{proof}
    See~\Cref{sec:thetaconvergesproof}.
\end{proof}

\begin{lemma}[{\cite[Theorem~4.1]{Borcherds98}}]\label{lem:gen1T}
    For any $\tau \in \calH$, we have:
    \begin{equation*}
        \Theta_{\calL}(\tau+1) = \rho_T \Theta_{\calL}(\tau),
    \end{equation*}
    for $\rho_T$ defined in~\Cref{def:weil}. In other words, noting that $G(T, \tau) = 1$, we have:
    \[
        \Theta_{\calL}(T \ast \tau) = G(T, \tau)^{r/2} \rho(T) \Theta_{\calL}(\tau).
    \]
\end{lemma}
\begin{proof}
    See~\Cref{sec:gen1Tproof}.
\end{proof}

\begin{lemma}[{\cite[Theorem~4.1]{Borcherds98}}]\label{lem:gen1S}
    For any $\tau \in \calH$, we have:
    \[
        \Theta_{\calL}(-1/\tau) = \tau^{r/2} \rho_S \Theta_{\calL}(\tau),
    \]
    for $\rho_S$ defined in~\Cref{def:weil}. In other words, noting that $G(S, \tau) = \tau$, we have:
    \[
        \Theta_{\calL}(S \ast \tau) = G(S, \tau)^{r/2} \rho(S) \Theta_{\calL}(\tau).
    \]
\end{lemma}

\begin{proof}
    See~\Cref{sec:gen1Sproof}.
\end{proof}

\Cref{lem:gen1T,lem:gen1S} together with~\Cref{prop:generatingisenough} imply that $\Theta_{\calL}$ is a weakly modular function of weight $r/2$ for $\SL_2(\ZZ)$ with the Weil representation $\rho$ of~\Cref{fact:weilrep}. It remains to address the cusps:
\begin{lemma}\label{lem:thetacusp}
    $\Theta_{\calL}$ is holomorphic at every cusp, and is therefore a modular form of weight $r/2$ for $\SL_2(\ZZ)$ with representation $\rho$. Moreover, $\order_\infty(\Theta_{\calL}) = 0$; more precisely, we have
    \begin{equation*}
        \lim_{\Im \tau \to \infty} \Theta_{\calL}(\tau) = \ev_0,
    \end{equation*}
    uniformly in $\Re \tau$.
\end{lemma}
\begin{proof}
    See~\Cref{sec:thetacuspproof}.
\end{proof}

Having constructed the vector-valued modular form, we next scalarize it, by combining it with the vector $\beta$ from Step 1. The polynomial below is engineered to meet three constraints at once. Its coefficients are the entries of $\beta$, arranged so that evaluating it on vectors of $q$th roots of unity produces sums of exactly the shape that~\eqref{eq:betafourier} controls; this is what will let us convert the duality conditions into vanishing Fourier coefficients in Step 3. It is homogeneous of degree $d$, so composing it with the weight-$r/2$ form $\Theta_{\calL}$ will produce a scalar form of integer weight $dr/2$. And it is averaged over $u \in \ZZ_q^*$: under the Weil action of $\Gamma_0(q)$, the coordinates of $\CC^q$ are rescaled in index by the (invertible) diagonal entry $d_0$ and multiplied by quadratic phases (\Cref{fact:weilrep}); the average absorbs the index rescaling, and the restriction of the support to the cone $\calQ$ makes the quadratic phases equal to $1$ --- which is why the composition will transform by a mere character.

\begin{definition}
    Define the polynomial $P_\beta: \CC^q \to \CC$ as follows:
    \begin{align*}
        P_{\beta}(X_0, \ldots, X_{q-1}) &:= \sum_{u \in \ZZ_q^*} \sum_{\xv \in \calQ} \beta_{\xv} \prod_{i = 1}^d X_{ux_i \bmod{q}}.
    \end{align*}
    By construction, $P_\beta$ is homogeneous with degree $d$.
\end{definition}

Next, we will show that $P_\beta$ satisfies~\eqref{eq:scalarisation} by appealing to the explicit formula of~\Cref{fact:weilrep}. Recall the definition of the subgroup $\Gamma_0(q)$ of $\SL_2(\ZZ)$ from~\Cref{def:subgroups}.

\begin{lemma}\label{lem:gen1scalarisation}
    For any $g \in \Gamma_0(q)$, we have $P_\beta(\rho(g)\vecv) \propto P_\beta(\vecv)$ as functions of $\vecv \in \CC^q$.
\end{lemma}
\begin{proof}
    See~\Cref{sec:gen1scalarisationproof}.
\end{proof}

Consequently, by~\Cref{lem:thetacusp,lem:restriction,lem:scalarisation} ($\Theta_{\calL}$ is a modular form for $\SL_2(\ZZ)$; restrict it to $\Gamma_0(q)$ and compose with $P_\beta$, which is not the zero polynomial since $P_\beta(\ev_0) = (q-1)\beta_0 \neq 0$, as computed in~\Cref{lem:gen1nonzero} below), the composed function $P_\beta \circ \Theta_{\calL}$ is a scalar modular form of weight $dr/2$ for $\Gamma_0(q)$, with the one-dimensional representation $\chi$ defined by~\eqref{eq:scalarisation} for $P = P_\beta$; moreover, $\chi^j = 1$ for some positive integer $j$. We will also need the classification of the cusp classes of $\Gamma_0(q)$:

\begin{proposition}[{\cite[Section~3.8]{DiamondShurman05}}]\label{prop:gamma0cusps}
    Under the action of $\Gamma_0(q)$, the cusps $\QQ \cup \{\infty\}$ split into two equivalence classes:
    \begin{itemize}
        \item The $\infty$ class contains $\infty$ and all $a/b \in \QQ$ where $\gcd(a, b) = 1$ and $b$ is divisible by $q$.
        \item The 0 class contains all $a/b \in \QQ$ where $\gcd(a, b) = 1$ and $b$ is not divisible by $q$.
    \end{itemize}
\end{proposition}

\subsection{Proof Step 3: The Order of the Cusp at 0}\label{sec:gen1step3}

This step is where we use the duality of Step~1. We will expand $P_\beta \circ \Theta_{\calL}$ at the cusp at $0$ and track its coefficients: each term of the expansion comes from a tuple of dual-lattice vectors, and the coefficient multiplying it is an average of sums of exactly the shape of~\eqref{eq:betafourier}, evaluated at frequency vectors built from the tuple's residue classes. The frequencies in $S_d$ are the ones on which the sums in~\eqref{eq:betafourier} vanish; and a frequency outside $S_d$ has, by the very definition of $S_d$, a large total cost $\sum_i w_\calL(t_i\av)$ --- which lower-bounds the Gaussian decay exponent of the corresponding terms. In this way the exclusion of $S_d$ from the support becomes a strict lower bound on the order of vanishing at the cusp.

Next, we analyze the behavior of $P_\beta \circ \Theta_{\calL}$ at the cusp at 0. To do this, we follow the procedure in~\Cref{sec:modularforms} and start by choosing a group element $g \in \SL_2(\ZZ)$ such that $g \ast \infty = 0$. We may simply take $g = S = \begin{bmatrix} 0 & -1 \\ 1 & 0 \end{bmatrix}$. Now, following~\eqref{eq:pullback}, we define:
\begin{equation}\label{eq:f0}
f_0(\tau) := \tau^{-dr/2} P_\beta(\Theta_{\calL}(S \ast \tau)) = \tau^{-dr/2} P_\beta(\Theta_{\calL}(-1/\tau)).
\end{equation}
This is modular of weight $dr/2$ with respect to the conjugate subgroup $S^{-1} \Gamma_{0}(q) S$ and the conjugate representation $\overline{\chi}$ defined by $\overline{\chi}(S^{-1} g S) = \chi(g)$. 
Our goal is to compute $\order_{\infty}(f_0)$. To do this, we follow the procedure in~\Cref{sec:cuspinfty}:

\begin{proposition}
    The width of the cusp at $\infty$ with respect to $S^{-1}\Gamma_0(q)S$ (equivalently, of the cusp at $0$ for $\Gamma_0(q)$) is $q$.
\end{proposition}
\begin{proof}
    The width is the minimal $h$ such that $T^h \in S^{-1} \Gamma_0(q) S \Leftrightarrow S T^h S^{-1} \in \Gamma_0(q)$. 
    To this end, we compute:
    \begin{align*}
        S T^h S^{-1} &= \begin{bmatrix} 0 & -1 \\ 1 & 0 \end{bmatrix} \begin{bmatrix} 1 & h \\ 0 & 1 \end{bmatrix} \begin{bmatrix} 0 & 1 \\ -1 & 0 \end{bmatrix} \\
        &= \begin{bmatrix} 1 & 0 \\ -h & 1 \end{bmatrix},
    \end{align*}
    which belongs to $\Gamma_0(q)$ if and only if $q \mid h$. Consequently, the width of the cusp is $q$ as claimed.
\end{proof}

\begin{lemma}
    We have $\order_0(P_\beta \circ \Theta_{\calL}) = \order_\infty(f_0) > r(q+1)d/24$.
\end{lemma}
\begin{proof}
Now continuing, we have:
\[
\overline{\chi}\left(\begin{bmatrix} 1 & q \\ 0 & 1 \end{bmatrix}\right) = \chi\left(\begin{bmatrix} 1 & 0 \\ -q & 1 \end{bmatrix}\right) = 1,
\]
because $\begin{bmatrix} 1 & 0 \\ -q & 1 \end{bmatrix} \in \Gamma(q)$, on which $\rho$ is trivial (\Cref{fact:weilrep}): for this element,~\eqref{eq:scalarisation} reads $P_\beta(\vecv) = \chi\left(\begin{bmatrix} 1 & 0 \\ -q & 1 \end{bmatrix}\right) P_\beta(\vecv)$ for all $\vecv$, and $P_\beta \neq 0$.
Plugging this into~\eqref{eq:qexpansion}, we may write:
\begin{align*}
    f_0(\tau) &= \sum_{n = -\infty}^\infty c_n \exp(2\pi i \tau n/q),
\end{align*}
and our goal is to show that $c_n = 0$ for all $n \leq r(q+1)d/24$. (In particular, this covers all negative $n$, so the computation below also establishes that $f_0$ is holomorphic at $\infty$, i.e., that $P_\beta \circ \Theta_{\calL}$ is holomorphic at the cusp at $0$.)
To do this, let us compute:
\begin{align*}
    f_0(\tau) &= \tau^{-dr/2} P_\beta(\Theta_{\calL}(-1/\tau)) \\
    &= \tau^{-dr/2} P_\beta(\tau^{r/2} \rho_S \Theta_{\calL}(\tau)) \quad\text{(\Cref{lem:gen1S})}\\
    &= P_\beta(\rho_S \Theta_{\calL}(\tau)) \quad\text{($P_\beta$ is homogeneous of degree $d$)}\\
    &= P_\beta\left(\sum_{t \in \ZZ_q} \left(\sum_{\yv \in t\av + \calL} \exp\left(\pi i \langle \yv, \yv \rangle \tau\right)\right) \rho_S \ev_t\right) \\
    &\propto P_\beta\left(\sum_{z \in \ZZ_q} \left(\sum_{t \in \ZZ_q} \sum_{\yv \in t\av+\calL} \exp(\pi i \langle \yv, \yv \rangle \tau)\, \zeta_q^{-2\mu tz}\right) \ev_z\right) \quad\text{(\Cref{def:weil})} \\
    &= \sum_{u \in \ZZ_q^*} \sum_{\xv \in \calQ} \beta_\xv \prod_{i = 1}^d \left(\sum_{t \in \ZZ_q} \sum_{\yv \in t\av+\calL} \exp(\pi i \langle \yv, \yv \rangle \tau)\, \zeta_q^{-2\mu tux_i}\right) \\
    &= \sum_{\tv \in \ZZ_q^d}\ \sum_{\substack{\yv_1, \ldots, \yv_d: \\ \yv_i \in t_i\av+\calL}} \exp\left(\pi i \sum_{i = 1}^d \langle \yv_i, \yv_i \rangle \tau\right) \sum_{u \in \ZZ_q^*} \sum_{\xv \in \calQ} \beta_\xv\, \zeta_q^{-2\mu u \langle \tv, \xv \rangle}.\numberthis \label{eq:f0expanded}
\end{align*}
In the third step from the end, we substituted the definition of $\rho_S$ and pulled its scalar prefactor, which does not depend on $t$, out of $P_\beta$ by homogeneity; so $f_0$ is a fixed nonzero multiple of~\eqref{eq:f0expanded}, which does not affect which of its coefficients vanish. In the last step, we expanded the product of the $d$ inner series and moved the (finite) sums over $u$ and $\xv$ inside; this rearrangement is justified because each inner series converges absolutely (\Cref{lem:latticegaussian}, exactly as in the proof of~\Cref{lem:thetaconverges}), so the product expands into an absolutely convergent sum over tuples. In the display, $\langle \tv, \xv \rangle := \sum_{i=1}^d t_i x_i \in \ZZ_q$.

We claim that the inner double sum in~\eqref{eq:f0expanded} vanishes whenever $\tv \in S_d$. First, observe that $S_d$ is closed under coordinatewise multiplication by any scalar $c \in \ZZ_q^*$: if $\zv$ satisfies the defining condition of $S_d$ for the index $u'$, then $c\zv$ satisfies it for the index $u'c^{-1}$, since $w_\calL\big((u'c^{-1})(cz_i)\av\big) = w_\calL(u'z_i\av)$. Now, for each fixed $u \in \ZZ_q^*$, the sum over $\xv$ in~\eqref{eq:f0expanded} is $\sum_{\xv \in \calQ} \beta_\xv\, \zeta_q^{\langle \xv, -2\mu u \tv \rangle}$, which vanishes by~\eqref{eq:betafourier} whenever $-2\mu u \tv \in S_d$; and since $-2\mu u \in \ZZ_q^*$ (here $\mu \not\equiv 0 \pmod q$ by~\Cref{prop:munonzero}, $u \neq 0$, and $q$ is odd), membership $\tv \in S_d$ implies $-2\mu u\tv \in S_d$ by the closure just observed. This proves the claim.

So the only tuples $(\tv, \yv_1, \ldots, \yv_d)$ that contribute nontrivially to~\eqref{eq:f0expanded} have $\tv \notin S_d$, which --- taking $u = 1$ in the definition of $S_d$ --- means:
\begin{align*}
    \frac{r(q+1)}{24}d &< \sum_{i = 1}^d w_{\calL}(t_i\av) = \frac{q}{2} \sum_{i = 1}^d \min_{\xv \in t_i\av + \calL} \langle \xv, \xv \rangle \leq \frac{q}{2} \sum_{i = 1}^d \langle \yv_i, \yv_i \rangle,
\end{align*}
where the final inequality uses $\yv_i \in t_i\av + \calL$. Each $\langle \yv_i, \yv_i \rangle$ lies in $\frac{2}{q}\ZZ$ by~\Cref{prop:evendistances}, and is nonnegative, so each contributing term of~\eqref{eq:f0expanded} equals its coefficient times $\exp(2\pi i n \tau/q)$ for the \emph{integer} $n := \frac{q}{2}\sum_{i=1}^d \langle \yv_i, \yv_i \rangle > r(q+1)d/24$. Collecting the terms of~\eqref{eq:f0expanded} by the value of $n$ (justified by absolute convergence, exactly as in the proof of~\Cref{lem:scalarisation}) and invoking the uniqueness in~\Cref{fact:fourierexp}, we conclude that $c_n = 0$ for every $n \leq r(q+1)d/24$. That is, $\order_0(P_\beta \circ \Theta_{\calL}) = \order_\infty(f_0) > r(q+1)d/24$, as desired.
\end{proof}

\subsection{Proof Step 4: Root-Counting Contradiction from the Valence Formula}\label{sec:gen1step4}

For the coup de grace, we will apply~\Cref{thm:valence} to $P_\beta \circ \Theta_{\calL}$ to obtain a contradiction. 
To do this, we must first verify that $P_\beta \circ \Theta_{\calL}$ is not identically 0. This will necessarily make use of~\eqref{eq:beta0} (it is the only thing preventing all the $\beta_\xv$'s from being 0).

\begin{lemma}\label{lem:gen1nonzero}
    As $\Im \tau \to \infty$, we have $P_\beta(\Theta_{\calL}(\tau)) \to (q-1)\beta_0 \neq 0$. Consequently, $P_\beta \circ \Theta_{\calL}$ is not identically 0.
\end{lemma}
\begin{proof}
    By~\Cref{lem:thetacusp}, $\Theta_{\calL}(\tau) \to \ev_0$ as $\Im \tau \to \infty$, so by continuity of the polynomial $P_\beta$ we get $P_\beta(\Theta_{\calL}(\tau)) \to P_\beta(\ev_0)$. Finally, evaluating $P_\beta$ at $\ev_0$ keeps exactly the terms with $ux_i \equiv 0 \pmod{q}$ for all $i$, i.e., with $\xv = \zerov$ (as $u$ is invertible); so $P_\beta(\ev_0) = \sum_{u \in \ZZ_q^*} \beta_0 = (q-1)\beta_0 \neq 0$ by~\eqref{eq:beta0}.
\end{proof}

\noindent
Finally, we can finish the proof of~\Cref{thm:gen1} by invoking the valence formula:
\begin{proof}[Proof of~\Cref{thm:gen1}]
    We have shown that $P_\beta \circ \Theta_{\calL}$ is a nonzero modular form of weight $rd/2$ that has order $> r(q+1)d/24$ at the cusp at 0. On the other hand,~\Cref{thm:valence} applies (we have $-I \in \Gamma_0(q)$, and the representation $\chi$ satisfies $\chi^j = 1$ for some positive integer $j$, by~\Cref{lem:scalarisation}); dropping the nonnegative contributions of the interior orders and of the cusp class at $\infty$ (by~\Cref{prop:gamma0cusps}, the classes of $\infty$ and $0$ are the only two), it implies that:
    \begin{align*}
        \frac{\order_0(P_\beta \circ \Theta_{\calL})}{2} &\leq \frac{rd}{48} [\SL_2(\ZZ) : \Gamma_0(q)] \\
        &= \frac{rd(q+1)}{48} \quad\text{(\Cref{lem:indices})},
    \end{align*}
    which is a direct contradiction, establishing the theorem.
\end{proof}

\section{Stronger Construction}\label{sec:gen4}

This section proves our main result, \Cref{thm:gen4aprime}: sparse quadratic-phase delta expansions, which power $s$-server PIR for $s$ below $q$, with the support exponents computed in~\Cref{sec:gen4exponent}. The proof follows the warm-up, with two new ingredients: an elliptic variable in the theta function, and a second scalarization; \Cref{sec:gen4overview} explains both, and~\Cref{tab:gen4comparison} compares the two proofs step by step.

The new ingredients are again classical in isolation. The components of the Jacobi theta function are, up to normalization, theta functions with rational characteristics~\cite[Chapter~I, \S3]{Mumford83}; composing it with the witness polynomial yields, up to roots of unity, a Jacobi form of weight $d/2$ (in the variable $z = qw$) in the sense of~\cite{EichlerZagier85} (\Cref{lem:bigF,lem:fgtranslation}); and the two scalar modular forms of Step~3 are the squares of, respectively, its first nonvanishing Taylor coefficient in the elliptic variable~\cite[Section~3]{EichlerZagier85} and the product of its restrictions to the nonzero 2-torsion points~\cite[Theorem~1.3]{EichlerZagier85}. Bounding orders of vanishing by combining such scalar forms with the valence formula is the strategy of~\cite{RichterWesterholtRaum15,Aoki22,LiWang25} (see the discussion after~\Cref{lem:valence2gen4}). What is new here is the argument that ties the cusp orders to the witness: the selected cusp classes, the per-class quantities $\ell_\frakc$ and $H_\frakc$, and the analysis of the edge polynomials in Step~4.

\subsection{Overview of the Proof}\label{sec:gen4overview}

This subsection explains how the proof of~\Cref{thm:gen4aprime} extends the warm-up (\Cref{sec:gen1overview}); we give the details in the rest of this section. \Cref{tab:gen4comparison} summarizes the changes, and~\Cref{fig:gen4flow} at the end of this subsection shows how the final inequalities fit together.

\input{tabgen4comparison}

\paragraph{The target and the witness.} A quadratic-phase delta expansion (\Cref{def:genconedelta}) with phases in a set $\calA \subseteq \ZZ_q$ of size $s$ expands the delta function at $\zerov$ in the characters $\xv \mapsto \zeta_q^{\langle \zv, \xv \rangle + a\langle \xv, \xv \rangle}$, for $\zv$ in the support and $a \in \calA$. Using all $q$ phases recovers cone-delta functions: if $f$ is a cone-delta function, then $\frac{1}{q}\sum_{a \in \ZZ_q} \zeta_q^{a\langle \xv, \xv \rangle} f(\xv)$ is the delta function at $\zerov$, since the average over $a$ is the indicator function of the cone. With $s < q$ phases, this average is no longer available. Dually (Step~1, \Cref{sec:gen4step1}), the witness $\alpha$ now satisfies $\alpha_\zerov = 1$ and
\begin{equation*}
    \sum_{\xv \in \ZZ_q^d} \alpha_\xv \zeta_q^{a\langle \xv, \xv \rangle + \langle \xv, \zv \rangle} = 0 \qquad\text{for all } a \in \calA \text{ and } \zv \text{ in the support},
\end{equation*}
and it need not be supported on the cone. We symmetrize it to a witness $\beta$ that is supported on $\{\xv : \sum_i x_i \equiv 0\}$ and invariant under $\xv \mapsto \xv + \bfone$ (\Cref{prop:symmetrization}), for a reason that will become clear below, and as in the warm-up we package $\beta$ as the homogeneous polynomial $P_\beta(\vecv) := \sum_{\xv} \beta_\xv\, v_{x_1} v_{x_2} \cdots v_{x_d}$ of degree $d$.

\paragraph{Where the phases go: the selected cusps.} Since $\beta$ is no longer supported on the cone, $P_\beta$ is not invariant under $\rho_T$, and the modularity for $\Gamma_0(q)$ used in the warm-up is lost. We instead work with the smaller subgroup $\Gamma(q) = \{g \in \SL_2(\ZZ) : g \equiv I \pmod{q}\}$, on which the Weil representation is trivial (\Cref{fact:weilrep};~\cite[Lemma~5.12]{Stromberg13}), so that modularity comes for free. In exchange, $\Gamma(q)$ has $\frac{q^2-1}{2}$ classes of cusps rather than two, and we must keep track of all of them. The class of a cusp is determined by its numerator and denominator in lowest terms (with $\infty = \frac{1}{0}$), taken modulo $q$ and up to a common sign, and every nonzero point of $\ZZ_q^2$, up to sign, arises in this way (\Cref{prop:gammaqcusps}; see, e.g.,~\cite[Section~3.8]{DiamondShurman05}). So the cusp classes lie on the $q + 1$ lines through the origin of $\ZZ_q^2$, with $\frac{q-1}{2}$ classes on each line. We label the line spanned by $(2a, 1)$ by $a \in \ZZ_q$, and the remaining line, spanned by $(1, 0)$, by $\infty$. A cusp class is then specified by its label and a multiplier $\lambda \in \{1, \ldots, \frac{q-1}{2}\}$, as the class $\pm\lambda(2a, 1)$ or $\pm\lambda(1, 0)$ (\Cref{fig:gen4cusps}); for instance, the cusps $\infty = \frac{1}{0}$ and $0 = \frac{0}{1}$ have labels $\infty$ and $0$, both with multiplier $1$. As in the warm-up, the behavior at a cusp $g \ast \infty$ is governed by $P_\beta \circ \rho(g)$. The phase $a \in \calA$ becomes visible at the cusp classes $\frakc_{a, \lambda} = \pm\lambda(2a, 1)$ with label $a$, for $\lambda \in \{1, \ldots, \frac{q-1}{2}\}$. For a suitable $g_{a, \lambda}$ sending $\infty$ into this class, $\rho(g_{a, \lambda})$ is a rescaled Fourier transform followed by multiplication by the quadratic phase $\zeta_q^{ay^2}$~\eqref{eq:rhogal}, so the coefficient of the monomial $v_{x_1} \cdots v_{x_d}$ in $P_\beta \circ \rho(g_{a, \lambda})$ is a constant multiple of
\begin{equation*}
    \sum_{\yv \in \ZZ_q^d} \beta_\yv\, \zeta_q^{a\langle \yv, \yv \rangle - \lambda^{-1}\langle \xv, \yv \rangle}.
\end{equation*}
This is a witness sum for the phase $a$, so it vanishes whenever $-\lambda^{-1}\xv$ lies in the support. Each phase $a \in \calA$ thus controls the $\frac{q-1}{2}$ cusp classes with label $a$, and the witness is visible at $s \cdot \frac{q-1}{2}$ \emph{selected} classes in all. In the warm-up, the cone made all phases agree, and a single cusp sufficed: the cusp $0 = S \ast \infty$, which has label $0$.

\input{figgen4cusps}

\paragraph{Tori and torsion points.} Recall from the technical overview that $\tau \in \calH$ stands for the lattice $\ZZ\tau + \ZZ$ in $\CC$, up to rotation and scaling (\Cref{fig:lattice}). The quotient $\CC/(\ZZ\tau + \ZZ)$ is a \emph{torus}: we draw it as the parallelogram with vertices $0$, $1$, $\tau$ and $1 + \tau$, with opposite sides glued (\Cref{fig:gen4twotorsion}). For an integer $N \geq 1$, its \emph{$N$-torsion points} are the points $w$ with $Nw \in \ZZ\tau + \ZZ$, that is, $w \equiv \frac{a\tau + b}{N}$ for $a, b \in \{0, \ldots, N - 1\}$; there are $N^2$ of them, forming a grid in the parallelogram. Two kinds will matter. Translations by the $q$-torsion points act on our new theta function by the Heisenberg operators (below), and the second scalarization evaluates at the three nonzero 2-torsion points $\frac{1}{2}$, $\frac{\tau}{2}$ and $\frac{1+\tau}{2}$.

\paragraph{The elliptic variable.} Our new theta function has an additional \emph{elliptic} variable $w \in \CC$, which we think of as a point of the torus; up to normalization, its coordinates are theta functions with rational characteristics~\cite[Chapter~I, \S3]{Mumford83}:
\begin{equation*}
    \Theta(\tau, w) := \sum_{k \in \ZZ} (-1)^k \exp\left(\frac{\pi i \tau}{q}\left(k - \frac{1}{2}\right)^2 + 2\pi i \left(k - \frac{1}{2}\right) w\right) \ev_{k + (q-1)/2 \bmod q} \in \CC^q.
\end{equation*}
It has two symmetries (\Cref{fig:gen4symmetries}). First, as in the warm-up, $\SL_2(\ZZ)$ acts, and $\Theta$ transforms by the Weil representation (now with $\mu = \frac{q+1}{2}$), up to an explicit factor (\Cref{lem:descended}). The action is now on pairs, by $g \ast (\tau, w) = \left(g \ast \tau, \frac{w}{c\tau + d_0}\right)$: it moves the point $w$ of the torus along with the lattice (\Cref{fig:gen4action}). This is the action of $\SL_2(\ZZ)$ that underlies Jacobi forms~\cite{EichlerZagier85}. Second, translating $w$ by a $q$-torsion point $\frac{a\tau + b}{q}$ multiplies $\Theta$ by the Heisenberg operator $\sfZ^b\sfX^a$ of~\Cref{def:heisenberg}, again up to an explicit factor (\Cref{lem:thetatranslation}). Both symmetries are instances of the classical transformation laws of theta functions with characteristics~\cite[Chapter~I, \S\S3 and~7]{Mumford83}. Conjugating by the Weil representation permutes the Heisenberg operators up to scalars (\Cref{lem:heisenbergweil}; see~\cite[n\textsuperscript{o}~10, Th\'eor\`eme~1]{Weil64}, \cite[Theorem~2.4(a)]{Gerardin77} and~\cite[Theorem~1.3.2 and Equation~(1.3.1)]{GurevichHadani07}), so the second symmetry persists after applying $\rho(g)$. The second symmetry is why we symmetrized the witness: the support condition and the shift-invariance of $\beta$ make $P_\beta$ invariant under $\sfZ$ and $\sfX$ (\Cref{prop:pproperties}), so the scalar function
\begin{equation*}
    F(\tau, z) := P_\beta\left(\Theta\left(\tau, \frac{z}{q}\right)\right)
\end{equation*}
is quasi-periodic in $z$ for the lattice $\ZZ\tau + \ZZ$. Up to roots of unity, $F$ is a Jacobi form of weight $\frac{d}{2}$ in the sense of~\cite{EichlerZagier85} (\Cref{lem:bigF,lem:fgtranslation}): a function of a modulus $\tau$ and of a point $z$ on the corresponding torus.

\input{figgen4symmetries}

\input{figgen4action}

\paragraph{Two scalarizations.} A Jacobi form gives rise to many scalar modular forms~\cite{EichlerZagier85}, and we use two of them (Step~3, \Cref{sec:gen4step3}). The first is a Taylor coefficient: expanding $F(\tau, z)$ in $z$ around $z = 0$, let $A_m(\tau)$ be the coefficient of $z^m$ for the least $m$ such that this coefficient is not identically zero. Its square $A_m^2$ is a modular form of weight $d + 2m$ for $\Gamma(q)$ (\Cref{lem:amsq}), as for the first nonvanishing Taylor coefficient of any Jacobi form~\cite[Section~3]{EichlerZagier85}. The second is the product $B(\tau)$ of the values of $F$, suitably normalized, at the three nonzero 2-torsion points $z = \frac{1}{2}, \frac{\tau}{2}, \frac{1+\tau}{2}$ of the torus. The action of $\SL_2(\ZZ)$ permutes these three points (\Cref{fig:gen4twotorsion}), so only their product is modular: $B^2$ is a modular form of weight $3d$ for $\Gamma(q)$ (\Cref{cor:bsqmodular}). Restrictions of Jacobi forms to torsion points are modular for smaller groups~\cite[Theorem~1.3]{EichlerZagier85}, and taking the product over an $\SL_2(\ZZ)$-orbit of torsion points is standard~\cite{RichterWesterholtRaum15,LiWang25}. The two scalarizations see the theta function through different cost functions. The $k$th term of a coordinate of $\Theta(\tau, z/q)$ has the factor $e^{\pi i \tau (k - 1/2)^2/q}$, which at $z = 0$ reads $e^{\pi i \tau(k(k-1) + 1/4)/q}$; at the 2-torsion point $z = \frac{\tau}{2}$, the factor $e^{2\pi i (k - 1/2)z/q}$ turns it into $e^{\pi i \tau (k^2 - 1/4)/q}$. So the Taylor coefficient sees the \emph{edge cost} $k(k-1)$, while the 2-torsion evaluations see the \emph{square cost} $k^2$, which is $w_q(k)$ for $|k| \leq \frac{q-1}{2}$ (\Cref{def:track3exponent}).

\input{figgen4twotorsion}

\paragraph{Root counting at each cusp.} At each cusp class $\frakc$, with a fixed matrix $g_\frakc$ sending $\infty$ into $\frakc$, we record two quantities (\Cref{sec:gen4cuspdata}). The first, $\ell_\frakc \geq 0$, records where the expansion of $F_{g_\frakc}(\tau, z) := P_\beta(\rho(g_\frakc)\Theta(\tau, z/q))$ in powers of $e^{\pi i \tau/(4q)}$ starts: its leading terms come from tuples $\kv \in \ZZ^d$ with $\sum_i k_i(k_i - 1) = 2\ell_\frakc$. The second, $H_\frakc$, is the least square cost $\sum_i w_q(x_i)$ of a monomial $v_{x_1} \cdots v_{x_d}$ of $P_\beta \circ \rho(g_\frakc)$. They bound the orders of our two forms at $\frakc$ from below: $\order_\frakc(A_m^2) \geq \frac{d}{4} + 2\ell_\frakc$ and $\order_\frakc(B^2) \geq \frac{d}{4} + 2\ell_\frakc + 2H_\frakc$ (\Cref{lem:amsq,lem:b2holomorphic}). Summing over all cusp classes, the valence formula (\Cref{thm:valence}; see, e.g.,~\cite[Theorem~4.1.4]{Rankin77}) then gives (\Cref{lem:valence1gen4,lem:valence2gen4})
\begin{equation*}
    \text{(V1)}\quad \sum_\frakc \ell_\frakc \leq \frac{d(q^2-1)(q-3)}{48} + \frac{m(q^3-q)}{24} \qquad\text{and}\qquad \text{(V2)}\quad \sum_\frakc \left(\ell_\frakc + H_\frakc\right) \leq \frac{d(q^2-1)(q-1)}{16}.
\end{equation*}
These take the place of the warm-up's single valence inequality, and both follow a known strategy: (V1) is essentially the case $t = 1$ of~\cite[Lemma~3.1]{LiWang25} (see also~\cite[Lemma~5.1]{Aoki22}), applied to a suitable norm of $F$, and (V2) applies the valence formula to the 2-torsion product in the same spirit as~\cite{RichterWesterholtRaum15,Aoki22,LiWang25}. What is specific to our setting is how the witness controls $\ell_\frakc$ and $H_\frakc$. It remains to show that the witness forces $H_\frakc$ and $\ell_\frakc$ to be large at the selected classes.

\paragraph{The square costs: the warm-up argument.} At a selected class $\frakc_{a, \lambda}$, the coefficients of $P_\beta \circ \rho(g_{a, \lambda})$ are the witness sums for the phase $a$ displayed above, which vanish on the support. So every surviving monomial has square cost more than $\nu d$, that is, $H_{\frakc_{a, \lambda}} > \nu d$ (\Cref{lem:Hgen4}). This is the warm-up's argument at the cusp $0$, now run at $s \cdot \frac{q-1}{2}$ cusps.

\paragraph{The edge polynomials.} The Taylor order $m$ gives a second lower bound on $\ell_\frakc$ at each selected class (Step~4, \Cref{sec:gen4step4}):
\begin{equation*}
    \gamma d < 2\ell_\frakc + \vartheta(d - qm)
\end{equation*}
(\Cref{lem:randomgen4}). 
The high-level idea of the proof of the above is the following: since $F$ vanishes to order $m$ at $z = 0$, some tuple $\kv$ in the leading term of the expansion at $\frakc$ has $\sum_i k_i$ within $\frac{d - qm}{2}$ of $0$ or of $d$. (These sums are the exponents of the \emph{edge polynomial} $p_\frakc$, which is divisible by $(X^q - 1)^m$; see~\Cref{lem:pc,lem:pcperiodic,lem:pcdivisible,cor:nonzeromonomial}.) Such a tuple, or its reflection $\bfone - \kv$, has low \emph{tilted edge cost} $w_{q, \vartheta}(j) = \min (k(k-1) + 2\vartheta k)$, the minimum over the integers $k \equiv j \pmod{q}$, so its coefficient vanishes unless $\ell_\frakc$ is large.

\paragraph{Combining the inequalities.} Summing the bound on $H_\frakc$ over the selected classes and inserting it into (V2), averaging the edge-polynomial bound over the selected classes, and using (V1) to eliminate $m$ yields
\begin{equation*}
    \gamma < \left(1 - \frac{6\vartheta s}{q+1}\right)\left(\frac{q^2-1}{4s} - 2\nu\right) + \frac{\vartheta(q-1)}{2},
\end{equation*}
which contradicts the definition of $\gamma = \gamma(\vartheta, \nu)$ in~\Cref{def:track3exponent} (Step~5, \Cref{sec:gen4step5}). We again work backwards: $\gamma(\vartheta, \nu)$ is the smallest budget for which this combination gives a contradiction, and a smaller budget means a smaller support. The constraints on $\vartheta$ and $\nu$ ensure that the coefficients in the combination have the right signs. When $\vartheta = 0$, the Taylor coefficient and the edge polynomials are not needed at all, and the 2-torsion product alone already gives a nontrivial exponent (\Cref{remark:thetazero}).

\input{figgen4flow}

\paragraph{The size of the support.} The support $S_d(\vartheta, \nu)$ consists of the frequency vectors on which the two arguments need the witness sums to vanish: the set $S_d^{\mathrm{sq}}(\nu)$ of vectors $\zv$ of total square cost $\sum_i w_q(z_i)$ at most $\nu d$, and the set $S_d^{\mathrm{edge}}(\vartheta, \nu)$ of vectors of total tilted edge cost $\sum_i w_{q, \vartheta}(z_i)$ at most $\gamma d$. These are closed under the affine maps $\zv \mapsto u\zv + b\bfone$ (the rescalings serve the selected classes $\frakc_{a, \lambda}$, and the shifts serve the symmetrization and the reflection above; for $q = 3$, the shifts let the support reach the far end of the weight range, as discussed in~\Cref{sec:q3elementary}) and then thickened by changing one coordinate (to absorb a prefactor from the symmetrization), which costs only a factor $q^{o(d)}$. The same moment-generating-function bound as in the warm-up (\Cref{lem:sublevel}), applied to these two sets, bounds $|S_d(\vartheta, \nu)|$ by $q^{(\eta_{q, s}(\vartheta, \nu) + o(1))d}$ (\Cref{lem:gen4supportsize}), and optimizing over $(\vartheta, \nu)$ numerically gives the exponents of~\Cref{cor:gen4exponent} (\Cref{lem:gen4numerics}).

\subsection{Main Result}\label{sec:gen4main}

We use the costs $w_{q, \vartheta}$ and $w_q = w_{q, 1/2}$, their generating functions $M_\vartheta$, the feasible pairs $(\vartheta, \nu)$, and the budget $\gamma(\vartheta, \nu)$ of~\Cref{def:track3exponent}. Explicitly, $w_q(j) = j^2$ for $j \in [0, (q-1)/2]$ and $w_q(j) = (q-j)^2$ for $j \in [(q+1)/2, q-1]$ (\Cref{lem:wqthetaexplicit} gives $w_{q, \vartheta}$ in the same way).

\begin{definition}[The set $S_d(\vartheta, \nu)$]\label{def:gen4aprimesupport}
Let $q$ be an odd prime, $s \in [2, q]$ an integer, $d \geq 1$ an integer dimension parameter that is divisible by $q$, and $(\vartheta, \nu)$ a feasible pair, with budget $\gamma = \gamma(\vartheta, \nu)$ (\Cref{def:track3exponent}).
Define the sets:
\begin{align*}
    S_d^{\mathrm{sq}}(\nu)
    &:=
    \left\{
        \zv\in\ZZ_q^d:
        \sum_{i=1}^d w_q(z_i)
        \leq
        \nu d
    \right\} \\
    S_d^{\mathrm{edge}}(\vartheta,\nu)
    &:=
    \left\{
        \zv\in\ZZ_q^d:
        \sum_{i=1}^d w_{q,\vartheta}(z_i)
        \leq
        \gamma d
    \right\}.
\end{align*}
Finally, define
\begin{align*}
    T_d(\vartheta, \nu) &:= \left\{u\zv + b\bfone: u \in \ZZ_q^*; b \in \ZZ_q; \zv \in S_d^{\mathrm{sq}}(\nu) \cup S_d^{\mathrm{edge}}(\vartheta, \nu)\right\}; \\
    S_d(\vartheta,\nu) &:= \left\{\zv+c\ev_i: c \in \ZZ_q; i \in [d]; \zv \in T_d(\vartheta, \nu)\right\}.
\end{align*}
\end{definition}

\noindent
We can finally state the main theorem of this section:

\begin{theorem}
\label{thm:gen4aprime}
For any $q, s, d, \vartheta, \nu$ as defined in~\Cref{def:gen4aprimesupport} (in particular, we require $q \mid d$) and any set $\calA \subseteq \ZZ_q$ of size $s$, there exists a quadratic-phase delta expansion over $\QQ(\zeta_q)$ on $\ZZ_q^d$, in the sense of~\Cref{def:genconedelta}, with support contained in $S_d(\vartheta, \nu) \times \calA$. In particular, this expansion is $(|S_d(\vartheta, \nu)|, s)$-sparse.
\end{theorem}

The set $S_d(\vartheta, \nu)$ is permutation-invariant. Indeed, the sets $S_d^{\mathrm{sq}}(\nu)$ and $S_d^{\mathrm{edge}}(\vartheta, \nu)$ are, as their definitions depend only on the multiset of coordinates; and permuting the coordinates of $u\zv + b\bfone + c\ev_i$ gives $u\zv' + b\bfone + c\ev_{i'}$, where $\zv'$ is obtained from $\zv$ by the same permutation and $i'$ is the new position of the $i$th coordinate.

\begin{corollary}[Generalizing to all $d$]\label{cor:roundd}
    For any $q, s, \vartheta, \nu$ as in~\Cref{def:gen4aprimesupport} and any positive integer $d$ (not necessarily divisible by $q$), there exists a $(|S_{q\lceil d/q\rceil}(\vartheta, \nu)|, s)$-sparse quadratic-phase delta expansion over $\QQ(\zeta_q)$ on $\ZZ_q^d$ with permutation-invariant support.
\end{corollary}
\begin{proof}
    See~\Cref{sec:rounddproof}.
\end{proof}

Bounding the support sizes $|S_d(\vartheta, \nu)|$ gives the following, which we prove in~\Cref{sec:gen4exponent}.
\begin{corollary}\label{cor:gen4exponent}
    Fix an odd prime $q$, an integer $s \in [2, q]$, and a feasible pair $(\vartheta, \nu)$, and let $d$ be a growing parameter. Then there exists a $(k, s)$-sparse quadratic-phase delta expansion over $\QQ(\zeta_q)$ on $\ZZ_q^d$ with permutation-invariant support and with $k \leq q^{(\eta_{q, s}(\vartheta, \nu) + o(1))d}$; the same holds with $\eta_{q, s}$ in place of $\eta_{q, s}(\vartheta, \nu)$. Here $\eta_{q, s}(\vartheta, \nu)$ and $\eta_{q, s}$ are the constants of~\Cref{def:track3exponent}.
    In particular, $\eta_{3, 2} \leq 0.820340$, $\eta_{3, 3} \leq 0.732672$, $\eta_{5, 4} \leq 0.702450$, $\eta_{5, 5} \leq 0.653434$, $\eta_{7, 6} \leq 0.652057$, $\eta_{7, 7} \leq 0.619701$, $\eta_{11, 10} \leq 0.607791$, and $\eta_{11, 11} \leq 0.590014$ (\Cref{lem:gen4numerics}).
    As $q$ grows, we have $\eta_{q, q} \leq \frac{1}{2} + \frac{\frac{1}{2}\ln(\pi e/6) + o(1)}{\ln q} \leq \frac{1}{2} + \frac{0.255 + o(1)}{\log_2 q}$ (\Cref{lem:gen4scaling}).
\end{corollary}

\subsection{Proof Step 1: Duality and Symmetrization}\label{sec:gen4step1}

Throughout this section, we fix an arbitrary subset $\calA \subseteq \ZZ_q$ of size $s$.

Here, we follow the first part of~\Cref{sec:gen1step1}. 
Unwinding~\Cref{def:genconedelta} as in the warm-up, \Cref{thm:gen4aprime} follows from the existence of coefficients $\{c_{\zv, a}\}$ supported on pairs $(\zv, a) \in S_d(\vartheta, \nu) \times \calA$ with
$\sum_{\zv, a} c_{\zv, a}\, \zeta_q^{\langle \zv, \xv \rangle + a \langle \xv, \xv \rangle} = \one[\xv = \zerov]$ for every $\xv \in \ZZ_q^d$: such a collection is a $(|S_d(\vartheta, \nu)|, s)$-sparse quadratic-phase delta expansion (its support is $\calW \times \calT$ in the sense of~\Cref{def:genconedelta}, with $\calW = S_d(\vartheta, \nu)$ and $\calT = \calA$), and since our $\calA \subseteq \ZZ_q$ of size $s$ is arbitrary, producing such coefficients for it proves the theorem. Suppose, for the sake of contradiction, that no such coefficients exist. 
By the duality argument of~\Cref{sec:gen1step1}, applied verbatim in the space $\CC^{\ZZ_q^d}$ with one spanning vector $\left(\zeta_q^{a\langle \xv, \xv \rangle + \langle \xv, \zv \rangle}\right)_{\xv \in \ZZ_q^d}$ per pair $(a, \zv) \in \calA \times S_d(\vartheta, \nu)$, there exists a vector $\alpha$ indexed by elements $\xv \in \ZZ_q^d$ such that:
\begin{align*}
    \alpha_0 &= 1 \\
    \sum_{\xv \in \ZZ_q^d} \alpha_{\xv} \zeta_q^{a\langle \xv, \xv \rangle + \langle \xv, \zv \rangle} &= 0 \quad \forall a \in \calA, \zv \in S_d(\vartheta, \nu).\numberthis\label{eq:alpha2}
\end{align*}
It will help us later to replace $\alpha$ by a symmetrized witness $\beta$ satisfying two additional properties: $\beta$ is supported on vectors $\xv$ with $\sum_{i \in [d]} x_i \equiv 0 \pmod{q}$, and $\beta_\xv = \beta_{\xv + \bfone}$ for all $\xv$. (These are what will make the polynomial of \Cref{prop:pproperties} below invariant under the Heisenberg operators $\sfZ$ and $\sfX$ respectively, which is how translations of the elliptic variable of our theta function will be absorbed, starting with~\Cref{lem:thetatranslation,lem:fgtranslation} in Step 2.)
These need not be true of $\alpha$ itself, so we will need to make some transformations first to ensure this. We do this using an averaging trick:
\begin{definition}
    For every $J \in \ZZ_q$ and $\xv \in \ZZ_q^d$, define:
    \begin{equation*}
        \alpha_\xv^{(J)} = \one\left[\sum_{i = 1}^d x_i = 0 \right] \cdot \zeta_q^{-Jx_1} \cdot \frac{1}{q} \sum_{b \in \ZZ_q} \zeta_q^{-Jb} \alpha_{\xv+b\bfone}.
    \end{equation*}
\end{definition}

\noindent Each ingredient buys one property: the indicator enforces the support condition~\eqref{eq:beta2} below; the weight $\zeta_q^{-Jb}$ inside the average provides the one degree of freedom, $J$, needed to make the normalization $\alpha^{(J)}_0 \neq 0$ achievable (it selects a Fourier coefficient along the line $\{b\bfone\}$, not all of which can vanish); and the matching prefactor $\zeta_q^{-Jx_1}$ is what preserves the shift-invariance~\eqref{eq:beta3} in the presence of that weight. The cost of the prefactor is a shift of the frequency vector by $-J\ev_1$, which is absorbed by the final thickening in the definition of $S_d(\vartheta, \nu)$ --- indeed, only the first-coordinate thickening $c\ev_1$ is consumed there; $S_d(\vartheta, \nu)$ thickens every coordinate only for symmetry of the definition.

\begin{proposition}
    There exists a $J \in \ZZ_q$ such that $\alpha_0^{(J)} \neq 0$.
\end{proposition}
\begin{proof}
    By definition, we have
    \begin{equation*}
        \alpha_0^{(J)} = \frac{1}{q} \sum_{b \in \ZZ_q} \zeta_q^{-Jb} \alpha_{b\bfone},
    \end{equation*}
    so this is just a Fourier-transform coefficient of $(\alpha_0, \alpha_{\bfone}, \alpha_{2\bfone}, \ldots, \alpha_{(q-1)\bfone}) \neq 0$. So one of the Fourier coefficients must be nonzero.
\end{proof}

\begin{proposition}\label{prop:symmetrization}
    Fix $J \in \ZZ_q$ such that $\alpha_0^{(J)} \neq 0$ and define $\beta$ by $\beta_{\xv} = \alpha_{\xv}^{(J)}/\alpha_0^{(J)}$. Then we have:
    \begin{align}
        \beta_0 &= 1 \label{eq:beta1}\\
        \beta_\xv &= 0 \quad \forall \xv \in \ZZ_q^d\text{ s.t. }\sum_{i = 1}^d x_i \neq 0 \label{eq:beta2} \\
        \beta_\xv &= \beta_{\xv + \bfone} \quad \forall \xv \in \ZZ_q^d.\label{eq:beta3}\\
        \sum_{\xv \in \ZZ_q^d} \beta_\xv \zeta_q^{a\langle \xv, \xv \rangle + \langle \xv, \zv \rangle} &= 0 \quad \forall a \in \calA, \zv \in T_d(\vartheta, \nu)\label{eq:beta4}.
    \end{align}
\end{proposition}
\begin{proof}
    See~\Cref{sec:symmetrizationproof}.
\end{proof}

As in the warm-up, we package the dual witness into a homogeneous polynomial. This time there is no average over $\ZZ_q^*$; instead, the symmetries~\eqref{eq:beta2} and~\eqref{eq:beta3} will make the polynomial invariant under the Heisenberg operators $\sfX$ and $\sfZ$ of~\Cref{def:heisenberg}, which will reappear as the effect of translating the elliptic variable of our theta function (\Cref{lem:thetatranslation} in Step 2).

\begin{definition}
    Define the polynomial $P_\beta: \CC^q \to \CC$ as follows:
    \begin{equation*}
        P_\beta(X_0, \ldots, X_{q-1}) = \sum_{\substack{\xv \in \ZZ_q^d}} \beta_\xv \prod_{i = 1}^d X_{x_i}.
    \end{equation*}
\end{definition}

Here $\sfX$ and $\sfZ$ are the Heisenberg operators of~\Cref{def:heisenberg}, and their conjugation by the Weil representation (\Cref{lem:heisenbergweil}) is the engine of the scalarization below.

\begin{proposition}\label{prop:pproperties}
    We have all of the following:
    \begin{itemize}
        \item $P_\beta$ is homogeneous of degree $d$.
        \item For all $x \in \ZZ_q$, $P_\beta(\ev_x) = 1$.
        \item For all $\yv \in \CC^q$, $P_\beta(\sfX \yv) = P_\beta(\sfZ \yv) = P_\beta(\yv)$.
        \item For all $g \in \SL_2(\ZZ)$ and every product $H$ of powers of $\sfX$ and $\sfZ$, we have $P_\beta(\rho(g) H \yv) \propto_\zeta P_\beta(\rho(g) \yv)$ as functions of $\yv \in \CC^q$.
    \end{itemize}
\end{proposition}
\begin{proof}
    See~\Cref{sec:pppropertiesproof}.
\end{proof}

\subsection{Proof Step 2: The Vector-Valued Jacobi Theta Modular Form}\label{sec:gen4step2}

We continue the proof of~\Cref{thm:gen4aprime}; recall the standing assumption that $d$ is divisible by $q$.

\begin{definition}[$\SL_2(\ZZ)$ action on $\calH \times \CC$]
    We define a group action by $\SL_2(\ZZ)$ on $\calH \times \CC$ as follows:
    \begin{align*}
        \begin{bmatrix} a & b \\ c & d_0 \end{bmatrix} \ast (\tau, w) := \left(\frac{a\tau+b}{c\tau+d_0}, \frac{w}{c\tau+d_0}\right).
    \end{align*}
\end{definition}

The intuition: the pair $(\tau, w)$ should be read as a point $w$ on the complex torus $\CC/(\ZZ\tau + \ZZ)$, together with the modulus $\tau$ of that torus. The action transports both consistently: if $\tau' = g \ast \tau$, then the lattice $\ZZ\tau' + \ZZ$ equals $(\ZZ\tau + \ZZ)/G(g, \tau)$ (the matrix $g$ is a change of basis of the lattice, followed by the rescaling), and dividing $w$ by the same factor $G(g, \tau)$ applies that rescaling to the marked point.

\begin{proposition}
    The above defines a valid group action.
\end{proposition}
\begin{proof}
    The validity on the first coordinate is automatic from the $\SL_2(\ZZ)$ action on $\calH$. For the second coordinate, we want to show for any $g_1, g_2 \in \SL_2(\ZZ)$ that:
    \begin{align*}
        \frac{w}{G(g_1g_2, \tau)} &= \frac{w}{G(g_1, g_2 \ast \tau) \cdot G(g_2, \tau)},
    \end{align*}
    which is in turn immediate from~\Cref{prop:chainrule}.
\end{proof}

For this proof, our vector-valued function is defined as follows:
\begin{definition}
    Define $\Theta: \calH \times \CC \to \CC^q$ by:
    \begin{equation*}
        \Theta(\tau, w) = \sum_{k \in \ZZ} (-1)^k \exp\left(\frac{\pi i \tau}{q} \left(k-\frac{1}{2}\right)^2 + 2\pi i \left(k-\frac{1}{2}\right) w\right) \ev_{k+(q-1)/2 \bmod{q}}.
    \end{equation*}
\end{definition}
An analogous argument to that in~\Cref{lem:thetaconverges} establishes that this is well-defined and also that it is holomorphic on $\calH \times \CC$ in the sense of~\Cref{def:jointholo}: each term is entire in $(\tau, w)$, and on a compact subset of $\calH \times \CC$ the real part of each exponent is at most $-c\left(k - \frac{1}{2}\right)^2 + C\left|k-\frac{1}{2}\right|$ for constants $c > 0$ and $C \geq 0$ (coming from $\Im \tau$ being bounded below and $|{\Im w}|$ bounded above), so~\Cref{lem:latticegaussian} (with $r = 1$ and the coset $\ZZ - \frac{1}{2}$ of $\ZZ$) and~\Cref{fact:weierstrass} apply.

At this point in the warm-up in~\Cref{sec:gen1step2}, we checked a few things about $\Theta_{\calL}$:
\begin{itemize}
    \item that it is weakly modular with respect to the Weil representation; and
    \item that it is holomorphic at the cusps, and therefore a vector-valued modular form.
\end{itemize}
We will not do either of these here, because these are not technically true. To begin with, the function now takes an element of $\calH \times \CC$ rather than just an element of $\calH$ as input. Moreover, it will turn out that $\Theta(\tau, w)$ can be thought of as a ``pseudo-modular form'' of weight $1/2$ (whereas a modular form's weight has to be an integer). This creates cumbersome technicalities in terms of appropriately defining the square root of $c\tau+d_0$ in the transformation law defined in~\eqref{eq:transformationlaw}.
Formally, $\Theta$ is a vector-valued Jacobi form of weight $1/2$ for a metaplectic cover of the Jacobi group, and its transformation laws are instances of the classical ones for theta functions with characteristics (see, e.g.,~\cite[Chapter~I, \S\S3 and~7]{Mumford83} and~\cite{EichlerZagier85}); but we will elide this formalism because we do not need it. In particular, we work up to roots of unity (the relation $\propto_\zeta$), much as the classical transformation formula for theta functions under $\SL_2(\ZZ)$ involves an eighth root of unity whose exact value is cumbersome to track~\cite[Chapter~I, Theorem~7.1]{Mumford83}.
Instead, we will proceed as follows:
\begin{itemize}
    \item In this subsection, we just explicitly calculate how $\Theta$ transforms under the action of $\SL_2(\ZZ)$. We do not address any limiting behavior at cusps.
    \item In~\Cref{sec:gen4step3}, we mirror~\Cref{sec:gen1step2} and compose $\Theta$ with a suitable homogeneous polynomial. The resulting object will lead us to two ordinary scalar-valued modular forms with integer weight; we will check this using the aforementioned transformation laws and we will also address the holomorphy at the cusps at this point.
    \item Finally, we will again finish with the valence formula: the scalarized forms' cusp orders, bounded from below in~\Cref{sec:gen4step3,sec:gen4step4}, feed the valence inequalities of~\Cref{sec:gen4step5}, which together contradict the existence of the witness.
\end{itemize}

First, we will once again use the Weil representation $\rho: \SL_2(\ZZ) \to \CC^{q \times q}$ of~\Cref{def:weil,fact:weilrep}, specializing to the parameter $\mu = (q+1)/2$ (a valid choice, as $(q+1)/2 \equiv 2^{-1} \not\equiv 0 \pmod{q}$). Thus $\rho_T$ maps $\ev_x \mapsto \zeta_q^{x^2(q+1)/2} \ev_x$, and $\rho_S$ maps $\ev_x \mapsto \frac{\epsilon_q}{\sqrt{q}} \sum_{y \in \ZZ_q} \zeta_q^{-xy} \ev_y$, where $\epsilon_q := \epsilon_{(q+1)/2,\, q} \in \{\pm 1, \pm i\}$ (\Cref{fact:gausssum}). With this in mind, we can now check the transformation laws. These are the classical transformation laws of theta functions with characteristics, in our normalization; since translating between normalizations is about as long as a direct verification, we include proofs. They closely mirror those in~\Cref{sec:gen1step2}, and we defer them to~\Cref{sec:gen4proofs}.

\begin{lemma}\label{lem:gen4T}
    As functions of $(\tau, w) \in \calH \times \CC$, we have:
    \begin{equation*}
        \Theta(T \ast (\tau, w)) = \Theta(\tau+1, w) \propto_\zeta \rho(T) \Theta(\tau, w).
    \end{equation*}
\end{lemma}
\begin{proof}
    See~\Cref{sec:gen4Tproof}.
\end{proof}

\begin{lemma}\label{lem:gen4S}
    As functions of $(\tau, w) \in \calH \times \CC$, we have:
    \begin{equation*}
        \Theta(S \ast (\tau, w)) = \Theta(-1/\tau, w/\tau) \propto_\zeta (-i\tau)^{1/2} e^{\pi i qw^2/\tau} \rho(S) \Theta(\tau, w).
    \end{equation*}
    In the above, $(-i\tau)^{1/2}$ denotes the principal branch, as in~\Cref{cor:gausspsf}; it is well-defined and holomorphic on $\calH$ because $\Re(-i\tau) = \Im \tau > 0$.
\end{lemma}
\begin{proof}
    See~\Cref{sec:gen4Sproof}.
\end{proof}

Finally, there is one more symmetry that we take a moment to highlight; this symmetry is enabled by the addition of the variable $w$. It says that $\Theta$ is \emph{quasi-periodic} in $w$ under translations by the scaled lattice $\frac{1}{q}(\ZZ\tau + \ZZ)$: shifting $w$ by $\frac{a\tau+b}{q}$ reproduces $\Theta$ up to an explicit nonvanishing factor, a root of unity, and an application of the Heisenberg operators $\sfZ^b \sfX^a$, which record the discrete data $(a, b)$ of the translation. This is what justifies regarding $w$ (equivalently $z = qw$) as a point on a torus, and --- once composed with the Heisenberg-invariant polynomial $P_\beta$ of~\Cref{prop:pproperties} --- it is the source of the elliptic transformation law~\Cref{lem:fgtranslation} below.

\begin{lemma}\label{lem:thetatranslation}
    For any $a, b \in \ZZ$, we have, as functions of $(\tau, w) \in \calH \times \CC$:
    \begin{equation}\label{eq:heisenbergtarget}
        \Theta\left(\tau, w+\frac{a\tau+b}{q}\right) \propto_\zeta e^{-\pi i a^2 \tau/q- 2\pi i aw} \cdot \sfZ^b \sfX^{a} \Theta(\tau, w).
    \end{equation}
\end{lemma}
\begin{proof}
    See~\Cref{sec:thetatranslationproof}.
\end{proof}

The transformation laws for the generators $T$ and $S$ (\Cref{lem:gen4T,lem:gen4S}) propagate to all of $\SL_2(\ZZ)$, up to a root of unity; this also absorbs the sign ambiguity in composing the square-root factors, whose branches multiply only up to sign:

\begin{lemma}\label{lem:descended}
    For any $g = \begin{bmatrix} a & b \\ c & d_0 \end{bmatrix} \in \SL_2(\ZZ)$, we have, as functions of $(\tau, w) \in \calH \times \CC$:
    \begin{equation}\label{eq:descendedtransformation}
        \Theta(g \ast (\tau, w)) \propto_\zeta G(g, \tau)^{1/2} \exp\left(\frac{\pi i q c w^2}{G(g, \tau)}\right) \rho(g) \Theta(\tau, w),
    \end{equation}
    where $G(g, \tau)^{1/2}$ denotes any holomorphic branch of the square root of $\tau \mapsto G(g, \tau)$. (Such a branch exists: the image of $\tau \mapsto c\tau + d_0$ on $\calH$ is an open half-plane when $c \neq 0$ and the single nonzero real $d_0$ when $c = 0$, in either case a simply connected set avoiding $0$ on which a branch can be chosen. Any two branches differ by a constant factor $\pm 1$, a root of unity, so the choice does not matter.)
\end{lemma}
\begin{proof}
    See~\Cref{sec:descendedproof}.
\end{proof}

\begin{definition}
    For any $g \in \SL_2(\ZZ)$, define $F_g: \calH \times \CC \to \CC$ by $F_g(\tau, z) := P_\beta(\rho(g) \Theta(\tau, z/q))$. Define $F := F_I$. Here and throughout, the two coordinates are related by $z = qw$.
\end{definition}

\begin{proposition}\label{prop:fg=f}
    For $g \in \Gamma(q)$, we have $F_g(\tau, z) = F(\tau, z)$.
\end{proposition}
\begin{proof}
    Immediate from~\Cref{fact:weilrep}, which tells us that $\rho(g) = I$ when $g \in \Gamma(q)$.
\end{proof}

\begin{lemma}\label{lem:bigF}
    For any $g = \begin{bmatrix} a & b \\ c & d_0 \end{bmatrix} \in \SL_2(\ZZ)$, we have, as functions of $(\tau, z) \in \calH \times \CC$:
    \begin{equation*}
        F(g \ast (\tau, z)) = F\left(g \ast \tau, \frac{z}{G(g, \tau)}\right) \propto_\zeta G(g, \tau)^{d/2} \exp\left(\frac{\pi i d c z^2}{q G(g, \tau)}\right) F_g(\tau, z),
    \end{equation*}
    where $G(g, \tau)^{d/2} := \left(G(g, \tau)^{1/2}\right)^d$ for a holomorphic branch as in~\Cref{lem:descended}.
\end{lemma}
\begin{proof}
    See~\Cref{sec:bigFproof}.
\end{proof}

\begin{lemma}\label{lem:fgtranslation}
    For all $a, b \in \ZZ$ and $g \in \SL_2(\ZZ)$, we have, as functions of $(\tau, z) \in \calH \times \CC$:
    \begin{equation*}
        F_g(\tau, z+a\tau+b) \propto_\zeta \exp\left(-\frac{\pi i d}{q}(a^2\tau + 2az)\right) F_g(\tau, z).
    \end{equation*}
\end{lemma}
\begin{proof}
    See~\Cref{sec:fgtranslationproof}.
\end{proof}

\subsection{Proof Step 3: Two Scalarizations}\label{sec:gen4step3}

\subsubsection{Cusp Data}\label{sec:gen4cuspdata}

Both scalarizations below follow the same pattern: compose $\Theta$ with the polynomial $P_\beta$ to obtain a scalar function of $\tau$, show that this function is a modular form for $\Gamma(q)$, and lower bound its orders at the cusps. Before carrying either out, we collect what the two constructions share: the fact that there is anything to scalarize at all ($F \not\equiv 0$), the classification of the cusps of $\Gamma(q)$, a matrix $g_\frakc$ fixed once and for all per cusp class, and two per-class quantities $\ell_\frakc$ and $H_\frakc$, in terms of which both cusp-order bounds will be expressed.

\begin{proposition}\label{prop:fnotzero}
    $F$ is not identically zero.
\end{proposition}
\begin{proof}
    See~\Cref{sec:fnotzeroproof}.
\end{proof}

\begin{corollary}\label{cor:fgnotzero}
    For every $g \in \SL_2(\ZZ)$, the function $F_g$ is not identically zero.
\end{corollary}
\begin{proof}
    By~\Cref{lem:bigF}, $F(g \ast (\tau, z)) \propto_\zeta G(g, \tau)^{d/2} \exp\left(\frac{\pi i d c z^2}{q G(g, \tau)}\right) F_g(\tau, z)$, where the constant and the other factors multiplying $F_g(\tau, z)$ never vanish. Since $(\tau, z) \mapsto g \ast (\tau, z)$ is a bijection of $\calH \times \CC$ and $F$ is not identically zero (\Cref{prop:fnotzero}), neither is $F_g$.
\end{proof}

Next, the cusps. We computed the cusp classes of $\Gamma_0(q)$ in~\Cref{prop:gamma0cusps}; the analogous statement for $\Gamma(q)$ is the following.

\begin{proposition}[{\cite[Proposition~3.8.3 and Section~3.8]{DiamondShurman05}}]\label{prop:gammaqcusps}
    Under the action of $\Gamma(q)$, the cusps $\QQ \cup \{\infty\}$ split into $(q^2-1)/2$ equivalence classes, each naturally represented by an element of $(\ZZ_q^2 \backslash \{(0, 0)\})/(\pm 1)$. Moreover, all these cusps have width $q$.
\end{proposition}

By~\Cref{lem:cuspwelldef}, the width of a cusp, holomorphy of a weakly modular function there, and its order there depend only on the $\Gamma(q)$-class of the cusp, and not on the matrix $g_\frakc$ used to compute them. The per-class quantities $\ell_\frakc$ and $H_\frakc$ defined below are different: they are defined \emph{in terms of} such a matrix. We therefore fix, once and for all, for each cusp class $\frakc$ a matrix $g_\frakc \in \SL_2(\ZZ)$ sending $\infty$ to a cusp in the class $\frakc$, and we define $\ell_\frakc$ and $H_\frakc$ relative to this fixed choice. (For the selected classes $\frakc_{a, \lambda}$ of~\Cref{sec:gen4step4} we will fix a specific, convenient $g_\frakc$; for all other classes the choice is arbitrary. No independence from the choice needs to be \emph{proved}: the cusp-order bounds of~\Cref{lem:amsq,lem:b2holomorphic} bound the choice-independent quantities $\order_\frakc(A_m^2)$ and $\order_\frakc(B^2)$ in terms of the fixed choice, and the per-class estimates of~\Cref{sec:gen4step4} (\Cref{lem:Hgen4,lem:randomgen4}) are proved for the same fixed lifts; using a single fixed choice consistently throughout is all that is required.)

The quantity $\ell_\frakc$ is extracted from an expansion of $F_{g_\frakc}$ in powers of $e^{\pi i \tau/(4q)}$, which we now construct. In the definition of $\Theta$, the $\tau$-dependence of the $k$-th term is through $\exp\left(\frac{\pi i \tau}{q}\left(k - \frac{1}{2}\right)^2\right) = \exp\left(\frac{\pi i \tau (2k-1)^2}{4q}\right)$, with an odd square $(2k-1)^2$ in the numerator; the possible $\tau$-frequencies of a degree-$d$ monomial in the coordinates of $\Theta$ are therefore sums of $d$ odd squares. Since every odd square is $\equiv 1 \pmod 8$, such a sum is at least $d$ and $\equiv d \pmod 8$. Formally:

\begin{lemma}\label{lem:ellexpansion}
    Fix $g \in \SL_2(\ZZ)$. For each integer $\ell \geq d$ with $\ell \equiv d \pmod 8$, there is an entire function $\Phi_{g, \ell}: \CC \to \CC$ such that:
    \begin{equation}\label{eq:ellexpansion}
        P_\beta(\rho(g)\, \Theta(\tau, w)) = \sum_{\substack{\ell \geq d \\ \ell \equiv d \pmod 8}} \Phi_{g, \ell}(w)\, e^{\pi i \tau \ell/(4q)},
    \end{equation}
    with the series converging absolutely and locally uniformly on $\calH \times \CC$. (The functions $\Phi_{g, \ell}$ are unique: for each fixed $w$,~\eqref{eq:ellexpansion} is the Fourier expansion of the $8q$-periodic holomorphic function $\tau \mapsto P_\beta(\rho(g)\Theta(\tau, w))$, whose coefficients are unique by~\Cref{fact:fourierexp}.) Moreover, writing $F_g(\tau, z) = \sum_{j \geq 0} A_{g, j}(\tau)\, z^j$ for the Taylor family of~\Cref{lem:taylorfamilies} (which applies: $F_g$ is holomorphic on $\calH \times \CC$, being the composition of $\Theta$ with the linear map $\rho(g)$, the polynomial $P_\beta$, and the rescaling $z \mapsto z/q$), and $\Xi_{g; j, \ell}$ for the $j$-th Taylor coefficient of $\Phi_{g, \ell}$, we have for every $j \geq 0$:
    \begin{equation}\label{eq:ellexpansioncoeffs}
        A_{g, j}(\tau) = \sum_{\substack{\ell \geq d \\ \ell \equiv d \pmod 8}} q^{-j}\, \Xi_{g; j, \ell}\, e^{\pi i \tau \ell/(4q)},
    \end{equation}
    again converging absolutely and locally uniformly on $\calH$.
\end{lemma}
\begin{proof}
    See~\Cref{sec:ellexpansionproof}.
\end{proof}

\begin{definition}\label{def:ella}
    Let $\frakc$ be a cusp class, with fixed matrix $g_\frakc$ as above. By~\Cref{cor:fgnotzero} we have $F_{g_\frakc} \not\equiv 0$, so by~\eqref{eq:ellexpansion} not all of the functions $\Phi_{g_\frakc, \ell}$ vanish identically; let $\ell^*$ be the minimal $\ell$ with $\Phi_{g_\frakc, \ell} \not\equiv 0$. Since every index $\ell$ in~\eqref{eq:ellexpansion} satisfies $\ell \geq d$ and $\ell \equiv d \pmod{8}$, the quantity
    \begin{equation*}
        \ell_\frakc := \frac{\ell^* - d}{8}
    \end{equation*}
    is a nonnegative integer.
\end{definition}

Finally, we define the second per-class quantity, which is used in the cusp-order bound of the second scalarization. Recall the weight function $w_q = w_{q, 1/2}$ of~\Cref{def:track3exponent}, and for a polynomial $P$ in the variables $X_0, \ldots, X_{q-1}$, write $[X_{x_1} X_{x_2} \cdots X_{x_d}]P$ for the coefficient of the monomial $X_{x_1} X_{x_2} \cdots X_{x_d}$ in $P$.

\begin{definition}\label{def:hc}
    Let $\frakc$ be a cusp class, with fixed matrix $g_\frakc$ as above. Define:
    \begin{equation*}
        H_\frakc := \min\left\{\sum_{i = 1}^d w_q(x_i): \xv \in \ZZ_q^d\text{ s.t. }[X_{x_1} X_{x_2} \ldots  X_{x_d}](P_\beta \circ \rho(g_\frakc)) \neq 0\right\}.
    \end{equation*}
    (The minimum is over a nonempty finite set: $P_\beta$ is not the zero polynomial by~\eqref{eq:beta1}, and $\rho(g_\frakc)$ is invertible, so $P_\beta \circ \rho(g_\frakc)$ is not the zero polynomial either; it is homogeneous of degree $d$, so all of its monomials have the displayed shape.)
\end{definition}

\subsubsection{Scalarization via Taylor Coefficients}

At long last, we can define the scalarized modular form that we will work with in this step. It is the first nonvanishing Taylor coefficient of $F$ in the elliptic variable, which, for a Jacobi form in the sense of~\cite{EichlerZagier85}, is classically a modular form~\cite[Section~3]{EichlerZagier85}; we verify the analogous statement directly for $F$, whose transformation laws hold only up to roots of unity.
\begin{definition}
    $F$ is holomorphic on $\calH \times \CC$ (it is the composition of $\Theta$, which is holomorphic by the discussion above, with the polynomial $P_\beta$ and the rescaling $z \mapsto z/q$), so by~\Cref{lem:taylorfamilies} there exist unique holomorphic functions $A_0, A_1, \ldots$ from $\calH \to \CC$ such that:
    \begin{equation*}
        F(\tau, z) = \sum_{j = 0}^\infty A_j(\tau) z^j.
    \end{equation*}
    By~\Cref{prop:fnotzero}, there exists a minimal integer $m \geq 0$ such that $A_m$ is not identically zero. Our scalarized function is $A_m: \calH \to \CC$. Equivalently, we can define:
    \begin{equation*}
        A_m(\tau) = \frac{1}{m!} \frac{\pp^m}{\pp z^m} F(\tau, z)\bigg|_{z = 0}.
    \end{equation*}
\end{definition}

\begin{proposition}\label{prop:mintaylorpreserved}
    For any $g \in \SL_2(\ZZ)$, there exist holomorphic functions $A_{g, m}, A_{g, m+1}, \ldots: \calH \to \CC$ such that:
    \begin{equation*}
        F_g(\tau, z) = P_\beta(\rho(g) \Theta(\tau, z/q)) = \sum_{j = m}^\infty A_{g, j}(\tau) z^j.
    \end{equation*}
    Moreover, we have $A_{g, m}(\tau) \propto_\zeta G(g, \tau)^{-d/2-m} A_m(g \ast \tau)$ as functions of $\tau$.
\end{proposition}
\begin{proof}
    See~\Cref{sec:mintaylorpreservedproof}.
\end{proof}

Here and below, for functions holomorphic on $\calH \times \CC$, the notation $O(z^{m+1})$ denotes a function of the form $z^{m+1} h(\tau, z)$ with $h$ holomorphic on $\calH \times \CC$ --- equivalently, by~\Cref{lem:taylorfamilies}, a holomorphic function whose Taylor coefficients (in $z$) of index at most $m$ vanish identically. The uniqueness in~\Cref{lem:taylorfamilies} then justifies equating the coefficients of $z^m$ between two such expansions.

We show that $A_m^2$ is a modular form, and bound its orders at the cusps in terms of the quantity $\ell_\frakc$ of~\Cref{def:ella}.

\begin{lemma}\label{lem:amsq}
    $A_m^2: \calH \to \CC$ is a modular form for $\Gamma(q)$ of weight $d+2m$, with a one-dimensional representation of finite order. Moreover, for all cusps $\frakc \in \QQ \cup \{\infty\}$, we have $\order_\frakc (A_m^2) \geq \frac{d}{4}+2\ell_\frakc$.
\end{lemma}
\begin{proof}
    See~\Cref{sec:amsqproof}.
\end{proof}

\subsubsection{Scalarization via 2-Torsion Evaluations}

The second scalarization fixes the elliptic variable at the three points $z = \frac{x\tau+y}{2}$ for $(x, y) \in \{(0, 1), (1, 0), (1, 1)\}$: representatives of the three nonzero \emph{2-torsion} points of the torus $\CC/(\ZZ\tau + \ZZ)$, i.e., of the points $z \notin \ZZ\tau + \ZZ$ with $2z \in \ZZ\tau + \ZZ$. By~\Cref{lem:fgtranslation}, moving $z$ by a lattice vector changes $F_g$ only by an explicit nonvanishing factor times a root of unity, so these evaluations are canonical up to such factors; and the action of $\SL_2(\ZZ)$ on $\tau$ permutes the three nonzero 2-torsion points (as the bijection $(x, y) \mapsto (\tilde{x}, \tilde{y})$ in~\Cref{lem:bxytransformation} below makes precise), which is why we work with the \emph{product} $B_g$ of all three evaluations: up to scalars it transforms like a weight-$3d/2$ form. The prefactor $\exp(\pi i d x^2 \tau/(4q))$ in the definition below normalizes the growth of $F_g(\tau, (x\tau+y)/2)$ as $\Im \tau \to \infty$ (see the proof of~\Cref{lem:b2holomorphic}); it is the standard normalizing factor with which restrictions of Jacobi forms to torsion points become modular forms~\cite[Theorem~1.3]{EichlerZagier85} (see also~\cite[Section~4.2]{Aoki22} and~\cite[Lemma~2.1]{RichterWesterholtRaum15}). Taking the product over an $\SL_2(\ZZ)$-orbit of torsion points, as we do here, is likewise standard~\cite{RichterWesterholtRaum15,LiWang25}.

\begin{definition}
    For $x, y \in \{0, 1\}$ and $g \in \SL_2(\ZZ)$, define:
    \begin{equation*}
        B_{g; x, y}(\tau) := \exp\left(\frac{\pi i d x^2 \tau}{4q}\right) F_g\left(\tau, \frac{x\tau+y}{2}\right),
    \end{equation*}
    and then define $B_g(\tau) := B_{g; 0, 1}(\tau) B_{g; 1, 0}(\tau) B_{g; 1, 1}(\tau)$. 
    When $g$ is omitted from notation, it is set to the identity. Note that this is consistent with~\Cref{prop:fg=f}.
\end{definition}

\begin{lemma}\label{lem:bxytransformation}
    Let $g \in \SL_2(\ZZ)$ and $(x, y) \in \{(0, 1), (1, 0), (1, 1)\}$. There exists $(\tilde{x}, \tilde{y}) \in \{(0, 1), (1, 0), (1, 1)\}$ --- explicitly, $(\tilde{x}, \tilde{y}) \equiv (ax + cy,\ bx + d_0 y) \pmod 2$ --- such that, as functions of $\tau \in \calH$:
    \begin{equation*}
        B_{x, y}(g \ast \tau) \propto_\zeta G(g, \tau)^{d/2} B_{g; \tilde{x}, \tilde{y}}(\tau),
    \end{equation*}
    with $G(g, \tau)^{d/2}$ as in~\Cref{lem:bigF}.
    Moreover, for each $g$ the mapping $(x, y) \mapsto (\tilde{x}, \tilde{y})$ is bijective.
\end{lemma}
\begin{proof}
    See~\Cref{sec:bxytransformationproof}.
\end{proof}

\begin{proposition}\label{prop:bnotzero}
    $B$ is not identically zero.
\end{proposition}
\begin{proof}
    First, each $B_{x, y}$ is holomorphic on $\calH$: writing $F(\tau, z) = \sum_j A_j(\tau) z^j$ (\Cref{lem:taylorfamilies}), we have $F(\tau, \varphi(\tau)) = \sum_j A_j(\tau) \varphi(\tau)^j$ for the holomorphic function $\varphi(\tau) := \frac{x\tau+y}{2}$, a locally uniform limit of holomorphic functions, which is holomorphic by~\Cref{fact:weierstrass}; multiplying by the nonvanishing factor $\exp(\pi i dx^2\tau/(4q))$ preserves this. Since $\calH$ is connected and the zeros of a nonzero holomorphic function are isolated (\Cref{fact:identitytheorem}), a product of finitely many holomorphic functions, none identically zero, is not identically zero; so it suffices to show $B_{x, y} \not\equiv 0$ for each of the three pairs.

    For $(x, y) = (1, 0)$: by definition, $B_{1, 0}(\tau) = e^{\pi i d \tau/(4q)}\, F(\tau, \tau/2) = e^{\pi i d\tau/(4q)}\, P_\beta(\Theta(\tau, \tau/(2q)))$. The proof of~\Cref{prop:fnotzero} shows precisely that $P_\beta(\Theta(iy, iy/(2q))) \neq 0$ for all large enough $y > 0$, so $B_{1, 0} \not\equiv 0$.

    For the other two pairs, we use~\Cref{lem:bxytransformation} as a pullback. In each case we exhibit a matrix $g \in \SL_2(\ZZ)$ with $g \equiv I \pmod q$ --- so that $\rho(g) = I$ (\Cref{fact:weilrep}), hence $F_g = F$ (\Cref{prop:fg=f}) and $B_{g; \tilde{x}, \tilde{y}} = B_{\tilde{x}, \tilde{y}}$ --- for which the lemma's bijection --- explicitly, $(\tilde{x}, \tilde{y}) \equiv (ax + cy,\ bx + d_0 y) \pmod 2$ --- sends the pair at hand to $(\tilde{x}, \tilde{y}) = (1, 0)$. The lemma then gives $B_{x, y}(g \ast \tau) \propto_\zeta G(g, \tau)^{d/2}\, B_{1, 0}(\tau)$; the right-hand side is not identically zero by the previous paragraph, and $\tau \mapsto g \ast \tau$ is a bijection of $\calH$, so $B_{x, y} \not\equiv 0$.
    \begin{itemize}
        \item For $(x, y) = (1, 1)$: take $g = T^q = \begin{bmatrix} 1 & q \\ 0 & 1 \end{bmatrix} \equiv I \pmod q$. Then $(\tilde{x}, \tilde{y}) \equiv (1 + 0,\ q + 1) \equiv (1, 0) \pmod 2$, since $q$ is odd.
        \item For $(x, y) = (0, 1)$: take $g = \begin{bmatrix} q^2-q+1 & q^2 \\ q & q+1 \end{bmatrix}$, which has determinant $(q^2-q+1)(q+1) - q^3 = 1$ and is $\equiv I \pmod q$. Then $(\tilde{x}, \tilde{y}) \equiv (c, d_0) = (q,\ q+1) \equiv (1, 0) \pmod 2$. \qedhere
    \end{itemize}
\end{proof}

\begin{corollary}\label{cor:bsqmodular}
    The function $\tau \mapsto (B(\tau))^2$ is a weakly modular function for $\Gamma(q)$ with weight $3d$, with a one-dimensional representation of finite order.
\end{corollary}
\begin{proof}
    Fix $g \in \Gamma(q)$. By~\Cref{lem:bxytransformation} and the bijectivity of $(x, y) \mapsto (\tilde{x}, \tilde{y})$:
    \begin{equation*}
        B(g \ast \tau) = \prod_{(x, y)} B_{x, y}(g \ast \tau) \propto_\zeta G(g, \tau)^{3d/2} \prod_{(\tilde{x}, \tilde{y})} B_{g; \tilde{x}, \tilde{y}}(\tau) = G(g, \tau)^{3d/2}\, B_g(\tau),
    \end{equation*}
    where all products range over $\{(0, 1), (1, 0), (1, 1)\}$ and $G(g, \tau)^{3d/2} := \left(G(g, \tau)^{1/2}\right)^{3d}$. Since $g \in \Gamma(q)$, we have $F_g = F$ (\Cref{prop:fg=f}) and hence $B_g = B$. Squaring gives $B^2(g \ast \tau) \propto_\zeta G(g, \tau)^{3d}\, B^2(\tau)$, now with an integer power of $G(g, \tau)$; since $B^2 \not\equiv 0$ (\Cref{prop:bnotzero}),~\Cref{lem:scalarcharacter} shows that $B^2$ is weakly modular of weight $3d$ for $\Gamma(q)$, with a one-dimensional representation of finite order.
\end{proof}

\begin{lemma}\label{lem:b2holomorphic}
    For all cusps $\frakc \in \QQ \cup \{\infty\}$, $B^2$ is holomorphic at $\frakc$ and moreover we have $\order_\frakc (B^2) \geq \frac{d}{4} + 2\ell_\frakc + 2H_\frakc$, with $\ell_\frakc$ and $H_\frakc$ as in~\Cref{def:ella,def:hc}.
\end{lemma}
\begin{proof}
    See~\Cref{sec:b2holomorphicproof}.
\end{proof}

Thus we have shown that $B^2$ is a modular form for $\Gamma(q)$ of weight $3d$, with a one-dimensional representation of finite order.

\subsection{Proof Step 4: Lower Bounding Cusp Orders}\label{sec:gen4step4}

Recall the classification of the cusp classes of $\Gamma(q)$ (\Cref{prop:gammaqcusps}) and the per-class matrices $g_\frakc$ and quantities $\ell_\frakc, H_\frakc$ fixed in~\Cref{sec:gen4cuspdata}. In this step we single out a family of \emph{selected} cusp classes and lower bound $H_\frakc$ on them, as well as upper bound the degrees to which the relevant expansions can be supported; these are the inputs that the final combination of inequalities in~\Cref{sec:gen4step5} will consume, alongside the valence bounds themselves.

By~\Cref{prop:gammaqcusps}, for each $a \in \calA$ and $\lambda \in \{1, \ldots, (q-1)/2\}$ we may speak of the cusp class $\pm(2a\lambda, \lambda)$; we write $\frakc_{a, \lambda} := \pm\lambda(2a, 1)$ for this class, and we freely identify a class with any cusp representing it. We let $g_{a, \lambda} = \begin{bmatrix} 2a\lambda & -\lambda^{-1} \\ \lambda & 0 \end{bmatrix} \in \SL_2(\ZZ_q)$ (a matrix of determinant $1$), and --- making the choice promised in~\Cref{sec:gen4cuspdata} specific for these classes --- we take the fixed matrix $g_{\frakc_{a, \lambda}}$ to be an integral lift $g \in \SL_2(\ZZ)$ with $g \equiv g_{a, \lambda} \pmod{q}$, which exists by the surjectivity in~\Cref{lem:indices}. Then $g \ast \infty$ is indeed a cusp in the class $\frakc_{a, \lambda}$: it is the ratio of the entries of the first column of $g$, which reduces to $(2a\lambda, \lambda)$ modulo $q$. By~\Cref{fact:weilrep}, the matrix $\rho(g)$ is determined by $g_{a, \lambda}$ alone. To compute it, factor $g_{a, \lambda} = p_{a, \lambda} \bar{S}$ in $\SL_2(\ZZ_q)$, where $\bar{S}$ is the reduction of $S$ and $p_{a, \lambda} := \begin{bmatrix} \lambda^{-1} & 2a\lambda \\ 0 & \lambda \end{bmatrix}$ (direct multiplication). Since $p_{a, \lambda}$ has lower-left entry $0$, any lift of it lies in $\Gamma_0(q)$, so~\eqref{eq:weilformula} (with $\mu = (q+1)/2 \equiv 2^{-1}$, so that $\mu \cdot 2a\lambda \cdot \lambda \equiv a\lambda^2$) gives $\rho(p_{a, \lambda})\ev_y = \left(\frac{\lambda^{-1}}{q}\right) \zeta_q^{a\lambda^2 y^2}\ev_{\lambda y}$. Combining this with $\rho(S)\ev_x = \frac{\epsilon_q}{\sqrt{q}} \sum_{y} \zeta_q^{-xy} \ev_y$, re-indexing $y \mapsto \lambda^{-1}y$, and noting $\left(\frac{\lambda^{-1}}{q}\right) = \left(\frac{\lambda}{q}\right)$ (\Cref{fact:legendre}), we obtain, for every $x \in \ZZ_q$:
\begin{equation}\label{eq:rhogal}
    \rho(g_{a, \lambda}) \ev_x = c_{a, \lambda} \sum_{y \in \ZZ_q} \zeta_q^{ay^2 - \lambda^{-1}xy}\, \ev_y, \qquad \text{where } c_{a, \lambda} := \frac{\epsilon_q}{\sqrt{q}} \left(\frac{\lambda}{q}\right) \neq 0.
\end{equation}
We define the set of selected cusp classes
\[
    \frakC := \left\{\frakc_{a, \lambda}: a \in \calA, \lambda \in \left\{1, \ldots, \frac{q-1}{2}\right\}\right\}.
\]
These classes are pairwise distinct: if $\pm(2a\lambda, \lambda) = \pm(2a'\lambda', \lambda')$, then comparing second coordinates gives $\lambda' = \pm\lambda$, and since $\lambda, \lambda' \in \{1, \ldots, (q-1)/2\}$ while $-\lambda \equiv q - \lambda \in \{(q+1)/2, \ldots, q-1\} \pmod{q}$, the sign must be $+$; so $\lambda' = \lambda$, and then $2a\lambda \equiv 2a'\lambda$ forces $a' = a$. Hence $|\frakC| = s \cdot \frac{q-1}{2}$.

In the following, we use the notation $\vartheta, \nu$ introduced in~\Cref{sec:gen4main}.

\begin{lemma}\label{lem:Hgen4}
    For every $a \in \calA$ and $\lambda \in \{1, \ldots, (q-1)/2\}$, we have $H_{\frakc_{a, \lambda}} > \nu d$.
\end{lemma}
\begin{proof}
    By~\Cref{fact:weilrep}, $\rho(g_{\frakc_{a, \lambda}}) = \rho(g_{a, \lambda})$. Substituting~\eqref{eq:rhogal} and using the degree-$d$ homogeneity of $P_\beta$ to pull out the nonzero constant $c_{a, \lambda}^d$, we have:
    \begin{align*}
        P_\beta \circ \rho(g_{a, \lambda}) \left(\sum_{x \in \ZZ_q} X_x \ev_x\right) &= c_{a, \lambda}^d\, P_\beta\left(\sum_{x \in \ZZ_q} \sum_{y \in \ZZ_q} X_x \zeta_q^{ay^2 - \lambda^{-1} xy} \ev_y \right) \\
        &= c_{a, \lambda}^d \sum_{\yv \in \ZZ_q^d} \beta_{\yv} \prod_{i = 1}^d \left(\sum_{x \in \ZZ_q} X_x \zeta_q^{ay_i^2 - \lambda^{-1} xy_i}\right) \\
        &= c_{a, \lambda}^d \sum_{\xv \in \ZZ_q^d} X_{x_1} \ldots X_{x_d} \left(\sum_{\yv \in \ZZ_q^d} \beta_{\yv} \zeta_q^{a \langle \yv, \yv \rangle - \lambda^{-1} \langle \xv, \yv \rangle}\right).
    \end{align*}
    By~\eqref{eq:beta4} (applied with $\zv = -\lambda^{-1}\xv$, noting $a \in \calA$), the bracketed coefficient vanishes whenever $-\lambda^{-1} \xv \in T_d(\vartheta, \nu)$; and since $T_d(\vartheta, \nu)$ is closed under coordinatewise multiplication by units (multiplying $u\zv + b\bfone$ by a unit $\mu$ gives $(\mu u)\zv + (\mu b)\bfone$), this happens if and only if $\xv \in T_d(\vartheta, \nu)$. In particular, the coefficient vanishes for every $\xv$ with $\sum_{i = 1}^d w_q(x_i) \leq \nu d$, as such $\xv$ lie in $S_d^{\mathrm{sq}}(\nu) \subseteq T_d(\vartheta, \nu)$. The conclusion is now immediate from~\Cref{def:hc}.
\end{proof}

\begin{lemma}\label{lem:pc}
    Let $\frakc$ be a cusp class, with fixed matrix $g_\frakc$ and expansion functions $\Phi_{g_\frakc, \ell}$ as in~\Cref{sec:gen4cuspdata}. There exists a polynomial $p_{\frakc} \in \CC[X]$ of degree $\leq d$ such that:
    \begin{equation}\label{eq:pc}
        F_{g_{\frakc}}(\tau, z) = e^{-2\pi i zd/(2q)}\, p_{\frakc}\left(e^{2\pi i z/q}\right) \exp\left(\frac{\pi i \tau (d+8\ell_{\frakc})}{4q}\right) + \sum_{\substack{\ell > d+8\ell_{\frakc} \\ \ell \equiv d \pmod 8}} \Phi_{g_\frakc, \ell}(z/q) \exp\left(\frac{\pi i \tau \ell}{4q}\right).
    \end{equation}
\end{lemma}
\begin{proof}
    See~\Cref{sec:pcproof}.
\end{proof}

\begin{lemma}\label{lem:pcperiodic}
    In the setting of~\Cref{lem:pc}, the exponents of the monomials with nonzero coefficient in $p_{\frakc}$ all lie in a single residue class modulo $q$. Consequently, there exist an integer $r \in \{0, \ldots, q-1\}$ and a univariate polynomial $f_{\frakc}$ of degree $\leq (d-r)/q$ such that $p_{\frakc}(X) = X^r f_{\frakc}(X^q)$.
\end{lemma}
\begin{proof}
    By~\Cref{lem:fgtranslation} (with $(a, b) = (0, 1)$), there is a nonzero constant $\omega$, independent of $\tau$ and $z$, such that:
    \begin{equation*}
        F_{g_{\frakc}}(\tau, z+1) = \omega F_{g_{\frakc}}(\tau, z).
    \end{equation*}
    Substituting~\eqref{eq:pc} into both sides and matching the coefficients of $\exp(\pi i \tau(d + 8\ell_\frakc)/(4q))$ --- which are determined, for each fixed $z$, by the uniqueness in~\Cref{fact:fourierexp} --- we obtain, for every $z \in \CC$:
    \begin{align*}
        e^{-2\pi i (z+1)d/(2q)} p_{\frakc}(e^{2\pi i (z+1)/q}) &= \omega e^{-2\pi i zd/(2q)} p_{\frakc}(e^{2\pi i z/q}) \\
        \Leftrightarrow p_{\frakc}(\zeta_q e^{2\pi i z/q}) &= \omega' p_{\frakc}(e^{2\pi i z/q}), \qquad \text{where } \omega' := \omega e^{\pi i d/q}.
    \end{align*}
    Since $e^{2\pi i z/q}$ ranges over all of $\CC^*$ as $z$ does, we conclude $p_\frakc(\zeta_q X) = \omega' p_\frakc(X)$ identically. Comparing coefficients, $\zeta_q^R = \omega'$ whenever the coefficient of $X^R$ is nonzero; so any two such exponents $R, R'$ satisfy $\zeta_q^{R - R'} = 1$, i.e., $R \equiv R' \pmod{q}$. Letting $r \in \{0, \ldots, q-1\}$ be their common residue (or $r := 0$ if $p_\frakc = 0$), and recalling $\deg p_\frakc \leq d$, this is exactly the displayed conclusion.
\end{proof}

\begin{lemma}\label{lem:pcdivisible}
    In the setting of~\Cref{lem:pc,lem:pcperiodic}, $f_{\frakc}$ is divisible by $(X-1)^m$. Consequently, $p_{\frakc}$ is divisible by $(X^q-1)^m$.
\end{lemma}
\begin{proof}
    By~\Cref{prop:mintaylorpreserved}, the Taylor coefficients of $F_{g_\frakc}$ in $z$ of index below $m$ vanish identically; by the uniqueness of the coefficients in~\eqref{eq:ellexpansioncoeffs} (\Cref{fact:fourierexp}), this forces $\Xi_{g_\frakc; j, \ell} = 0$ for all $j < m$ and all $\ell$. In particular, the entire function $\Phi_{g_\frakc, \ell^*}$ (whose Taylor coefficients are the $\Xi_{g_\frakc; j, \ell^*}$) vanishes to order at least $m$ at $0$; and by the construction of $p_\frakc$ (see~\Cref{sec:pcproof}), $\Phi_{g_\frakc, \ell^*}(z/q) = e^{-2\pi i zd/(2q)}\, p_\frakc(e^{2\pi i z/q})$. Consequently, we have:
    \begin{align*}
        e^{-2\pi i zd/(2q)} p_{\frakc}(e^{2\pi i z/q}) &= O(z^m) \\
        \Rightarrow p_{\frakc}(e^{2\pi i z/q}) &= e^{2\pi i zd/(2q)} \cdot O(z^m) \\
        &= O(z^m) \\
        \Rightarrow e^{2\pi i rz/q} f_{\frakc}(e^{2\pi i z}) &= O(z^m) \quad\text{(\Cref{lem:pcperiodic}: $p_\frakc(X) = X^r f_\frakc(X^q)$)} \\
        \Rightarrow f_{\frakc}(e^{2\pi i z}) &= O(z^m).
    \end{align*}
    To see that this implies the conclusion, first note that if $f_{\frakc} \equiv 0$ then the conclusion holds trivially; so assume $f_\frakc \not\equiv 0$ and write $f_{\frakc}(X) = (X-1)^t h(X)$ for some $t \geq 0$ and a polynomial $h$ with $h(1) \neq 0$.
    Then we have:
    \begin{align*}
        f_{\frakc}(e^{2\pi i z}) &= (e^{2\pi i z} - 1)^t h(e^{2\pi i z}) \\
        &= (2\pi i z + O(z^2))^t (h(1) + O(z)).
    \end{align*}
    The lowest-order term on the RHS is $(2\pi i)^t h(1) z^t$, so since $h(1) \neq 0$ this forces $t \geq m$ as desired. Finally, substituting $X^q$ for $X$, the polynomial $(X^q-1)^m$ divides $f_\frakc(X^q)$, and hence divides $p_\frakc(X) = X^r f_\frakc(X^q)$.
\end{proof}

\begin{corollary}\label{cor:nonzeromonomial}
    In the setting of~\Cref{lem:pc}, we have $qm \leq d$. Moreover, there exists an integer $R \in [0, d]$ such that:
    \begin{itemize}
        \item the coefficient of $X^R$ in $p_\frakc(X)$ is nonzero; and
        \item $\min(R, d-R) \leq \frac{d-qm}{2}$.
    \end{itemize}
\end{corollary}
\begin{proof}
    First, $p_\frakc$ is not the zero polynomial: by the construction of $p_\frakc$ (see~\Cref{sec:pcproof}), $e^{-2\pi i zd/(2q)}\, p_\frakc(e^{2\pi i z/q}) = \Phi_{g_\frakc, \ell^*}(z/q)$, which is not identically zero by the minimality of $\ell^*$ in~\Cref{def:ella}.
    Call the \emph{spread} of a nonzero univariate polynomial the difference of its largest and smallest exponents with a nonzero coefficient. Then the spread of $f_1f_2$ is the sum of the spreads of $f_1$ and $f_2$, as long as neither $f_1$ nor $f_2$ is the zero polynomial: the lowest terms of $f_1$ and $f_2$ multiply to the (nonzero) lowest term of $f_1 f_2$, and likewise for the highest terms.
    Since $(X^q-1)^m$ divides $p_\frakc$ (\Cref{lem:pcdivisible}) and has spread $qm$, the spread of $p_\frakc$ is at least $qm$. As the exponents of $p_\frakc$ lie in $[0, d]$ (\Cref{lem:pc}), this already forces $qm \leq d$. Moreover, the exponents that appear in $p_{\frakc}$ with nonzero coefficient cannot all lie in the open interval $\left(\frac{d-qm}{2}, \frac{d+qm}{2}\right)$, whose width is exactly $qm$; so some exponent $R$ satisfies $R \leq \frac{d-qm}{2}$ or $R \geq \frac{d+qm}{2}$, i.e., $\min(R, d-R) \leq \frac{d-qm}{2}$.
\end{proof}

\begin{lemma}\label{lem:randomgen4}
    For every cusp $\frakc \in \frakC$, we have:
    \begin{equation*}
        \gamma d < 2\ell_{\frakc} + \vartheta(d-qm).
    \end{equation*}
\end{lemma}
\begin{proof}
    Write $\frakc = \frakc_{a, \lambda}$, and let $R \in [0, d]$ be the integer guaranteed by~\Cref{cor:nonzeromonomial} for this cusp class. Let us unwrap the definition of $F_{g_{\frakc}}(\tau, z)$ a bit, using $\rho(g_\frakc) = \rho(g_{a, \lambda})$ (\Cref{fact:weilrep}):
    \begin{align*}
        &F_{g_{\frakc}}(\tau, z) \\
        ={}&P_\beta(\rho(g_{\frakc})\Theta(\tau, z/q)) \\
        ={}& P_\beta\left(\rho(g_{a, \lambda}) \sum_{k \in \ZZ} (-1)^k \exp\left(\frac{\pi i \tau}{q} \left(k-\frac{1}{2}\right)^2 + 2\pi i \left(k-\frac{1}{2}\right) \frac{z}{q}\right) \ev_{k+(q-1)/2 \bmod{q}}\right) \\
        \propto_\zeta{} &P_\beta\left(\rho(g_{a, \lambda}) \sum_{k \in \ZZ} (-1)^k \exp\left(\frac{\pi i \tau}{q} \left(k-\frac{1}{2}\right)^2 + 2\pi i \left(k-\frac{1}{2}\right) \frac{z}{q}\right) \ev_{k\bmod{q}}\right) \quad\text{(\Cref{prop:pproperties}, with $\sfX^{(q+1)/2}$)} \\
        \propto{} &P_\beta\left(\sum_{k \in \ZZ} \sum_{y \in \ZZ_q} (-1)^k \exp\left(\frac{\pi i \tau}{q} \left(k-\frac{1}{2}\right)^2 + 2\pi i \left(k-\frac{1}{2}\right) \frac{z}{q}\right) \exp\left(\frac{2\pi i}{q}\left(ay^2 - \lambda^{-1}ky\right)\right) \ev_{y}\right) \\
        ={} &\sum_{\xv \in \ZZ_q^d} \beta_{\xv} \prod_{i = 1}^d \left(\sum_{k \in \ZZ} (-1)^k \exp\left(\frac{\pi i \tau}{q} \left(k-\frac{1}{2}\right)^2 + 2\pi i \left(k-\frac{1}{2}\right) \frac{z}{q}\right) \exp\left(\frac{2\pi i}{q}\left(ax_i^2 - \lambda^{-1}kx_i\right)\right)\right) \\
        ={} &\sum_{k_1, \ldots, k_d \in \ZZ}  (-1)^{k_1 + \ldots + k_d} \exp\left(\frac{\pi i \tau}{q} \sum_i \left(k_i - \frac{1}{2}\right)^2 + 2\pi i \sum_i \left(k_i - \frac{1}{2}\right) \frac{z}{q}\right) \left[\sum_{\xv \in \ZZ_q^d} \beta_{\xv} \zeta_q^{a \langle \xv, \xv \rangle - \lambda^{-1} \langle \kv, \xv \rangle}\right].
    \end{align*}
    (In the third line we used the $\sfX$-invariance, up to a root of unity, of $P_\beta \circ \rho(g)$ --- the fourth bullet of~\Cref{prop:pproperties}; in the fourth we substituted~\eqref{eq:rhogal} and pulled out the nonzero constant $c_{a, \lambda}^d$ by homogeneity --- this constant, like the root of unity hidden in $\propto_\zeta$, is immaterial for what follows; the rearrangements are justified by absolute convergence, exactly as in~\Cref{sec:ellexpansionproof}.)

    A tuple $(k_1, \ldots, k_d)$ above contributes to the coefficient of $X^R$ in $p_\frakc$ exactly when $\sum_i (k_i - 1/2)^2 = d/4 + 2\ell_\frakc$ and $\sum_i k_i = R$: such a term has the $\tau$-exponential of level $d + 8\ell_\frakc$ and the $z$-exponential $e^{2\pi i (R - d/2)z/q}$, matching the leading part of~\eqref{eq:pc}. Moreover, by~\eqref{eq:beta4} (with $\zv = -\lambda^{-1}\kv \bmod q$, noting $a \in \calA$), the bracketed factor vanishes whenever $-\lambda^{-1}\kv \bmod q \in T_d(\vartheta, \nu)$, which --- $T_d(\vartheta, \nu)$ being closed under multiplication by units, as in the proof of~\Cref{lem:Hgen4} --- happens if and only if $\kv \bmod q \in T_d(\vartheta, \nu)$. Since the coefficient of $X^R$ in $p_\frakc$ is \emph{nonzero} (\Cref{cor:nonzeromonomial}), at least one contributing tuple has a nonvanishing bracket. Fix such a tuple $\kv \in \ZZ^d$; it satisfies:
    \begin{itemize}
        \item $\kv \bmod q \notin T_d(\vartheta, \nu)$;
        \item $\sum_{i = 1}^d (k_i - 1/2)^2 = d/4 + 2\ell_{\frakc}$ --- equivalently, $\sum_{i = 1}^d k_i(k_i - 1) = 2\ell_\frakc$;
        \item $\sum_{i = 1}^d k_i = R$.
    \end{itemize}
    At this point, we use the fact from~\Cref{cor:nonzeromonomial} that either $R$ or $d-R$ is $\leq \frac{d-qm}{2}$, and address these two cases separately:
    \begin{itemize}
        \item If $R \leq \frac{d-qm}{2}$:
        taking $(u, b) = (1, 0)$ in the definition of $T_d(\vartheta, \nu)$ (\Cref{def:gen4aprimesupport}), the first bullet implies $\kv \bmod q \notin S_d^{\mathrm{edge}}(\vartheta, \nu)$, i.e., $\sum_i w_{q, \vartheta}(k_i \bmod q) > \gamma d$. Bounding each summand by its integer representative $k_i$ (\Cref{def:track3exponent}), we have:
        \begin{align*}
            \gamma d &< \sum_{i = 1}^d w_{q, \vartheta}(k_i \bmod q) \\
            &\leq \sum_{i = 1}^d \left(k_i(k_i-1) + 2\vartheta k_i\right) \\
            &= 2\ell_{\frakc} + 2\vartheta R \\
            &\leq 2\ell_{\frakc} + \vartheta(d-qm).
        \end{align*}
        \item If $d-R \leq \frac{d-qm}{2}$: taking $(u, b) = (-1, 1)$ instead, the first bullet implies $(\bfone - \kv) \bmod q \notin S_d^{\mathrm{edge}}(\vartheta, \nu)$ (if $\bfone - \kv \equiv \zv$ for some $\zv \in S_d^{\mathrm{edge}}(\vartheta, \nu)$, then $\kv \equiv -\zv + \bfone \in T_d(\vartheta, \nu)$), i.e., $\sum_i w_{q, \vartheta}((1-k_i) \bmod q) > \gamma d$. Bounding each summand by the integer representative $1 - k_i$, whose value is $(1-k_i)((1-k_i)-1) + 2\vartheta(1-k_i) = k_i(k_i-1) + 2\vartheta(1-k_i)$, we have:
        \begin{align*}
            \gamma d &< \sum_{i = 1}^d w_{q, \vartheta}((1-k_i) \bmod q) \\
            &\leq \sum_{i = 1}^d \left(k_i(k_i-1) + 2\vartheta(1-k_i)\right) \\
            &= 2\ell_{\frakc} + 2\vartheta (d-R) \\
            &\leq 2\ell_{\frakc} + \vartheta(d-qm).
        \end{align*}
    \end{itemize}
    Either way, we are done.
\end{proof}

\subsection{Proof Step 5: Applying the Valence Formula}\label{sec:gen4step5}

We now apply the valence formula to the two modular forms constructed in Step 3 and combine the resulting inequalities with Step 4's per-class bounds. Recall the standing hypothesis of this proof (\Cref{sec:gen4step1}): for the sake of contradiction, we assumed that the coefficients required by~\Cref{thm:gen4aprime} do not exist, and from this assumption constructed the symmetrized witness $\beta$; the function $F$, the integer $m$, the modular forms $A_m^2$ and $B^2$, and the per-class quantities $\ell_\frakc$ and $H_\frakc$ were all built from that $\beta$. In the two lemmas below, the sums range over all $(q^2-1)/2$ cusp classes of $\Gamma(q)$ (\Cref{prop:gammaqcusps}).

\begin{lemma}\label{lem:valence1gen4}
    Summing over all cusp classes of $\Gamma(q)$, we have:
    \begin{equation*}
        \sum_\frakc \ell_\frakc \leq \frac{d(q^2-1)(q-3)}{48} + \frac{m(q^3-q)}{24}.
    \end{equation*}
\end{lemma}
\begin{proof}
    We apply~\Cref{lem:valencegammaq} to $A_m^2$, which is a nonzero modular form of weight $d+2m$ for $\Gamma(q)$ (\Cref{lem:amsq}; it is nonzero since $A_m \not\equiv 0$ by the definition of $m$) whose one-dimensional representation has finite order (\Cref{lem:amsq}), as~\Cref{lem:valencegammaq} requires. Since $A_m^2$ is holomorphic, its interior orders are nonnegative, so dropping the interior sum in~\eqref{eq:valencegammaq} and inserting the cusp-order bound of~\Cref{lem:amsq} gives:
    \begin{align*}
        \frac{d+2m}{24} \cdot \frac{q^3-q}{2} &\geq \frac{1}{2} \sum_{\frakc} \order_\frakc(A_m^2) \\
        &\geq \frac{1}{2} \sum_\frakc \left(2\ell_\frakc + \frac{d}{4}\right) \\
        &= \sum_\frakc \ell_\frakc + \frac{d}{8} \cdot \frac{q^2-1}{2} \\
        \Leftrightarrow \sum_\frakc \ell_\frakc &\leq \frac{d(q^2-1)(q-3)}{48} + \frac{m(q^3-q)}{24},
    \end{align*}
    as desired.
\end{proof}

\begin{lemma}\label{lem:valence2gen4}
    Summing over all cusp classes of $\Gamma(q)$, we have:
    \begin{equation*}
        \sum_\frakc \ell_\frakc + \sum_\frakc H_\frakc \leq \frac{d(q^2-1)(q-1)}{16}.
    \end{equation*}
\end{lemma}
\begin{proof}
    We apply~\Cref{lem:valencegammaq} to $B^2$, which is a nonzero modular form of weight $3d$ for $\Gamma(q)$ (\Cref{cor:bsqmodular,lem:b2holomorphic,prop:bnotzero}) whose one-dimensional representation has finite order (\Cref{cor:bsqmodular}), as~\Cref{lem:valencegammaq} requires. Since $B^2$ is holomorphic, its interior orders are nonnegative, so dropping the interior sum in~\eqref{eq:valencegammaq} and inserting the cusp-order bound of~\Cref{lem:b2holomorphic} gives:
    \begin{align*}
        \frac{3d}{24} \cdot \frac{q^3-q}{2} &\geq \frac{1}{2} \sum_\frakc \order_\frakc(B^2) \\
        &\geq \frac{1}{2} \sum_\frakc \left(\frac{d}{4} + 2\ell_\frakc + 2H_\frakc\right) \\
        &= \sum_\frakc \ell_\frakc + \sum_\frakc H_\frakc + \frac{d}{8} \cdot \frac{q^2-1}{2} \\
        \Leftrightarrow \sum_\frakc \ell_\frakc + \sum_\frakc H_\frakc &\leq \frac{d(q^2-1)(q-1)}{16}.
    \end{align*}
\end{proof}

Both valence bounds are instances of a known strategy. \Cref{lem:valence1gen4} applies the valence formula to (a power of) the first nonvanishing Taylor coefficient, which is the approach of~\cite[Lemma~3.1]{LiWang25} (see also~\cite[Lemma~5.1]{Aoki22}); indeed, it is essentially the case $t = 1$ of~\cite[Lemma~3.1]{LiWang25}, applied to a suitable norm of $F$. \Cref{lem:valence2gen4} applies it to the product of the restrictions to the nonzero 2-torsion points, in the same spirit as~\cite{RichterWesterholtRaum15,Aoki22,LiWang25}. What is particular to our setting is how the cusp orders entering these bounds are controlled by the witness, through $\ell_\frakc$ and $H_\frakc$ (\Cref{lem:amsq,lem:b2holomorphic,lem:Hgen4,lem:randomgen4}).

We can finally complete the proof of~\Cref{thm:gen4aprime}:
\begin{proof}[Proof of~\Cref{thm:gen4aprime}]
    We collect here the four inequalities we will use, from~\Cref{lem:valence1gen4,lem:valence2gen4,lem:Hgen4,lem:randomgen4} respectively. In the first two, the sums range over all cusp classes; the third holds for every selected class $\frakc_{a, \lambda}$; the fourth holds for every $\frakc \in \frakC$. (We will also freely use that $\ell_\frakc \geq 0$ and $H_\frakc \geq 0$ for \emph{every} class --- the former by~\Cref{def:ella}, the latter from~\Cref{def:hc} since $w_q \geq 0$ --- and that $|\frakC| = s \cdot \frac{q-1}{2}$.)
    \begin{align}
        \sum_\frakc \ell_\frakc &\leq \frac{d(q^2-1)(q-3)}{48} + \frac{m(q^3-q)}{24}\nonumber \\
        \Leftrightarrow d-qm &\leq \frac{d(q-1)}{2} - \frac{24}{q^2-1} \sum_\frakc \ell_\frakc \quad\text{(multiply by $\tfrac{24}{q^2-1}$ and rearrange)}\label{eq:valence1gen4} \\
        \sum_\frakc \ell_\frakc + \sum_\frakc H_\frakc &\leq \frac{d(q^2-1)(q-1)}{16}\label{eq:valence2gen4} \\
        H_{\frakc_{a, \lambda}} &> \nu d\label{eq:Hgen4} \\
        \gamma d &< 2\ell_{\frakc} + \vartheta(d-qm).\label{eq:randomgen4}
    \end{align}
    First, summing~\eqref{eq:Hgen4} over the $s \cdot \frac{q-1}{2}$ selected classes and using $H_\frakc \geq 0$ for the others gives $\sum_\frakc H_\frakc \geq \sum_{\frakc \in \frakC} H_\frakc > \frac{\nu d s(q-1)}{2}$; plugging this into~\eqref{eq:valence2gen4} yields:
    \begin{equation}\label{eq:plugin}
        \sum_{\frakc} \ell_{\frakc} + \frac{\nu ds(q-1)}{2} < \frac{d(q^2-1)(q-1)}{16}.
    \end{equation}
    Now summing~\eqref{eq:randomgen4} over all $\frakc \in \frakC$ and dividing by $|\frakC| = s \cdot \frac{q-1}{2}$ yields the first line below:
    \begin{align*}
        \gamma d &< \frac{4}{s(q-1)}\sum_{\frakc \in \frakC} \ell_{\frakc} + \vartheta(d-qm) \\
        &\leq \frac{4}{s(q-1)}\sum_{\frakc \in \frakC} \ell_{\frakc} + \frac{\vartheta d(q-1)}{2} - \frac{24\vartheta}{q^2-1} \sum_{\frakc} \ell_{\frakc} \quad\text{(plugging in~\eqref{eq:valence1gen4}, using $\vartheta \geq 0$)} \\
        &\leq \frac{4}{s(q-1)}\underbrace{\left(1 - \frac{6\vartheta s}{q+1}\right)}_{\geq 0}\sum_{\frakc \in \frakC} \ell_{\frakc} + \frac{\vartheta d(q-1)}{2} \\
        &\leq \frac{4}{s(q-1)}\left(1 - \frac{6\vartheta s}{q+1}\right) \left(\frac{d(q^2-1)(q-1)}{16} - \frac{\nu ds(q-1)}{2}\right) + \frac{\vartheta d(q-1)}{2} \quad\text{(plugging in~\eqref{eq:plugin})} \\
        &= \left(1 - \frac{6\vartheta s}{q+1}\right) \left(\frac{d(q^2-1)}{4s} - 2\nu d\right) + \frac{\vartheta d(q-1)}{2} \\
        \Leftrightarrow \gamma &< \left(1 - \frac{6\vartheta s}{q+1}\right) \left(\frac{q^2-1}{4s} - 2\nu \right) + \frac{\vartheta (q-1)}{2}.
    \end{align*}
    Here the third line drops the non-selected classes from $\sum_\frakc \ell_\frakc$ --- valid since $\ell_\frakc \geq 0$ and the coefficient $-\frac{24\vartheta}{q^2-1}$ is nonpositive --- and then uses the identity $\frac{24\vartheta}{q^2-1} = \frac{4}{s(q-1)} \cdot \frac{6\vartheta s}{q+1}$; the underbraced factor is nonnegative because $\vartheta \leq \frac{q+1}{6s}$ (\Cref{def:gen4aprimesupport}), which also justifies the fourth line, together with $\sum_{\frakc \in \frakC} \ell_\frakc \leq \sum_\frakc \ell_\frakc$ and~\eqref{eq:plugin}; and the final line divides by $d > 0$. But the right-hand side of the final line is \emph{exactly} the definition of $\gamma = \gamma(\vartheta, \nu)$ in~\Cref{def:gen4aprimesupport}, so we have derived $\gamma < \gamma$ --- a contradiction. Hence the assumed coefficients exist after all, proving~\Cref{thm:gen4aprime}.
\end{proof}
\begin{remark}[The case $\vartheta = 0$]\label{remark:thetazero}
    For the boundary parameters $\vartheta = 0$ and $\nu = \frac{q^2-1}{8s}$, the proof above simplifies considerably. Here $\gamma(0, \nu) = 0$, and summing~\eqref{eq:Hgen4} over the $s \cdot \frac{q-1}{2}$ selected classes gives $\sum_\frakc H_\frakc > \frac{d(q^2-1)(q-1)}{16}$, which, together with $\ell_\frakc \geq 0$, already contradicts~\eqref{eq:valence2gen4}. So in this case only the scalarization via 2-torsion evaluations and a single application of the valence formula are needed --- not the scalarization via Taylor coefficients, \Cref{lem:valence1gen4}, or the edge-polynomial analysis of Step~4 (\Cref{lem:pc} through~\Cref{lem:randomgen4}). This case already gives a nontrivial support exponent for every odd prime $q$ and every $s \in [2, q]$: the edge set $S_d^{\mathrm{edge}}(0, \nu)$ is $\{0, 1\}^d$ (as $w_{q, 0}(j) = 0$ exactly when $j \in \{0, 1\}$), so $E^{\mathrm{edge}}_{q, s}(0, \nu) = \log_q 2$, while $E^{\mathrm{sq}}_q(\nu) < 1$ because $\nu$ is smaller than the average $\frac{q^2-1}{12}$ of $w_q$ over $\ZZ_q$ (for $s \geq 2$). It is, however, weaker than the full result: for $(q, s) = (5, 4), (7, 6), (7, 7), (11, 10)$, the case $\vartheta = 0$ gives exponents of approximately $0.791$, $0.729$, $0.690$, and $0.676$, respectively, compared with the bounds of~\Cref{lem:gen4numerics}. The scalarization via Taylor coefficients and the edge-polynomial analysis of Step~4 make positive tilts $\vartheta > 0$ available, and all of the bounds in~\Cref{lem:gen4numerics} use such tilts.
\end{remark}

%% file: figcuspinfty.tex
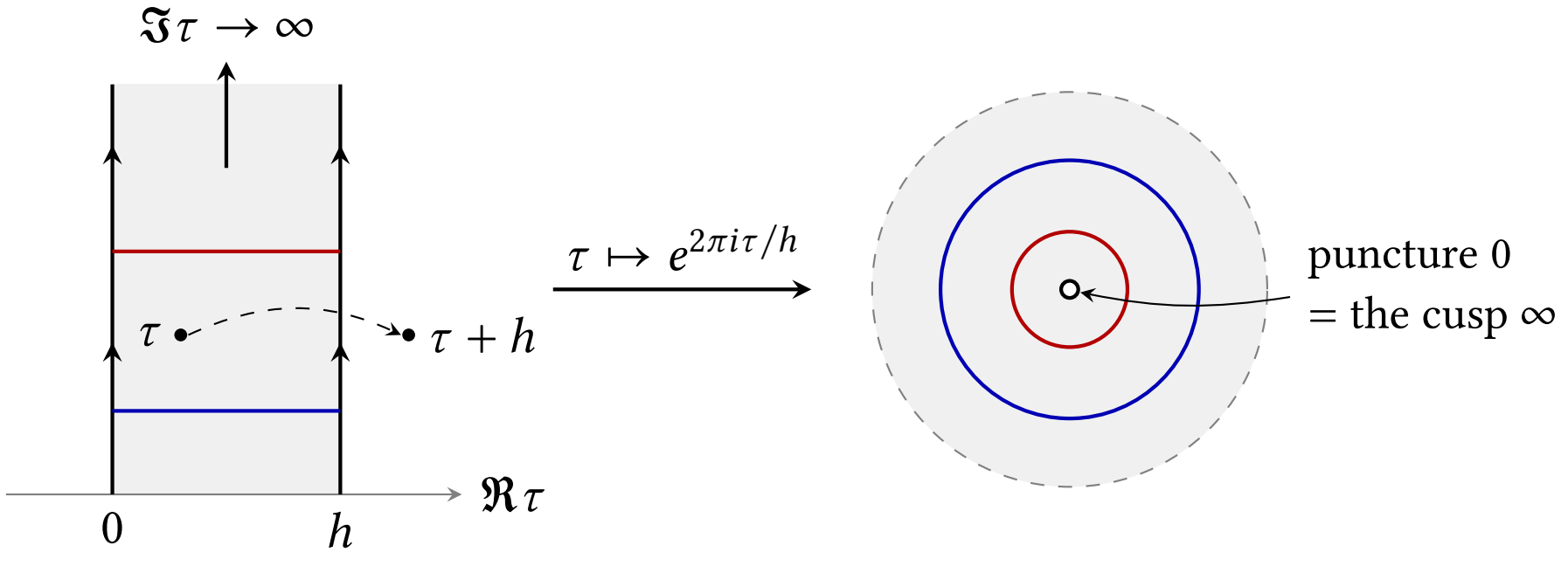
\begin{figure}[H]
    \centering
    \begin{tikzpicture}[>=stealth, scale=1.15]
        \fill[gray!12] (0,0) rectangle (1.5,2.7);
        \draw[->, gray] (-0.7,0) -- (2.3,0) node[right, black] {$\Re \tau$};
        \draw[thick] (0,0) -- (0,2.7);
        \draw[thick] (1.5,0) -- (1.5,2.7);
        \foreach \x in {0,1.5} {
            \draw[thick, ->] (\x,0.95) -- (\x,1.0);
            \draw[thick, ->] (\x,2.25) -- (\x,2.3);
        }
        \node[below] at (0,0) {$0$};
        \node[below] at (1.5,0) {$h$};
        \draw[blue!70!black, thick] (0,0.55) -- (1.5,0.55);
        \draw[red!70!black, thick] (0,1.6) -- (1.5,1.6);
        \draw[->, thick] (0.75,2.15) -- (0.75,2.85) node[above] {$\Im \tau \to \infty$};
        \fill (0.45,1.05) circle (1.2pt) node[left] {$\tau$};
        \fill (1.95,1.05) circle (1.2pt) node[right] {$\tau + h$};
        \draw[->, dashed] (0.5,1.05) to[bend left=25] (1.9,1.05);
        \draw[->, thick] (2.9,1.35) -- node[above] {$\tau \mapsto e^{2\pi i \tau/h}$} (4.6,1.35);
        \begin{scope}[shift={(6.3,1.35)}]
            \fill[gray!12] (0,0) circle (1.3);
            \draw[dashed, gray] (0,0) circle (1.3);
            \draw[blue!70!black, thick] (0,0) circle (0.85);
            \draw[red!70!black, thick] (0,0) circle (0.38);
            \filldraw[fill=white, draw=black, thick] (0,0) circle (1.6pt);
            \node[anchor=west, align=left, font=\small] at (1.45,0) {puncture $0$\\$=$ the cusp $\infty$};
            \draw[->, thin] (1.45,-0.05) to[bend left=10] (0.07,-0.02);
        \end{scope}
    \end{tikzpicture}
    \caption{The cusp at $\infty$, drawn schematically. A function $f$ with $f(\tau + h) = f(\tau)$ is a function on the strip $0 \leq \Re \tau \leq h$ with its two vertical edges glued (left). The map $\tau \mapsto e^{2\pi i \tau/h}$ identifies this glued strip with the punctured unit disk (right): it sends horizontal lines to circles around $0$ (higher lines to smaller circles), and $\Im \tau \to \infty$ to the puncture. So the cusp $\infty$ becomes the point $0$: $f$ is holomorphic at $\infty$ when $0$ is a removable singularity, and its order at $\infty$ is its order of vanishing at $0$, the least $n$ with a nonzero coefficient in the expansion $f(\tau) = \sum_n c_n e^{2\pi i n\tau/h}$. For a vector-valued $F$, the periodicity is twisted by $\rho(T^h)$, and untwisting it shifts the exponents by the numbers $\alpha_j$ (\Cref{def:holoinfty}).}
    \label{fig:cuspinfty}
\end{figure}

%% file: figgen1flow.tex
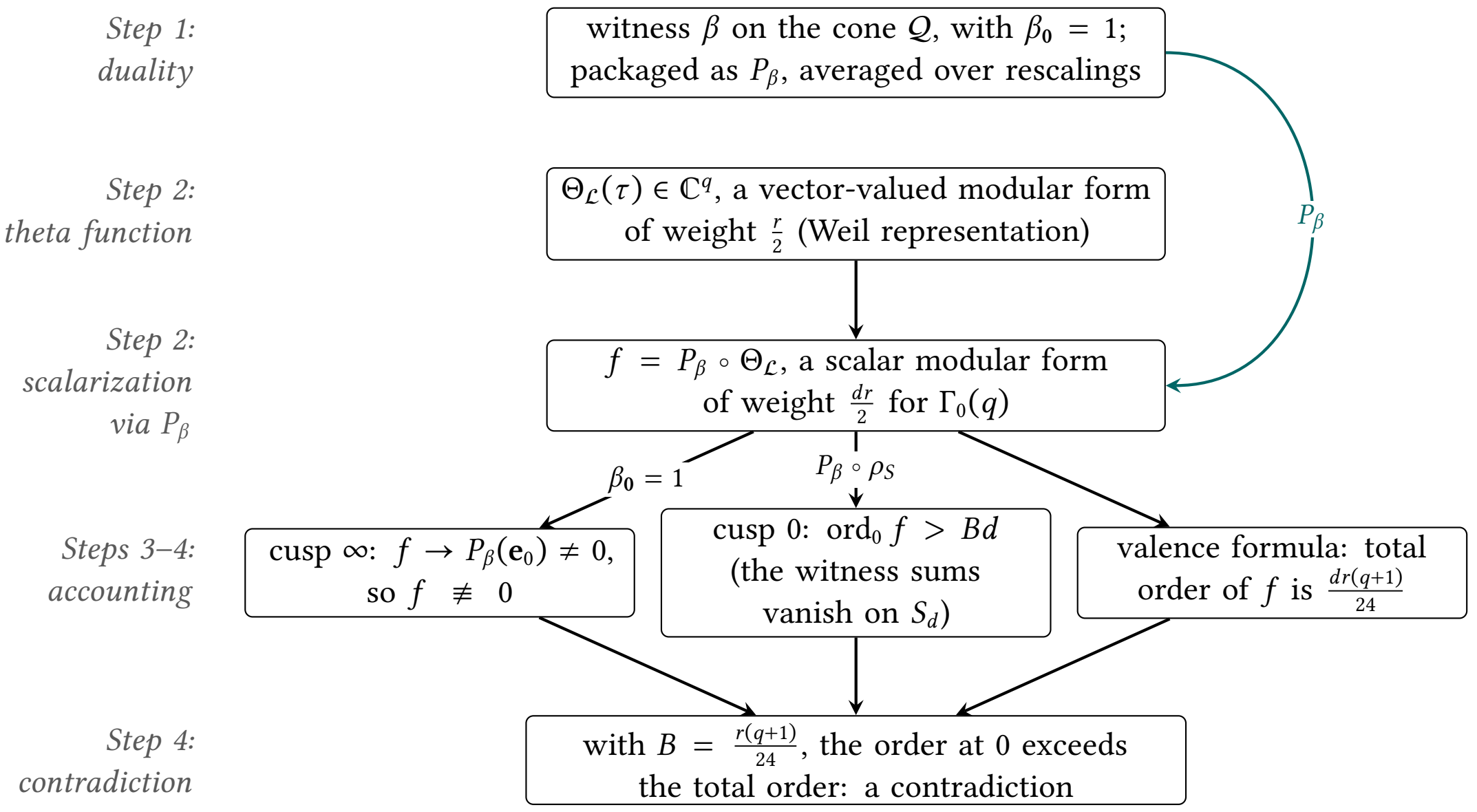
\begin{figure}[H]
    \centering
    \begin{tikzpicture}[>=stealth, x=1cm, y=1cm,
        box/.style={draw, rounded corners=2pt, align=center, font=\footnotesize, inner sep=2pt, text width=3.6cm, fill=white},
        layer/.style={anchor=east, align=right, font=\footnotesize\itshape, text=black!65, text width=1.9cm},
        lab/.style={font=\scriptsize, fill=white, inner sep=1pt}]
        \node[layer] at (-6.25,0) {Step~1:\\duality};
        \node[layer] at (-6.25,-1.55) {Step~2:\\theta function};
        \node[layer] at (-6.25,-3.2) {Step~2:\\scalarization\\via $P_\beta$};
        \node[layer] at (-6.25,-5.0) {Steps~3--4:\\accounting};
        \node[layer] at (-6.25,-6.8) {Step~4:\\contradiction};
        \node[box, text width=5.8cm] (W) at (0,0) {witness $\beta$ on the cone $\calQ$, with $\beta_\zerov = 1$;\\packaged as $P_\beta$, averaged over rescalings};
        \node[box, text width=5.8cm] (T) at (0,-1.55) {$\Theta_\calL(\tau) \in \CC^q$, a vector-valued modular form\\of weight $\frac{r}{2}$ (Weil representation)};
        \node[box, text width=5.8cm] (f) at (0,-3.2) {$f = P_\beta \circ \Theta_\calL$, a scalar modular form\\of weight $\frac{dr}{2}$ for $\Gamma_0(q)$};
        \node[box] (I) at (-4.0,-5.0) {cusp $\infty$: $f \to P_\beta(\ev_0) \neq 0$,\\so $f \not\equiv 0$};
        \node[box] (Z) at (0,-5.0) {cusp $0$: $\order_0 f > Bd$\\(the witness sums vanish on $S_d$)};
        \node[box] (V) at (4.0,-5.0) {valence formula: total\\order of $f$ is $\frac{dr(q+1)}{24}$};
        \node[box, text width=6.2cm] (C) at (0,-6.8) {with $B = \frac{r(q+1)}{24}$, the order at $0$ exceeds\\the total order: a contradiction};
        \draw[->, thick, teal!80!black] (W.east) to[out=0, in=0, looseness=1.5] node[lab, pos=0.5, text=teal!80!black] {$P_\beta$} (f.east);
        \draw[->, thick] (T) -- (f);
        \draw[->, thick] (f) -- node[lab, pos=0.5] {$\beta_\zerov = 1$} (I);
        \draw[->, thick] (f) -- node[lab, pos=0.5] {$P_\beta \circ \rho_S$} (Z);
        \draw[->, thick] (f) -- (V);
        \draw[->, thick] (I) -- (C);
        \draw[->, thick] (Z) -- (C);
        \draw[->, thick] (V) -- (C);
    \end{tikzpicture}
    \caption{How the proof of~\Cref{thm:gen1} fits together, one row per layer of the argument (numbered by the warm-up's steps). The witness and the theta function are built independently, and the witness enters through the polynomial $P_\beta$ (teal), which turns $\Theta_\calL$ into a single scalar modular form $f$. Each property of the witness plays a role: the cone (with the averaging over rescalings) makes $f$ modular for $\Gamma_0(q)$; the normalization $\beta_\zerov = 1$ makes $f$ nonzero at the cusp $\infty$ (left); and the orthogonality to $S_d$, which reaches the cusp $0$ through $P_\beta \circ \rho_S$, makes $f$ vanish there to order more than $Bd$ (middle), while the valence formula fixes its total order (right). Looking ahead, the stronger construction has a chart of the same shape (\Cref{fig:gen4flow}), in which this chain reappears as the right-hand branch, run at $s \cdot \frac{q-1}{2}$ cusps instead of one.}
    \label{fig:gen1flow}
\end{figure}

%% file: tabgen4comparison.tex
\begin{table}[t]
    \centering
    \small
    \renewcommand{\arraystretch}{1.2}
    \begin{tabular}{>{\raggedright\arraybackslash}p{0.2\textwidth} >{\raggedright\arraybackslash}p{0.34\textwidth} >{\raggedright\arraybackslash}p{0.36\textwidth}}
        \toprule
        & Warm-up (\Cref{sec:gen1}) & Stronger construction (\Cref{sec:gen4}) \\
        \midrule
        Goal & cone-delta function ($q$ servers) & quadratic-phase delta expansion with $s \leq q$ phases ($s$ servers) \\
        Witness & $\beta$ on the cone $\calQ = \{\xv : \langle \xv, \xv \rangle \equiv 0 \pmod{q}\}$, with $\beta_\zerov = 1$ & $\beta$ on $\{\xv : \sum_i x_i \equiv 0\}$, invariant under $\xv \mapsto \xv + \bfone$, with $\beta_\zerov = 1$ \\
        Theta function & lattice theta function $\Theta_\calL(\tau) \in \CC^q$, of weight $\frac{r}{2}$ & Jacobi theta function $\Theta(\tau, w) \in \CC^q$, of weight $\frac{1}{2}$ \\
        Symmetries absorbed by $P_\beta$ & $\rho(g)$ for $g \in \Gamma_0(q)$ (after averaging) & the Heisenberg operators $\sfX$ and $\sfZ$ \\
        Group & $\Gamma_0(q)$, with $2$ cusp classes & $\Gamma(q)$, with $\frac{q^2-1}{2}$ cusp classes \\
        Cusps where we force high vanishing order & the cusp $0$ & the $s \cdot \frac{q-1}{2}$ selected classes $\frakc_{a, \lambda}$ \\
        Scalar modular form(s) & $P_\beta \circ \Theta_\calL$, of weight $\frac{dr}{2}$ & $A_m^2$ and $B^2$, of weights $d + 2m$ and $3d$ \\
        Costs & shortest squared lengths of the cosets $c$ of $\calL$ in $\calL^*$, scaled: $w_\calL(c) = \frac{q}{2} \min_{\yv \in c} \langle \yv, \yv \rangle$ (\Cref{def:prices}) & square cost $w_q(j) = \min_{k \equiv j} k^2$ and tilted edge cost $w_{q, \vartheta}(j) = \min_{k \equiv j} (k(k-1) + 2\vartheta k)$ (\Cref{def:track3exponent}) \\
        Root counting & one valence inequality & two valence inequalities and two per-cusp bounds \\
        Support & vectors $\zv$ of total cost $\sum_i w_\calL(z_i\av)$ at most $\frac{r(q+1)}{24}d$, and their scalar multiples (\Cref{thm:gen1}) & vectors $\zv$ of total square cost $\sum_i w_q(z_i)$ at most $\nu d$ or total tilted edge cost $\sum_i w_{q, \vartheta}(z_i)$ at most $\gamma d$; their images under $\zv \mapsto u\zv + b\bfone$; and these with one coordinate changed arbitrarily (\Cref{def:gen4aprimesupport}) \\
        \bottomrule
    \end{tabular}
    \caption{The warm-up and the stronger construction side by side.}
    \label{tab:gen4comparison}
\end{table}

%% file: figgen4cusps.tex
\begin{figure}[H]
    \centering
    \begin{tikzpicture}[x=1.62cm, y=0.62cm]
        \fill[blue!10] (-0.5,1.5) rectangle (2.5,-2.5);
        \foreach \c/\lab in {0/{a = 0}, 1/{a = 1}, 2/{a = 2}, 3/{a = 3}, 4/{a = 4}, 5/{a = 5}, 6/{a = 6}, 7/{\infty}} {
            \node at (\c,1) {$\lab$};
        }
        \node[anchor=east] at (-0.55,1) {label};
        \draw (-1.6,0.5) -- (7.5,0.5);
        \foreach \l in {1,2,3} {
            \node[anchor=east] at (-0.55,1-\l) {$\lambda = \l$};
        }
        \foreach \c/\l/\x/\y in {
            0/1/0/1, 0/2/0/2, 0/3/0/3,
            1/1/2/1, 1/2/-3/2, 1/3/-1/3,
            2/1/-3/1, 2/2/1/2, 2/3/-2/3,
            3/1/-1/1, 3/2/-2/2, 3/3/-3/3,
            4/1/1/1, 4/2/2/2, 4/3/3/3,
            5/1/3/1, 5/2/-1/2, 5/3/2/3,
            6/1/-2/1, 6/2/3/2, 6/3/1/3,
            7/1/1/0, 7/2/2/0, 7/3/3/0} {
            \node at (\c,1-\l) {$\pm(\x, \y)$};
        }
        \draw[thick, rounded corners=4pt] (6.6,-0.4) rectangle (7.4,0.4);
        \draw[thick, rounded corners=4pt] (-0.4,-0.4) rectangle (0.4,0.4);
        \draw (-0.45,-2.7) -- (-0.45,-2.9) -- (2.45,-2.9) -- (2.45,-2.7);
        \node[below] at (1,-2.9) {selected classes $\frakc_{a, \lambda}$, for $a \in \calA = \{0, 1, 2\}$};
    \end{tikzpicture}
    \caption{The $\frac{q^2-1}{2} = 24$ cusp classes of $\Gamma(q)$ for $q = 7$, as nonzero points of $\ZZ_q^2$ up to sign (with coordinates in $\{-3, \ldots, 3\}$), arranged by label and multiplier: the class with label $a \in \ZZ_q$ and multiplier $\lambda$ is $\pm\lambda(2a, 1)$, and the class with label $\infty$ and multiplier $\lambda$ is $\pm\lambda(1, 0)$. For $\calA = \{0, 1, 2\}$, the selected classes are the $s \cdot \frac{q-1}{2} = 9$ entries in the shaded columns. The boxed entries are the warm-up's cusps $0 = \frac{0}{1}$ and $\infty = \frac{1}{0}$.}
    \label{fig:gen4cusps}
\end{figure}

%% file: figgen4symmetries.tex
\begin{figure}[H]
    \centering
    \begin{minipage}[b]{0.44\textwidth}
        \centering
        \begin{tikzcd}[column sep=large, row sep=large]
            w \arrow[r, mapsto, "\Theta(\tau{,}\, \cdot)"] \arrow[d, mapsto, "\cdot\, + \frac{a\tau + b}{q}"'] & \Theta(\tau, w) \arrow[d, mapsto, "\sfZ^b \sfX^a"] \\
            w + \frac{a\tau+b}{q} \arrow[r, mapsto, "\Theta(\tau{,}\, \cdot)"'] & \Theta\left(\tau, w + \frac{a\tau+b}{q}\right)
        \end{tikzcd}

        \vspace{0.5em}
        {\small (a) Translations by $q$-torsion points;\\ the Heisenberg operators}
    \end{minipage}%
    \hfill
    \begin{minipage}[b]{0.54\textwidth}
        \centering
        \begin{tikzcd}[column sep=large, row sep=large]
            (\tau, w) \arrow[r, mapsto, "\Theta"] \arrow[d, mapsto, "g \ast \cdot"'] & \Theta(\tau, w) \arrow[d, mapsto, "\rho(g)"] \\
            g \ast (\tau, w) \arrow[r, mapsto, "\Theta"'] & \Theta\big(g \ast (\tau, w)\big)
        \end{tikzcd}

        \vspace{0.5em}
        {\small (b) The action of $\SL_2(\ZZ)$ on $(\tau, w)$;\\ the Weil representation}
    \end{minipage}
    \caption{The two symmetries of the Jacobi theta function $\Theta$, each holding up to an explicit nonvanishing factor and a root of unity (\Cref{lem:thetatranslation,lem:descended}); both are classical for theta functions with characteristics~\cite[Chapter~I, \S\S3 and~7]{Mumford83}. (a) For fixed $\tau$, translating $w$ by a $q$-torsion point $\frac{a\tau+b}{q}$ multiplies $\Theta(\tau, w)$ by the Heisenberg operator $\sfZ^b\sfX^a$: this is the abelian, Fourier-analytic side of the analogy in~\Cref{fig:fourieranalogy}, now present inside the proof. (b) The action of $g \in \SL_2(\ZZ)$ on $(\tau, w)$ multiplies $\Theta$ by $\rho(g)$, as for the lattice theta functions of the warm-up. The witness polynomial $P_\beta$ is invariant under $\sfX$ and $\sfZ$, so it erases the symmetry (a): the scalar function $P_\beta \circ \Theta$ is quasi-periodic in $w$.}
    \label{fig:gen4symmetries}
\end{figure}

%% file: figgen4action.tex
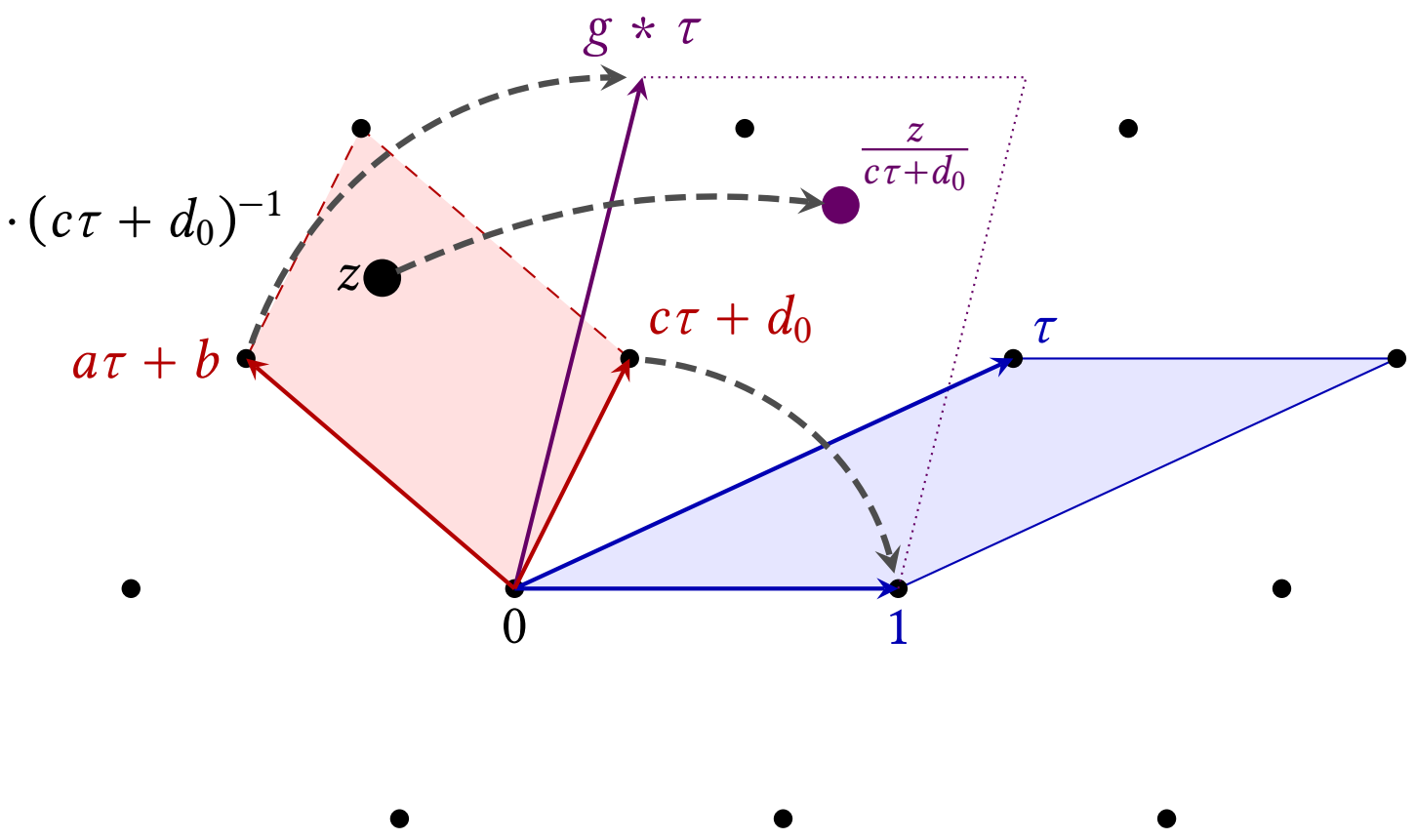
\begin{figure}[H]
    \centering
    \pgfmathsetmacro{\tx}{1.3}
    \pgfmathsetmacro{\ty}{0.6}
    \pgfmathsetmacro{\ga}{1}
    \pgfmathsetmacro{\gb}{-2}
    \pgfmathsetmacro{\gc}{1}
    \pgfmathsetmacro{\gd}{-1}
    \pgfmathsetmacro{\zx}{-0.345}
    \pgfmathsetmacro{\zy}{0.81}
    \pgfmathsetmacro{\ax}{\gc*\tx + \gd}
    \pgfmathsetmacro{\ay}{\gc*\ty}
    \pgfmathsetmacro{\bx}{\ga*\tx + \gb}
    \pgfmathsetmacro{\by}{\ga*\ty}
    \pgfmathsetmacro{\nrm}{\ax*\ax + \ay*\ay}
    \pgfmathsetmacro{\sx}{(\bx*\ax + \by*\ay)/\nrm}
    \pgfmathsetmacro{\sy}{(\by*\ax - \bx*\ay)/\nrm}
    \pgfmathsetmacro{\wx}{(\zx*\ax + \zy*\ay)/\nrm}
    \pgfmathsetmacro{\wy}{(\zy*\ax - \zx*\ay)/\nrm}
    \begin{tikzpicture}[scale=2.6, >=stealth]
        \fill[blue!10] (0,0) -- (1,0) -- (1+\tx,\ty) -- (\tx,\ty) -- cycle;
        \fill[red!12] (0,0) -- (\ax,\ay) -- (\ax+\bx,\ay+\by) -- (\bx,\by) -- cycle;
        \draw[blue!70!black] (1,0) -- (1+\tx,\ty) -- (\tx,\ty);
        \draw[red!70!black, dashed] (\ax,\ay) -- (\ax+\bx,\ay+\by) -- (\bx,\by);
        \foreach \m in {-4,...,4} {
            \foreach \n in {-2,...,3} {
                \pgfmathsetmacro{\px}{\m+\n*\tx}
                \pgfmathsetmacro{\py}{\n*\ty}
                \pgfmathparse{(\px > -1.3) && (\px < 2.5) && (\py > -0.75) && (\py < 1.5) ? 1 : 0}
                \ifdim\pgfmathresult pt>0.5pt \fill (\px,\py) circle (0.7pt); \fi
            }
        }
        \draw[violet!80!black, densely dotted] (1,0) -- (1+\sx,\sy) -- (\sx,\sy);
        \draw[->, thick, violet!80!black] (0,0) -- (\sx,\sy) node[above] {$g \ast \tau$};
        \draw[->, thick, blue!70!black] (0,0) -- (1,0) node[below] {$1$};
        \draw[->, thick, blue!70!black] (0,0) -- (\tx,\ty) node[above right=-1pt] {$\tau$};
        \draw[->, thick, red!70!black] (0,0) -- (\ax,\ay) node[above right=-1pt] {$c\tau + d_0$};
        \draw[->, thick, red!70!black] (0,0) -- (\bx,\by) node[left] {$a\tau + b$};
        \draw[->, very thick, black!70, dash pattern=on 4pt off 2pt, shorten <=3pt, shorten >=3pt] (\ax,\ay) to[bend left=35] (1,0);
        \draw[->, very thick, black!70, dash pattern=on 4pt off 2pt, shorten <=3pt, shorten >=3pt] (\bx,\by) to[bend left=35] node[pos=0.3, left=3pt, black] {$\cdot\, (c\tau + d_0)^{-1}$} (\sx,\sy);
        \fill (\zx,\zy) circle (1.4pt) node[left] {$z$};
        \fill[violet!80!black] (\wx,\wy) circle (1.4pt) node[above right=-1pt] {$\frac{z}{c\tau + d_0}$};
        \draw[->, very thick, black!70, dash pattern=on 4pt off 2pt, shorten <=3pt, shorten >=3pt] (\zx,\zy) to[bend left=15] (\wx,\wy);
        \node[below] at (0,0) {$0$};
    \end{tikzpicture}
    \caption{The action of $\SL_2(\ZZ)$ on pairs $(\tau, z)$, where $z$ is a point of the torus $\CC/(\ZZ\tau + \ZZ)$. As in~\Cref{fig:lattice}, the matrix $g$ gives another basis $(c\tau+d_0, a\tau+b)$ of the lattice $\ZZ + \ZZ\tau$ (red), and rotating and scaling the plane by $(c\tau+d_0)^{-1}$ (dashed arrows) turns it into the basis $(1, g \ast \tau)$ (purple). The same similarity sends the point $z$ to $\frac{z}{c\tau + d_0}$, a point of the torus $\CC/(\ZZ(g \ast \tau) + \ZZ)$. This is the action $g \ast (\tau, z) = \left(g \ast \tau, \frac{z}{c\tau + d_0}\right)$ on $\calH \times \CC$ that underlies Jacobi forms~\cite{EichlerZagier85}.}
    \label{fig:gen4action}
\end{figure}

%% file: figgen4twotorsion.tex
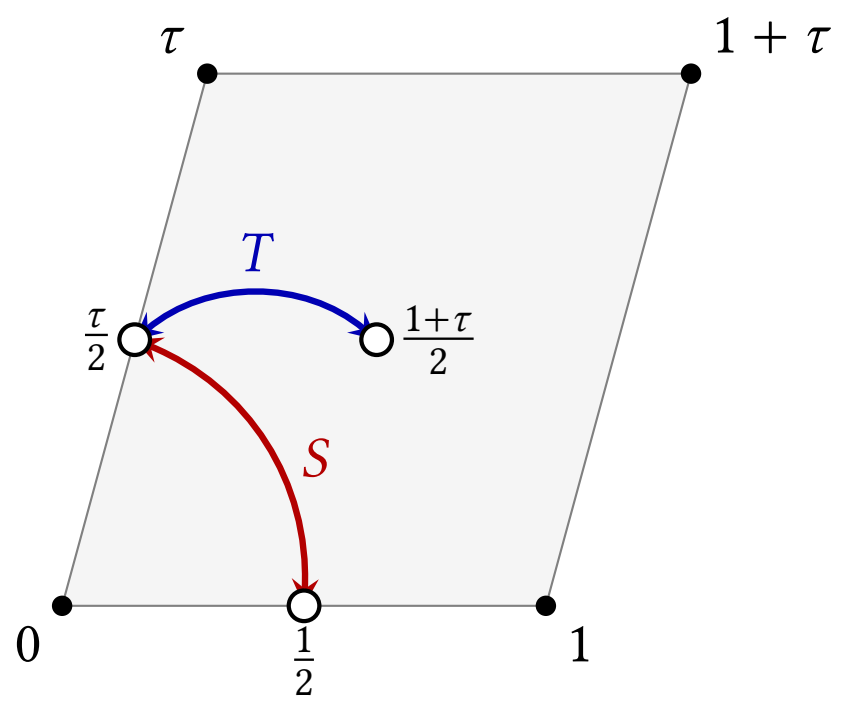
\begin{figure}[H]
    \centering
    \pgfmathsetmacro{\ux}{0.3}
    \pgfmathsetmacro{\uy}{1.1}
    \begin{tikzpicture}[scale=3.2, >=stealth]
        \fill[gray!8] (0,0) -- (1,0) -- (1+\ux,\uy) -- (\ux,\uy) -- cycle;
        \draw[gray] (0,0) -- (1,0) -- (1+\ux,\uy) -- (\ux,\uy) -- cycle;
        \foreach \p in {(0,0), (1,0), (\ux,\uy), (1+\ux,\uy)} {
            \fill \p circle (0.6pt);
        }
        \node[below left] at (0,0) {$0$};
        \node[below right] at (1,0) {$1$};
        \node[above left] at (\ux,\uy) {$\tau$};
        \node[above right] at (1+\ux,\uy) {$1 + \tau$};
        \coordinate (H1) at (0.5,0);
        \coordinate (Ht) at (\ux/2,\uy/2);
        \coordinate (H1t) at (0.5+\ux/2,\uy/2);
        \draw[<->, very thick, red!70!black] (H1) to[bend right=35] node[pos=0.45, right=1pt] {$S$} (Ht);
        \draw[<->, very thick, blue!70!black] (Ht) to[bend left=40] node[midway, above] {$T$} (H1t);
        \foreach \p in {(H1), (Ht), (H1t)} {
            \filldraw[fill=white, draw=black, thick] \p circle (0.9pt);
        }
        \node[below] at (H1) {$\frac{1}{2}$};
        \node[left] at (Ht) {$\frac{\tau}{2}$};
        \node[right] at (H1t) {$\frac{1+\tau}{2}$};
    \end{tikzpicture}
    \caption{The three nonzero 2-torsion points $\frac{1}{2}$, $\frac{\tau}{2}$ and $\frac{1+\tau}{2}$ of the torus $\CC/(\ZZ\tau + \ZZ)$, in the variable $z = qw$. Rewriting them in the coordinates of a new basis (\Cref{fig:gen4action}) permutes them: $S$ swaps $\frac{1}{2}$ and $\frac{\tau}{2}$ and fixes $\frac{1+\tau}{2}$, while $T$ swaps $\frac{\tau}{2}$ and $\frac{1+\tau}{2}$ and fixes $\frac{1}{2}$ (all modulo the lattice). As $S$ and $T$ generate $\SL_2(\ZZ)$, no single nonzero 2-torsion point is preserved, which is why the second scalarization uses the product of the evaluations at all three, as is standard~\cite{RichterWesterholtRaum15,LiWang25}.}
    \label{fig:gen4twotorsion}
\end{figure}

%% file: figgen4flow.tex
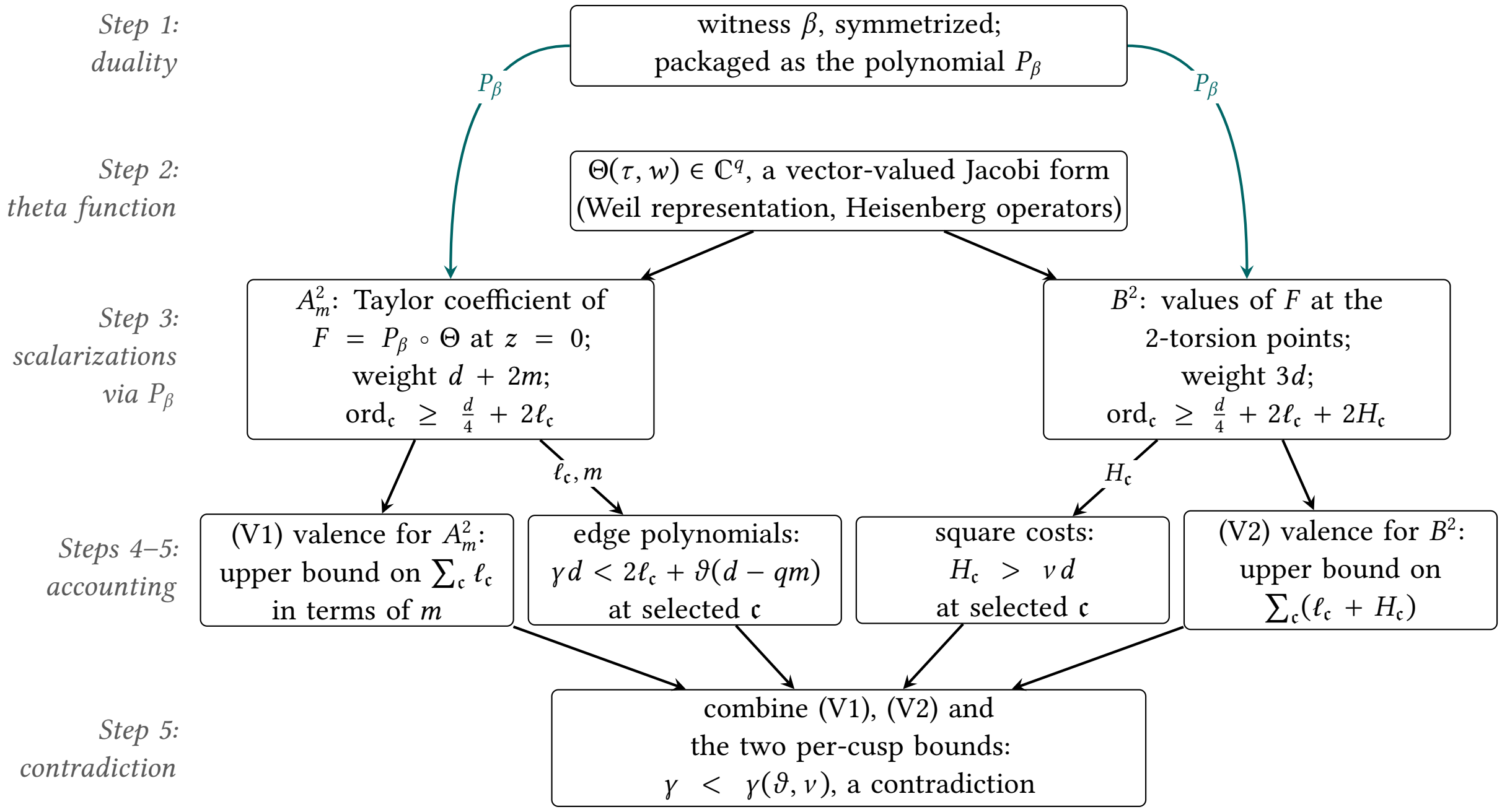
\begin{figure}[H]
    \centering
    \begin{tikzpicture}[>=stealth, x=1cm, y=1cm,
        box/.style={draw, rounded corners=2pt, align=center, font=\footnotesize, inner sep=2pt, text width=3.2cm, fill=white},
        layer/.style={anchor=east, align=right, font=\footnotesize\itshape, text=black!65, text width=1.9cm},
        lab/.style={font=\scriptsize, fill=white, inner sep=1pt}]
        \node[layer] at (-7.05,0) {Step~1:\\duality};
        \node[layer] at (-7.05,-1.55) {Step~2:\\theta function};
        \node[layer] at (-7.05,-3.35) {Step~3:\\scalarizations\\via $P_\beta$};
        \node[layer] at (-7.05,-5.6) {Steps~4--5:\\accounting};
        \node[layer] at (-7.05,-7.5) {Step~5:\\contradiction};
        \node[box, text width=5.8cm] (W) at (0,0) {witness $\beta$, symmetrized;\\packaged as the polynomial $P_\beta$};
        \node[box, text width=5.8cm] (T) at (0,-1.55) {$\Theta(\tau, w) \in \CC^q$, a vector-valued Jacobi form\\(Weil representation, Heisenberg operators)};
        \node[box, text width=4.2cm] (A) at (-4.25,-3.35) {$A_m^2$: Taylor coefficient of\\$F = P_\beta \circ \Theta$ at $z = 0$;\\weight $d + 2m$;\\$\order_\frakc \geq \frac{d}{4} + 2\ell_\frakc$};
        \node[box, text width=4.2cm] (B) at (4.25,-3.35) {$B^2$: values of $F$ at the\\2-torsion points;\\weight $3d$;\\$\order_\frakc \geq \frac{d}{4} + 2\ell_\frakc + 2H_\frakc$};
        \node[box] (V1) at (-5.25,-5.6) {(V1) valence for $A_m^2$:\\upper bound on $\sum_\frakc \ell_\frakc$\\in terms of $m$};
        \node[box] (E) at (-1.75,-5.6) {edge polynomials:\\$\gamma d < 2\ell_\frakc + \vartheta(d - qm)$\\at selected $\frakc$};
        \node[box] (H) at (1.75,-5.6) {square costs:\\$H_\frakc > \nu d$\\at selected $\frakc$};
        \node[box] (V2) at (5.25,-5.6) {(V2) valence for $B^2$:\\upper bound on\\$\sum_\frakc (\ell_\frakc + H_\frakc)$};
        \node[box, text width=6.2cm] (C) at (0,-7.5) {combine (V1), (V2) and the two per-cusp bounds:\\$\gamma < \gamma(\vartheta, \nu)$, a contradiction};
        \draw[->, thick, teal!80!black] (W.west) to[out=180, in=90] node[lab, pos=0.35, text=teal!80!black] {$P_\beta$} (A.north);
        \draw[->, thick, teal!80!black] (W.east) to[out=0, in=90] node[lab, pos=0.35, text=teal!80!black] {$P_\beta$} (B.north);
        \draw[->, thick] (T) -- (A);
        \draw[->, thick] (T) -- (B);
        \draw[->, thick] (A) -- (V1);
        \draw[->, thick] (A) -- node[lab, pos=0.45] {$\ell_\frakc, m$} (E);
        \draw[->, thick] (B) -- node[lab, pos=0.45] {$H_\frakc$} (H);
        \draw[->, thick] (B) -- (V2);
        \draw[->, thick] (V1) -- (C);
        \draw[->, thick] (E) -- (C);
        \draw[->, thick] (H) -- (C);
        \draw[->, thick] (V2) -- (C);
    \end{tikzpicture}
    \caption{How the proof of~\Cref{thm:gen4aprime} fits together, one row per layer of the argument. The witness and the theta function are built independently (Steps~1 and~2). The witness enters through the polynomial $P_\beta$ (teal, as in~\Cref{fig:fourieranalogy}), which turns $\Theta$ into the Jacobi form $F = P_\beta \circ \Theta$ and from it into two scalar modular forms for $\Gamma(q)$ (Step~3). Their orders at each cusp class $\frakc$ are bounded below in terms of $\ell_\frakc$ (where the expansion of $F$ at $\frakc$ starts), the Taylor order $m$, and $H_\frakc$ (the least square cost of a surviving monomial). In the accounting layer, the valence formula bounds the sums of these quantities over all cusp classes from above (outer boxes), while the witness's orthogonality, which reaches the cusps through $P_\beta$, bounds them from below at the $s \cdot \frac{q-1}{2}$ selected classes (inner boxes). Combining the four inequalities gives a contradiction, by the choice of $\gamma(\vartheta, \nu)$ in~\Cref{def:gen4aprimesupport}. Compare with the warm-up (\Cref{fig:gen1flow}): the right-hand branch ($B^2$, the square costs and (V2)) is its argument, now run at the $s \cdot \frac{q-1}{2}$ selected classes; for $\vartheta = 0$ it suffices on its own (\Cref{remark:thetazero}), and the left-hand branch is what the Taylor coefficient adds.}
    \label{fig:gen4flow}
\end{figure}

%% file: depirx.tex
\section{Proofs of Formal Theorem Statements}
\label{sec:depirx}

\paragraph{Proof of \cref{mainthm:track1}.} \begin{proof}
    Let $n$ be sufficiently large, and let $d$ be the largest integer with $16^d \leq n$. By~\Cref{lem:goodprime} with $q = 2$ and $M = n$, there are an integer $B = \polylog(n)$ and an odd prime $p > n$ with $p \geq n^4$ and $2^B < p \leq 2^B(1 + \frac{1}{4n})$. Let $j \in [d]$. A database of $16^j$ records in $\F_p$ splits into four databases of $2^{4j-2}$ records each, where $4j - 2 \equiv 2 \pmod 4$ and $4j - 2 < n < p$. Running the scheme of~\Cref{thm:2-server-rate-one} on each of the four, with independent randomness and with a query for an arbitrary fixed record to each of the three that do not contain the desired record, gives a $2$-server scheme over databases of $16^j$ records in $\F_p$. In it, each server stores $16^j$ elements of $\F_p$, the user uploads $4(4j-2)$ bits, and each server reads and sends back at most $4 \cdot 2^{H_2(1/4)(4j-1)} \leq 4 \cdot 16^{H_2(1/4) d} \leq 4 n^{H_2(1/4)}$ elements of $\F_p$. So~\Cref{lem:bit-compiler}, applied with $b = 16$, gives a $2$-server scheme with exponent $H_2(1/4)$.

    \emph{Three servers.} Let $n$ be sufficiently large, and let $d$ be the largest integer with $3^d \leq n$. By~\Cref{lem:goodprime} with $q = 3$ and $M = n$, there are an integer $B = \polylog(n)$ and a prime $p \equiv 1 \pmod 3$ with $p > n$, $p \geq n^4$ and $2^B < p \leq 2^B(1 + \frac{1}{4n})$. Then $\F_p$ contains a primitive cube root of unity, and its characteristic exceeds $\max\{\lfloor d/3 \rfloor, 2\}$. So~\Cref{lem:q3ends}, with $\ell = \lfloor \alpha^* d \rfloor$, gives a cone-delta function on $\ZZ_3^d$ over $\F_p$ with Fourier sparsity $k \leq 3^{(\eta_{\mathrm{ends}} + o(1))d} \leq n^{\eta_{\mathrm{ends}} + o(1)}$ (\Cref{cor:q3ends}). By~\Cref{lem:conedelta-qpde}, this yields a $(k, 3)$-sparse quadratic-phase delta expansion over $\F_p$, which~\Cref{thm:s-server-rate-one} (with $s = q = 3$) and~\Cref{rem:smallerdims} turn into $3$-server schemes over databases of $3^j$ records in $\F_p$, for every $j \in [d]$, in which each server reads and sends back at most $k$ elements of $\F_p$. Applying~\Cref{lem:bit-compiler} with $b = 3$ gives a $3$-server scheme with exponent $\eta_{\mathrm{ends}}$.
\end{proof}

\paragraph{Proof of \cref{mainthm:track2}.} \begin{proof}
    Fix a generator $\av + \calL$ of $\calL^*/\calL$. Let $n$ be sufficiently large, let $d$ be the largest integer with $q^d \leq n$, and let $S_d \subseteq \ZZ_q^d$ be the set of~\Cref{thm:gen1}. That theorem gives a cone-delta function on $\ZZ_q^d$ over $\QQ(\zeta_q)$ whose Fourier support is contained in $S_d$, and $|S_d| \leq q^{(\eta_\calL + o(1))d} \leq n^{\eta_\calL + o(1)}$ by~\Cref{cor:gen1general}. By~\Cref{lem:conedelta-qpde}, this yields a quadratic-phase delta expansion over $\QQ(\zeta_q)$ with support $S_d \times \ZZ_q$. The set $S_d$ is invariant under permutations of the $d$ coordinates, since its definition depends only on the multiset of coordinates, so by~\Cref{lem:symmetrize}, there is also a symmetric such expansion. By~\Cref{lem:goodprime} with $M = M_d$, the explicit bound of~\Cref{lem:finite-field-specialization}, there are an integer $B = \polylog(n)$ and a prime $p \equiv 1 \pmod q$ with $p > M$, $p \geq n^4$ and $2^B < p \leq 2^B(1 + \frac{1}{4n})$. Then~\Cref{lem:finite-field-specialization} gives a quadratic-phase delta expansion over $\F_p$ supported on $S_d \times \ZZ_q$, whose coefficients the user can compute in $\polylog(n)$ time, and which~\Cref{thm:s-server-rate-one} (with $s = q$) and~\Cref{rem:smallerdims} turn into $q$-server schemes over databases of $q^j$ records in $\F_p$, for every $j \in [d]$, in which each server reads and sends back at most $|S_d|$ elements of $\F_p$. Applying~\Cref{lem:bit-compiler} with $b = q$ gives a $q$-server scheme with exponent $\eta_\calL$.

    For primes $q \equiv 3 \pmod 4$, the rank-$2$ lattice of~\Cref{def:gen1concrete} has $\eta_\calL = \eta_q$, as shown in the proof of~\Cref{cor:gen1specific}, and the bounds on $\eta_q$ are~\Cref{cor:gen1specific}. Finally, given $s \geq 3$, let $q$ be the largest prime with $q \equiv 3 \pmod 4$ and $q \leq s$, and run the $q$-server scheme on $q$ of the $s$ servers. By the prime number theorem in arithmetic progressions~\cite[Chapter~22, p.~133]{Davenport80}, $q \geq s/2$ for all sufficiently large $s$, so $\log_2 q \geq \log_2 s - 1$ and $\frac{0.467 + o(1)}{\log_2 q} = \frac{0.467 + o(1)}{\log_2 s}$.
\end{proof}

\paragraph{Proof of \cref{mainthm:track3}.} \begin{proof}
    Let $\eta$ be either $\eta_{q, s}(\vartheta, \nu)$ for a fixed feasible pair $(\vartheta, \nu)$, or $\eta_{q, s}$. Let $n$ be sufficiently large, and let $d$ be the largest integer with $q^d \leq n$. By~\Cref{cor:gen4exponent}, there is a $(k, s)$-sparse quadratic-phase delta expansion over $\QQ(\zeta_q)$ on $\ZZ_q^d$ with permutation-invariant support, where $k \leq q^{(\eta + o(1))d} \leq n^{\eta + o(1)}$; by~\Cref{lem:symmetrize}, there is also a symmetric such expansion, with the same support. By~\Cref{lem:goodprime} with $M = M_d$, the explicit bound of~\Cref{lem:finite-field-specialization}, there are an integer $B = \polylog(n)$ and a prime $p \equiv 1 \pmod q$ with $p > M$, $p \geq n^4$ and $2^B < p \leq 2^B(1 + \frac{1}{4n})$. Then~\Cref{lem:finite-field-specialization} gives a $(k, s)$-sparse quadratic-phase delta expansion over $\F_p$, whose coefficients the user can compute in $\polylog(n)$ time, and which~\Cref{thm:s-server-rate-one} and~\Cref{rem:smallerdims} turn into $s$-server schemes over databases of $q^j$ records in $\F_p$, for every $j \in [d]$, in which each server reads and sends back at most $k$ elements of $\F_p$. Applying~\Cref{lem:bit-compiler} with $b = q$ gives an $s$-server scheme with exponent $\eta$.

    The bounds on $\eta_{q, s}$ are~\Cref{cor:gen4exponent}. Finally, given $s \geq 3$, let $q$ be the largest prime with $q \leq s$, and run the scheme with exponent $\eta_{q, q}$ on $q$ of the $s$ servers. By the prime number theorem~\cite[Chapter~18]{Davenport80}, $q \geq s/2$ for all sufficiently large $s$, so $\log_2 q \geq \log_2 s - 1$ and $\frac{0.255 + o(1)}{\log_2 q} = \frac{0.255 + o(1)}{\log_2 s}$.
\end{proof}

%% file: mathapp.tex
\section{Background for Part II}\label{sec:pt2appendix}

\subsection{Complex Analysis}\label{sec:complexanalysis}

This subsection collects the facts from complex analysis that we use, in the forms in which we use them. Throughout, ``locally uniformly'' means uniformly on every compact subset of the domain in question. For the one-variable facts, see, e.g.,~\cite[Chapter~2]{SteinShakarchi03}.

\begin{definition}[Holomorphic functions]\label{def:holomorphic}
    Let $U \subseteq \CC$ be open. A function $f: U \to \CC$ is \emph{holomorphic} if it is complex differentiable at every point of $U$, i.e., for every $z_0 \in U$ the limit
    \begin{equation*}
        f'(z_0) := \lim_{z \to z_0} \frac{f(z) - f(z_0)}{z - z_0}
    \end{equation*}
    exists. A function $F: U \to \CC^N$ is holomorphic if each of its coordinates is. Equivalently, $f$ is holomorphic if and only if, near each point $z_0 \in U$, it is given by a convergent power series $f(z) = \sum_{n \geq 0} a_n (z - z_0)^n$, its \emph{Taylor expansion} at $z_0$; see, e.g.,~\cite[Chapter~2]{SteinShakarchi03}.
\end{definition}

\begin{fact}[Identity theorem]\label{fact:identitytheorem}
    Let $U \subseteq \CC$ be a connected open set and let $f: U \to \CC$ be holomorphic and not identically zero. Then the zeros of $f$ are isolated, and the Taylor expansion of $f$ at every point of $U$ has a first nonzero coefficient. Consequently, if two holomorphic functions on $U$ agree on a subset of $U$ with an accumulation point in $U$, then they agree on all of $U$.
\end{fact}

\begin{fact}[Weierstrass convergence theorem]\label{fact:weierstrass}
    Let $U \subseteq \CC$ be open and let $f_1, f_2, \ldots: U \to \CC^N$ be holomorphic functions converging locally uniformly to $f: U \to \CC^N$. Then $f$ is holomorphic, and $f'_n \to f'$ locally uniformly. In particular, a series of holomorphic functions that converges locally uniformly (e.g., because the terms admit a convergent series of uniform bounds on each compact set) has a holomorphic sum.
\end{fact}

\begin{fact}[Fourier expansion of periodic holomorphic functions]\label{fact:fourierexp}
    Let $h > 0$ and let $G: \calH \to \CC^N$ be a holomorphic function satisfying $G(\tau + h) = G(\tau)$ for all $\tau \in \calH$. Then there exist unique $\cv_n \in \CC^N$ (for $n \in \ZZ$) such that
    \begin{equation*}
        G(\tau) = \sum_{n = -\infty}^{\infty} \cv_n e^{2\pi i \tau n/h},
    \end{equation*}
    with the series converging absolutely and locally uniformly on $\calH$. (Applying the scalar case coordinatewise gives the vector-valued statement, and the scalar case is standard: $G$ descends through the holomorphic covering map $\tau \mapsto e^{2\pi i \tau/h}$ from $\calH$ onto the punctured unit disk, where the expansion above is the Laurent expansion. See, e.g.,~\cite[Section~1.1]{DiamondShurman05} (stated there for $h = 1$; the general case follows by rescaling $\tau \mapsto \tau/h$).)
\end{fact}

We will also deal with holomorphic functions of two complex variables, for which the following definition suffices for our purposes.

\begin{definition}\label{def:jointholo}
    Let $U \subseteq \CC \times \CC$ be open. A function $F: U \to \CC^N$ is \emph{holomorphic} if it is continuous and, for each fixed value of either variable, holomorphic in the other. (By a theorem of Hartogs, the continuity requirement is automatic, but we will not need this.)
\end{definition}

With this definition, sums, products, and compositions with polynomials of holomorphic functions are holomorphic, and the convergence statement of \Cref{fact:weierstrass} extends verbatim to two variables: continuity passes to locally uniform limits, and holomorphy in each variable follows from the one-variable case applied on slices. The two-variable fact we need is that Taylor expansion in one variable behaves well in the other:

\begin{lemma}[{Taylor coefficients of a holomorphic family~\cite[Theorem~1.2.2 and Corollary~1.2.4]{Krantz01}}]\label{lem:taylorfamilies}
    Let $U \subseteq \CC$ be open and let $F: U \times \CC \to \CC^N$ be holomorphic. Then there exist unique holomorphic functions $A_0, A_1, \ldots: U \to \CC^N$ such that:
    \begin{equation*}
        F(\tau, z) = \sum_{j = 0}^\infty A_j(\tau)\, z^j,
    \end{equation*}
    with the series converging absolutely and locally uniformly on $U \times \CC$. Moreover, $A_j(\tau) = \frac{1}{j!} \frac{\pp^j}{\pp z^j} F(\tau, z)\big|_{z = 0}$.
\end{lemma}

\subsection{Poisson Summation Formula}

We use the Poisson summation formula only through the following identity for Gaussian sums over a lattice (\Cref{cor:gausspsf}). Throughout this subsection, $r$ is an arbitrary positive integer (we do \emph{not} impose the evenness convention of~\Cref{sec:highdimlattices}; one application below has $r = 1$), $\langle \cdot, \cdot \rangle$ is the standard inner product on $\RR^r$ together with its bilinear extension to $\CC^r$, $\norm{\xv} := \langle \xv, \xv \rangle^{1/2}$ for $\xv \in \RR^r$, and $\calL = \BB \ZZ^r \subseteq \RR^r$ is a full-rank lattice with dual $\calL^*$ and determinant $\det \calL = |\det \BB|$ as in~\Cref{sec:highdimlattices} (so $\det \calL^* = 1/\det \calL$).

We first record a crude bound that makes every sum in this subsection (and our theta functions) converge:

\begin{lemma}[Gaussian-dominated lattice sums]\label{lem:latticegaussian}
    For any $\vecv \in \RR^r$ and constants $c > 0$ and $C \geq 0$:
    \begin{equation*}
        \sum_{\yv \in \vecv + \calL} \exp\left(-c \norm{\yv}^2 + C \norm{\yv}\right) < \infty.
    \end{equation*}
\end{lemma}
\begin{proof}
    Distinct points of $\vecv + \calL$ differ by a nonzero element of $\calL$, and every nonzero $\yv = \BB\xv \in \calL$ (where $\xv \in \ZZ^r \setminus \{\zerov\}$) satisfies $\norm{\yv} \geq \sigma \norm{\xv} \geq \sigma$, where $\sigma > 0$ is the smallest singular value of $\BB$. So the open balls of radius $\sigma/2$ centered at the points of $\vecv + \calL$ are pairwise disjoint. For an integer $n \geq 0$, each such ball whose center has norm less than $n+1$ lies inside the ball of radius $n + 1 + \sigma/2$ around the origin, so comparing volumes, the number of $\yv \in \vecv + \calL$ with $n \leq \norm{\yv} < n+1$ is at most $\left(\frac{2(n+1)}{\sigma} + 1\right)^r$. Each such $\yv$ contributes at most $\exp(-cn^2 + C(n+1))$, so the sum is at most $\sum_{n = 0}^\infty \left(\frac{2(n+1)}{\sigma} + 1\right)^r \exp(-cn^2 + C(n+1))$, which converges because the summand decays superexponentially in $n$.
\end{proof}

\begin{fact}[{\cite[Chapter~II, Equation~(5.6)]{Mumford83}}]\label{cor:gausspsf}
    For any $\uv \in \CC^r$ (the shift may be complex) and $\tau \in \calH$, we have:
    \[
        \sum_{\yv \in \calL} \exp\left(-\pi i \langle \yv+\uv, \yv+\uv \rangle/\tau\right) = (-i\tau)^{r/2} \det(\calL^*) \sum_{\yv \in \calL^*} \exp(\pi i \tau \langle \yv, \yv \rangle - 2\pi i \langle \yv, \uv \rangle),
    \]
    with both sums converging absolutely. Here and throughout, $(-i\tau)^{r/2} := \exp\left(\frac{r}{2} \operatorname{Log}(-i\tau)\right)$ with $\operatorname{Log}$ the principal branch of the logarithm, which is well-defined and holomorphic on $\calH$ because $\Re(-i\tau) = \Im \tau > 0$.
\end{fact}
This is the functional equation $\vartheta(\Omega^{-1} z, -\Omega^{-1}) = \det(\Omega/i)^{1/2} \exp(\pi i\, z^\top \Omega^{-1} z)\, \vartheta(z, \Omega)$ of the Riemann theta function~\cite[Chapter~II, Equation~(5.6) and the paragraph after it]{Mumford83}, with $\Omega = \tau (\BB^\top \BB)^{-1}$ and $z = -\BB^{-1} \uv$, after cancelling the factor $\exp(\pi i \langle \uv, \uv \rangle/\tau)$ from both sides. Mumford's branch of $\det(\Omega/i)^{1/2}$ is positive for purely imaginary $\Omega$, so it agrees with $(-i\tau)^{r/2} \det(\calL^*)$. Both sums converge absolutely by~\Cref{lem:latticegaussian}.

\begin{remark}
    For purely imaginary $\tau = is^2$, with $s > 0$, and real $\uv$,~\Cref{cor:gausspsf} reads
    \[
        \sum_{\xv \in \calL + \uv} e^{-\pi \norm{\xv}^2/s^2} = s^r \det(\calL^*) \sum_{\yv \in \calL^*} e^{-\pi s^2 \norm{\yv}^2 - 2\pi i \langle \yv, \uv \rangle}.
    \]
    This is the form of the Poisson summation formula behind Banaszczyk's Gaussian measures on lattices~\cite[Lemma~1.1(i)]{Ban93} and the analysis of discrete Gaussians in lattice-based cryptography (the proofs of~\cite[Lemma~4.1]{MR07} and~\cite[Claim~3.8]{R09}).
\end{remark}

\subsection{Gaussian Riemann Sums}

\begin{lemma}[Gaussian Riemann sums]\label{lem:riemannsum}
    Let $N$ be a positive integer, $h, \lambda > 0$ real numbers, and $\alpha \in [0, 1]$ some fixed constant offset. Then:
    \[
    \left|h\sum_{j = 1}^N e^{-\lambda (j-\alpha)^2 h^2} - \frac{1}{2}\sqrt{\frac{\pi}{\lambda}}\right| = \left|h\sum_{j = 1}^N e^{-\lambda (j-\alpha)^2 h^2} - \int_0^\infty e^{-\lambda x^2} dx\right| \leq h + \frac{e^{-\lambda h^2N^2}}{2\lambda hN}.
    \]
\end{lemma}
\begin{proof}
    The equality is the Gaussian integral $\int_0^\infty e^{-\lambda x^2}\, dx = \frac{1}{2}\sqrt{\pi/\lambda}$. For the inequality, split the integral as $\int_0^{Nh} + \int_{Nh}^\infty$. The function $g(x) := e^{-\lambda x^2}$ is nonincreasing on $[0, \infty)$, and $h \sum_{j=1}^N g((j-\alpha)h)$ is a Riemann sum for $\int_0^{Nh} g(x)\, dx$ with one sample point $(j - \alpha)h \in [(j-1)h,\, jh]$ in each cell; for a nonincreasing function, the error in each cell is at most $h\left(g((j-1)h) - g(jh)\right)$, and these telescope:
    \[
    \left|h\sum_{j = 1}^N g((j-\alpha)h) - \int_0^{Nh} g(x)\, dx\right| \leq h\,(g(0) - g(Nh)) \leq h.
    \]
    For the tail, $\int_{Nh}^\infty e^{-\lambda x^2}\, dx \leq \int_{Nh}^\infty \frac{x}{Nh}\, e^{-\lambda x^2}\, dx = \frac{e^{-\lambda h^2 N^2}}{2\lambda h N}$.
\end{proof}

\section{Deferred Proofs from~\Cref{sec:pt2bg}}

\subsection*{Proof of~\Cref{lem:cuspwelldef}}\label{sec:cuspwelldefproof}

First consider two choices $g_\fraka, g'_\fraka \in \SL_2(\ZZ)$ with $g_\fraka \ast \infty = g'_\fraka \ast \infty = \fraka$. Then $g_\fraka^{-1} g'_\fraka$ fixes $\infty$, and the stabilizer of $\infty$ in $\SL_2(\ZZ)$ is $\{\pm T^n : n \in \ZZ\}$: a matrix fixes $\infty$ if and only if its lower-left entry is $0$, and then its diagonal entries are integers with product $1$, hence both $1$ or both $-1$. So $g'_\fraka = \pm g_\fraka T^n$ for some choice of sign and some $n \in \ZZ$.
    The width is unaffected: for every $t$ we have $g'_\fraka T^t (g'_\fraka)^{-1} = (\pm g_\fraka T^n) T^t (\pm g_\fraka T^n)^{-1} = g_\fraka T^t g_\fraka^{-1}$ (the signs cancel and powers of $T$ commute), so both choices yield the same set $\{t : g_\fraka T^t g_\fraka^{-1} \in \Gamma\}$ and hence the same width $h$, as well as the same matrix $A := \rho(g_\fraka T^h g_\fraka^{-1})$ and thus the same matrix $M$ from~\Cref{eq:eigendecomp}.
    Next, writing $F'_\fraka(\tau) := G(g'_\fraka, \tau)^{-k} F(g'_\fraka \ast \tau)$, we claim that $F'_\fraka(\tau) = (\pm 1)^k F_\fraka(\tau + n)$: indeed, $(\pm g_\fraka T^n) \ast \tau = g_\fraka \ast (\tau + n)$, while $G(\pm g_\fraka T^n, \tau) = \pm G(g_\fraka T^n, \tau) = \pm G(g_\fraka, \tau+n)$ by~\Cref{prop:chainrule} and $G(T^n, \tau) = 1$ (and $(\pm 1)^{-k} = (\pm 1)^k$).
    Now let $\cv_m$ and $\cv'_m$ be the expansion coefficients (\Cref{eq:qexpansion}) of $\exp(-2\pi i M \tau/h)F_\fraka(\tau)$ and $\exp(-2\pi i M \tau/h)F'_\fraka(\tau)$ respectively. We compute:
    \begin{equation*}
        \exp(-2\pi i M \tau/h)F'_\fraka(\tau) = (\pm 1)^k \exp(2\pi i M n/h) \cdot \exp(-2\pi i M (\tau+n)/h) F_\fraka(\tau + n),
    \end{equation*}
    so by the uniqueness in~\Cref{fact:fourierexp}, $\cv'_m = (\pm 1)^k e^{2\pi i mn/h} \exp(2\pi i Mn/h)\, \cv_m$. The matrix $\exp(2\pi i Mn/h)$ acts on the range of each spectral projector $P_\alpha$ of $A$ as the nonzero scalar $e^{2\pi i \alpha n/h}$, so $P_\alpha \cv'_m \neq 0$ if and only if $P_\alpha \cv_m \neq 0$. By~\Cref{remark:orderwelldef}, holomorphy at $\fraka$ and $\order_\fraka(F)$ are determined by exactly these conditions, so both choices give the same answers.

    Now consider $\frakb = \gamma \ast \fraka$ with $\gamma \in \Gamma$. By the first part we are free to choose the lifts, so take $g_\frakb := \gamma g_\fraka$ (indeed $g_\frakb \ast \infty = \frakb$). The widths agree because $g_\frakb T^t g_\frakb^{-1} = \gamma (g_\fraka T^t g_\fraka^{-1}) \gamma^{-1}$, and conjugation by $\gamma \in \Gamma$ preserves membership in $\Gamma$. Next, by~\Cref{eq:pullback}, \Cref{def:weakmodular}, and~\Cref{prop:chainrule}:
    \begin{align*}
        F_\frakb(\tau) &= G(\gamma g_\fraka, \tau)^{-k} F(\gamma g_\fraka \ast \tau) \\
        &= G(\gamma g_\fraka, \tau)^{-k} G(\gamma, g_\fraka \ast \tau)^k \rho(\gamma) F(g_\fraka \ast \tau) \\
        &= G(\gamma g_\fraka, \tau)^{-k} G(\gamma, g_\fraka \ast \tau)^k G(g_\fraka, \tau)^k \rho(\gamma) F_\fraka(\tau) \\
        &= \rho(\gamma) F_\fraka(\tau).
    \end{align*}
    The matrices $A$ at the two cusps are conjugate: $A_\frakb := \rho(g_\frakb T^h g_\frakb^{-1}) = \rho(\gamma) A_\fraka \rho(\gamma)^{-1}$. Hence $A_\frakb$ has the same eigenvalues as $A_\fraka$, with spectral projectors $P^\frakb_\alpha = \rho(\gamma) P^\fraka_\alpha \rho(\gamma)^{-1}$, and $M_\frakb = \rho(\gamma) M_\fraka \rho(\gamma)^{-1}$. Therefore:
    \begin{equation*}
        \exp(-2\pi i M_\frakb \tau/h)\, F_\frakb(\tau) = \rho(\gamma) \exp(-2\pi i M_\fraka \tau/h)\, F_\fraka(\tau),
    \end{equation*}
    so the expansion coefficients satisfy $\cv^\frakb_m = \rho(\gamma)\, \cv^\fraka_m$ by uniqueness, and $P^\frakb_\alpha \cv^\frakb_m = \rho(\gamma)\, P^\fraka_\alpha \cv^\fraka_m$, which is nonzero if and only if $P^\fraka_\alpha \cv^\fraka_m$ is. As before, by~\Cref{remark:orderwelldef}, holomorphy and order at the two cusps coincide.
    \qed

\subsection*{Proof of~\Cref{lem:scalarisation}: Holomorphy at the Cusps}\label{sec:scalarisationcuspsproof}

Let $F$, $P$, $\chi$ and $f = P \circ F$ be as in~\Cref{lem:scalarisation}. Recall from the main-text part of the proof that $\chi$ has finite order and that $f$ is weakly modular of weight $kD$ for $\Gamma$ with representation $\chi$. It remains to show that $f$ is holomorphic at every cusp $\fraka \in \QQ \cup \{\infty\}$. Fix a lift $g_\fraka$ (taking $g_\infty := I$ for the cusp at infinity), and let $h$ be the width of the cusp --- the width is a property of $\Gamma$ and $\fraka$ alone, so it is the same for $f$ and $F$. Homogeneity gives:
    \begin{equation*}
        f_\fraka(\tau) = G(g_\fraka, \tau)^{-kD}\, P(F(g_\fraka \ast \tau)) = P\left(G(g_\fraka, \tau)^{-k} F(g_\fraka \ast \tau)\right) = P(F_\fraka(\tau)).
    \end{equation*}
    Let $A := \rho(g_\fraka T^h g_\fraka^{-1})$ be the matrix describing how $F_\fraka$ transforms over one period, with eigenbasis $\vecv_j$, dual basis $\uv_j$, and eigenvalues $e^{2\pi i \alpha_j}$ ($\alpha_j \in [0, 1)$) as in~\Cref{sec:cuspinfty}; and let $e^{2\pi i \beta}$ with $\beta \in [0, 1)$ be the scalar $\chi(g_\fraka T^h g_\fraka^{-1})$ (well-defined this way since $\chi$ has finite order), which is the corresponding phase picked up by $f_\fraka$ over one period.

    Write the coordinates of $F_\fraka$ in the eigenbasis as $y_j(\tau) := \uv_j^* F_\fraka(\tau)$. Since $F$ is holomorphic at $\fraka$, the expansion from~\Cref{sec:cuspinfty} gives:
    \begin{equation*}
        y_j(\tau) = \sum_{n \geq 0} (\uv_j^* \cv_n)\, e^{2\pi i (n + \alpha_j) \tau/h},
    \end{equation*}
    converging absolutely and locally uniformly. Also expand $P$ in these coordinates: there are coefficients $\tilde{p}_\kappa$, indexed by multi-indices $\kappa = (\kappa_1, \ldots, \kappa_N)$ with $\kappa_1 + \ldots + \kappa_N = D$, such that $P(\wv) = \sum_\kappa \tilde{p}_\kappa \prod_j (\uv_j^* \wv)^{\kappa_j}$ for all $\wv \in \CC^N$.
    The covariance hypothesis~\eqref{eq:scalarisation}, applied to the element $g_\fraka T^h g_\fraka^{-1} \in \Gamma$, pins down which $\kappa$ can occur: since $A$ scales the coordinate $\uv_j^* \wv$ by $e^{2\pi i \alpha_j}$, the identity $P(A\wv) = e^{2\pi i \beta} P(\wv)$ reads
    \begin{equation*}
        \sum_\kappa \tilde{p}_\kappa\, e^{2\pi i \langle \kappa, \alpha \rangle} \prod_j (\uv_j^* \wv)^{\kappa_j} = \sum_\kappa \tilde{p}_\kappa\, e^{2\pi i \beta} \prod_j (\uv_j^* \wv)^{\kappa_j} \qquad \text{for all } \wv \in \CC^N,
    \end{equation*}
    where $\langle \kappa, \alpha \rangle := \sum_j \kappa_j \alpha_j$. Distinct monomials are linearly independent polynomials of $\wv$, so every $\kappa$ with $\tilde{p}_\kappa \neq 0$ satisfies $\langle \kappa, \alpha \rangle \equiv \beta \pmod{1}$.

    Now substitute the series for the $y_j$ into $f_\fraka(\tau) = \sum_\kappa \tilde{p}_\kappa \prod_j y_j(\tau)^{\kappa_j}$, and expand each product of (finitely many) absolutely convergent series into an absolutely convergent sum of products. Every resulting term is a constant times $e^{2\pi i t \tau/h}$, where $t$ is a sum of $D$ exponents $n_i + \alpha_{j_i}$ with each $n_i \geq 0$ and with the multiset $\{j_i\}$ prescribed by some $\kappa$ with $\tilde{p}_\kappa \neq 0$. Hence every occurring $t$ satisfies:
    \begin{equation*}
        t = \sum_i n_i + \langle \kappa, \alpha \rangle \ \in\ \beta + \ZZ \qquad\text{and}\qquad t \geq 0,
    \end{equation*}
    so $t = m + \beta$ for an integer $m$ with $m \geq -\beta > -1$, i.e., $m \geq 0$. Collecting terms (justified by absolute convergence), we conclude that $e^{-2\pi i \beta \tau/h} f_\fraka(\tau) = \sum_{m \geq 0} b_m\, e^{2\pi i m \tau/h}$ for some $b_m \in \CC$; the collected series converges absolutely at every point of $\calH$, so $\sum_m |b_m| r^m < \infty$ for every $r < 1$, and the convergence is therefore also locally uniform. By the uniqueness in~\Cref{fact:fourierexp}, these are precisely the cusp-expansion coefficients of $f$ at $\fraka$, and they vanish for all $m < 0$; so $f$ is holomorphic at $\fraka$.
    \qed

\subsection*{Proof of~\Cref{lem:restriction}}\label{sec:restrictionproof}

Weak modularity of $F$ for $\Gamma'$ with $\rho|_{\Gamma'}$ is immediate, since~\eqref{eq:transformationlaw} holds for all $g \in \Gamma \supseteq \Gamma'$. Fix a cusp $\fraka$ with lift $g_\fraka$ (taking $g_\infty := I$ as before), and let $h, h'$ be its widths with respect to $\Gamma, \Gamma'$ respectively. The sets $\{t \in \ZZ : g_\fraka T^t g_\fraka^{-1} \in \Gamma'\} \subseteq \{t \in \ZZ : g_\fraka T^t g_\fraka^{-1} \in \Gamma\}$ are subgroups of $\ZZ$ (closure follows from $g_\fraka T^s g_\fraka^{-1} \cdot g_\fraka T^t g_\fraka^{-1} = g_\fraka T^{s+t} g_\fraka^{-1}$), and they are nonzero because $g_\fraka T^q g_\fraka^{-1} \in \Gamma(q) \subseteq \Gamma'$ (by normality of $\Gamma(q)$ and because $\Gamma'$ has level $q$); so they equal $h'\ZZ \subseteq h\ZZ$, and $h \mid h'$. Write $e := h'/h \in \ZZ_{\geq 1}$.

    The matrix $A$ of $F$ at $\fraka$ with respect to $\Gamma'$ is the $e$th power of the one with respect to $\Gamma$: $A' := \rho(g_\fraka T^{h'} g_\fraka^{-1}) = \rho\left((g_\fraka T^{h} g_\fraka^{-1})^e\right) = A^e$. In particular, the eigenbasis $\vecv_j$ of $A$ (with dual basis $\uv_j$) also diagonalizes $A'$, with eigenvalues $e^{2\pi i e \alpha_j}$; the representatives in $[0, 1)$ are $\alpha'_j := e\alpha_j - \lfloor e\alpha_j \rfloor$, and $M' := \sum_j \alpha'_j \vecv_j \uv_j^*$.

    Let $\sum_n \cv_n e^{2\pi i n \tau/h}$ be the expansion of $\exp(-2\pi i M \tau/h) F_\fraka(\tau)$ from~\Cref{eq:qexpansion}. We compute the corresponding $h'$-periodic function directly:
    \begin{align*}
        \exp(-2\pi i M' \tau/h')\, F_\fraka(\tau) &= \sum_{j} e^{-2\pi i \alpha'_j \tau/h'}\, \vecv_j \uv_j^* F_\fraka(\tau) = \sum_{n, j} (\uv_j^* \cv_n)\, e^{2\pi i (e(n+\alpha_j) - \alpha'_j)\tau/h'}\, \vecv_j \\
        &= \sum_{n, j} (\uv_j^* \cv_n)\, e^{2\pi i m_{n,j} \tau/h'}\, \vecv_j, \qquad\text{where } m_{n,j} := en + \lfloor e \alpha_j \rfloor \in \ZZ.
    \end{align*}
    By the uniqueness in~\Cref{fact:fourierexp}, the $\Gamma'$-expansion coefficients of $F_\fraka$ are obtained by grouping this (absolutely convergent) sum by the value of $m_{n,j}$: $\cv'_m = \sum_{(n, j):\ m_{n,j} = m} (\uv_j^* \cv_n)\, \vecv_j$. Now suppose $F$ is holomorphic at $\fraka$ with respect to $\Gamma$, i.e., $\cv_n = 0$ for all $n < 0$. Then every nonzero term has $n \geq 0$, hence $m_{n,j} = en + \lfloor e\alpha_j \rfloor \geq 0$ (as $\lfloor e\alpha_j \rfloor \geq 0$), so $\cv'_m = 0$ for all $m < 0$: $F$ is holomorphic at $\fraka$ with respect to $\Gamma'$. Since $\fraka$ was arbitrary, $F$ is a modular form for $\Gamma'$.
    \qed

\section{Deferred Proofs from~\Cref{sec:gen1}}\label{sec:gen1proofs}

\subsection*{Bounding the Support Size}\label{sec:instantiationgen1}

In this subsection, we derive the two corollaries of~\Cref{thm:gen1} stated in~\Cref{sec:gen1}: first~\Cref{cor:gen1general}, for an arbitrary lattice, and then~\Cref{cor:gen1specific}, by instantiating the lattice of~\Cref{def:gen1concrete}. The restriction to primes $q \equiv 3 \pmod{4}$ in~\Cref{cor:gen1specific} enters only in the second step: \Cref{thm:gen1,cor:gen1general} hold for every odd prime $q$ and every even lattice of determinant $\sqrt{q}$, but the rank-$2$ instantiation below exists only in this residue class (\Cref{remark:qmod4}).

The first step is the Chernoff bound, in the form of a bound on the number of vectors $\zv \in \ZZ_q^d$ whose total weight $\sum_i w(z_i)$ is at most a given budget, which we will use again in~\Cref{sec:gen4exponent}.

\begin{lemma}[{Chernoff bound~\cite[Section~2.2]{BLM13}}]\label{lem:sublevel}
    Let $w: \ZZ_q \to \RR$ be a function, let $B \in \RR$, and let $d$ be a positive integer. Then for every $0 < t \leq 1$, we have:
    \begin{equation*}
        \left|\left\{\zv \in \ZZ_q^d : \sum_{i=1}^d w(z_i) \leq Bd\right\}\right| \leq \left(t^{-B} \sum_{j \in \ZZ_q} t^{w(j)}\right)^d.
    \end{equation*}
\end{lemma}
(This is the bound $\Pr\{Z \geq a\} \leq e^{-\lambda a}\, \mathbb{E}\, e^{\lambda Z}$ for $\lambda \geq 0$ of~\cite[Section~2.2]{BLM13}, applied to $Z = -\sum_i w(Z_i)$ for $(Z_1, \ldots, Z_d)$ uniform on $\ZZ_q^d$, with $a = -Bd$ and $t = e^{-\lambda}$, and multiplied by $q^d$.)

\begin{proof}[Proof of~\Cref{cor:gen1general}]
    Fix a generator $\av + \calL$ of $\calL^*/\calL$. \Cref{thm:gen1} provides a cone-delta function $f: \ZZ_q^d \to \QQ(\zeta_q)$ with $\hat{f}$ supported on $S_d$, so its Fourier sparsity is at most $|S_d|$, and it suffices to show that $|S_d| \leq q^{(\eta_\calL + o(1))d}$. By definition, $S_d$ is the union, over $u \in \ZZ_q \backslash \{0\}$, of the sets $\left\{\zv \in \ZZ_q^d : \sum_{i=1}^d w_u(z_i) \leq \frac{r(q+1)}{24} d\right\}$, where $w_u(x) := w_\calL(ux\av)$. For each such $u$, the map $x \mapsto ux\av + \calL$ is a bijection from $\ZZ_q$ onto $\calL^*/\calL$ (as $u$ is invertible modulo $q$, and $\av + \calL$ generates this group of order $q$), so $\sum_{x \in \ZZ_q} t^{w_u(x)} = M_\calL(t)$ for every $0 < t \leq 1$. Applying~\Cref{lem:sublevel} to each of these $q - 1$ sets, with $B = \frac{r(q+1)}{24}$, thus gives:
    \begin{equation*}
        |S_d| \leq (q-1) \left(t^{-r(q+1)/24} M_\calL(t)\right)^d \qquad \text{for every } 0 < t \leq 1.
    \end{equation*}
    Taking the infimum over $t$ gives $|S_d| \leq (q-1)\, q^{\eta_\calL d}$, by the definition~\eqref{eq:gen1generalexp} of $\eta_\calL$; and $q - 1 = q^{o(d)}$, as $q$ is fixed while $d$ grows.
\end{proof}

The second step uses the rank-$2$ lattice of~\Cref{def:gen1concrete}.

\begin{lemma}\label{lem:gen1weights}
    Fix a prime $q \equiv 3 \pmod{4}$, and define the lattice $\calL$ and generator $\av$ as in~\Cref{def:gen1concrete}. Then $w_{\calL}(u\av) = w_{\calL}(-u\av)$ for every $u \in \ZZ_q$, and for $u \in [0, (q-1)/2]$, we have:
    \[
    w_{\calL}(u\av) = \begin{cases}
        0,\text{ if }u = 0;\\
        u^2,\text{ if }1 \leq u \leq K;\\
        u^2-qu+qK,\text{ if }K+1 \leq u \leq (q-1)/2.
    \end{cases}
    \]
    Consequently, for $0 < t \leq 1$, we have $M_\calL(t) = 1 + 2\sum_{u = 1}^K t^{u^2} + 2\sum_{u = K+1}^{(q-1)/2} t^{u^2-qu+qK}$, and the exponent $\eta_\calL$ of~\eqref{eq:gen1generalexp} equals the exponent $\eta_q$ of~\eqref{eq:gen1exp}.
\end{lemma}
\begin{proof}
We first compute the prices; throughout, $K = (q+1)/4$, as in~\Cref{cor:gen1specific}.
    If $u = 0$, then $u\av + \calL$ contains 0 so the assertion is immediate. We also have $u\av+\calL = -(-u\av + \calL)$ so the symmetry claim is immediate. It remains to handle $u \in [1, (q-1)/2]$. 
    For such $u$, recall from~\Cref{def:prices} that:
    \begin{align*}
        w_{\calL}(u\av) &= \frac{q}{2} \min_{\xv \in u\av+\calL} \langle \xv, \xv \rangle \\
        &= \frac{q}{2} \min_{a, b \in \ZZ} \norm{u(0, \sqrt{2/q}) + a(\sqrt{2}, 0) + b(1/\sqrt{2}, \sqrt{q/2})}^2 \\
        &= \frac{q}{2} \min_{a, b \in \ZZ} \left((a\sqrt{2} + b/\sqrt{2})^2 + (u\sqrt{2/q} + b\sqrt{q/2})^2\right) \\
        &= \min_{a, b \in \ZZ} \left(q(a + b/2)^2 + (u + bq/2)^2\right).
    \end{align*} 
    Now we have two cases:
    \begin{itemize}
        \item If $b$ is even, then minimizing over $a$ will set $a+b/2 = 0$, so the value is $(u+bq/2)^2$. For $u \in [1, (q-1)/2]$, the choice of even $b$ that minimizes this is $b = 0$. Thus the minimum when we restrict to $b$ being even is $u^2$.
        \item If $b$ is odd, then minimizing over $a$ will set $a+b/2 = \pm 1/2$, so the value is $q/4 + (u+bq/2)^2$. For $u \in [1, (q-1)/2]$, the choice of odd $b$ that minimizes this is $b = -1$. Thus the minimum when we restrict to $b$ being odd is $q/4 + (u-q/2)^2 = u^2 - uq + (q+q^2)/4 = u^2-uq+qK$.
    \end{itemize}
    To minimize between these two branches, we simply note that $u^2 \leq u^2-qu+qK$ if and only if $u \leq K$.

It remains to deduce the consequences. Since $\av + \calL$ generates $\calL^*/\calL$, the map $x \mapsto x\av + \calL$ is a bijection from $\ZZ_q$ onto $\calL^*/\calL$, so $M_\calL(t) = \sum_{x \in \ZZ_q} t^{w_\calL(x\av)}$. In this sum, the term $x = 0$ contributes $t^0 = 1$, and the symmetry $w_\calL(x\av) = w_\calL(-x\av)$ pairs each $x \in [1, (q-1)/2]$ with $-x$; so the formula above gives the stated expression for $M_\calL(t)$. Finally, here $r = 2$, so the coefficient $\frac{r(q+1)}{24}$ in~\eqref{eq:gen1generalexp} equals $\frac{q+1}{12}$, and $\ln M_\calL(t) - \frac{q+1}{12} \ln t = \ln\left(t^{-(q+1)/12} M_\calL(t)\right)$ is the logarithm of the quantity whose infimum over $0 < t \leq 1$ defines $\eta_q$ in~\eqref{eq:gen1exp}. Hence $\eta_\calL = \eta_q$.
\end{proof}

\begin{lemma}\label{lem:gen1asymptotic}
    As $q \to \infty$ along primes $q \equiv 3 \pmod{4}$ (so that $\eta_q$ is defined), we have
    $$\eta_q \leq \frac{1}{2} + \frac{C_{\mathrm{warmup}}}{\ln q} + o\left(\frac{1}{\ln q}\right),$$
    where we define:
    $$C_{\mathrm{warmup}} := \inf_{\lambda > 0} \left(\lambda/12 + \ln(1+e^{-\lambda/4}) + \ln(\sqrt{\pi/\lambda})\right) \leq 0.3234.$$
\end{lemma}
\begin{proof}
In~\eqref{eq:gen1exp}, we will set $t = e^{-\lambda/q}$ for a constant $\lambda > 0$. The rough idea is to show using a Riemann sum argument that for fixed $\lambda$, the expression in~\eqref{eq:gen1exp} scales like $1/2 + C_\lambda/\ln q + o(1/\ln q)$ as $q \to \infty$. Then we will minimize over $\lambda$ at the end.
    
    Plugging in this setting of $t$ and replacing $u$ with $(q-1)/2 - u$ in the second sum, we have:
    \begin{align*}
        &t^{-(q+1)/12}\left(1 + 2\sum_{u = 1}^K t^{u^2} + 2\sum_{u = K+1}^{(q-1)/2} t^{u^2-qu+qK}\right) \\
        = &e^{\lambda(q+1)/(12q)} \left(1 + 2\sum_{u = 1}^K e^{-\lambda u^2/q} + 2\sum_{u = K+1}^{(q-1)/2} e^{-\lambda (u^2-qu+qK)/q}\right) \\
        = &e^{\lambda(q+1)/(12q)} \left(1 + 2\sum_{u = 1}^{(q+1)/4} e^{-\lambda u^2/q} + 2\sum_{u = 0}^{(q-7)/4} e^{-\lambda ((2u+1)^2+q)/(4q)}\right) \\
        = &e^{\lambda(q+1)/(12q)} \left(1 + 2\sum_{u = 1}^{(q+1)/4} e^{-\lambda u^2/q} + 2e^{-\lambda/4}\sum_{u = 0}^{(q-7)/4} e^{-\lambda (u+1/2)^2/q}\right).
    \end{align*}
    Now by~\Cref{lem:riemannsum} --- applied with $h = 1/\sqrt{q}$, the same $\lambda$, and $(N, \alpha) = \left(\frac{q+1}{4}, 0\right)$ for the first sum, respectively $(N, \alpha) = \left(\frac{q-3}{4}, \frac{1}{2}\right)$ for the second (reindexing $j = u+1$; here $q \geq 7$, as we may since $q \to \infty$) --- we have:
    \begin{align*}
        \sum_{u = 1}^{(q+1)/4} e^{-\lambda u^2/q} &= \frac{1}{2} \sqrt{\frac{q\pi}{\lambda}} + O(1) \\
        \sum_{u = 0}^{(q-7)/4} e^{-\lambda (u+1/2)^2/q} &= \frac{1}{2} \sqrt{\frac{q\pi}{\lambda}} + O(1).
    \end{align*}
    (Both $O(1)$'s hide constants depending only on $\lambda$: multiplying the lemma's error bound by $\sqrt{q}$ gives at most $1 + \frac{q\, e^{-\lambda N^2/q}}{2\lambda N}$, which is bounded as $q \to \infty$ with $\lambda$ fixed, since $N \geq (q-3)/4$.)
    Plugging these in, we may continue as follows:
    \begin{align*}
        &e^{\lambda(q+1)/(12q)} \left(1 + 2\sum_{u = 1}^{(q+1)/4} e^{-\lambda u^2/q} + 2e^{-\lambda/4}\sum_{u = 0}^{(q-7)/4} e^{-\lambda (u+1/2)^2/q}\right) \\
        = &e^{\lambda(q+1)/(12q)} \left(\sqrt{\frac{q\pi}{\lambda}} + e^{-\lambda/4} \sqrt{\frac{q\pi}{\lambda}} + O(1)\right) \\
        = &\exp\left(\frac{\lambda}{12} + \ln(1+e^{-\lambda/4})+\ln(\sqrt{\pi/\lambda})+\frac{1}{2}\ln q + o(1)\right),
    \end{align*}
    so the expression in~\eqref{eq:gen1exp} for this particular setting of $t$ is at most:
    $$\frac{1}{2} + \frac{1}{\ln q}\left(\lambda/12 + \ln(1+e^{-\lambda/4}) + \ln(\sqrt{\pi/\lambda})\right) + o\left(\frac{1}{\ln q}\right).$$
    Since this holds for every fixed $\lambda > 0$, and the infimum defining $C_{\mathrm{warmup}}$ is attained (the expression is continuous in $\lambda$ and tends to $+\infty$ as $\lambda \to 0^+$ and as $\lambda \to \infty$), we may run the argument at a minimizing $\lambda$, which gives the statement. Finally, the numerical bound $C_{\mathrm{warmup}} \leq 0.3234$ is witnessed by $\lambda = 8.67$: numerically, $\frac{8.67}{12} + \ln\left(1+e^{-8.67/4}\right) + \ln\sqrt{\pi/8.67} = 0.32330\ldots \leq 0.3234$.
\end{proof}

\begin{lemma}\label{lem:gen1numerics}
    We have $\eta_3 \leq 0.789691$, $\eta_7 \leq 0.709410$, and $\eta_{11} \leq 0.660807$.
\end{lemma}
\begin{proof}
    Each bound follows by evaluating the expression inside the infimum in~\eqref{eq:gen1exp} at one point $t$: at $t = 1/4$ for $q = 3$ (giving $0.7896901$), at $t = 0.351795$ for $q = 7$ (giving $0.7094098$), and at $t = 0.481707$ for $q = 11$ (giving $0.6608066$); rounding up in the sixth decimal gives the stated values.
\end{proof}

With these two lemmas, \Cref{cor:gen1specific} reduces to~\Cref{cor:gen1general}:

\begin{proof}[Proof of~\Cref{cor:gen1specific}]
    Instantiate~\Cref{cor:gen1general} with the rank-$2$ lattice $\calL$ of~\Cref{def:gen1concrete} (available exactly for the primes $q \equiv 3 \pmod{4}$ under consideration, \Cref{remark:qmod4}). This produces a cone-delta function $f: \ZZ_q^d \to \QQ(\zeta_q)$ with Fourier sparsity at most $q^{(\eta_\calL + o(1))d}$, and $\eta_\calL = \eta_q$ by~\Cref{lem:gen1weights}. Note the order of the limits: $q$ is held fixed while $d \to \infty$ (this is the $o(1)$ in the exponent); the final claim of the corollary concerns the subsequent limit $q \to \infty$ along primes $q \equiv 3 \pmod{4}$, and is~\Cref{lem:gen1asymptotic}, together with $\frac{C_{\mathrm{warmup}}}{\ln q} = \frac{C_{\mathrm{warmup}} / \ln 2}{\log_2 q}$ and $0.3234 / \ln 2 < 0.467$. The numerical bounds on $\eta_3$, $\eta_7$ and $\eta_{11}$ are~\Cref{lem:gen1numerics}.
\end{proof}

\subsection*{Proof of~\Cref{prop:evendistances}}\label{sec:evendistancesproof}

The first bullet is immediate by direct expansion: the difference between the two sides is $\langle \xv, \yv \rangle + \frac{1}{2} \langle \yv, \yv \rangle$.
    The first term is an integer by definition of $\calL^*$ and the second term is an integer since $\yv \in \calL$ and $\calL$ is even.

    Now for the second bullet, start by noting that we have $|\calL^* / \calL| = q$, so for any $\xv \in \calL^*$ we have $q\xv \in \calL$.
    Since $q\xv \in \calL$ and $\xv \in \calL^*$, their inner product must be an integer, hence $q \langle \xv, \xv \rangle \in \ZZ \Rightarrow \langle \xv, \xv \rangle \in \frac{1}{q} \ZZ$.
    Secondly, by the evenness of $\calL$ we have $\langle q\xv, q\xv \rangle \in 2\ZZ \Rightarrow \langle \xv, \xv \rangle \in \frac{2}{q^2} \ZZ$.
    Therefore we have $\langle \xv, \xv \rangle \in \frac{1}{q} \ZZ \cap \frac{2}{q^2}\ZZ = \frac{2}{q} \ZZ$, as claimed.
    \qed

\subsection*{Proof of~\Cref{prop:munonzero}}\label{sec:munonzeroproof}

Suppose instead that $q \mid \mu$, so that $\langle \av, \av \rangle = 2\mu/q$ is an integer. Every element of $\calL^*$ has the form $x\av + \vecv$ with $x \in \ZZ$ and $\vecv \in \calL$ (as $\av + \calL$ generates $\calL^*/\calL$), and expanding
    $$\langle x\av + \vecv,\ y\av + \wv \rangle = xy \langle \av, \av \rangle + x\langle \av, \wv \rangle + y\langle \av, \vecv \rangle + \langle \vecv, \wv \rangle$$
    shows that any two elements of $\calL^*$ would then have integer inner product: the first term is an integer by assumption, and the remaining three because $\av \in \calL^*$ and $\vecv \in \calL \subseteq \calL^*$ (\Cref{prop:evencontainment}) pair integrally with the lattice vectors $\wv, \vecv \in \calL$. In other words, we would have $\calL^* \subseteq (\calL^*)^*$. But writing $\calL = \BB\ZZ^r$, we have $(\calL^*)^* = (\BB^{-\top})^{-\top}\ZZ^r = \BB\ZZ^r = \calL$, so this would give $\calL^* \subseteq \calL$ --- contradicting $|\calL^*/\calL| = q > 1$.
    \qed

\subsection*{Proof of~\Cref{lem:thetaconverges}}\label{sec:thetaconvergesproof}

Fix a coordinate $x \in \ZZ_q$. Then it suffices to check that the sum $\sum_{\yv \in x\av+\calL} \exp(\pi i\langle \yv, \yv \rangle \tau)$ is locally uniformly absolutely convergent: each term therein is holomorphic, so the sum is then holomorphic by~\Cref{fact:weierstrass}. To do this, note that:
    $$|\exp(\pi i \langle \yv, \yv \rangle \tau)| = \exp(-\pi \langle \yv, \yv \rangle \Im \tau).$$ 
    Now consider any compact subset $\Omega$ of $\calH$; we want to show that $\sum_{\yv \in x\av + \calL} \exp(-\pi \langle \yv, \yv \rangle \Im \tau)$ converges uniformly on $\Omega$. On $\Omega$, any $\tau$ has imaginary part at least some $h_0 > 0$.
    Consequently, we just need to check that
    $$\sum_{\yv \in x\av + \calL} \exp(-\pi h_0 \langle \yv, \yv \rangle)$$
    converges. This is exactly~\Cref{lem:latticegaussian}, applied with $\vecv$ any representative of the coset $x\av + \calL$, $c = \pi h_0$, and $C = 0$.
    \qed

\subsection*{Proof of~\Cref{lem:gen1T}}\label{sec:gen1Tproof}

    Expanding the RHS, we have:
    \begin{align*}
        \rho_T \Theta_{\calL}(\tau) &= \sum_{x \in \ZZ_q} \left(\sum_{\yv \in x\av + \calL} \exp\left(\pi i \langle \yv, \yv \rangle \tau\right)\right) \rho_T \ev_x \\
        &= \sum_{x \in \ZZ_q} \left(\sum_{\yv \in x\av + \calL} \exp\left(\pi i \langle \yv, \yv \rangle \tau\right)\right) \exp\left(\frac{2\pi i \mu x^2}{q}\right) \ev_x.
    \end{align*}
    As we argued in~\Cref{def:phases}, every $\yv \in x\av + \calL$ satisfies $\frac{1}{2} \langle \yv, \yv \rangle \equiv \phi(x) \equiv \frac{\mu x^2}{q} \pmod{1}$, so $\exp\left(\frac{2\pi i \mu x^2}{q}\right) = \exp(\pi i \langle \yv, \yv \rangle)$ for every $\yv$ in the coset. Moving this factor inside the inner sum, we have:
    \begin{align*}
        \rho_T \Theta_{\calL}(\tau) &= \sum_{x \in \ZZ_q} \left(\sum_{\yv \in x\av+\calL} \exp(\pi i \langle \yv, \yv \rangle \tau) \exp(\pi i \langle \yv, \yv \rangle)\right) \ev_x \\
        &= \sum_{x \in \ZZ_q} \left(\sum_{\yv \in x\av+\calL} \exp(\pi i \langle \yv, \yv \rangle (\tau+1))\right) \ev_x \\
        &= \Theta_{\calL}(\tau+1),
    \end{align*}
    as claimed.
    \qed

\subsection*{Proof of~\Cref{lem:gen1S}}\label{sec:gen1Sproof}

    We have:
    \begin{align*}
        \Theta_{\calL}(-1/\tau) &= \sum_{x \in \ZZ_q} \sum_{\yv \in x\av+\calL} \exp(-\pi i \langle \yv, \yv \rangle/\tau) \ev_x,\text{ and}\\
        \tau^{r/2} \rho_S \Theta_{\calL}(\tau) &= \tau^{r/2} \sum_{x \in \ZZ_q} \sum_{\yv \in x\av+\calL} \exp(\pi i \langle \yv, \yv \rangle \tau) \rho_S \ev_x \\
        &= \frac{\epsilon_{\mu, q}\tau^{r/2}}{\sqrt{q}} \sum_{x \in \ZZ_q} \sum_{\yv \in x\av+\calL} \sum_{z \in \ZZ_q} \exp\left(\pi i \langle \yv, \yv \rangle \tau - \frac{4\pi i \mu xz}{q}\right) \ev_z.
    \end{align*}
    Consequently, our goal is to show for every $z \in \ZZ_q$ that:
    \begin{equation}\label{eq:psftarget}
        \sum_{\yv \in z\av+\calL} \exp\left(-\pi i \langle \yv, \yv \rangle/\tau\right) = \frac{\epsilon_{\mu, q} \tau^{r/2}}{\sqrt{q}} \sum_{x \in \ZZ_q} \sum_{\yv \in x\av+\calL} \exp\left(\pi i \langle \yv, \yv \rangle \tau - \frac{4\pi i \mu xz}{q}\right).
    \end{equation}
    We further unpack each side of~\eqref{eq:psftarget} as follows:
    \begin{align*}
        \sum_{\yv \in z\av+\calL} \exp(-\pi i \langle \yv, \yv \rangle/\tau) &= \sum_{\yv \in \calL} \exp(-\pi i \langle z\av+\yv, z\av+\yv \rangle/\tau),\text{ and} \\
        \sum_{x \in \ZZ_q} \sum_{\yv \in x\av+\calL} \exp\left(\pi i \langle \yv, \yv \rangle \tau - \frac{4\pi i \mu xz}{q}\right) &= \sum_{x \in \ZZ_q} \sum_{\yv \in \calL} \exp\left(\pi i \langle \yv+x\av, \yv+x\av \rangle \tau - \frac{4\pi i \mu xz}{q}\right) \\
        &= \sum_{x \in \ZZ_q} \sum_{\yv \in \calL} \exp\left(\pi i \langle \yv+x\av, \yv+x\av \rangle \tau - 2\pi i \langle x\av+\yv, z\av\rangle\right) \\
        &= \sum_{\yv \in \calL^*} \exp(\pi i \langle \yv, \yv \rangle \tau - 2\pi i \langle \yv, z\av \rangle).
    \end{align*}
    In the above, we have used: a change of variables to sum over $\calL$ instead of one of its cosets; the observation from~\Cref{def:phases} that $2\mu xz/q \equiv \langle x\av+\yv, z\av \rangle \pmod{1}$; and the observation that as $x, \yv$ range over $\ZZ_q$ and $\calL$ respectively, the vector $\yv+x\av$ ranges over $\calL^*$ (each element exactly once). All the series here converge absolutely (\Cref{lem:latticegaussian}), so these rearrangements are valid. In summary,~\eqref{eq:psftarget} is equivalent to:
    \begin{equation}
        \sum_{\yv \in \calL} \exp(-\pi i \langle z\av+\yv, z\av+\yv \rangle/\tau) = \frac{\epsilon_{\mu, q} \tau^{r/2}}{\sqrt{q}} \sum_{\yv \in \calL^*} \exp(\pi i \langle \yv, \yv \rangle \tau - 2\pi i \langle \yv, z\av \rangle).
    \end{equation}
    This in turn follows directly from the Poisson summation formula (\Cref{cor:gausspsf}), setting $\uv = z\av$. To explain the normalization factor on the RHS, note that:
    \[
    (-i\tau)^{r/2} \det(\calL^*) = \frac{(-i\tau)^{r/2}}{\sqrt{q}} = \frac{\epsilon_{\mu, q} \tau^{r/2}}{\sqrt{q}},
    \]
    where $\det(\calL^*) = 1/\det(\calL) = 1/\sqrt{q}$, the factor splits as $(-i\tau)^{r/2} = (-i)^{r/2} \tau^{r/2}$ (these are ordinary integer powers, as $r$ is even), and $\epsilon_{\mu, q} = (-i)^{r/2}$ by Milgram's formula (\Cref{fact:milgram}). This completes the proof.
    \qed

\subsection*{Proof of~\Cref{lem:thetacusp}}\label{sec:thetacuspproof}

First consider the cusp at $\infty$. Its width with respect to $\SL_2(\ZZ)$ is $1$ (we have $T \in \SL_2(\ZZ)$), and the associated matrix of~\Cref{sec:cuspinfty} is $A = \rho(T) = \rho_T$, which is diagonal with entries $\zeta_q^{\mu x^2} = e^{2\pi i \alpha_x}$ for $\alpha_x := (\mu x^2 \bmod q)/q \in [0, 1)$, indexed by $x \in \ZZ_q$. We may therefore apply the results of~\Cref{sec:cuspinfty} in the standard basis $\{\ev_x\}$, with $M$ the diagonal matrix with entries $\alpha_x$. The $x$-th coordinate of $\exp(-2\pi i M \tau)\, \Theta_{\calL}(\tau)$ is:
\begin{equation*}
    e^{-2\pi i \alpha_x \tau} \sum_{\yv \in x\av + \calL} e^{\pi i \langle \yv, \yv \rangle \tau} = \sum_{\yv \in x\av + \calL} e^{2\pi i \left(\frac{1}{2}\langle \yv, \yv \rangle - \alpha_x\right) \tau}.
\end{equation*}
By~\Cref{def:phases}, every $\yv \in x\av + \calL$ has $\frac{1}{2}\langle \yv, \yv \rangle \equiv \phi(x) \equiv \alpha_x \pmod{1}$, so every exponent $\frac{1}{2}\langle \yv, \yv \rangle - \alpha_x$ above is an integer; and since $\frac{1}{2}\langle \yv, \yv \rangle \geq 0$ and $\alpha_x < 1$, every exponent exceeds $-1$, hence is nonnegative. Collecting terms by the value of the exponent (justified by absolute convergence, exactly as in the proof of~\Cref{lem:scalarisation}) and invoking the uniqueness in~\Cref{fact:fourierexp}, we conclude that the coefficients $\cv_n$ of~\eqref{eq:qexpansion} vanish for all $n < 0$, i.e., $\Theta_{\calL}$ is holomorphic at $\infty$ (\Cref{def:holoinfty}).

Next, every cusp $\fraka \in \QQ$ is $\SL_2(\ZZ)$-equivalent to $\infty$: by~\Cref{def:cusps}, there is $g_\fraka \in \SL_2(\ZZ)$ with $g_\fraka \ast \infty = \fraka$. So holomorphy at every other cusp follows from~\Cref{lem:cuspwelldef}, and $\Theta_{\calL}$ is a modular form.

Finally, we identify the behavior at $\infty$. The exponent $\frac{1}{2}\langle \yv, \yv \rangle$ vanishes if and only if $\yv = \zerov$, which lies in the coset $x\av + \calL$ exactly when $x = 0$; so the unique term with $n + \alpha_x = \frac{1}{2}\langle \yv, \yv \rangle = 0$ is the $\yv = \zerov$ term, which occurs in the coordinate $x = 0$ with coefficient $1$; since all other exponents $n + \alpha_x$ are positive, this gives $\order_\infty(\Theta_{\calL}) = 0$. For the limit, let $2\delta := \min\{\langle \yv, \yv \rangle : \yv \in \calL^* \setminus \{\zerov\}\}$, which is attained and positive because every ball contains only finitely many points of $\calL^*$ (as in the proof of~\Cref{lem:latticegaussian}). For $\Im \tau \geq 1$, every coordinate of $\Theta_{\calL}(\tau) - \ev_0$ is bounded in absolute value by:
\begin{equation*}
    \sum_{\yv \in \calL^* \setminus \{\zerov\}} e^{-\pi \langle \yv, \yv \rangle \Im \tau} \leq e^{-2\pi \delta (\Im \tau - 1)} \sum_{\yv \in \calL^* \setminus \{\zerov\}} e^{-\pi \langle \yv, \yv \rangle},
\end{equation*}
using $e^{-\pi \langle \yv, \yv \rangle \Im \tau} = e^{-\pi \langle \yv, \yv \rangle (\Im \tau - 1)} e^{-\pi \langle \yv, \yv \rangle} \leq e^{-2\pi\delta(\Im \tau - 1)} e^{-\pi \langle \yv, \yv \rangle}$. The remaining sum is finite by~\Cref{lem:latticegaussian}, so the bound tends to $0$ as $\Im \tau \to \infty$, and it does not depend on $\Re \tau$. Hence $\Theta_{\calL}(\tau) \to \ev_0$, uniformly in $\Re \tau$.
\qed

\subsection*{Proof of~\Cref{lem:gen1scalarisation}}\label{sec:gen1scalarisationproof}

    Write $g = \begin{bmatrix} a & b \\ c & d_0 \end{bmatrix}$ and $\vecv = (v_0, \ldots, v_{q-1}) = \sum_{x \in \ZZ_q} v_x \ev_x$. Since $g \in \Gamma_0(q)$, \Cref{fact:weilrep} (specifically,~\eqref{eq:weilformula}) gives $\rho(g)\ev_x = \left(\frac{a}{q}\right) \zeta_q^{\mu bd_0x^2}\ev_{d_0x \bmod{q}}$. Hence, as functions of $\vecv$:
    \begin{align*}
        \rho(g)\vecv &= \left(\frac{a}{q}\right) \sum_{x \in \ZZ_q} v_x \zeta_q^{\mu bd_0x^2} \ev_{d_0x \bmod{q}} \\
        &= \left(\frac{a}{q}\right) \sum_{x \in \ZZ_q} v_{d_0^{-1}x \bmod{q}} \zeta_q^{\mu bd_0^{-1}x^2} \ev_x \\
        \Rightarrow P_\beta(\rho(g) \vecv) &\propto \sum_{u \in \ZZ_q^*} \sum_{\xv \in \calQ} \beta_\xv \prod_{i = 1}^d \left(v_{d_0^{-1} ux_i \bmod{q}} \zeta_q^{\mu b d_0^{-1} (ux_i)^2}\right) \\
        &= \sum_{u \in \ZZ_q^*} \sum_{\xv \in \calQ} \beta_\xv \underbrace{\zeta_q^{\mu b d_0^{-1} u^2 \langle \xv, \xv \rangle}}_{= 1\text{ since }\xv \in \calQ \Rightarrow \langle \xv, \xv \rangle \equiv 0 \pmod{q}} \prod_{i = 1}^d v_{d_0^{-1} ux_i \bmod{q}} \\
        &= \sum_{u \in \ZZ_q^*} \sum_{\xv \in \calQ} \beta_\xv \prod_{i = 1}^d v_{d_0^{-1} ux_i \bmod{q}} \\
        &= \sum_{u \in \ZZ_q^*} \sum_{\xv \in \calQ} \beta_\xv \prod_{i = 1}^d v_{ux_i \bmod{q}} \\
        &= P_\beta(\vecv),
    \end{align*}
    as desired. (The implication uses the degree-$d$ homogeneity of $P_\beta$ to pull out the sign $\left(\frac{a}{q}\right)^d$, and the penultimate step substitutes $u \mapsto d_0 u$, a bijection of $\ZZ_q^*$.)
    \qed

\section{Deferred Proofs from~\Cref{sec:gen4}}\label{sec:gen4proofs}

\subsection*{Calculating the Work Exponent}\label{sec:gen4exponent}

In this subsection, we bound the support sizes $|S_d(\vartheta, \nu)|$ that~\Cref{thm:gen4aprime} delivers, and so prove~\Cref{cor:gen4exponent}. We emphasize that this subsection bounds a \emph{support size} only; the conversion of such a bound into PIR server work and communication is the proof of~\Cref{mainthm:track3}, which combines~\Cref{thm:s-server-rate-one} with~\Cref{lem:bit-compiler}: a $(k, s)$-sparse quadratic-phase delta expansion on $\ZZ_q^d$ with $k \leq q^{(\eta + o(1))d}$ gives an $s$-server scheme with exponent $\eta$ on $n$-bit databases. Throughout this subsection, $q$ and $s$ are fixed and $d \to \infty$ along multiples of $q$; all $o(1)$ terms refer to this regime.

The support-size bound is again the moment-generating-function bound of~\Cref{lem:sublevel}, as in the proof of~\Cref{cor:gen1general}; the new ingredients are the two sets $S_d^{\mathrm{sq}}(\nu)$ and $S_d^{\mathrm{edge}}(\vartheta, \nu)$ of vectors of small total cost, their union, and the affine closure and coordinate thickening in the definitions of $T_d$ and $S_d$, which cost only a $q^{o(d)}$ factor.

\begin{lemma}\label{lem:wqthetaexplicit}
    Let $\vartheta \in [0, 1]$. For $j \in \ZZ_q$ identified with its representative in $[0, q-1]$:
    \begin{equation*}
        w_{q, \vartheta}(j) = \begin{cases}
            j(j-1) + 2\vartheta j, & \text{if } 0 \leq j \leq \frac{q-1}{2}; \\
            (q-j)(q-j+1) - 2\vartheta(q-j), & \text{if } \frac{q+1}{2} \leq j \leq q-1.
        \end{cases}
    \end{equation*}
    In particular, $w_{q, \vartheta} \geq 0$ and $w_{q, \vartheta}(0) = 0$, and the generating function $M_\vartheta$ of~\Cref{def:track3exponent} is given, for $t \in (0, 1]$, by:
    \begin{equation*}
        M_\vartheta(t) = 1 + \sum_{u = 1}^{(q-1)/2} \left(t^{u(u-1) + 2\vartheta u} + t^{u(u+1) - 2\vartheta u}\right); \qquad\text{in particular, } M_{1/2}(t) = 1 + 2\sum_{u=1}^{(q-1)/2} t^{u^2}.
    \end{equation*}
\end{lemma}
\begin{proof}
    The function $k \mapsto k(k-1) + 2\vartheta k = k^2 + (2\vartheta - 1)k$ is a parabola in $k$ with vertex at $k_0 = \frac{1 - 2\vartheta}{2} \in [-\frac{1}{2}, \frac{1}{2}]$, so among the representatives $\{j + aq : a \in \ZZ\}$ of $j$, its minimum is attained at the representative closest to $k_0$ (a tie gives equal values, by the symmetry of the parabola about $k_0$). For $0 \leq j \leq \frac{q-1}{2}$ that representative is $j$ itself: $|j - k_0| \leq \frac{q-1}{2} + \frac{1}{2} = \frac{q}{2}$, while every other representative is at distance at least $q - \frac{q}{2} = \frac{q}{2}$, with both equalities only in the tie case. For $\frac{q+1}{2} \leq j \leq q-1$ the closest representative is $j - q \in [-\frac{q-1}{2}, -1]$, and $(j-q)(j-q-1) + 2\vartheta(j-q) = (q-j)(q-j+1) - 2\vartheta(q-j)$. Nonnegativity: for any integer $k$, the factors $k$ and $k - 1 + 2\vartheta$ are both $\geq 0$ (if $k \geq 1$) or both $\leq 0$ (if $k \leq -1$, using $\vartheta \leq 1$), while for $k = 0$ the product vanishes. The formula for $M_\vartheta$ substitutes $u := j$ in the first branch and $u := q - j$ in the second, and the special case sets $\vartheta = \frac{1}{2}$ (both exponents become $u^2$).
\end{proof}

\begin{lemma}\label{lem:gen4supportsize}
    Fix an odd prime $q$, an integer $s \in [2, q]$, and a feasible pair $(\vartheta, \nu)$ (\Cref{def:track3exponent}). Then, as $d \to \infty$ along multiples of $q$, we have $|S_d(\vartheta, \nu)| \leq q^{(\eta_{q, s}(\vartheta, \nu) + o(1))\, d}$, with $\eta_{q, s}(\vartheta, \nu)$ as defined in~\eqref{eq:gen4exp}.
\end{lemma}
\begin{proof}
    By~\Cref{def:gen4aprimesupport}, every element of $S_d(\vartheta, \nu)$ has the form $u\zv + b\bfone + c\ev_i$ with $u \in \ZZ_q^*$, $b, c \in \ZZ_q$, $i \in [d]$, and $\zv \in S^{\mathrm{sq}}_d(\nu) \cup S^{\mathrm{edge}}_d(\vartheta, \nu)$; hence:
    \begin{equation*}
        |S_d(\vartheta, \nu)| \leq q(q-1) \cdot qd \cdot \left(|S^{\mathrm{sq}}_d(\nu)| + |S^{\mathrm{edge}}_d(\vartheta, \nu)|\right),
    \end{equation*}
    and the prefactor $q^2(q-1)d$ is $q^{o(d)}$. It remains to bound $|S^{\mathrm{sq}}_d(\nu)|$ and $|S^{\mathrm{edge}}_d(\vartheta, \nu)|$. Both sets have the form $\{\zv \in \ZZ_q^d : \sum_{i} w(z_i) \leq Bd\}$ for a nonnegative weight function $w$ on $\ZZ_q$ and a budget $B \geq 0$: take $(w, B) = (w_{q, 1/2}, \nu)$ for the square set and $(w, B) = (w_{q, \vartheta}, \gamma(\vartheta, \nu))$ for the edge set. (Here $\gamma(\vartheta, \nu) \geq 0$ under feasibility: both summands in its definition are products of nonnegative factors, since $\vartheta \leq \frac{q+1}{6s}$ and $\nu \leq \frac{q^2-1}{8s}$.) By~\Cref{lem:sublevel}, for any fixed $0 < t \leq 1$, each of them satisfies:
    \begin{equation*}
        \left|\left\{\zv \in \ZZ_q^d : \sum_{i=1}^d w(z_i) \leq Bd\right\}\right| \leq \left(t^{-B} \sum_{j \in \ZZ_q} t^{w(j)}\right)^d.
    \end{equation*}
    Taking the infimum over $t$ for each set, and bounding the sum of the two cardinalities by twice the larger, gives the claim.
\end{proof}

Optimizing this bound over the feasible parameters is a finite-dimensional problem, which we solve numerically:

\begin{lemma}\label{lem:gen4numerics}
    We have $\eta_{3, 2} \leq 0.820340$, $\eta_{3, 3} \leq 0.732672$, $\eta_{5, 4} \leq 0.702450$, $\eta_{5, 5} \leq 0.653434$, $\eta_{7, 6} \leq 0.652057$, $\eta_{7, 7} \leq 0.619701$, $\eta_{11, 10} \leq 0.607791$, and $\eta_{11, 11} \leq 0.590014$.
\end{lemma}
\begin{proof}
    Any feasible pair $(\vartheta, \nu)$, together with one evaluation point $t$ for each of the two infima in~\eqref{eq:gen4exp}, certifies an upper bound on $\eta_{q, s}$ by direct evaluation. The following witnesses certify the eight bounds, which are the exponents of~\Cref{mainthm:track3}; rounding the last column up in the sixth decimal gives the stated values.
    \begin{center}
    \begin{tabular}{ccccccc}
        $q$ & $s$ & $\vartheta$ & $\nu$ & $t^{\mathrm{sq}}$ & $t^{\mathrm{edge}}$ & $\eta_{q,s} \leq$ \\
        \hline
        $3$ & $2$ & $1/3$ & $0.358650$ & $0.279606$ & $0.286169$ & $0.8203393$ \\
        $3$ & $3$ & $0.169085$ & $0.291108$ & $0.205326$ & $0.201386$ & $0.7326712$ \\
        $5$ & $4$ & $1/4$ & $0$ & --- & $0.414520$ & $0.7024495$ \\
        $5$ & $5$ & $0.100050$ & $0.479668$ & $0.354308$ & $0.354830$ & $0.6534337$ \\
        $7$ & $6$ & $0.166638$ & $0.740704$ & $0.509311$ & $0.509527$ & $0.6520565$ \\
        $7$ & $7$ & $0.071420$ & $0.653042$ & $0.465067$ & $0.465248$ & $0.6197008$ \\
        $11$ & $10$ & $0.1$ & $1.08$ & $0.629376$ & $0.629376$ & $0.6077910$ \\
        $11$ & $11$ & $0.045454$ & $0.991735$ & $0.604009$ & $0.604010$ & $0.5900131$ \\
    \end{tabular}
    \end{center}
    (For $(q, s) = (5, 4)$ the optimum lies on the boundary of the feasible region: $\nu = 0$ makes $S^{\mathrm{sq}}_d(0) = \{\zerov\}$, so $E^{\mathrm{sq}}_q(0) = 0$ and no $t^{\mathrm{sq}}$ is needed, while $\vartheta = \frac{q+1}{6s} = \frac{1}{4}$ zeroes the second term of $\gamma$, leaving $\gamma = \frac{(q-1)\vartheta}{2} = \frac{1}{2}$.) In each row, the pair $(\vartheta, \nu)$ is feasible for~\Cref{def:gen4aprimesupport}, and evaluating the two expressions of~\eqref{eq:gen4exp} at the listed $t$'s and taking the maximum yields a value at most the stated bound; finer numerical optimization does not improve these bounds in the sixth decimal, so we believe them to be essentially optimal for this analysis. The row $(q, s) = (7, 7)$ is the ``seven-server'' exponent that Part~I also deploys for $s \in \{8, 9\}$ (running the seven-server construction with idle servers).
\end{proof}

\begin{lemma}\label{lem:gen4scaling}
    We have $\eta_{q, q} \leq \frac{1}{2} + \frac{1}{2} \frac{\ln(\pi e/6)}{\ln q} + o\left(\frac{1}{\ln q}\right)$ as $q \to \infty$.
\end{lemma}
\begin{proof}
    We will take $\vartheta = 1/6$ and $\nu = 0$ as our feasible values, so that:
    \begin{align*}
        \gamma &= \frac{q-1}{2} \cdot \frac{1}{6} + \left(1 - \frac{q}{q+1}\right) \left(\frac{q^2-1}{4q}\right) \\
        &= \frac{q-1}{12} + \frac{q-1}{4q} \\
        &= \frac{q^2+2q-3}{12q}.
    \end{align*}
    Now note that $E_q^{\mathrm{sq}}(\nu) = 0$, as can be seen by taking the limit as $t \to 0$. So we need only address $E_{q, q}^{\mathrm{edge}}(1/6, 0)$. To this end, write $t = e^{-\lambda/q}$ for $\lambda > 0$, then we have:
    \begin{align*}
        M_{1/6}(t) &= 1 + \sum_{u = 1}^{(q-1)/2} \exp\left(-\frac{\lambda}{q}(u^2-2u/3)\right) + \sum_{u = 1}^{(q-1)/2} \exp\left(-\frac{\lambda}{q}(u^2+2u/3)\right) \\
        &= 1 + e^{\lambda/(9q)} \sum_{u = 1}^{(q-1)/2} \exp\left(-\frac{\lambda}{q}(u-1/3)^2\right) + e^{\lambda/(9q)} \sum_{u = 1}^{(q-1)/2} \exp\left(-\frac{\lambda}{q}(u+1/3)^2\right).
    \end{align*}
    Now by~\Cref{lem:riemannsum}, applied with $h = 1/\sqrt{q}$, we have:
    \begin{align*}
        \sum_{u = 1}^{(q-1)/2} \exp\left(-\frac{\lambda}{q}(u-1/3)^2\right) &= \frac{1}{2} \sqrt{\frac{q\pi}{\lambda}} + O(1) \\
        \sum_{u = 1}^{(q-1)/2} \exp\left(-\frac{\lambda}{q}(u+1/3)^2\right) &= -e^{-\lambda/(9q)} + \sum_{u = 1}^{(q+1)/2} \exp\left(-\frac{\lambda}{q}(u-2/3)^2\right) \\
        &= \frac{1}{2} \sqrt{\frac{q\pi}{\lambda}} + O(1).
    \end{align*}
    Plugging these in, we get:
    \begin{align*}
        M_{1/6}(e^{-\lambda/q}) &= e^{\lambda/(9q)} \sqrt{\frac{q\pi}{\lambda}} + O(1) \\
        &= \sqrt{\frac{q\pi}{\lambda}}(1+o(1)).
    \end{align*}
    Thus we have:
    \begin{align*}
        E_{q, q}^{\mathrm{edge}}(\vartheta, \nu) &= \frac{1}{\ln q} \inf_{\lambda > 0} (\ln M_{1/6}(e^{-\lambda/q}) + \gamma \lambda/q) \\
        &= \frac{1}{\ln q} \inf_{\lambda > 0} \left(\ln \sqrt{\frac{q\pi}{\lambda}} + \frac{\lambda}{12}+o(1)\right) \\
        &\leq \frac{1}{\ln q} \left(\ln \sqrt{\frac{q\pi e}{6}} +o(1)\right)&&\text{(setting $\lambda = 6$)} \\
        &= \frac{1}{2} + \frac{1}{2} \frac{\ln(\pi e/6)}{\ln q} + o\left(\frac{1}{\ln q}\right),
    \end{align*}
    as desired.
\end{proof}

With these lemmas, the corollary reduces to the theorem:

\begin{proof}[Proof of~\Cref{cor:gen4exponent}]
    First fix a feasible pair $(\vartheta, \nu)$. For any $d$, \Cref{cor:roundd} (which rests on~\Cref{thm:gen4aprime}) provides a $(|S_D(\vartheta, \nu)|, s)$-sparse quadratic-phase delta expansion over $\QQ(\zeta_q)$ on $\ZZ_q^d$ with permutation-invariant support, where $D := q\lceil d/q \rceil$ satisfies $d \leq D < d + q$. By~\Cref{lem:gen4supportsize}, $|S_D(\vartheta, \nu)| \leq q^{(\eta_{q, s}(\vartheta, \nu) + o(1))D} = q^{(\eta_{q, s}(\vartheta, \nu) + o(1))d}$, which proves the first statement.

    For the second, fix $\epsilon > 0$; it suffices to show that for all sufficiently large $d$ there is a $(k, s)$-sparse quadratic-phase delta expansion on $\ZZ_q^d$ with $k \leq q^{(\eta_{q, s} + \epsilon)d}$. Pick feasible $(\vartheta, \nu)$ with $\eta_{q, s}(\vartheta, \nu) \leq \eta_{q, s} + \epsilon/3$. For any $d$, \Cref{cor:roundd} (which rests on~\Cref{thm:gen4aprime}) provides a $(|S_D(\vartheta, \nu)|, s)$-sparse quadratic-phase delta expansion over $\QQ(\zeta_q)$ on $\ZZ_q^d$ with permutation-invariant support, where $D := q\lceil d/q \rceil$ satisfies $d \leq D < d + q$. By~\Cref{lem:gen4supportsize}, $|S_D(\vartheta, \nu)| \leq q^{(\eta_{q, s} + 2\epsilon/3)D} \leq q^{(\eta_{q, s} + 2\epsilon/3)(d + q)}$ for all sufficiently large $d$, and the right-hand side is at most $q^{(\eta_{q, s} + \epsilon)d}$ once $(\eta_{q, s} + 2\epsilon/3)\, q \leq \epsilon d/3$. The particular numerical bounds are~\Cref{lem:gen4numerics}, and the asymptotic bound is given by~\Cref{lem:gen4scaling}, together with $\frac{\frac{1}{2}\ln(\pi e/6)}{\ln q} = \frac{\frac{1}{2}\log_2(\pi e/6)}{\log_2 q}$ and $\frac{1}{2}\log_2(\pi e/6) = 0.2546\ldots \leq 0.255$.
\end{proof}

\subsection*{Proof of~\Cref{cor:roundd}}\label{sec:rounddproof}

Let $D := q \lceil d/q \rceil \geq d$, so that $q \mid D$ and $D - d < q$. By~\Cref{thm:gen4aprime}, there exist coefficients $\{c_{\zv, a}\}$ supported on $\calW \times \calA$, where $\calW := S_D(\vartheta, \nu) \subseteq \ZZ_q^D$ and $\calA \subseteq \ZZ_q$ has size $s$, satisfying:
    \begin{equation*}
        \sum_{a \in \calA} \sum_{\zv \in \ZZ_q^D} c_{\zv, a}\, \zeta_q^{\langle \zv, \xv \rangle + a \langle \xv, \xv \rangle} = \one[\xv = \zerov] \qquad \text{for all } \xv \in \ZZ_q^D.
    \end{equation*}
    Restrict this identity to the padded vectors $\xv = (\xv', \zerov)$ with $\xv' \in \ZZ_q^d$: then $\langle \xv, \xv \rangle = \langle \xv', \xv' \rangle$, $\langle \zv, \xv \rangle = \langle \pi(\zv), \xv' \rangle$ for the projection $\pi: \ZZ_q^D \to \ZZ_q^d$ onto the first $d$ coordinates, and $\xv = \zerov$ if and only if $\xv' = \zerov$. Grouping the left-hand side by the value of $\zv' := \pi(\zv)$, the coefficients:
    \begin{equation*}
        c'_{\zv', a} := \sum_{\zv \in \ZZ_q^D:\ \pi(\zv) = \zv'} c_{\zv, a}
    \end{equation*}
    form a quadratic-phase delta expansion on $\ZZ_q^d$ (\Cref{def:genconedelta}). Their support satisfies $\supp\{c'_{\zv', a}\} \subseteq \pi(\calW) \times \calA$, and $|\pi(\calW)| \leq |\calW| = |S_D(\vartheta, \nu)|$: distinct frequencies may merge under the projection $\pi$, which only shrinks the support, never grows it. So this expansion is $(|S_D(\vartheta, \nu)|, s)$-sparse. Its support $\pi(\calW)$ is permutation-invariant: a permutation $\sigma$ of the first $d$ coordinates extends to a permutation $\tilde\sigma$ of the $D$ coordinates that fixes the last $D - d$, and $\sigma(\pi(\calW)) = \pi(\tilde\sigma(\calW)) = \pi(\calW)$, as $\calW$ is permutation-invariant (\Cref{thm:gen4aprime}).
    \qed

\subsection*{Proof of~\Cref{prop:symmetrization}}\label{sec:symmetrizationproof}

\eqref{eq:beta1} and~\eqref{eq:beta2} are apparent from definition. To see~\eqref{eq:beta3}, note that:
    \begin{align*}
        \beta_{\xv+\bfone} \alpha_0^{(J)} &= \alpha_{\xv+\bfone}^{(J)} \\
        &= \one\left[\sum_{i = 1}^d (x_i+1) = 0\right] \cdot \zeta_q^{-J(x_1+1)} \cdot \frac{1}{q} \sum_{b \in \ZZ_q} \zeta_q^{-Jb} \alpha_{\xv+(b+1)\bfone} \\
        &= \one\left[\sum_{i = 1}^d x_i = 0\right] \cdot \zeta_q^{-Jx_1} \cdot \frac{1}{q} \sum_{b \in \ZZ_q} \zeta_q^{-J(b+1)} \alpha_{\xv+(b+1)\bfone} \\
        &= \alpha_{\xv}^{(J)} \\
        &= \beta_{\xv} \alpha_0^{(J)}.
    \end{align*}
    In the above, we have used the fact that $d \equiv 0 \pmod{q}$, so that $\sum_i (x_i + 1) \equiv \sum_i x_i \pmod{q}$. Finally, we check~\eqref{eq:beta4}:
    \begin{align*}
        \alpha_0^{(J)} \sum_{\xv \in \ZZ_q^d} \beta_{\xv} \zeta_q^{a\langle \xv, \xv \rangle + \langle \xv, \zv \rangle} &= \sum_{\xv \in \ZZ_q^d} \alpha_{\xv}^{(J)} \zeta_q^{a\langle \xv, \xv \rangle + \langle \xv, \zv \rangle} \\
        &= \frac{1}{q} \sum_{\xv \in \ZZ_q^d} \sum_{b \in \ZZ_q} \one\left[\sum_{i = 1}^d x_i = 0\right] \cdot \zeta_q^{-Jx_1} \cdot \zeta_q^{-Jb} \alpha_{\xv+b\bfone} \zeta_q^{a\langle \xv, \xv \rangle + \langle \xv, \zv \rangle} \\
        &= \frac{1}{q} \sum_{\xv \in \ZZ_q^d}\one\left[\sum_{i = 1}^d x_i = 0\right] \cdot \left(\sum_{b \in \ZZ_q} \zeta_q^{-Jx_1} \alpha_{\xv} \zeta_q^{a\langle \xv-b\bfone, \xv-b\bfone \rangle + \langle \xv-b\bfone, \zv \rangle}\right) \\
        &= \frac{1}{q^2 } \sum_{\xv \in \ZZ_q^d} \sum_{c \in \ZZ_q} \zeta_q^{c\langle \xv, \bfone \rangle} \cdot \left(\sum_{b \in \ZZ_q} \zeta_q^{-Jx_1} \alpha_{\xv} \zeta_q^{a\langle \xv-b\bfone, \xv-b\bfone \rangle + \langle \xv-b\bfone, \zv \rangle}\right).
    \end{align*}
    (The substitution $\xv \mapsto \xv - b\bfone$ in the third line uses $q \mid d$ once more: the indicator $\one[\sum_i x_i = 0]$ is unchanged by it because $\sum_i b = bd \equiv 0$, and the factor $\zeta_q^{Jb}$ it produces cancels the $\zeta_q^{-Jb}$ of the average. The final line replaced the indicator by the character sum $\frac{1}{q}\sum_c \zeta_q^{c \langle \xv, \bfone \rangle}$.)
    To unpack this, note that:
    \begin{align*}
        a\langle \xv-b\bfone, \xv - b\bfone \rangle + \langle \xv - b\bfone, \zv \rangle + c\langle \xv, \bfone \rangle - Jx_1 &= a\langle \xv, \xv \rangle + \langle \xv, -2ab\bfone+\zv+c\bfone-J\ev_1 \rangle + ab^2\langle \bfone, \bfone \rangle - b \langle \bfone, \zv \rangle \\
        &= a\langle \xv, \xv \rangle + \langle \xv, \zv+(c-2ab) \bfone - J\ev_1 \rangle - b\langle \bfone, \zv \rangle,
    \end{align*}
    where the second equality uses $\langle \bfone, \bfone \rangle = d \equiv 0 \pmod{q}$ --- the third and final place where $q \mid d$ enters this step.
    Note that $\zv \in T_d(\vartheta, \nu) \Rightarrow \zv+(c-2ab)\bfone \in T_d(\vartheta, \nu) \Rightarrow \zv + (c-2ab)\bfone - J\ev_1 \in S_d(\vartheta, \nu)$. Thus we have:
    \begin{align*}
        \frac{1}{q^2} \sum_{\xv \in \ZZ_q^d} \sum_{c \in \ZZ_q} \zeta_q^{c\langle \xv, \bfone \rangle} \cdot \left(\sum_{b \in \ZZ_q} \zeta_q^{-Jx_1} \alpha_{\xv} \zeta_q^{a\langle \xv-b\bfone, \xv-b\bfone \rangle + \langle \xv-b\bfone, \zv \rangle}\right) &= \frac{1}{q^2} \sum_{b, c \in \ZZ_q} \zeta_q^{-b\langle \bfone, \zv\rangle} \underbrace{\left(\sum_{\xv \in \ZZ_q^d} \alpha_{\xv} \zeta_q^{a\langle \xv, \xv \rangle + \langle \xv, \zv+(c-2ab)\bfone - J\ev_1 \rangle}\right)}_{= 0\text{ by }\eqref{eq:alpha2}}.
    \end{align*}
    This is hence zero as claimed.
    \qed

\subsection*{Proof of~\Cref{prop:pproperties}}\label{sec:pppropertiesproof}

    Homogeneity is clear. For the second bullet, we have $P_\beta(\ev_x) = \beta_{x\bfone} = \beta_0 = 1$ by~\eqref{eq:beta1} and~\eqref{eq:beta3}. 
    For the invariance properties under $\sfX$ and $\sfZ$, recall the coordinate descriptions $(\sfX\yv)_j = y_{j+1}$ and $(\sfZ\yv)_j = \zeta_q^j y_j$ (indices modulo $q$). Then:
    \begin{align*}
        P_\beta(\sfX \yv) &= P_\beta(y_1, y_2, \ldots, y_{q-1}, y_0) \\
        &= \sum_{\xv \in \ZZ_q^d} \beta_{\xv} \prod_{i = 1}^d y_{x_i+1} \\
        &= \sum_{\xv \in \ZZ_q^d} \beta_{\xv-\bfone} \prod_{i = 1}^d y_{x_i} \\
        &= \sum_{\xv \in \ZZ_q^d} \beta_{\xv} \prod_{i = 1}^d y_{x_i} \text{ (by~\eqref{eq:beta3})} \\
        &= P_\beta(\yv),\text{ and} \\
        P_\beta(\sfZ \yv) &= P_\beta(y_0, \zeta_q y_1, \zeta_q^2 y_2, \ldots, \zeta_q^{q-1} y_{q-1}) \\
        &= \sum_{\xv \in \ZZ_q^d} \beta_{\xv} \prod_{i = 1}^d \zeta_q^{x_i} y_{x_i} \\
        &= \sum_{\xv \in \ZZ_q^d} \beta_{\xv} \zeta_q^{\langle \xv, \bfone \rangle} \prod_{i = 1}^d y_{x_i} \\
        &= \sum_{\xv \in \ZZ_q^d} \beta_{\xv} \prod_{i = 1}^d y_{x_i} \text{ (by~\eqref{eq:beta2}, $\zeta_q^{\langle \xv, \bfone \rangle} = 1$ on the support of $\beta$)} \\
        &= P_\beta(\yv),
    \end{align*}
    as desired. 
    For the final assertion, it suffices to treat $H \in \{\sfX, \sfZ\}$: a general $H$ is a product of such factors (negative powers are positive powers, as $\sfX^q = \sfZ^q = I$), which we may strip off one at a time, as $\propto_\zeta$ is transitive. For $H \in \{\sfX, \sfZ\}$, \Cref{lem:heisenbergweil} (with $(m, n) = (1, 0)$ or $(0, 1)$) gives integers $a, b, c$ such that $\rho(g) H = \zeta_q^c \sfX^a \sfZ^b \rho(g)$. This then tells us that:
    \begin{align*}
        P_\beta(\rho(g)H\yv) &= P_\beta(\zeta_q^c \sfX^a \sfZ^b \rho(g) \yv) \\
        &= \zeta_q^{cd} P_\beta(\sfX^a \sfZ^b \rho(g) \yv) \\
        &\propto_\zeta P_\beta(\rho(g)\yv).
    \end{align*}
    In the last step, we used the invariance $P_\beta(\sfX^a \sfZ^b \wv) = P_\beta(\wv)$ (for all $\wv \in \CC^q$) that follows from the third bullet, and that $\zeta_q^{cd}$ is a root of unity.
    \qed

\subsection*{Proof of~\Cref{lem:gen4T}}\label{sec:gen4Tproof}

We prove the exact identity $\Theta(\tau+1, w) = e^{\pi i q/4} \rho(T) \Theta(\tau, w)$; since $e^{\pi i q/4}$ is a root of unity, this gives the statement.

    Firstly, note that:
    \begin{align*}
        (k+(q-1)/2)^2(q+1) &= (k-1/2)^2 + k^2q + kq^2 + \frac{q^3-q^2-q}{4} \\
        &\equiv (k-1/2)^2 - \frac{q^2}{4} \pmod{2q} \\
        \Rightarrow \exp\left(\frac{\pi i(k+(q-1)/2)^2(q+1)}{q}\right) &= \exp\left(\frac{\pi i (k-1/2)^2}{q}\right) \exp\left(\frac{-\pi i q}{4}\right).
    \end{align*}
    Bearing this in mind, we have:
    \begin{align*}
        \rho(T) \Theta(\tau, w) &= \sum_{k \in \ZZ} (-1)^k \exp\left(\frac{\pi i \tau}{q} \left(k-\frac{1}{2}\right)^2 + 2\pi i \left(k-\frac{1}{2}\right) w\right) \rho(T) \ev_{k+(q-1)/2 \bmod{q}} \\
        &= \sum_{k \in \ZZ} (-1)^k \exp\left(\frac{\pi i \tau}{q} \left(k-\frac{1}{2}\right)^2 + 2\pi i \left(k-\frac{1}{2}\right) w\right) \exp\left(\frac{\pi i (k+(q-1)/2)^2(q+1)}{q}\right) \ev_{k+(q-1)/2 \bmod{q}} \\
        &= \exp\left(\frac{-\pi i q}{4}\right) \sum_{k \in \ZZ} (-1)^k \exp\left(\frac{\pi i \tau}{q} \left(k-\frac{1}{2}\right)^2 + 2\pi i \left(k-\frac{1}{2}\right) w\right) \exp\left(\frac{\pi i (k-1/2)^2}{q}\right) \ev_{k+(q-1)/2 \bmod{q}} \\
        &= \exp\left(\frac{-\pi i q}{4}\right) \sum_{k \in \ZZ} (-1)^k \exp\left(\frac{\pi i (\tau+1)}{q} \left(k-\frac{1}{2}\right)^2 + 2\pi i \left(k-\frac{1}{2}\right) w\right) \ev_{k+(q-1)/2 \bmod{q}} \\
        &= e^{-\pi i q/4} \Theta(\tau+1, w),
    \end{align*}
    which implies the conclusion.\qed

\subsection*{Proof of~\Cref{lem:gen4S}}\label{sec:gen4Sproof}

We prove the exact identity $\Theta(-1/\tau, w/\tau) = \epsilon_q^{-1} (-i)^q (-i\tau)^{1/2} e^{\pi i qw^2/\tau} \rho(S) \Theta(\tau, w)$; since $\epsilon_q \in \{\pm 1, \pm i\}$, the constant $\epsilon_q^{-1}(-i)^q$ is a root of unity, so this gives the statement. As in the proof of~\Cref{lem:gen1S}, the identity is a consequence of the Poisson summation formula (\Cref{cor:gausspsf}).
    We have:
    \begin{align*}
        \Theta(-1/\tau, w/\tau) &= \sum_{k \in \ZZ} (-1)^k \exp\left(\frac{-\pi i}{q\tau} \left(k-\frac{1}{2}\right)^2 + 2\pi i \left(k-\frac{1}{2}\right) \frac{w}{\tau}\right) \ev_{k+(q-1)/2 \bmod{q}}, \\
        &= \sum_{y \in \ZZ_q} \sum_{k \in y-(q-1)/2+q\ZZ} (-1)^k \exp\left(\frac{-\pi i}{q\tau} \left(k-\frac{1}{2}\right)^2 + 2\pi i \left(k-\frac{1}{2}\right) \frac{w}{\tau}\right) \ev_{y} \\
        &= \sum_{y \in \ZZ_q} \sum_{k \in \ZZ} (-1)^{y-(q-1)/2+qk} \exp\left(\frac{-\pi i}{q\tau} \left(y-\frac{q}{2}+qk\right)^2 + 2\pi i \left(y-\frac{q}{2}+qk\right) \frac{w}{\tau}\right) \ev_{y} \\
        &= \sum_{y \in \ZZ_q} \sum_{k \in \ZZ} (-1)^{y-(q-1)/2} \exp\left(\frac{-\pi i}{q\tau} \left(y-\frac{q}{2}+qk\right)^2 + 2\pi i \left(y-\frac{q}{2}+qk\right) \frac{w}{\tau} + \pi i k\right) \ev_{y}\text{ and}\\
        \rho(S) \Theta(\tau, w) &= \frac{\epsilon_q}{\sqrt{q}} \sum_{k \in \ZZ} (-1)^k \exp\left(\frac{\pi i \tau}{q} \left(k-\frac{1}{2}\right)^2 + 2\pi i \left(k-\frac{1}{2}\right) w\right) \sum_{y \in \ZZ_q} \exp\left(\frac{-2\pi i (k+(q-1)/2)y}{q}\right) \ev_{y}.
    \end{align*}
    Hence our goal is to show for every $y \in \ZZ_q$ that:
    \begin{align}
        &(-1)^{y - (q-1)/2} \sum_{k \in \ZZ} \exp\left(\frac{-\pi i}{q\tau} \left(y-\frac{q}{2}+qk\right)^2 + 2\pi i \left(y-\frac{q}{2}+qk\right) \frac{w}{\tau} - \pi i k\right) \label{eq:poissongen4lhs}\\
        = &\frac{(-i)^q (-i\tau)^{1/2} e^{\pi i qw^2/\tau}}{\sqrt{q}} \sum_{k \in \ZZ} (-1)^k \exp\left(\frac{\pi i \tau}{q} \left(k-\frac{1}{2}\right)^2 + 2\pi i \left(k-\frac{1}{2}\right) w\right) \exp\left(\frac{-2\pi i (k+(q-1)/2)y}{q}\right).\label{eq:poissongen4rhs}
    \end{align}
    To do this, note that:
    \begin{align*}
        &\frac{-\pi i}{q\tau} \left(y-\frac{q}{2}+qk\right)^2 + 2\pi i \left(y-\frac{q}{2}+qk\right) \frac{w}{\tau} - \pi i k \\
        = &\frac{-\pi i q}{\tau} \left(\left(k + \frac{2y-q}{2q}\right)^2 - \frac{w(2y-q)}{q}-2wk+\frac{k\tau}{q}\right) \\
        = &\frac{-\pi i q}{\tau} \left(k + \frac{2y-q-2qw+\tau}{2q}\right)^2 - \pi i w + \frac{\pi i qw^2}{\tau} + \frac{\pi i(2y-q)}{2q} + \frac{\pi i \tau}{4q}.
    \end{align*}
    This allows us to rewrite~\eqref{eq:poissongen4lhs} and apply~\Cref{cor:gausspsf} to it --- with $r = 1$, lattice $\calL = \ZZ$, modular variable $\tau/q \in \calH$, and the \emph{complex} shift $\uv = \frac{2y-q-2qw+\tau}{2q}$, which~\Cref{cor:gausspsf} allows --- using $(-i\tau/q)^{1/2} = (-i\tau)^{1/2}/\sqrt{q}$ (valid for principal branches, as $q > 0$):
    \begin{align*}
        &(-1)^{y - (q-1)/2} \exp\left(-\pi i w + \frac{\pi i qw^2}{\tau} + \frac{\pi i (2y-q)}{2q} + \frac{\pi i \tau}{4q} \right)\sum_{k \in \ZZ} \exp\left(\frac{-\pi i q}{\tau} \left(k + \frac{2y-q-2qw+\tau}{2q}\right)^2\right) \\
        = &(-1)^{y - (q-1)/2} \frac{(-i\tau)^{1/2} e^{\pi i qw^2/\tau}}{\sqrt{q}} \exp\left(-\pi i w + \frac{\pi i (2y-q)}{2q} + \frac{\pi i \tau}{4q} \right) \sum_{k \in \ZZ} \exp\left(\frac{\pi i \tau k^2}{q} - 2\pi i k\left(\frac{2y-q-2qw+\tau}{2q}\right)\right)\\
        = &(-1)^{y - (q-1)/2} \frac{(-i\tau)^{1/2} e^{\pi i qw^2/\tau}}{\sqrt{q}} \exp\left(-\pi i w + \frac{\pi i (2y-q)}{2q} + \frac{\pi i \tau}{4q} \right) \sum_{k \in \ZZ} \exp\left(\frac{\pi i \tau k^2}{q} - 2\pi i k\left(\frac{y}{q} - \frac{1}{2} - w + \frac{\tau}{2q}\right)\right) \\
        = &-i(-1)^{y - (q-1)/2} \frac{(-i\tau)^{1/2} e^{\pi i qw^2/\tau}}{\sqrt{q}} \exp\left(-\pi i w + \frac{\pi i y}{q} + \frac{\pi i \tau}{4q} \right) \sum_{k \in \ZZ} \exp\left(\frac{\pi i \tau k^2}{q} - 2\pi i k\left(\frac{y}{q} - \frac{1}{2} - w + \frac{\tau}{2q}\right)\right).\numberthis \label{eq:22intermediate}
    \end{align*}
    On the other hand, working from~\eqref{eq:poissongen4rhs}, we have:
    \begin{align*}
        &(-1)^k \exp\left(\frac{\pi i \tau}{q} \left(k-\frac{1}{2}\right)^2 + 2\pi i \left(k-\frac{1}{2}\right) w - \frac{2\pi i (k+(q-1)/2)y}{q}\right) \\
        = &\exp\left(\frac{\pi i \tau}{q} \left(k-\frac{1}{2}\right)^2 + 2\pi i \left(k-\frac{1}{2}\right) w - \frac{2\pi i (k+(q-1)/2)y}{q} + \pi i k\right) \\
        = &\exp\left(\frac{\pi i \tau k^2}{q} - 2\pi i k \left(\frac{y}{q} - \frac{1}{2} - w +\frac{\tau}{2q}\right) + \frac{\pi i \tau}{4q} - \pi i w - \frac{\pi i y(q-1)}{q}\right),
    \end{align*}
    so~\eqref{eq:poissongen4rhs} simplifies as follows:
    \begin{align*}
        &\frac{(-i)^q (-i\tau)^{1/2} e^{\pi i qw^2/\tau}}{\sqrt{q}} \sum_{k \in \ZZ} \exp\left(\frac{\pi i \tau k^2}{q} - 2\pi i k \left(\frac{y}{q} - \frac{1}{2} - w +\frac{\tau}{2q}\right) + \frac{\pi i \tau}{4q} - \pi i w - \frac{\pi i y(q-1)}{q}\right) \\
        = &\frac{(-i)^q (-i\tau)^{1/2} e^{\pi i qw^2/\tau}}{\sqrt{q}} \exp\left(\frac{\pi i \tau}{4q} - \pi i w + \frac{\pi i y}{q} - \pi i y\right) \sum_{k \in \ZZ} \exp\left(\frac{\pi i \tau k^2}{q} - 2\pi i k \left(\frac{y}{q} - \frac{1}{2} - w +\frac{\tau}{2q}\right)\right) \\
        = &\frac{(-i)^q (-i\tau)^{1/2} e^{\pi i qw^2/\tau} (-1)^y}{\sqrt{q}} \exp\left(\frac{\pi i \tau}{4q} - \pi i w + \frac{\pi i y}{q}\right) \sum_{k \in \ZZ} \exp\left(\frac{\pi i \tau k^2}{q} - 2\pi i k \left(\frac{y}{q} - \frac{1}{2} - w +\frac{\tau}{2q}\right)\right).
    \end{align*}
    To show that this is equal to~\eqref{eq:22intermediate} (and hence~\eqref{eq:poissongen4lhs}), it suffices to show that:
    \begin{align*}
        -i(-1)^{y-(q-1)/2} \frac{(-i\tau)^{1/2} e^{\pi i qw^2/\tau}}{\sqrt{q}} &= \frac{(-i)^q (-i\tau)^{1/2} e^{\pi i qw^2/\tau} (-1)^y}{\sqrt{q}} \\
        \Leftrightarrow -i(-1)^{-(q-1)/2} &= (-i)^q,
    \end{align*}
    which is evident as $(-1)^{-(q-1)/2} = (-1)^{(q-1)/2} = (-i)^{q-1}$.
    \qed

\subsection*{Proof of~\Cref{lem:thetatranslation}}\label{sec:thetatranslationproof}

All proportionality constants below are roots of unity independent of $(\tau, w)$. First, we unpack the RHS of~\eqref{eq:heisenbergtarget}, using $\sfZ^b \sfX^a \ev_j = \zeta_q^{b(j-a)} \ev_{j-a}$:
    \begin{align*}
        &e^{-\pi i a^2 \tau/q- 2\pi i aw} \cdot \sfZ^b \sfX^{a} \Theta(\tau, w) \\
        = &e^{-\pi i a^2 \tau/q- 2\pi i aw} \sum_{k \in \ZZ} (-1)^{k} \exp\left(\frac{\pi i \tau}{q} \left(k-\frac{1}{2}\right)^2 + 2\pi i \left(k-\frac{1}{2}\right) w + \frac{2\pi i b(k+(q-1)/2-a)}{q}\right) \ev_{k+(q-1)/2-a \bmod{q}} \\
        \propto_\zeta &e^{-\pi i a^2 \tau/q- 2\pi i aw} \sum_{k \in \ZZ} (-1)^k\exp\left(\frac{\pi i \tau}{q} \left(k-\frac{1}{2}\right)^2 + 2\pi i \left(k-\frac{1}{2}\right) w + \frac{2\pi i bk}{q}\right) \ev_{k+(q-1)/2-a \bmod{q}} \\
        \propto_\zeta &e^{-\pi i a^2 \tau/q- 2\pi i aw} \sum_{k \in \ZZ} (-1)^k \exp\left(\frac{\pi i \tau}{q} \left(k+a-\frac{1}{2}\right)^2 + 2\pi i \left(k+a-\frac{1}{2}\right) w + \frac{2\pi i bk}{q}\right) \ev_{k+(q-1)/2 \bmod{q}}.\numberthis \label{eq:heisenbergrhs}
    \end{align*}
    In the second step we dropped the constant $\zeta_q^{b((q-1)/2-a)}$; in the third we substituted $k \mapsto k+a$ and dropped the resulting constant $(-1)^a \zeta_q^{ab}$.
    Now let us unpack the LHS of~\eqref{eq:heisenbergtarget}:
    \begin{align*}
        \Theta\left(\tau, w + \frac{a\tau+b}{q}\right) = &\sum_{k \in \ZZ} (-1)^k \exp\left(\frac{\pi i \tau}{q} \left(k-\frac{1}{2}\right)^2 + 2\pi i \left(k-\frac{1}{2}\right) \left(w + \frac{a\tau+b}{q}\right)\right) \ev_{k+(q-1)/2 \bmod{q}} \\
        = &\exp\left(-\frac{\pi i \tau a^2}{q}\right) \sum_{k \in \ZZ} (-1)^k \exp\left(\frac{\pi i \tau}{q} \left(k+a-\frac{1}{2}\right)^2 + 2\pi i \left(k-\frac{1}{2}\right) \left(w + \frac{b}{q}\right)\right) \ev_{k+(q-1)/2 \bmod{q}} \\
        = &\exp\left(-\frac{\pi i \tau a^2}{q} - \frac{\pi i b}{q}\right) \cdot \\
        &\sum_{k \in \ZZ} (-1)^k \exp\left(\frac{\pi i \tau}{q} \left(k+a-\frac{1}{2}\right)^2 + 2\pi i \left(k-\frac{1}{2}\right) w + \frac{2\pi i kb}{q}\right) \ev_{k+(q-1)/2 \bmod{q}} \\
        = &\exp\left(-\frac{\pi i \tau a^2}{q} - \frac{\pi i b}{q} - 2\pi i aw\right) \cdot \\
        &\sum_{k \in \ZZ} (-1)^k \exp\left(\frac{\pi i \tau}{q} \left(k+a-\frac{1}{2}\right)^2 + 2\pi i \left(k+a-\frac{1}{2}\right) w + \frac{2\pi i kb}{q}\right) \ev_{k+(q-1)/2 \bmod{q}},
    \end{align*}
    which matches~\eqref{eq:heisenbergrhs} up to the constant $e^{-\pi i b/q}$, a root of unity, as desired.
    \qed

\subsection*{Proof of~\Cref{lem:descended}}\label{sec:descendedproof}

All constants below are independent of $(\tau, w)$, and we work with $\propto_\zeta$ throughout. We repeatedly use the following observation: if two nowhere-vanishing holomorphic functions on the connected set $\calH$ have the same square, then their ratio is continuous and squares to $1$, so it is a constant $\pm 1$ --- a root of unity. First we address the generators $T, S$:
    \begin{itemize}
        \item When $g = T$, the claim is~\Cref{lem:gen4T} (here $c = 0$ and $G(T, \tau) = 1$).
        \item When $g = S$, we have $c = 1$ and $G(S, \tau) = \tau$, and~\Cref{lem:gen4S} gives the claim with the square root $(-i\tau)^{1/2}$ in place of a branch $\tau^{1/2}$ of $G(S, \tau)^{1/2}$. These differ by a root of unity: the ratio $(-i\tau)^{1/2}/\tau^{1/2}$ is continuous on $\calH$ and squares to the constant $-i$, so it is a constant square root of $-i$, i.e., an eighth root of unity.
    \end{itemize}
    Since $S$ and $T$ generate $\SL_2(\ZZ)$ (\Cref{fact:generation}), it remains to show that if the claim is true for $g_1, g_2 \in \SL_2(\ZZ)$, then it is also true for $g_1^{-1}$ and $g_1g_2$. First we address $g_1^{-1}$: the claim for $g_1^{-1}$ at the point $(\tau', w') := g_1 \ast (\tau, w)$ --- which ranges over all of $\calH \times \CC$ as $(\tau, w)$ does --- follows by rearranging~\eqref{eq:descendedtransformation} (the inverse of a root of unity is again one):
    \begin{align*}
        \Theta(\tau, w) &\propto_\zeta G(g_1, \tau)^{-1/2} \exp\left(-\frac{\pi i qcw^2}{G(g_1, \tau)}\right) \rho(g_1^{-1}) \Theta(g_1 \ast (\tau, w)).
    \end{align*}
    To address the $G(g_1, \tau)^{-1/2}$ term, note by~\Cref{prop:chainrule} that:
    \begin{equation*}
        1 = G(I, \tau) = G(g_1^{-1}, g_1 \ast \tau) G(g_1, \tau) \Rightarrow G(g_1, \tau)^{-1/2} = \pm\, G(g_1^{-1}, g_1 \ast \tau)^{1/2},
    \end{equation*}
    with a constant sign by the observation above.
    To address the exponential term, note that:
    \begin{align*}
        \frac{(w/G(g_1, \tau))^2}{G(g_1^{-1}, g_1 \ast \tau)} &= \frac{w^2}{G(g_1, \tau)^2 G(g_1^{-1}, g_1 \ast \tau)} \\
        &= \frac{w^2}{G(g_1, \tau)} \text{ (\Cref{prop:chainrule})},
    \end{align*}
    and also that $\begin{bmatrix} a & b \\ c & d_0 \end{bmatrix}^{-1} = \begin{bmatrix} d_0 & -b \\ -c & a \end{bmatrix}$, so this accounts for the negative sign.
    Finally, we address $g_1g_2$:
    \begin{align*}
        \Theta(g_1g_2 \ast (\tau, w)) &\propto_\zeta G(g_1, g_2 \ast \tau)^{1/2} \exp\left(\frac{\pi i q c(g_1) (w/G(g_2, \tau))^2}{G(g_1, g_2 \ast \tau)}\right) \rho(g_1) \Theta(g_2 \ast (\tau, w)) \\
        &\propto_\zeta G(g_1, g_2 \ast \tau)^{1/2} G(g_2, \tau)^{1/2} \exp\left(\frac{\pi i q c(g_1) (w/G(g_2, \tau))^2}{G(g_1, g_2 \ast \tau)}\right) \times \\
        &\qquad\quad \exp\left(\frac{\pi i q c(g_2) w^2}{G(g_2, \tau)}\right) \rho(g_1g_2) \Theta(\tau, w).
    \end{align*}
    By~\Cref{prop:chainrule}, the two square-root factors satisfy $G(g_1, g_2 \ast \tau)^{1/2}\, G(g_2, \tau)^{1/2} = \pm\, G(g_1g_2, \tau)^{1/2}$, again with a constant sign. The remaining task is to handle the exponential terms. Pulling out the factors of $\pi i qw^2$, we want to show that:
    \begin{align*}
        \frac{c(g_1)}{G(g_2, \tau)^2 G(g_1, g_2 \ast \tau)} + \frac{c(g_2)}{G(g_2, \tau)} &= \frac{c(g_1g_2)}{G(g_1g_2, \tau)} \\
        \Leftrightarrow c(g_1) + c(g_2) G(g_1g_2, \tau) &= c(g_1g_2) G(g_2, \tau).
    \end{align*}
    To open this up, write:
    \begin{equation*}
        g_1 = \begin{bmatrix} a_1 & b_1 \\ c_1 & d_1 \end{bmatrix} \qquad g_2 = \begin{bmatrix} a_2 & b_2 \\ c_2 & d_2 \end{bmatrix} \qquad g_1g_2 = \begin{bmatrix} \cdot & \cdot \\ c_1a_2 + d_1c_2 & c_1b_2 + d_1d_2 \end{bmatrix}.
    \end{equation*}
    Then we have:
    \begin{align*}
        c(g_1) + c(g_2) G(g_1g_2, \tau) &= c_1 + c_2 (c_1a_2 \tau + d_1c_2 \tau + c_1b_2 + d_1d_2) \\
        &= c_2\tau (d_1c_2 + c_1a_2) + (c_1 + c_1c_2b_2 + d_1d_2c_2) \\
        &= c_2\tau (d_1c_2 + c_1a_2) + (c_1(a_2d_2-b_2c_2) + c_1c_2b_2 + d_1d_2c_2) \\
        &= c_2 \tau(d_1c_2 + c_1a_2) + (c_1a_2d_2 + d_1d_2c_2) \\
        &= (c_1a_2 + d_1c_2)(c_2\tau+d_2) \\
        &= c(g_1g_2) G(g_2, \tau),
    \end{align*}
    as desired.
    \qed

\subsection*{Proof of~\Cref{lem:bigF}}\label{sec:bigFproof}

Writing $w = z/q$, so that $F(g \ast (\tau, z)) = P_\beta(\Theta(g \ast (\tau, w)))$ and $F_g(\tau, z) = P_\beta(\rho(g)\, \Theta(\tau, w))$, we apply~\Cref{lem:descended} and the degree-$d$ homogeneity of $P_\beta$ (a $d$-th power of a root of unity is again one):
    \begin{align*}
        P_\beta(\Theta(g \ast (\tau, w))) &\propto_\zeta P_\beta\left(G(g, \tau)^{1/2} \exp\left(\frac{\pi i q c w^2}{G(g, \tau)}\right) \rho(g) \Theta(\tau, w)\right) \\
        &= G(g, \tau)^{d/2} \exp\left(\frac{\pi i q d c w^2}{G(g, \tau)}\right) P_\beta(\rho(g) \Theta(\tau, w)).
    \end{align*}
    Since $\pi i q d c w^2/G(g, \tau) = \pi i d c z^2/(q G(g, \tau))$, this is the claim.
    \qed

\subsection*{Proof of~\Cref{lem:fgtranslation}}\label{sec:fgtranslationproof}

We have, as functions of $(\tau, z)$:
    \begin{align*}
        F_g(\tau, z+a\tau+b) &= P_\beta\left(\rho(g) \Theta\left(\tau, \frac{z+a\tau+b}{q}\right)\right) \\
        &\propto_\zeta P_\beta\left(\rho(g)\, e^{-\pi i a^2 \tau/q- 2\pi i aw} \cdot \sfZ^b \sfX^{a} \Theta\left(\tau, \frac{z}{q}\right)\right) \quad\text{(\Cref{lem:thetatranslation}, writing $w = z/q$)} \\
        &= \exp\left(-\frac{\pi i a^2 d \tau}{q} - \frac{2\pi i da z}{q}\right) P_\beta\left(\rho(g) \sfZ^b \sfX^a \Theta\left(\tau, \frac{z}{q}\right)\right) \\
        &\propto_\zeta \exp\left(-\frac{\pi i a^2 d \tau}{q} - \frac{2\pi i da z}{q}\right) P_\beta\left(\rho(g) \Theta\left(\tau, \frac{z}{q}\right)\right) \quad{\text{(\Cref{prop:pproperties})}} \\
        &= \exp\left(-\frac{\pi i a^2 d \tau}{q} - \frac{2\pi i da z}{q}\right) F_g(\tau, z),
    \end{align*}
    as desired. (The second and third steps use the degree-$d$ homogeneity of $P_\beta$; a $d$-th power of a root of unity is again one.)
    \qed

\subsection*{Proof of~\Cref{prop:fnotzero}}\label{sec:fnotzeroproof}

Since $F(\tau, z) = P_\beta(\Theta(\tau, z/q))$ and $z \mapsto z/q$ is a bijection of $\CC$, it suffices to find a point where $P_\beta(\Theta(\tau, w)) \neq 0$. Set $(\tau, w) = (iy, iy/(2q))$ for a parameter $y > 0$. The exponent of the $k$-th term of $\Theta$ then becomes:
    \begin{equation*}
        \frac{\pi i \tau}{q}\left(k - \frac{1}{2}\right)^2 + 2\pi i \left(k-\frac{1}{2}\right) w = -\frac{\pi y}{q}\left(\left(k-\frac{1}{2}\right)^2 + \left(k - \frac{1}{2}\right)\right) = -\frac{\pi y}{q}\left(k^2 - \frac{1}{4}\right),
    \end{equation*}
    so that, normalizing by the $k = 0$ term:
    \begin{equation*}
        \vecv_y := e^{-\pi y/(4q)}\, \Theta\left(iy, \frac{iy}{2q}\right) = \sum_{k \in \ZZ} (-1)^k e^{-\pi y k^2/q}\, \ev_{k + (q-1)/2 \bmod q} = \ev_{(q-1)/2} + E_y,
    \end{equation*}
    where the error term $E_y$ collects the terms with $k \neq 0$. Each coordinate of $E_y$ is bounded in absolute value by:
    \begin{equation*}
        \sum_{k \neq 0} e^{-\pi y k^2/q} \leq 2\sum_{k = 1}^\infty e^{-\pi y k/q} = \frac{2e^{-\pi y/q}}{1 - e^{-\pi y /q}} \xrightarrow{\ y \to \infty\ } 0,
    \end{equation*}
    so $\vecv_y \to \ev_{(q-1)/2}$ as $y \to \infty$. Since $P_\beta$ is a polynomial, hence continuous, $P_\beta(\vecv_y) \to P_\beta(\ev_{(q-1)/2}) = 1$ (\Cref{prop:pproperties}). So $P_\beta(\vecv_y) \neq 0$ for all large enough $y$, and then by the degree-$d$ homogeneity of $P_\beta$ (\Cref{prop:pproperties} again):
    \begin{equation*}
        P_\beta\left(\Theta\left(iy, \frac{iy}{2q}\right)\right) = e^{\pi d y/(4q)}\, P_\beta(\vecv_y) \neq 0.
    \end{equation*}
    \qed

\subsection*{Proof of~\Cref{lem:ellexpansion}}\label{sec:ellexpansionproof}

Since $P_\beta(\vecv) = \sum_{\xv \in \ZZ_q^d} \beta_\xv \prod_{i=1}^d v_{x_i}$ and $\rho(g)$ is linear, the composition $P_\beta \circ \rho(g)$ is again a homogeneous polynomial of degree $d$: expanding $(\rho(g)\vecv)_{x_i} = \sum_{r \in \ZZ_q} \rho(g)_{x_i, r}\, v_r$ and multiplying out,
    \begin{equation*}
        P_\beta(\rho(g) \vecv) = \sum_{\rv \in \ZZ_q^d} \beta^{(g)}_{\rv} \prod_{i = 1}^d v_{r_i}, \qquad \beta^{(g)}_{\rv} := \sum_{\xv \in \ZZ_q^d} \beta_\xv \prod_{i = 1}^d \rho(g)_{x_i, r_i}.
    \end{equation*}
    Now substitute $\vecv = \Theta(\tau, w)$. By definition, the $r$-th coordinate of $\Theta(\tau, w)$ is:
    \begin{equation*}
        \Theta_r(\tau, w) = \sum_{\substack{k \in \ZZ \\ k + (q-1)/2 \equiv r \pmod q}} (-1)^k \exp\left(\frac{\pi i \tau (2k-1)^2}{4q} + \pi i (2k-1) w\right),
    \end{equation*}
    using $\left(k - \frac{1}{2}\right)^2 = \frac{(2k-1)^2}{4}$ and $2\left(k - \frac{1}{2}\right) = 2k-1$. Each such series converges absolutely, locally uniformly on $\calH \times \CC$ (see the discussion following the definition of $\Theta$), so the $d$-fold products in $P_\beta(\rho(g)\Theta(\tau, w))$ expand term by term (Tonelli's theorem for sums: on each compact set, the iterated sums of absolute values are finite, being bounded by products of the one-dimensional bounds). Writing, for a tuple $\kv \in \ZZ^d$,
    \begin{equation*}
        \rv(\kv) := \left(k_i + \tfrac{q-1}{2} \bmod q\right)_{i \in [d]}, \qquad \ell(\kv) := \sum_{i=1}^d (2k_i - 1)^2, \qquad n(\kv) := \sum_{i = 1}^d (2k_i - 1),
    \end{equation*}
    we obtain the absolutely, locally uniformly convergent multiple series:
    \begin{equation}\label{eq:ellexpansionraw}
        P_\beta(\rho(g)\Theta(\tau, w)) = \sum_{\kv \in \ZZ^d} \beta^{(g)}_{\rv(\kv)}\, (-1)^{k_1 + \cdots + k_d} \exp\left(\frac{\pi i \tau\, \ell(\kv)}{4q} + \pi i\, n(\kv)\, w\right).
    \end{equation}
    We record three facts about these quantities:
    \begin{itemize}
        \item $\ell(\kv) \geq d$, since each $(2k_i - 1)^2 \geq 1$;
        \item $\ell(\kv) \equiv d \pmod 8$, since each odd square satisfies $(2k-1)^2 = 4k(k-1) + 1 \equiv 1 \pmod 8$ (as $k(k-1)$ is even);
        \item $|n(\kv)| \leq \sqrt{d \cdot \ell(\kv)}$, by the Cauchy--Schwarz inequality applied to the vectors $(2k_i - 1)_{i}$ and $(1, \ldots, 1)$.
    \end{itemize}
    In particular, for each fixed value $\ell$ of $\ell(\kv)$ there are finitely many contributing tuples (each $|2k_i - 1| \leq \sqrt{\ell}$, so there are at most $(\sqrt{\ell}+1)^d$ of them), so:
    \begin{equation*}
        \Phi_{g, \ell}(w) := \sum_{\kv \in \ZZ^d:\ \ell(\kv) = \ell} \beta^{(g)}_{\rv(\kv)}\, (-1)^{k_1 + \cdots + k_d}\, e^{\pi i\, n(\kv)\, w}
    \end{equation*}
    is a finite sum of exponentials, hence entire, and satisfies:
    \begin{equation}\label{eq:phibound}
        |\Phi_{g, \ell}(w)| \leq B_g\, (\sqrt{\ell} + 1)^d\, e^{\pi \sqrt{d \ell}\, |{\Im w}|}, \qquad\text{where } B_g := \max_{\rv \in \ZZ_q^d} \left|\beta^{(g)}_{\rv}\right|.
    \end{equation}
    Grouping the absolutely convergent series~\eqref{eq:ellexpansionraw} by the value of $\ell(\kv)$ yields~\eqref{eq:ellexpansion}. For the asserted convergence, fix a compact subset of $\calH \times \CC$, on which $\Im \tau \geq y_0 > 0$ and $|{\Im w}| \leq W$; then by~\eqref{eq:phibound}:
    \begin{equation*}
        \sum_{\ell} \left|\Phi_{g, \ell}(w)\, e^{\pi i \tau \ell/(4q)}\right| \leq B_g \sum_{\ell \geq d} (\sqrt{\ell}+1)^d\, e^{\pi \sqrt{d\ell}\, W}\, e^{-\pi \ell y_0/(4q)} < \infty,
    \end{equation*}
    since the exponent $-\pi \ell y_0 /(4q)$ eventually dominates $\pi\sqrt{d\ell}\, W + d \log(\sqrt{\ell}+1)$, which is $o(\ell)$.

    For the second claim, fix $\tau \in \calH$ and $j \geq 0$, and recall from~\Cref{lem:taylorfamilies} (and its proof) that $A_{g, j}(\tau) = \frac{1}{2\pi i} \oint_{|z| = q} F_g(\tau, z)\, z^{-j-1}\, dz$. Substituting $F_g(\tau, z) = P_\beta(\rho(g)\Theta(\tau, z/q))$ and the expansion~\eqref{eq:ellexpansion} --- which converges uniformly on the circle $\{z : |z| = q\}$ for our fixed $\tau$ --- we may integrate term by term:
    \begin{equation*}
        A_{g, j}(\tau) = \sum_{\ell} e^{\pi i \tau \ell/(4q)} \cdot \frac{1}{2\pi i}\oint_{|z| = q} \Phi_{g, \ell}(z/q)\, z^{-j-1}\, dz = \sum_{\ell} e^{\pi i \tau \ell/(4q)}\, q^{-j}\, \Xi_{g; j, \ell},
    \end{equation*}
    where the last step substitutes $z = qw$ and applies Cauchy's coefficient formula to the entire function $\Phi_{g, \ell}$ on the circle $|w| = 1$. This proves~\eqref{eq:ellexpansioncoeffs}; its absolute, locally uniform convergence in $\tau$ follows from the same majorant as above, since the same computation bounds $|q^{-j}\, \Xi_{g; j, \ell}| \leq q^{-j} \max_{|w| = 1} |\Phi_{g, \ell}(w)| \leq B_g\, (\sqrt{\ell}+1)^d\, e^{\pi\sqrt{d\ell}}$ by~\eqref{eq:phibound}.
    \qed

\subsection*{Proof of~\Cref{prop:mintaylorpreserved}}\label{sec:mintaylorpreservedproof}

    We have:
    \begin{align*}
        \sum_{j = m}^\infty \frac{A_j(g \ast \tau)}{G(g, \tau)^j} z^j &= F\left(g \ast (\tau, z)\right) \\
        &\propto_\zeta G(g, \tau)^{d/2} \exp\left(\frac{\pi i dc z^2}{qG(g, \tau)}\right) F_g(\tau, z) \quad\text{(\Cref{lem:bigF})} \\
        \Leftrightarrow F_g(\tau, z) &\propto_\zeta G(g, \tau)^{-d/2} \exp\left(-\frac{\pi i dcz^2}{qG(g, \tau)}\right) \left(\sum_{j = m}^\infty \frac{A_j(g \ast \tau)}{G(g, \tau)^j} z^j\right),
    \end{align*}
    and now the conclusion follows after expanding the RHS as a power series in $z$ (replacing the exponential term with its power series as well, and noting that it has constant term $1$): the function $F_g$ is holomorphic on $\calH \times \CC$ (like $F$, it is a composition of $\Theta$ with a polynomial and a rescaling), so~\Cref{lem:taylorfamilies} applies to it, and the displayed identity (whose implicit constant does not depend on $z$, so applies to every Taylor coefficient) shows that its Taylor coefficients of index below $m$ vanish while its coefficient of $z^m$ is $\propto_\zeta G(g, \tau)^{-d/2-m} A_m(g \ast \tau)$.
    \qed

\subsection*{Proof of~\Cref{lem:amsq}}\label{sec:amsqproof}

We first show that $A_m^2$ is weakly modular of weight $d+2m$ for $\Gamma(q)$, with a one-dimensional representation of finite order.
    By~\Cref{lem:bigF} (and~\Cref{prop:fg=f}), for any $g \in \Gamma(q)$ we have:
    \begin{align*}
        F\left(g \ast \tau, \frac{z}{G(g, \tau)}\right) &\propto_\zeta G(g, \tau)^{d/2} \exp\left(\frac{\pi i d c z^2}{q G(g, \tau)}\right) F(\tau, z) \\
        &= G(g, \tau)^{d/2} \left(\sum_{j = 0}^\infty \frac{1}{j!} \left(\frac{\pi i d c}{q G(g, \tau)}\right)^j z^{2j}\right) \left(\sum_{j = m}^\infty A_j(\tau) z^j\right) \\
        &= G(g, \tau)^{d/2} A_m(\tau) z^m + O(z^{m+1}).\numberthis\label{eq:taylorRHS}
    \end{align*}
    On the other hand, we have:
    \begin{align*}
        F\left(g \ast \tau, \frac{z}{G(g, \tau)}\right) &= \sum_{j = m}^\infty A_j(g \ast \tau) \left(\frac{z}{G(g, \tau)}\right)^j \\
        &= \frac{A_m(g \ast \tau)}{G(g, \tau)^m} z^m + O(z^{m+1}).\numberthis\label{eq:taylorLHS}
    \end{align*}
    Equating coefficients of $z^m$ in~\eqref{eq:taylorRHS} and~\eqref{eq:taylorLHS} (the constant implicit in $\propto_\zeta$ does not depend on $z$) implies:
    \begin{align*}
        \frac{A_m(g \ast \tau)}{G(g, \tau)^m} &\propto_\zeta G(g, \tau)^{d/2} A_m(\tau) \\
        \Leftrightarrow A_m(g \ast \tau) &\propto_\zeta G(g, \tau)^{d/2+m} A_m(\tau).
    \end{align*}
    Squaring both sides gives $A_m^2(g \ast \tau) \propto_\zeta G(g, \tau)^{d+2m} A_m^2(\tau)$ for all $g \in \Gamma(q)$, now with an integer power of $G(g, \tau)$. Since $A_m^2$ is not identically zero (as $A_m \not\equiv 0$ by the definition of $m$),~\Cref{lem:scalarcharacter} shows that $A_m^2$ is weakly modular of weight $d+2m$ for $\Gamma(q)$, with a one-dimensional representation of finite order.

Next, we show that $A_m^2$ is holomorphic at every cusp $\frakc \in \QQ \cup \{\infty\}$, with $\order_\frakc (A_m^2) \geq \frac{d}{4}+2\ell_\frakc$.
    By~\Cref{lem:cuspwelldef}, holomorphy at $\frakc$ and $\order_\frakc(A_m^2)$ depend only on the class of $\frakc$ and may be computed using the matrix $g_\frakc$ fixed in~\Cref{sec:gen4cuspdata}; also, the cusp has width $q$ by~\Cref{prop:gammaqcusps}. Following~\Cref{sec:modularforms}, we therefore want to bound the exponents appearing in the expansion at $\infty$ of $\tau \mapsto G(g_\frakc, \tau)^{-d-2m} (A_m(g_\frakc \ast \tau))^2$. By~\Cref{prop:mintaylorpreserved}, we have $G(g_\frakc, \tau)^{-d-2m} (A_m(g_\frakc \ast \tau))^2 \propto_\zeta (A_{g_\frakc, m}(\tau))^2$, and a nonzero constant factor does not affect the exponents in the expansion.

    The function $A_{g_\frakc, m}$ is the $m$-th Taylor coefficient of $F_{g_\frakc}$ (\Cref{prop:mintaylorpreserved} and the uniqueness in~\Cref{lem:taylorfamilies}), so~\eqref{eq:ellexpansioncoeffs} and the minimality of $\ell^* = d + 8\ell_\frakc$ in~\Cref{def:ella} give the absolutely and locally uniformly convergent expansion:
    \begin{equation*}
        A_{g_\frakc, m}(\tau) = \sum_{\substack{\ell \geq d + 8\ell_\frakc \\ \ell \equiv d \pmod 8}} q^{-m}\, \Xi_{g_\frakc; m, \ell}\, e^{\pi i \tau \ell/(4q)}.
    \end{equation*}
    Multiplying this series by itself (justified by absolute convergence), every term of $(A_{g_\frakc, m}(\tau))^2$ has the form $c\, e^{\pi i \tau \ell/(4q)} = c\, e^{2\pi i (\ell/8) \tau/q}$ for an integer $\ell \geq 2d + 16\ell_\frakc$ with $\ell \equiv 2d \pmod 8$. Collecting terms by the value of $\ell$ (justified by absolute convergence, as in the proof of~\Cref{lem:scalarisation}) and invoking the uniqueness in~\Cref{fact:fourierexp}, the cusp expansion of $A_m^2$ at $\frakc$ --- taken in the local variable $e^{2\pi i \tau/q}$, as the width is $q$ --- has all its exponents at least $\frac{2d+16\ell_\frakc}{8} = \frac{d}{4} + 2\ell_\frakc \geq 0$. So $A_m^2$ is holomorphic at $\frakc$, with $\order_\frakc(A_m^2) \geq \frac{d}{4} + 2\ell_\frakc$.
    \qed

\subsection*{Proof of~\Cref{lem:bxytransformation}}\label{sec:bxytransformationproof}

It will be helpful for us to write $\begin{bmatrix} X \\ Y \end{bmatrix} := \begin{bmatrix} a & c \\ b & d_0 \end{bmatrix} \begin{bmatrix} x \\ y \end{bmatrix}$ i.e., $X = ax+cy$ and $Y = bx+d_0 y$. 
    We have:
    \begin{align*}
        B_{x, y}(g \ast \tau) = &\exp\left(\frac{\pi i d x^2(g\ast \tau)}{4q}\right) F\left(g\ast \tau, \frac{x(g\ast\tau)+y}{2}\right) \\
        = &\exp\left(\frac{\pi i d x^2(g\ast \tau)}{4q}\right) F\left(g \ast \left(\tau, G(g, \tau) \frac{x(g \ast \tau)+y}{2}\right)\right) \\
        = &\exp\left(\frac{\pi i d x^2(g\ast \tau)}{4q}\right) F\left(g \ast \left(\tau, \frac{x(a\tau+b) +y(c\tau+d_0)}{2}\right)\right) \\
        \propto_\zeta &\exp\left(\frac{\pi i d x^2(g\ast \tau)}{4q}\right) G(g, \tau)^{d/2} \exp\left(\frac{\pi i dc}{qG(g, \tau)} \left(\frac{x(a\tau+b) + y(c\tau+d_0)}{2}\right)^2\right) \\
        &F_g\left(\tau, \frac{x(a\tau+b) + y(c\tau+d_0)}{2}\right) \quad\text{(\Cref{lem:bigF})} \\
        = &\exp\left(\frac{\pi i dx^2(a\tau+b)}{4q(c\tau+d_0)}\right) G(g, \tau)^{d/2} \exp\left(\frac{\pi i dc}{4q(c\tau+d_0)} (X\tau+Y)^2\right) F_g\left(\tau, \frac{X\tau+Y}{2}\right).\numberthis\label{eq:Bxystep1}
    \end{align*}
    Now write $X = \tilde{x}+2e_x$ and $Y = \tilde{y}+2e_y$ for $e_x, e_y \in \ZZ$. Then~\Cref{lem:fgtranslation} tells us that:
    \begin{align*}
        F_g\left(\tau, \frac{X\tau + Y}{2}\right) &= F_g\left(\tau, \frac{\tilde{x}\tau+\tilde{y}}{2} + e_x \tau + e_y\right) \\
        &\propto_\zeta \exp\left(-\frac{\pi i d}{q}(e_x^2 \tau + e_x(\tilde{x} \tau + \tilde{y}))\right) F_g\left(\tau, \frac{\tilde{x}\tau+\tilde{y}}{2}\right) \\
        &= \exp\left(-\frac{\pi i d}{q}(e_x^2 \tau + e_x(\tilde{x} \tau + \tilde{y}))\right) \exp\left(-\frac{\pi i d\tilde{x}^2 \tau}{4q}\right) B_{g; \tilde{x}, \tilde{y}}(\tau).\numberthis\label{eq:Bxystep2}
    \end{align*}
    Plugging this into~\eqref{eq:Bxystep1} implies that $B_{x, y}(g \ast \tau) \propto_\zeta \Psi(\tau)\, B_{g; \tilde{x}, \tilde{y}}(\tau)$, where:
    \begin{align*}
        \Psi(\tau) := G(g, \tau)^{d/2} \exp\left(\frac{\pi i dx^2(a\tau+b)}{4q(c\tau+d_0)} + \frac{\pi i dc}{4q(c\tau+d_0)} (X\tau+Y)^2-\frac{\pi i d}{q}(e_x^2 \tau + e_x(\tilde{x} \tau + \tilde{y})) -\frac{\pi i d\tilde{x}^2 \tau}{4q}\right).
    \end{align*}
    Therefore, isolating the exponential term and pulling out a factor of $\pi i d/(4q)$, it suffices to show that the following is independent of $\tau$ and rational (the exponential is then $e^{\pi i r}$ for a rational $r$, a root of unity):
    \begin{equation}\label{eq:indepoftau}
        \frac{x^2(a\tau+b)+c(X\tau+Y)^2}{(c\tau+d_0)} -4(e_x^2 \tau + e_x(\tilde{x} \tau + \tilde{y})) -\tilde{x}^2 \tau
    \end{equation}
    This is routine but rather tedious to check. We make use of an intermediate claim.
    \begin{claim}\label{clm:weirdcalc}
        We have:
        \begin{equation*}
            \frac{x^2(a\tau+b) + c(X\tau+Y)^2}{(c\tau+d_0)} = X^2 \tau + XY - xy.
        \end{equation*}
    \end{claim}
    \begin{proof}
        We have $\begin{bmatrix} x \\ y \end{bmatrix} = \begin{bmatrix} d_0 & -c \\ -b & a \end{bmatrix} \begin{bmatrix} X \\ Y \end{bmatrix} \Rightarrow x = d_0X - cY, y = -bX + aY$. Plugging these in, we want to show that:
        \begin{align*}
            \frac{(d_0X-cY)^2(a\tau+b) + c(X\tau+Y)^2}{(c\tau+d_0)} &= X^2 \tau + XY + (d_0X-cY)(bX-aY).
        \end{align*}
        Indeed, we have:
        \begin{align*}
            &(c\tau+d_0)(X^2 \tau + XY + (d_0X-cY)(bX-aY)) \\
            = &(c\tau+d_0)(X^2 \tau + XY + bd_0X^2 - (ad_0+bc)XY + acY^2) \\
            = &\tau^2(cX^2) + \tau(d_0X^2 + cXY + bcd_0X^2-(acd_0+bc^2)XY + ac^2Y^2) + (d_0XY+bd_0^2X^2 - (ad_0^2 + bcd_0)XY + acd_0Y^2) \\
            = &\tau^2(cX^2) + \tau(2cXY + ad_0^2X^2 - cXY(\underbrace{1+ad_0+bc}_{=2ad_0}) + ac^2Y^2) + (bd_0^2X^2 + c(1+bc)Y^2 +d_0XY(\underbrace{1-ad_0-bc}_{= -2bc})) \\
            = &\tau^2 (cX^2) + \tau(2cXY + a(d_0X-cY)^2) + (cY^2 + b(d_0X - cY)^2) \\
            = &(d_0X-cY)^2(a\tau+b) + c(X\tau+Y)^2.
        \end{align*}
    \end{proof}
    Plugging Claim~\ref{clm:weirdcalc} into~\eqref{eq:indepoftau} yields an affine function in $\tau$, so it suffices to focus on its linear coefficient, which is:
    \begin{equation*}
        X^2 - 4e_x^2 - 4e_x \tilde{x} - \tilde{x}^2 = X^2 - (2e_x + \tilde{x})^2 = 0,
    \end{equation*}
    so~\eqref{eq:indepoftau} equals its constant term $XY - xy - 4e_x\tilde{y}$, an integer, and the conclusion follows.
    
    Finally, the bijectivity claim follows from noting that $\begin{bmatrix} \tilde{x} \\ \tilde{y} \end{bmatrix} = \begin{bmatrix} a & c \\ b & d_0 \end{bmatrix} \begin{bmatrix} x \\ y \end{bmatrix} \bmod{2}$, and that the matrix is by definition invertible.
    \qed

\subsection*{Proof of~\Cref{lem:b2holomorphic}}\label{sec:b2holomorphicproof}

By~\Cref{lem:cuspwelldef}, holomorphy at $\frakc$ and $\order_\frakc(B^2)$ depend only on the class of $\frakc$ and may be computed using the matrix $g_\frakc$ fixed in~\Cref{sec:gen4cuspdata}; also, the cusp has width $q$ by~\Cref{prop:gammaqcusps}. Following~\Cref{sec:modularforms} (as in the proof of~\Cref{lem:amsq}), we want to bound the exponents in the expansion at $\infty$ of $\tau \mapsto G(g_\frakc, \tau)^{-3d} (B(g_\frakc \ast \tau))^2$. Exactly as in the proof of~\Cref{cor:bsqmodular} --- but now for the single matrix $g_\frakc \in \SL_2(\ZZ)$, which need not lie in $\Gamma(q)$ --- \Cref{lem:bxytransformation} and the bijectivity of $(x, y) \mapsto (\tilde{x}, \tilde{y})$ give:
    \begin{equation*}
        G(g_\frakc, \tau)^{-3d}\, (B(g_\frakc \ast \tau))^2 \propto_\zeta (B_{g_\frakc}(\tau))^2,
    \end{equation*}
    so it suffices to show that $(B_{g_\frakc}(\tau))^2$ is a sum of terms $e^{\pi i \tau \ell/(4q)}$ with $\ell \geq 2d + 16\ell_\frakc + 16 H_\frakc$. We bound the three factors of $B_{g_\frakc}$ in turn.
    First, we have:
    \begin{align*}
        B_{g_\frakc; 0, 1}(\tau) &= F_{g_\frakc}\left(\tau, \frac{1}{2}\right) \\
        &= P_\beta\left(\rho(g_\frakc)\Theta\left(\tau, \frac{1}{2q}\right)\right) = \sum_{\ell \geq d + 8\ell_\frakc} \Phi_{g_\frakc, \ell}\left(\frac{1}{2q}\right) e^{\pi i \tau \ell/(4q)},
    \end{align*}
    a sum of terms of the form $e^{\pi i \tau\ell/(4q)}$ for $\ell \geq d+8\ell_\frakc$, by~\Cref{lem:ellexpansion} (at $w = \frac{1}{2q}$) and the minimality of $\ell^* = d + 8\ell_\frakc$ in~\Cref{def:ella}.
    Next, we have:
    \begin{align*}
        \exp\left(-\frac{\pi i d \tau}{4q}\right) B_{g_\frakc; 1, 0}(\tau) &= F_{g_\frakc}\left(\tau, \frac{\tau}{2}\right) \\
        &= P_\beta\left(\rho(g_\frakc) \Theta\left(\tau, \frac{\tau}{2q}\right)\right) \\
        &= P_\beta \circ \rho(g_\frakc)\left(\sum_{k \in \ZZ} (-1)^k \exp\left(\frac{\pi i \tau}{q} \left(k-\frac{1}{2}\right)^2 + 2\pi i \left(k-\frac{1}{2}\right) \frac{\tau}{2q}\right) \ev_{k+(q-1)/2 \bmod{q}}\right) \\
        &= P_\beta \circ \rho(g_\frakc)\left(\sum_{k \in \ZZ} (-1)^k \exp\left(\frac{\pi i \tau}{q}k^2 - \frac{\pi i \tau}{4q} \right) \ev_{k+(q-1)/2 \bmod{q}}\right) \\
        &\propto_\zeta P_\beta \circ \rho(g_\frakc)\left(\sum_{k \in \ZZ} (-1)^k \exp\left(\frac{\pi i \tau}{q}k^2 - \frac{\pi i \tau}{4q} \right) \ev_{k\bmod{q}}\right) \quad\text{(\Cref{prop:pproperties})} \\
        &= \exp\left(-\frac{\pi i d\tau}{4q}\right) P_\beta \circ \rho(g_\frakc)\left(\sum_{k \in \ZZ} (-1)^k \exp\left(\frac{\pi i \tau}{q}k^2\right) \ev_{k\bmod{q}}\right) \\
        \Rightarrow B_{g_\frakc; 1, 0}(\tau) &\propto_\zeta P_\beta \circ \rho(g_\frakc)\left(\sum_{k \in \ZZ} (-1)^k \exp\left(\frac{\pi i \tau}{q}k^2\right) \ev_{k\bmod{q}}\right).
    \end{align*}
    (The step invoking~\Cref{prop:pproperties} uses its fourth bullet --- the $\sfX$-invariance, up to a root of unity, of $P_\beta \circ \rho(g_\frakc)$ --- applied with $H = \sfX^{(q+1)/2}$.)
    Now let us consider the expansion of $P_\beta \circ \rho(g_\frakc)$ on the RHS. The $x$-th coordinate of the vector in the argument is $\sum_{k \equiv x \pmod q} (-1)^k \exp\left(\frac{\pi i \tau}{4q} \cdot 4k^2\right)$, whose exponents satisfy $4k^2 \geq 4 w_q(x)$, since $w_q(x) = \min_{k \equiv x \pmod q} k^2$ (\Cref{def:track3exponent}). Hence a monomial $X_{x_1} \ldots X_{x_d}$ with nonzero coefficient in $P_\beta \circ \rho(g_\frakc)$ expands out to terms of the form $\exp\left(\frac{\pi i \tau}{4q} \ell\right)$ for $\ell \geq 4\sum_{i \in [d]} w_q(x_i) \geq 4H_\frakc$, where the last inequality is~\Cref{def:hc}; expanding the products term by term is justified by absolute convergence, as in the proof of~\Cref{lem:ellexpansion}.
    Finally, we have:
    \begin{align*}
        \exp\left(-\frac{\pi i d \tau}{4q}\right) B_{g_\frakc; 1, 1}(\tau) &= F_{g_\frakc}\left(\tau, \frac{1+\tau}{2}\right) \\
        &= P_\beta\left(\rho(g_\frakc) \Theta\left(\tau, \frac{1+\tau}{2q}\right)\right) \\
        &= P_\beta \circ \rho(g_\frakc)\left(\sum_{k \in \ZZ} (-1)^k \exp\left(\frac{\pi i \tau}{q} \left(k-\frac{1}{2}\right)^2 + 2\pi i \left(k-\frac{1}{2}\right) \frac{1+\tau}{2q}\right) \ev_{k+(q-1)/2 \bmod{q}}\right) \\
        &= P_\beta \circ \rho(g_\frakc)\left(\sum_{k \in \ZZ} (-1)^k \exp\left(\frac{\pi i \tau}{q}k^2 - \frac{\pi i \tau}{4q} + \frac{\pi i k}{q} - \frac{\pi i}{2q} \right) \ev_{k+(q-1)/2 \bmod{q}}\right) \\
        &\propto_\zeta P_\beta \circ \rho(g_\frakc)\left(\sum_{k \in \ZZ} (-1)^k \exp\left(\frac{\pi i \tau}{q}k^2 - \frac{\pi i \tau}{4q} + \frac{\pi i k}{q} - \frac{\pi i}{2q} \right) \ev_{k\bmod{q}}\right) \quad\text{(\Cref{prop:pproperties})} \\
        &\propto_\zeta \exp\left(-\frac{\pi i d\tau}{4q}\right) P_\beta \circ \rho(g_\frakc)\left(\sum_{k \in \ZZ} (-1)^k \exp\left(\frac{\pi i k}{q}\right) \exp\left(\frac{\pi i \tau}{q}k^2\right) \ev_{k\bmod{q}}\right),
    \end{align*}
    where the last step pulls out, by degree-$d$ homogeneity, both the $\tau$-exponential and the constant $e^{-\pi i/(2q)}$ common to every coordinate (the latter contributes only a root of unity); and now an analogous argument (the extra $k$-dependent phase $e^{\pi i k/q}$ does not affect the $\tau$-exponents) implies that this is also a sum of terms of the form $\exp\left(\frac{\pi i \tau}{4q}\ell\right)$ for $\ell \geq 4H_\frakc$.

    Multiplying the three factors together (Cauchy products of absolutely convergent series, so the minimal exponents add), we infer that $B_{g_\frakc}(\tau)$ is a sum of terms of the form $\exp\left(\frac{\pi i \tau}{4q}\ell\right)$ for $\ell \geq (d+8\ell_\frakc) + 4H_\frakc + 4H_\frakc = d+8\ell_\frakc + 8H_\frakc$, and its square is a sum of terms of the same form with $\ell \geq 2d + 16\ell_\frakc + 16H_\frakc$. (Unlike in~\Cref{lem:amsq}, the exponents of the individual factors here are not confined to a single residue class modulo $8$; this does not matter for the argument, which uses only the lower bound on the exponents and the uniqueness of the coefficients.) Collecting terms by the value of $\ell$ and invoking the uniqueness in~\Cref{fact:fourierexp} (as in the proof of~\Cref{lem:amsq}), the cusp expansion of $B^2$ at $\frakc$ --- in the local variable $e^{2\pi i \tau/q}$, as the width is $q$ --- has all its exponents at least $\frac{2d + 16\ell_\frakc + 16H_\frakc}{8} = \frac{d}{4} + 2\ell_\frakc + 2H_\frakc \geq 0$. So $B^2$ is holomorphic at $\frakc$, with $\order_\frakc(B^2) \geq \frac{d}{4} + 2\ell_\frakc + 2H_\frakc$.
    \qed

\subsection*{Proof of~\Cref{lem:pc}}\label{sec:pcproof}

Throughout, write $g := g_\frakc$ and $\ell^* := d + 8\ell_\frakc$, and adopt the convention $\Phi_{g, \ell} := 0$ for the indices $\ell \not\equiv d \pmod 8$ not covered by~\Cref{lem:ellexpansion}, so that sums over $\ell$ may range over all integers. By~\Cref{lem:ellexpansion} (with $w = z/q$) and the minimality of $\ell^*$ in~\Cref{def:ella}:
    \begin{equation}\label{eq:pcexpansion}
        F_g(\tau, z) = \sum_{\ell \geq \ell^*} \Phi_{g, \ell}(z/q)\, e^{\pi i \tau \ell/(4q)}, \qquad \Phi_{g, \ell}(z/q) = \sum_{\xi \in \ZZ - d/2} c_{\ell, \xi}\, e^{2\pi i \xi z/q},
    \end{equation}
    where the second display follows from the explicit formula for $\Phi_{g, \ell}$ derived in the proof of~\Cref{lem:ellexpansion}: each contributing tuple $\kv$ has $w$-exponential $e^{\pi i n(\kv) w} = e^{2\pi i \xi(\kv) z/q}$ with $\xi(\kv) := \frac{n(\kv)}{2} = \left(\sum_i k_i\right) - \frac{d}{2}$ --- an element of the coset $\ZZ - \frac{d}{2}$ (half-integers when $d$ is odd) --- so that:
    \begin{equation*}
        c_{\ell, \xi} := \sum_{\kv \in \ZZ^d:\ \ell(\kv) = \ell,\ \xi(\kv) = \xi} \beta^{(g)}_{\rv(\kv)}\, (-1)^{k_1 + \cdots + k_d}.
    \end{equation*}
    For each fixed $\ell$ the inner sum in~\eqref{eq:pcexpansion} is finite: $c_{\ell, \xi} = 0$ unless $|\xi| \leq \frac{1}{2}\sqrt{d\ell}$, by the Cauchy--Schwarz bound $|n(\kv)| \leq \sqrt{d\, \ell(\kv)}$ from the proof of~\Cref{lem:ellexpansion}. We also adopt the convention $c_{\ell, \xi} := 0$ for $\ell < \ell^*$, consistent with~\eqref{eq:pcexpansion}.

    Define $p_\frakc(X) := \sum_{\xi} c_{\ell^*, \xi}\, X^{\xi + d/2}$ --- a Laurent polynomial a priori, with integer exponents $\xi + \frac{d}{2} = \sum_i k_i$. By construction, the coefficient of $e^{\pi i \tau \ell^*/(4q)}$ in~\eqref{eq:pcexpansion} is exactly $\Phi_{g, \ell^*}(z/q) = e^{-2\pi i z d/(2q)}\, p_\frakc(e^{2\pi i z/q})$, so the identity~\eqref{eq:pc} holds. It remains to prove that the exponents of $p_\frakc$ lie in $[0, d]$ --- i.e., that $c_{\ell^*, \xi} = 0$ whenever $|\xi| > d/2$ --- so that $p_\frakc$ is a genuine polynomial of degree at most $d$.

    By~\Cref{lem:fgtranslation} (with $(a, b) = (n, 0)$), for any $n \in \ZZ$ there is a nonzero constant $\omega_n$, independent of $\tau$ and $z$, such that:
    \begin{equation*}
        F_g(\tau, z+n\tau) = \omega_n \exp\left(-\frac{\pi i d}{q}(n^2\tau+2nz)\right) F_g(\tau, z).
    \end{equation*}
    We expand both sides using~\eqref{eq:pcexpansion}; all rearrangements below are of absolutely convergent series, by the majorants from the proof of~\Cref{lem:ellexpansion}. On the left, $e^{2\pi i \xi(z+n\tau)/q} = e^{2\pi i \xi z/q}\, e^{\pi i \tau (8n\xi)/(4q)}$, so, reindexing by $\ell' := \ell + 8n\xi$:
    \begin{equation*}
        F_g(\tau, z+n\tau) = \sum_{\ell, \xi} c_{\ell, \xi}\, e^{2\pi i \xi z/q}\, e^{\pi i \tau(\ell + 8n\xi)/(4q)} = \sum_{\ell', \xi} c_{\ell' - 8n\xi,\ \xi}\, e^{2\pi i \xi z/q}\, e^{\pi i \tau \ell'/(4q)}.
    \end{equation*}
    On the right, reindexing by $\ell' := \ell - 4n^2 d$ and $\xi' := \xi - nd$:
    \begin{equation*}
        \omega_n e^{-\pi i d n^2 \tau/q}\, e^{-2\pi i dnz/q} \sum_{\ell, \xi} c_{\ell, \xi}\, e^{2\pi i \xi z/q}\, e^{\pi i \tau \ell/(4q)} = \sum_{\ell', \xi'} \omega_n\, c_{\ell' + 4n^2 d,\ \xi' + nd}\, e^{2\pi i \xi' z/q}\, e^{\pi i \tau \ell'/(4q)}.
    \end{equation*}
    Now fix $\ell' \in \ZZ$ and compare the two sides. By the uniqueness in~\Cref{fact:fourierexp} (applied in $\tau$ with period $8q$, for each fixed $z$), the coefficients of $e^{\pi i \tau \ell'/(4q)}$ agree as functions of $z$. Both are \emph{finite} sums of the characters $z \mapsto e^{2\pi i \xi z/q}$: on the left, $c_{\ell'-8n\xi, \xi} \neq 0$ requires $\ell' - 8n\xi \geq \ell^*$ and $4\xi^2 \leq d(\ell' - 8n\xi)$, which has finitely many solutions $\xi$; similarly on the right. Since these characters (for distinct $\xi$ in the coset $\ZZ - \frac{d}{2}$) are linearly independent --- after factoring out $e^{-\pi i dz/q}$, they become distinct integer powers of $e^{2\pi i z/q}$ --- the coefficients match term by term:
    \begin{equation}\label{eq:pcrelation}
        c_{\ell - 8n\xi,\ \xi} = \omega_n\, c_{\ell + 4n^2 d,\ \xi + nd} \qquad \text{for all } n, \ell \in \ZZ \text{ and } \xi \in \ZZ - \tfrac{d}{2}.
    \end{equation}

    Now suppose for the sake of contradiction that $c_{\ell^*, \xi^*} \neq 0$ for some $\xi^*$ with $|\xi^*| > d/2$. We have two cases:
    \begin{itemize}
        \item If $\xi^* > d/2$: apply~\eqref{eq:pcrelation} with $n = 1$, $\ell = -3d+8\ell_{\frakc}$, and $\xi = \xi^*-d$. The right-hand side is $\omega_1 c_{\ell + 4d,\ \xi + d} = \omega_1 c_{d + 8\ell_\frakc,\ \xi^*} = \omega_1 c_{\ell^*, \xi^*} \neq 0$, so the left-hand side $c_{\ell - 8\xi,\ \xi}$ is nonzero. But $\ell-8\xi = \ell-8(\xi^*-d) = 5d + 8\ell_{\frakc} - 8\xi^* < 5d + 8\ell_\frakc - 4d = \ell^*$, contradicting the convention $c_{\ell, \xi} = 0$ for $\ell < \ell^*$ (the minimality in~\Cref{def:ella}).

        \item If $\xi^* < -d/2$: apply~\eqref{eq:pcrelation} with $n = -1$, $\ell = -3d+8\ell_{\frakc}$, and $\xi = \xi^*+d$. The right-hand side is $\omega_{-1} c_{\ell+4d,\ \xi - d} = \omega_{-1} c_{\ell^*, \xi^*} \neq 0$, so $c_{\ell + 8\xi,\ \xi} \neq 0$. But $\ell + 8\xi = -3d + 8\ell_\frakc + 8\xi^* + 8d = 5d+8\ell_{\frakc}+8\xi^* < \ell^*$, again a contradiction.
    \end{itemize}
    So the exponents $\xi + d/2$ of $p_\frakc$ all lie in $[0, d]$, completing the proof.
    \qed